\documentstyle[aps,eqsecnum]{revtex}
\newcommand{\psibar}{\bar{\psi}}
\newcommand{\qqb}{q\bar{q}}
\newcommand{\vecr}{\vec{r}}
\newcommand{\vp}{\vec{p}}
\newcommand{\vn}{\vec{n}}
\newcommand{\nhat}{\hat{\eta}}
\newcommand{\dslasha}{\not \! \! D}
\newcommand{\dslash}{\not \! \! \! D}
\newcommand{\intd}[1]{\int {\rm d}^4#1 \,}
\newcommand{\intoned}[1]{\int {\rm d}#1 \,}
\newcommand{\intD}[1]{\int {\rm d}[#1] \,}

\begin{document}
\draft
\title{QCD Inequalities}
\author{Shmuel Nussinov\thanks{electronic mail:nussinov@ccsg.tau.ac.il}
and Melissa A. Lampert\thanks{electronic mail:melissa@albert.tau.ac.il}}
\address{School of Physics and Astronomy \\
Raymond and Beverly Sackler Faculty of Exact Sciences \\
Tel Aviv University \\
69978 Tel Aviv, Israel
}
\date{\today}
\maketitle
\begin{abstract}
{We review the subject of QCD inequalities, using both a Hamiltonian
  variational approach, and a rigorous Euclidean path integral
  approach.}
\end{abstract}
\pacs{PACS number(s): 12.35.Eq,12.70.+q}

\tableofcontents
\newpage
%
\section{Introduction}
\label{sec:intro}
The crucial steps in the evolution of any scientific discipline are the
identification of the underlying degrees of freedom and the dynamics
governing them. For the theory of strongly interacting particles these
degrees of freedom are the quarks and gluons, and the elegant quantum
chromodynamics (QCD) Lagrangian
\begin{equation}
   {\cal L}_{\text{QCD}} = \sum_{i=1}^{N_f} \psibar_i 
      (\dslasha + m_i) \psi_i + {\rm tr} (F^a_{\mu\nu} \lambda_a)^2
\label{eq:lqcd}
\end{equation}
prescribes the dynamics.

Here we would like to review how one can deduce directly from ${\cal
  L}_{\text{QCD}}$, and from its Hamiltonian counterpart (with
possible additional assumptions), various inequalities between
hadronic masses and/or other hadronic matrix elements (observables).

The euclidean correlation functions of color singlet (gauge invariant)
local operators $O_{a_i}(x)$ are given by the functional path integral
\cite{ref:iandz,ref:abers}
\begin{eqnarray}
   W_{a_1 \ldots a_n}(x_1 \ldots x_n) &=& \langle 0 
      | O_{a_1}(x_1) \ldots O_{a_n}(x_n) | 0 \rangle
   \nonumber \\
   &=& \intD{A_\mu} \intD{\psi} \intD{\psibar} 
      O_{a_1}(x_1) \ldots O_{a_n}(x_n) \, {\rm e}^{-\intd{x} {\cal L}(x)}
\label{eq:correlator}
\>,
\end{eqnarray}
with $\intD{A_\mu} \, \intD{\psi} \, \intD{\psibar}$ indicating the
functional integral over the ordinary gauge field and fermionic
(Grassman) degrees of freedom. By analytically continuing the
corresponding momentum space correlations $W_{a_1 \ldots a_n}(p_1
\ldots p_n)$ all hadronic scattering amplitudes can be determined.

The simplest two-point functions are particularly useful. The spectral
representation for such functions
\begin{equation}
   W_a(x,y) = \langle 0 | J_a(x) J^{\dag}_a(y) | 0 \rangle
      = \int {\rm d}(\mu^2) \sigma_a(\mu^2) {\rm e}^{-\mu |x-y|}
\end{equation}
yields information on the hadronic states in the channel with $J_a$
quantum numbers, {\em i.e.} the energy-momentum eigenstates
$|n\rangle$ with non-vanishing $\langle 0 | J_a | n \rangle$ matrix
elements. Thus a lowest state of mass $m_a^{(0)}$ implies an
asymptotic behavior which, up to powers of $|x-y|$, is
\begin{equation}
    \langle 0 | J_a(x) J^{\dag}_a(y) | 0 \rangle \stackrel
      {{\textstyle \longrightarrow}}
      {{\scriptscriptstyle \lim_{|x-y| \rightarrow \infty}}}
      {\rm e}^{- m_a^{(0)} |x-y|}
\label{eq:asymp}   
\>.
\end{equation}

The hadronic spectrum can also be directly obtained via the
Schr\"{o}dinger equation
\begin{equation}
   H_{\text{QCD}} \Psi = m \Psi
\label{eq:sch}
\>,
\end{equation}
with $\Psi$ a wave functional describing the degrees of freedom of the
valence quarks and any number of additional $\qqb$ pairs and/or
gluons. The complexity of the physical states in Eq.~(\ref{eq:sch}) or
the richness of field configurations in the functional integral
equation (\ref{eq:correlator}) impede quantitative computations of
hadronic matrix elements and the hadronic spectrum, a goal pursued
over more than two decades, utilizing in particular lattice
calculations \cite{ref:creutzetal,ref:creutz,ref:montvay}.

The QCD inequalities are derived by comparing expressions for
different correlation functions (or the energies of different hadronic
systems) {\em without} requiring explicit evaluation. We only need to
assume that an appropriate regularization scheme and gauge fixing have
been devised to make the path integral (or the Schr\"{o}dinger
problem) well defined.

A key ingredient in deriving relations between correlation functions is the
positivity of the functional path integration measure ${\rm d}\mu(A)$
obtained after integrating out the fermionic degrees of freedom. The
bilinear $ \sum_{i=1}^{N_f} \psibar_i (\dslasha + m_i) \psi_i$ part of
${\cal L}_{{\rm QCD}}$ then yields the determinantal factor
\begin{equation}
   {\rm Det} =  \prod_{i=1}^{N_f} {\rm Det} (\dslasha + m_i)
\label{eq:det}
\>,
\end{equation}
which for any vectorial (non-chiral) theory can be shown to be positive for
any $A_\mu(x)$ (see Sec.~\ref{sec:quarkbi})
\cite{ref:weingarten,ref:vafanp,ref:vafaprl,ref:vafacmp,ref:witt83}.

If the integrand in the path integral expression for one correlation
function is greater than the integrand in another correlation function for
all $A_\mu(x)$, then the positivity of the path integration measure
guarantees that this feature persists for the integrated values. A rigorous
inequality between the two correlation functions for all possible
(euclidean) locations of the external currents will then follow.

For the particular case of two-point functions an inequality of the
form
\begin{equation}
   \langle 0 | J_a(x) J^{\dag}_a(y) | 0 \rangle
      \geq \langle 0 | J_b(x) J^{\dag}_b(y) | 0 \rangle
\label{eq:2ptineq}
\end{equation}
implies, via Eq.~(\ref{eq:asymp}), the reversed inequality for the lowest
mass physical states with the quantum numbers of $J_a(x), J_b(x)$:
\begin{equation}
   m_a^{(0)} \leq m_b^{(0)}
\label{eq:massineq}
\>.
\end{equation}
Most of the inequalities (\ref{eq:2ptineq}) involve the pseudoscalar
currents ($J_a = \psibar_i \gamma_5 \psi_j$) and the corresponding
mass inequalities (\ref{eq:massineq}) the pion ({\em i.e.} the lowest
pseudoscalar states in the $\bar{u} \gamma_5 d, \bar{u} \gamma_5 u -
\bar{d} \gamma_5 d, \bar{d} \gamma_5 u$ channels) \cite{ref:weingarten}:
\begin{mathletters}
   \label{eq:pionineqs}
   \begin{equation}
      m^{(0)} \mbox{(any meson)} \geq m_\pi,
   \label{eq:meson}
   \end{equation}
   \begin{equation}
      m^{(0)} \mbox{(any baryon)} \geq m_\pi,
   \label{eq:baryon}
   \end{equation}
   \begin{equation}
      m_{\pi^+} \geq m_{\pi^0}.
   \label{eq:pion}
   \end{equation}
\end{mathletters}

The efforts to obtain inequalities in the Hamiltonian approach follow a
similar general pattern. Rather than attempting to solve the QCD
Schr\"{o}dinger equation (\ref{eq:sch}) for a particular channel, relations
are sought between baryonic and/or mesonic sectors of different flavors
$B_{ijk}$, $M_{i\bar{\jmath}}$, and different spins.

Flavor enters the Lagrangian (\ref{eq:lqcd}) only via the bilinear,
local mass term. Comparison of masses (or other features) of mesons or
baryons differing just by flavor may be easier than {\em ab initio}
computations. The additive form of the mass term implies a
relationship between the Hamiltonians obtained by restricting the full
$H_{\text{QCD}}$ to different flavor sectors. Using a variational
principle for the ground state masses and {\em assuming flavor
  symmetric} ground state wave functions, these relations imply the
inequalities
\begin{equation}
   m^{(0)}_{ij} \geq \frac{1}{2} \left( m^{(0)}_{i\bar{\imath}} 
      + m^{(0)}_{j\bar{\jmath}} \right)
\label{eq:interflavor}
\>,
\end{equation}
with $m^{(0)}_{ij}$ the ground state mass in the mesonic sector
$M_{i\bar{\jmath}}^{(J^{PC})}$. 

Physical color singlet states can be achieved in a variety of ways:
the {\em mesonic} -- $\psibar_i^a \psi_{ja}$; {\em baryonic} --
$\epsilon_{abc} \psi_i^a \psi_j^b \psi_k^c$; {\em exotic} --
$\epsilon_{de}^c \epsilon_{abc} \psi_i^a \psi_j^b \psi_k^d \psi_l^e$
(or {\em hybrid} -- $\psibar_a \psi_b G_r \lambda^{r(ab)}$)
configurations; or {\em glueballs} -- $G^r G_r, f_{rst}$ (or $d_{rst}$)
$G^r G^s G^t$ [Here $G_r = G_{\mu\nu}$ represents the
chromo-electromagnetic antisymmetric field tensor. We use $abc \,
(rst)$ for triplet (octet) color indices; $\lambda^{r(ab)}$ are the
Gell-Mann matrices; and $ijkl$ refer to flavors. We reserve $\alpha
\beta \gamma$ for spinor indices, and $\lambda \mu \kappa$ for vector
Lorentz indices.]  These configurations (mesonic, baryonic, {\em
  etc.}) differ dynamically by having different ``color networks''.
However, the different (non-glueball or hybrid) sectors contain quarks
and/or antiquarks which are sources or sinks of chromoelectric flux of
the same universal strength. This suggests that the QCD Hamiltonians
in the different sectors may be related and mass relations of the type
\cite{ref:nussprl51}
\begin{equation}
   m_{{\rm baryon}} \geq \frac{3}{2} \, m_{{\rm meson}}
\label{eq:mesbarineq}
\end{equation}
can be obtained.

The rigorous inequalities (\ref{eq:meson}) and (\ref{eq:baryon})
derived via the euclidean path integral formulation amount to the
well-known fact that the pion is the lightest hadron. These
inequalities have, however, profound implications for the phase
structure of QCD. Equation (\ref{eq:meson}) implies no spontaneous
breaking of vectorial global (isospin) symmetries in QCD. Equation
(\ref{eq:baryon}), along with the 't Hooft anomaly matching
condition, proves that the axial global flavor symmetry must be
spontaneously broken.

It is important to note that the basic feature of positivity of the
determinant factor (\ref{eq:det}) and the functional path integral
measure ${\rm d} \mu(A)$ is common to all gauge, QCD-like vectorial
theories with Dirac fermions.  This has far-reaching implications for
composite models for quarks and leptons. A basic puzzle facing such
models is the smallness of the masses of the composite quarks and
leptons: $m_e \simeq m_u \simeq m_d \simeq$ MeV in comparison with
$\Lambda_p \geq$ TeV, the compositeness (``preonic'') scale. A natural
mechanism for protecting (almost) massless composite fermions is an
unbroken chiral symmetry. 't Hooft \cite{ref:thooftcarg} and others
\cite{ref:frishman}, using the anomaly matching constraint, formulated
some {\em necessary} conditions for such a realization of an
underlying global chiral symmetry in the spectrum of the theory.
Together with the mass inequalities, these conditions rule out all
vectorial composite models for which fermion-boson mass inequalities
like Eq.~(\ref{eq:baryon}) [or (\ref{eq:mesbarineq})] can be proven.

In this work we will mention, at one stage or another, most of the papers
written on QCD inequalities, or on the related subject of inequalities
in potential models. Particular attention will be paid to the seminal
works of Weingarten \cite{ref:weingarten}, Vafa and Witten
\cite{ref:vafanp,ref:vafaprl,ref:vafacmp}, and Witten 
\cite{ref:witt83} -- all utilizing the euclidean path integral approach. 
Weingarten proved the inequalities (\ref{eq:meson}) and (\ref{eq:baryon})
and pointed out the relevance of (\ref{eq:baryon}) to spontaneous chiral
symmetry breaking (S$\chi$SB) in QCD. Vafa and Witten directly used the
measure positivity to prove that parity and global vectorial symmetries
like isospin do not break spontaneously 
\cite{ref:vafanp,ref:vafaprl}. Finally Witten \cite{ref:witt83} 
proved (\ref{eq:pion}) and the interflavor relation 
(\ref{eq:interflavor}) which holds rigorously for the case of
pseudoscalars, with no need for the flavor symmetry assumption.

To date, QCD inequalities have been mentioned in approximately 600
papers. Most authors were concerned with symmetry breaking patterns,
the motivation for the Vafa-Witten paper \cite{ref:vafanp}, which is
cited most often. Here we equally emphasize the other facet of the
inequalities, which constitute useful, testable, constraints on
observed (and yet to be discovered) hadrons. This is why we elaborate
on the baryon-meson mass inequalities and related inequalities for the
exotic sector despite the fact that we have not been able to prove
them via the rigorous euclidean path integral approach; and on the
inequalities between mesons of different flavors, which cannot be
justified without the additional specific assumption of flavor
symmetric mesonic wave function(al)s.

In general we will follow a didactic rather than a chronological
approach.  We start in Sec.~\ref{sec:mmineq} by deriving
Eq.~(\ref{eq:interflavor}) using a simple potential model
\cite{ref:nussplb136} and the flavor symmetry assumption.  This is
followed up in Sec.~\ref{sec:mbineq} by a potential model derivation
of the baryon-meson mass inequalities (\ref{eq:mesbarineq})
\cite{ref:nussprl51,ref:richplb139,ref:ader}, and a discussion of
baryon-baryon mass inequalities in Sec.~\ref{sec:bbineq}.  In
Secs.~\ref{sec:interflavor} and \ref{sec:nonpertmbineq} and
App.~\ref{app:eq42} we show that these relations may be valid far
beyond the simple potential model \cite{ref:nussprl51,ref:nussprl52}.
The elegant analysis of E. Lieb \cite{ref:lieb} of the potential model
approach is also reproduced in some detail in App.~\ref{app:lieb2}.
Section~\ref{sec:compdata} concludes the first part of this review
which utilizes the Hamiltonian variational approach by verifying that
the inequalities indeed hold in the approximately forty cases where
they can be tested, and by presenting lower bounds to the masses of
new, yet to be discovered, mesonic and baryonic states.

The middle part of the review focuses on the rigorous euclidean path
integral approach. We start in Sec.~\ref{sec:quarkbi} by proving the
positivity of the measure and the ensuing inequalities
(\ref{eq:pionineqs}), stating that the pseudoscalar pion is the
lightest meson. We illustrate the power of these inequalities in
Sec.~\ref{sec:vecsym} where $m_{u\bar{d}}^{(0^+)} \geq
m_{u\bar{d}}^{(0^-)}$ is used as a shortcut to motivate the
Vafa-Witten theorem on non-breaking of isospin (which is then
presented in some detail). We proceed in Sec.~\ref{sec:mbineqcorr}
with Weingarten's proof of the pion-nucleon mass inequality
[Eq.~(\ref{eq:baryon})] and in Sec.~\ref{sec:sxsb} indicate how it can
be utilized to prove S$\chi$SB and discuss its implication for
composite models of quarks and leptons.  Section~\ref{sec:pseumeson}
presents Witten's proofs of Eq.~(\ref{eq:pion}) and of the flavor mass
inequality for pseudoscalars. We present also an alternate derivation
of Eq.~(\ref{eq:pion}) \cite{ref:nussplb139}.  The beautiful
Vafa-Witten argument \cite{ref:vafaprl} for non-spontaneous breaking
of parity in QCD is presented in Sec.~\ref{sec:parity}.
Section~\ref{sec:largenc} is concerned with the inequalities in the
large $N_c$ limit of QCD \cite{ref:nusssath}. Section~\ref{sec:glue}
is devoted to the glueball sector \cite{ref:muzinich}, and
Sec.~\ref{sec:exotic} discusses inequalities in the continuum
meson-meson sector and for exotic states; in particular we also
discuss extensions to two-point functions involving local quark
combinations of a quartic degree \cite{ref:espriu,ref:goodyear}. In
Sec.~\ref{sec:finite} we discuss extensions to finite temperature,
finite chemical potential, and external electromagnetic fields.  In
Sec.~\ref{sec:qqbar} we discuss the constraints implied by QCD for the
chiral Lagrangian approach, and also discuss the utilization of QCD
inequalities to constrain the $\bar{Q}Q$ potential $V(R)$ in heavy
quarkonium states, quark mass ratios, and weak matrix elements.
Finally Sec.~\ref{sec:beyond2pt} discusses extensions beyond two-point
functions.

Towards the end of the review we adopt a more heuristic approach,
applying QCD inequality-like relations to four and five particle
states in App.~\ref{app:fourfive} and to electromagnetic effects on
scattering lengths in App.~\ref{app:scatt}.  We discuss some
applications of the inequalities in atomic, chemical, and biological
contexts in App.~\ref{app:bio}. Also in Sec.~\ref{sec:largenc}, we use
the large $N$ (planar) limit to extend the interflavor meson mass
inequalities which were rigorously proven only for the pseudoscalar
channel \cite{ref:witt83} to other cases.  Sections~\ref{sec:largenc},
\ref{sec:glue}, \ref{sec:exotic}, and also Apps.~\ref{app:eq42},
\ref{app:bmineq}, \ref{app:fourfive}, \ref{app:scatt}, and
\ref{app:bio} constitute mostly new, unpublished material.
Section~\ref{sec:conc} includes a short summary.  We also present two
new conjectures concerning the possible utilization of the
ferromagnetic character of the QCD euclidean Lagrangian, and a
possible monotonic behavior of mass ratios with the number of quark
flavors $N_f$, as well as inequalities for quantities other than
hadronic masses.

%
\section{Derivation of flavor mass relations in a simple potential model}
\label{sec:mmineq}

We first discuss the inequalities in a simple potential ``toy model''
\cite{ref:nussplb136,ref:nussprl52,ref:lieb,ref:bertlmann,ref:richtaxil,ref:richphysrep},
which contains some features of the full-fledged QCD problem.
Specifically the interactions -- represented here by the potentials --
are flavor independent, and all flavor dependence is manifested only
via the masses in the kinetic term.

Let us consider a two-body system described by the Hamiltonian
\begin{equation}
   H_{12} = T_1 + T_2 + V_{12}
\>.
\end{equation}
For a nonrelativistic Schr\"{o}dinger equation, the kinetic terms are
\begin{equation}
   T_1 = \frac{\vec{p}_1^{\, 2}}{2m_1}, \qquad
   T_2 = \frac{\vec{p}_2^{\, 2}}{2m_2}
\end{equation}
with $m_{1,2}$ the masses of particles 1 and 2. We assume that the
potential $V$ depends only on the relative coordinate $\vecr = \vecr_1
- \vecr_2$ (translational invariance), and we also take $V =
V(|\vecr|) = V(r)$ only to ensure rotational invariance.

We can separate the motion of the center of mass:
\begin{eqnarray*}
   \psi &=& {\rm e}^{i \vec{P} \cdot \vec{R}} \psi_{12}(\vecr) \\
   \vec{R} &=& \frac{m_1 \vecr_1 + m_2 \vecr_2}{m_1 + m_2} \\
   \vecr &=& \vecr_1 - \vecr_2
\>,
\end{eqnarray*}
and write $H_{12}$ as
\begin{equation}
   H_{12} = \frac{\vec{P}^2}{2M} + \frac{\vec{p}^{\, 2}}{2\mu} + V_{12}(r)
      \equiv \frac{\vec{P}^2}{2M} + h_{12}
\>,
\end{equation}
with $\vec{P} = \vp_1 + \vp_2, \vp = \vp_1 - \vp_2, M = m_1 + m_2,$
and $\mu = \frac{m_1 m_2}{m_1 + m_2}$. For the subsequent discussion
we will specialize to $\vec{P} = 0$, {\em i.e.} to the center of mass
system (CMS).

We will be interested in the bound states of $h_{12}$ satisfying
\begin{eqnarray}
   h_{12} \psi_{12} &=& \epsilon_{12} \psi_{12}
   \nonumber \\
   h_{12} &=& \frac{\vec{p}^{\, 2}}{2\mu} + V_{12}(r)
      = \frac{\vec{p}^{\, 2}}{2m_1} + \frac{\vec{p}^{\, 2}}{2m_2} + V_{12}(r)
      \equiv \frac{\vec{p}^{\, 2}}{2m_1} + \frac{\vec{p}^{\, 2}}{2m_2} + V(r)
\label{eq:hpsi}
\>,
\end{eqnarray}
with $\psi_{12} = \psi_{12}(\vecr)$ a normalized ($\int {\rm d}^3 r \,
\psi_{12}^2(\vecr) = 1$) state. (We assume that such bound states exist.)

We can next consider two additional systems with identical potentials
$V_{11}(r) = V_{22}(r) = V(r)$, but made of two particles with the
same mass $m_1$ (or $m_2$)
\begin{eqnarray}
   h_{11} &=& \vec{p}^{\, 2} \left( \frac{1}{2m_1} + \frac{1}{2m_1} \right)
      + V(r)
   \nonumber \\
   h_{22} &=& \vec{p}^{\, 2} \left( \frac{1}{2m_2} + \frac{1}{2m_2} \right)
      + V(r)
\label{eq:hdef}
\>.
\end{eqnarray}
To mimic the QCD problem, with quark-antiquark bound states, we still
take the particles as non-identical so as to avoid issues of
statistics.

Let $\epsilon_{12}^{(0)}, \epsilon_{11}^{(0)}, \epsilon_{22}^{(0)}$ be the
ground state energies for the three Hamiltonians
\begin{equation} 
   h_{ij} \psi_{ij}^{(0)}(\vecr) = 
      \epsilon_{ij}^{(0)} \psi_{ij}^{(0)}(\vecr)
\>.
\end{equation}
We wish to derive the relation
\begin{equation}
   \epsilon_{12}^{(0)} \geq \frac{1}{2}
      \left( \epsilon_{11}^{(0)} + \epsilon_{22}^{(0)} \right)
\label{eq:engineq}
\>.
\end{equation}
Equations (\ref{eq:hpsi}) and (\ref{eq:hdef}) imply the operator identity
\begin{equation}
   h_{12} = \frac{1}{2} \left( h_{11} + h_{22} \right)
\label{eq:hamineq}
\>.
\end{equation}
Let us take the diagonal matrix element of both sides of this equation
with $\psi_{12}^{(0)}$, the ground-state wave function of $h_{12}$. We
then have
\begin{equation}
   \langle \psi_{12}^{(0)} | h_{12} |\psi_{12}^{(0)} \rangle
      = \epsilon_{12}^{(0)} = \frac{1}{2} \left(
      \langle \psi_{12}^{(0)} | h_{11} |\psi_{12}^{(0)} \rangle +    
      \langle \psi_{12}^{(0)} | h_{22} |\psi_{12}^{(0)} \rangle \right)
\>.
\end{equation}
By the variational principle \cite{ref:shankar} each of the
expectation values on the right-hand side exceeds
$\epsilon_{11}^{(0)}$ (or $\epsilon_{22}^{(0)}$) respectively, which
are minima of $\langle \psi | h_{11} | \psi \rangle$ and $\langle \psi
| h_{22} | \psi \rangle$ with $\psi = \psi_{11}^{(0)}$ and $\psi =
\psi_{22}^{(0)}$.  Thus Eq.~(\ref{eq:engineq}) is obtained.

The previous discussion has been carried out in the CMS frame,
$\vec{P}_{{\rm total}} = 0$. In this frame, upon adding the rest masses
$m_1 + m_2$ to $\epsilon_{12}^{(0)}$, {\em etc.}, the inequality
(\ref{eq:engineq}) translates into an inequality for the total masses
of the bound states:
\begin{equation} 
   m_{1\bar{2}}^{(0)} \geq \frac{1}{2}
      \left( m_{1\bar{1}}^{(0)} + m_{2\bar{2}}^{(0)} \right)
\>.
\end{equation}
Also in this frame total angular momenta are the spins of the
composites.

Let us now make the following observations which will also be very useful
for the discussion of Sec.~\ref{sec:interflavor}:
\begin{trivlist}
\item[ 1.] The addition of the two Hamiltonians $h_{11}$ and $h_{22}$ and
  their comparison with $h_{12}$ may appear to present some formal
  difficulties. In nonrelativistic physics we have a super-selection
  rule for different masses. Hence strictly speaking $h_{11}, h_{22},$
  and $h_{12}$ operate in different Hilbert spaces consisting of
  states with the (1,1), (2,2), and (1,2) pairs of particles
  respectively. However, the crucial observation is that these are
  {\em identical} spaces. All the Hamiltonians $h_{11}, h_{22},
  h_{12}$ (or any other $h_{ij}$) which are written in terms of
  $\vecr$ and $\vp$ can, therefore, be made to operate on the common
  generic Hilbert space of wave functions $\psi(\vecr)$. The mass
  dependence is explicit in $h_{ij}$. It happens to be additive and
  therefore Eq.~(\ref{eq:hamineq}) is a meaningful and useful operator
  identity.
  
\item[ 2.] The above generic space can be block diagonalized by using the
  symmetries of the Hamiltonians $h_{ij}$. Thus for the case of
  central potentials, we consider separately subspaces with given
  total angular momentum $l$ and solve
\begin{equation}
   h_{ij} \psi_{ij}^{(l)} = \epsilon_{ij}^{(l)} \psi_{ij}^{(l)}   
\>.
\end{equation}
The projection operator $P_l$ used in $P_l h_{ij} (P_l)^{\dag} \equiv
h_{ij}^{(l)}$ can be expressed in terms of $\vecr$ and $\vp$ only, and
is independent of the masses. Thus $P_l$ simultaneously projects into
the $l$ subspace every term in Eq.~(\ref{eq:hamineq}):
\begin{equation}
   h_{12}^{(l)} = \frac{1}{2} \left( h_{11}^{(l)} 
      + h_{22}^{(l)} \right)
\>.
\end{equation}
Using the same arguments as above, we can deduce that
\begin{equation}
   \epsilon_{12}^{l(0)} \geq \frac{1}{2}
      \left( \epsilon_{11}^{l(0)} + \epsilon_{22}^{l(0)} \right)
\label{eq:englineq}
\end{equation}
holds for the ground states in each $l$-wave separately.

\item[ 3.] To the extent that heavy quarkonium systems can be treated via a
  nonrelativistic Schr\"{o}dinger equation approximation, with
  negligible spin effects, the above derivation would indeed directly
  suggest mass relations such as $m_{c\bar{b}}^{(0)} \geq \frac{1}{2}
  \left( m_{c\bar{c}}^{(0)} + m_{b\bar{b}}^{(0)} \right)$.
  
\item[ 4.] The inequalities (\ref{eq:engineq}) and (\ref{eq:englineq}) are
  obtained by taking just one particular matrix element of the
  operator relation (\ref{eq:hamineq}). Evidently the latter contains
  far more information. In particular one may wonder if the
  inequalities hold for excited states in each given $l$ channel ({\em
    i.e.}  for radial excitations).
  
  The simple generalization, {\em e.g.} $\epsilon_{12}^{(1)} \geq
  \frac{1}{2} \left( \epsilon_{11}^{(1)} + \epsilon_{22}^{(1)}
  \right)$ for the first radially excited state is, in general,
  incorrect. The point is that $\psi_{11}^{(1)}, \psi_{22}^{(1)}$ and
  $\psi_{12}^{(1)}$ should minimize the expectation values $\langle
  \psi | H_{11} | \psi \rangle$, $\langle \psi | H_{22} | \psi
  \rangle$ and $\langle \psi | H_{12} | \psi \rangle$ respectively,
  subject to {\em different} constraints: $\psi_{11}^{(1)}$ should be
  orthogonal to $\psi_{11}^{(0)}$ {\em etc.} and the three ground
  state wave functions are in general different. However, such
  constraints can be avoided if we consider the two-dimensional space
  $v_2$ spanned by the first {\em and} second excited states. The
  variational principal (for $h_{ij}$) tells us that $\langle
  \psi^{(0)} | h_{ij} | \psi^{(0)} \rangle + \langle \psi^{(1)} |
  h_{ij} | \psi^{(1)} \rangle = \epsilon_{ij}^{(0)} +
  \epsilon_{ij}^{(1)}$ is the minimal value that ${\rm tr}_{v_2}
  h_{ij}$ can achieve when we consider all possible two-dimensional
  subspaces $\epsilon_{ij}^{(0)} + \epsilon_{ij}^{(1)} = {\rm
    min}_{v_2} {\rm tr} h_{ij}$. More generally we have for the sum of
  the ground state and the first $n-1$ excited states:
  $\epsilon_{ij}^{(0)} + \epsilon_{ij}^{(1)} + \ldots +
  \epsilon_{ij}^{(n-1)} = {\rm min}_{v_n} {\rm tr} h_{ij}$, where the
  trace is now to be minimized over all $n$-dimensional subspaces
  $v_n$.  Considering then the operator relation $h_{12} = \frac{1}{2}
  \left( h_{11} + h_{22} \right)$ and taking the trace of both sides
  in the space $v_n^{1,2}$ which minimizes ${\rm tr}_{v_n}h_{12}$, we
  conclude that
\begin{eqnarray}
   \epsilon_{12}^{(0)} + \epsilon_{12}^{(1)} 
      + \ldots + \epsilon_{12}^{(n-1)} &\geq& \frac{1}{2} \left[ (
      \epsilon_{11}^{(0)} + \epsilon_{11}^{(1)} 
      + \ldots + \epsilon_{11}^{(n-1)}) \right. \nonumber \\
      &+& \left. (\epsilon_{22}^{(0)} + \epsilon_{22}^{(1)}
      + \ldots + \epsilon_{22}^{(n-1)}) \right]
\>,
\label{eq:sumeng}
\end{eqnarray}
with the summation extending up to any one of the radially excited states. 
   
\item[ 5.] It is amusing to see how the inequalities work in familiar
  cases.  For the Coulomb potential the ground state bindings are
\begin{eqnarray}
   B_{ij}^{(0)} &=& \frac{1}{2} \mu_{ij} \alpha^2 
      = \frac{1}{2} \left( \frac{m_i m_j}{m_i + m_j} \right)\alpha^2
   \nonumber \\
   B_{ii}^{(0)} &=& \frac{1}{2} \left(\frac{m_i}{2}\right) \alpha^2
   \nonumber \\
   B_{jj}^{(0)} &=& \frac{1}{2} \left(\frac{m_j}{2}\right) \alpha^2
\>.
\end{eqnarray}
Obviously, $B_{ij}^{(0)} \leq \frac{1}{2} \left( B_{ii}^{(0)} +
B_{jj}^{(0)} \right)$ so that the masses of the corresponding states
satisfy the desired inequality
\begin{equation}
    m_{ij}^{(0)} = m_i + m_j - B_{ij}^{(0)}
      > \frac{1}{2} \left( m_{ii}^{(0)} + m_{jj}^{(0)} \right)
\>.
\end{equation}

Next, consider the harmonic oscillator potential, $\frac{1}{2} K r^2$,
as a prototype of a confining potential. The ground state energies,
measured with respect to the minimum of the potential well, are
\begin{displaymath}
   \epsilon_{ij}^{(0)} = \frac{\hbar}{2} \omega_{ij}^{(0)}
      = \frac{\hbar}{2} \sqrt{\frac{K}{\mu_{ij}}}
      = \frac{\hbar}{2} \sqrt{K\left(\frac{1}{m_i} + \frac{1}{m_j}\right)}
\>.
\end{displaymath}
We therefore have $\left(\epsilon_{ij}^{(0)}\right)^2 = \frac{1}{2}
\left[(\epsilon_{ii}^{(0)})^2 + (\epsilon_{jj}^{(0)})^2 \right]$ which
implies by simple algebra that
\begin{equation}
   \epsilon_{ij}^{(0)} \geq \frac{1}{2}
      \left( \epsilon_{ii}^{(0)} + \epsilon_{jj}^{(0)} \right)
\>,
\label{eq:epsij}
\end{equation}
and upon adding the rest masses:
\begin{equation}
   m_{ij}^{(0)} \geq \frac{1}{2}
      \left( m_{ii}^{(0)} + m_{jj}^{(0)} \right)
\>.
\label{eq:mijineq}
\end{equation}

We note in passing that the energies for these two systems happen to
have quantization rules of the form $\epsilon^{(n)} = f(n)
\epsilon^{(0)}$, where $f(n)$ is independent of $\mu$, the reduced
mass.\footnote{This feature stems from the fact that these potentials
  have a power law behavior, $r^\alpha$, in which case $\epsilon^{(n)}
  = f(n)$ is clearly manifest in the semiclassical WKB limit.}
Specifically, $f(n) = - \frac{1}{(n+1)^2}$ and $(2n+1)$ for the two
cases. Hence all the above inequalities happen to hold separately for
each of the excited states $\epsilon_{ij}^{(n)}$,
$\epsilon_{ii}^{(n)}$, and $\epsilon_{jj}^{(n)}$.

\item[ 6.] While the above discussion was in the framework of a
  nonrelativistic Schr\"{o}dinger equation, the result
  (\ref{eq:epsij}) holds in a far more general context.  The
  particular form of the kinetic energy $T_i = \frac{\vec{p}^{\,
      2}}{2m_i}$ did not play any role in deriving the operator
  relation (\ref{eq:hamineq}). Hence $T_i$ could have an arbitrary $p$
  dependence. In particular, we can take $T_i = \sqrt{\vec{p}^{\, 2} +
    m_i^2}$, the expression appropriate for relativistic motion.
  
  The key ingredient was that the potentials $V_{12}, V_{11}$, and
  $V_{22}$ are all the {\em same}, {\em i.e.} that we have ``flavor
  independent interactions''. However, besides the requirement of
  translational (and preferably also rotational) invariance, the
  ``potential'' is not restricted. Thus, $V$ need not depend on
  $\vecr$ alone, but could have arbitrary dependence on $\vecr$ and
  $\vp$. The only aspect we have to preserve is that $h_{ij} = T_i +
  T_j + V_{ij}$ so that $h_{ij} = \frac{1}{2} (h_{ii} + h_{jj})$.

Nonrelativistic quark models have potentials with some explicit flavor
(quark mass) dependence, {\em e.g.} in the ``hyperfine interaction'' 
\cite{ref:rujula}:
\begin{equation}
   V_{ij}^{{\rm HF}} = \frac{(\lambda_i \cdot \lambda_j)}{m_i m_j} 
      (\sigma_i \cdot \sigma_j) \delta^3(\vecr)
\>.
\end{equation}

In Sec.~\ref{sec:interflavor} we show that in the full-fledged theory we still
have an operator relation
\begin{equation}
   H_{i\bar{\jmath}} + H_{k\bar{l}} = H_{i\bar{l}} + H_{k\bar{\jmath}}
\>,
\end{equation}
with $H_{ij}$ being the QCD Hamiltonian restricted to a particular flavor
sector. With $i = k, j = l$ the last relation reduces to
\begin{equation}
   H_{i\bar{\jmath}} + H_{j\bar{\imath}} 
      = H_{i\bar{\imath}} + H_{j\bar{\jmath}}
\>,
\end{equation}  
which is very reminiscent of Eq.~(\ref{eq:hamineq}) and, to the extent that
the wave functions $\psi_{i\bar{\jmath}}$ are {\em symmetric} in flavor,
again leads to the same conclusion.

One may also avoid the mass dependent $\vec{s}_1 \cdot \vec{s}_2$ or
$\vec{s} \cdot \vec{L}$ interactions by considering the ``centers of
mass'' of the various multiplets, {\em e.g.} $(3m_\rho + m_\pi)/4$,
representing the mass values prior to hyperfine splittings and
applying the inequalities to these combinations
\cite{ref:richplb139,ref:ader}.

We note that the hyperfine interaction is strongly attractive for the
spin singlet pseudoscalar meson case. Since furthermore $m_1^{-2} +
m_2^{-2} \geq 2 (m_1 m_2)^{-1}$, this extra binding enhances the
inequality (\ref{eq:engineq}) derived for the spin independent part of
the interaction.  Indeed, the only case for which relation
(\ref{eq:engineq}) was proved by utilizing the rigorous euclidean
correlation function approach, is that of the pseudoscalar mesons
\cite{ref:witt83}.

\end{trivlist}

%
\section{Baryon-meson mass inequalities in the gluon exchange model}
\label{sec:mbineq}

The simple ``quark counting rules'' relating meson and baryon total
cross-sections \cite{ref:levin,ref:lip66} and masses \cite{ref:freund}
were early indications for the relevance of the quark model
\cite{ref:lip73,ref:hendry}. If the mass of a ground state hadron is
just the sum of the masses of its quark constituents, then
\begin{equation}
   2m_{ijk}^{(0)} = m_{i\bar{\jmath}}^{(0)} + m_{j\bar{k}}^{(0)}
      + m_{k\bar{\imath}}^{(0)}
\>,
\end{equation}
with $m_{ijk}^{(0)}$ the mass of the ground state baryon consisting of
quarks $q_i, q_j, q_k$; and $m_{i\bar{\jmath}}^{(0)}$ the mass of the
lowest lying $q_i \bar{q}_j$ meson. The following discussion motivates a
related QCD inequality \cite{ref:nussprl51,ref:richplb139,ref:lieb,ref:bmr}
\begin{equation}
   2m_{ijk}^{(0)} \geq m_{i\bar{\jmath}}^{(0)} + m_{j\bar{k}}^{(0)}
      + m_{k\bar{\imath}}^{(0)}
\>.
\label{eq:mbmassineq}
\end{equation}
The particular variant $m_N \geq m_\pi$ was derived by Weingarten
\cite{ref:weingarten} from the correlation function inequalities, and
will be discussed in Sec.~\ref{sec:mbineqcorr}.

If $qq$ ($\qqb$) interactions are generated via one gluon
exchange, then the color structure is $V_{12} \propto \vec{\lambda}_1 \cdot
\vec{\lambda}_2$, with $\vec{\lambda}$ a vector consisting of the $N^2 - 1$ 
Gell-Mann matrices of the fundamental SU(N) representation. In a meson the
quarks $q_1 \bar{q}_2$ couple to a singlet and, using $\langle \cdots
\rangle$ to indicate expectation values,
\begin{equation}
   0 = \langle (\vec{\lambda}_1 +\vec{\lambda}_2 )^2 
      \rangle_{\rm meson} = 2 \vec{\lambda}^2 + 2 \langle
      \vec{\lambda}_1 \cdot \vec{\lambda}_2 \rangle_{\rm meson}
\>,
\end{equation}
so that
\begin{equation}
   \langle \vec{\lambda}_1 \cdot \vec{\lambda}_2 \rangle_M =
      - \vec{\lambda}^{\, 2}
\>.
\label{eq:lambdam}
\end{equation}
Here $\vec{\lambda}^2 = \sum_{n=0}^{N^2 - 1} \, (\lambda_n)^2$, is a
fixed diagonal $N \times N$ matrix proportional to the unit matrix.
The SU(N) baryon is constructed as the color singlet, completely
antisymmetric combination of $N$ quarks
\begin{equation}
   \epsilon_{a_1 \ldots a_N} q^{a_1 \ldots a_N}
\>.
\end{equation}
Thus
\begin{equation}
   0 = \left\langle \left(\sum_i^N \lambda_i\right)^2 \right\rangle_B 
     = N \vec{\lambda}^2 + \sum_{i \neq j}^N \langle \lambda_i 
     \lambda_j \rangle_B = N \vec{\lambda}^2 + N(N-1) \langle 
     \lambda_1 \cdot \lambda_2 \rangle_B
\>,
\end{equation}
where we used the fact that all the $N(N-1)$ expectation values $
\langle \lambda_i \cdot \lambda_j \rangle_B$ are equal. Thus using
Eq.~(\ref{eq:lambdam}) we find that
\begin{equation}
   \langle \lambda_1 \cdot \lambda_2 \rangle_B
      = - \frac{1}{(N-1)} \vec{\lambda}^2 = \frac{1}{(N-1)}
      \langle \lambda_1 \cdot \lambda_2 \rangle_M
\>.
\label{eq:lblm}
\end{equation}
This implies that the strength of the one gluon exchange interaction
in the meson $q_i \bar{q}_j$ is, for QCD ({\em i.e.} $N=3$), precisely
twice the corresponding gluon exchange interaction for a $q_i q_j$
pair in a baryon.

Since $\langle \lambda_1 \cdot \lambda_2 \rangle_B = \frac{1}{2}
\langle \lambda_1 \cdot \lambda_2 \rangle_M$ is just an overall
relation of the color factors, we have the corresponding ratio of the
pairwise interactions in the meson and in the baryon:
\begin{equation}
   V_{i\bar{\jmath}}(\cdots)_M = 2 V_{ij}(\cdots)_B
\>,
\end{equation}
as long as the $q_i \bar{q}_j$ are in the same angular momentum, spin,
radial state, {\em etc.} as the $q_i q_j$ in the baryon (or in the
appropriate mixture of states).

For $N=3$ and a general two-body interaction, we write the Hamiltonian
describing the baryon as:
\begin{equation}
   H_B = H_{ijk} = T_i + T_j + T_k + V_{ij}^B + V_{jk}^B + V_{ki}^B
\>,
\label{eq:hbaryon}
\end{equation}
with $T_i$ the kinetic, single-particle operators.

The $q_i q_j q_k$ baryonic system can be partitioned into (2 + 1)
subsystems in three different ways: $((ij),k), (i,(jk)), (j,(ki))$.
Let us consider the three mesonic $\qqb$ systems corresponding to
these partitionings: $M_{i\bar{\jmath}}, M_{j\bar{k}},
M_{k\bar{\imath}}$, composed of $(q_i \bar{q}_j), \, (q_j \bar{q}_k),
\, (q_k \bar{q}_i)$.  The Hamiltonians describing these mesons are
\begin{eqnarray}
   H_{i\bar{\jmath}} &=& T_i + T_j + V_{i\bar{\jmath}}^M
   \nonumber \\
   H_{j\bar{k}} &=& T_j + T_k + V_{j\bar{k}}^M
   \nonumber \\  
   H_{k\bar{\imath}} &=& T_k + T_i + V_{k\bar{\imath}}^M
\>.
\label{eq:hmeson}
\end{eqnarray}
From $V_{i\bar{\jmath}}^M = 2 V_{ij}^B$ {\em etc.} it is easy to
verify the key relation
\begin{equation}
   2 H_{ijk} =  H_{i\bar{\jmath}} + H_{j\bar{k}} + H_{k\bar{\imath}}
\label{eq:mbham}
\end{equation}
between the Hamiltonian describing the baryon $B_{ijk}$ and the three
mesonic Hamiltonians corresponding to its diquark subsystems. Let us next
take the expectation value of the last operator relation in the normalized
ground state $\psi_{ijk}^{(0)}$ of the baryon at rest [to simplify the
notation we suppress Lorentz ($J^{PC}$) quantum numbers]. The left hand
side yields
\begin{equation}
   2 \langle \psi_{ijk}^{(0)} | H_{ijk} | \psi_{ijk}^{(0)} \rangle
      = 2 m_{ijk}^{(0)}
\>,
\label{eq:lhs}
\end{equation}
with $ m_{ijk}^{(0)}$ the mass of the ground state baryon. The right hand
side is
\begin{equation}
   \langle \psi_{ijk}^{(0)} | H_{i\bar{\jmath}} | \psi_{ijk}^{(0)} \rangle
   + \langle \psi_{ijk}^{(0)} | H_{j\bar{k}} | \psi_{ijk}^{(0)} \rangle
   + \langle \psi_{ijk}^{(0)} | H_{k\bar{\imath}} | \psi_{ijk}^{(0)} \rangle
\>.
\label{eq:rhs}
\end{equation}
The matrix elements of the two-body operators in the three-body wave
function, {\em e.g.} $ \langle \psi_{ijk}^{(0)} | H_{i\bar{\jmath}} |
\psi_{ijk}^{(0)} \rangle$, are evaluated by considering $H_{i\bar{\jmath}}$
as a three-body operator which is just the identity for quark $k$. Thus we
view $\psi_{ijk}$, for fixed coordinates of the quark $q_k$, as a two-body
$(q_i q_j)$ wave function $\tilde{\psi}_{(i\bar{\jmath})}^{(k)}$, and
compute $ \langle \tilde{\psi}_{i\bar{\jmath}}^{(k)} | H_{i\bar{\jmath}}
| \tilde{\psi}_{i\bar{\jmath}}^{(k)} \rangle$. Finally, this is integrated 
over all values of the coordinates of $q_k$. 

The key observation is that $\tilde{\psi}_{(i\bar{\jmath})}^{(k)}$,
the two-body wave function prescribed by $\psi_{ijk}^{(0)}$, is, in
general, different from $\psi_{(i\bar{\jmath})}^{(0)}$, the ground
state wave function of the meson $q_i q_{\bar{\jmath}}$. By the
variational principle the latter wave function minimizes the
expectation value of $H_{i\bar{\jmath}}$, the mesonic Hamiltonian:
\begin{equation}
   {\rm min}_{\tilde{\psi}_{i\bar{\jmath}}} \langle
      \tilde{\psi}_{i\bar{\jmath}} | H_{i\bar{\jmath}} | 
      \tilde{\psi}_{i\bar{\jmath}} \rangle
      = \langle \tilde{\psi}_{i\bar{\jmath}}^{(0)} | H_{i\bar{\jmath}} | 
      \tilde{\psi}_{i\bar{\jmath}}^{(0)} \rangle = m_{i\bar{\jmath}}^{(0)}
\>,
\end{equation}
where $ m_{i\bar{\jmath}}^{(0)}$, the energy of the state at rest, is the
mass of the ground state $q_i q_{\bar{\jmath}}$ meson. Hence we have
\begin{mathletters}
   \begin{equation}
       \langle \tilde{\psi}_{i\bar{\jmath}}^{(k)} | H_{i\bar{\jmath}} | 
       \tilde{\psi}_{i\bar{\jmath}}^{(k)} \rangle 
       \geq m_{i\bar{\jmath}}^{(0)} 
   \label{eq:mepsiij}
   \end{equation}
and likewise
   \begin{equation}
       \langle \tilde{\psi}_{j\bar{k}}^{(i)} | H_{j\bar{k}} | 
       \tilde{\psi}_{j\bar{k}}^{(i)} \rangle \geq m_{j\bar{k}}^{(0)} 
   \label{eq:mepsijk}
   \end{equation}
   \begin{equation}
       \langle \tilde{\psi}_{k\bar{\imath}}^{(j)} | H_{k\bar{\imath}} | 
       \tilde{\psi}_{k\bar{\imath}}^{(j)} \rangle 
       \geq m_{k\bar{\imath}}^{(0)} .
   \label{eq:mepsiki}
   \end{equation}
\end{mathletters}
Since each of these inequalities persists after the normalized integration
over the coordinates of the third quark [{\em e.g.} $q_k$ for
Eq.~(\ref{eq:mepsiij})], we have,
\begin{equation}
   \langle \psi_{ijk}^{(0)} | H_{i\bar{\jmath}} | \psi_{ijk}^{(0)} \rangle
      \geq m_{i\bar{\jmath}}^{(0)}, \, \, {\em etc.}
\end{equation}
Equating (\ref{eq:lhs}) and (\ref{eq:rhs}), the matrix elements of the
left hand side and right hand side of the original operator relation,
we arrive at the desired inequality (\ref{eq:mbmassineq})
\cite{ref:nussprl51,ref:richplb139}. The following remarks elaborate on
these inequalities and the conditions for their applicability.

\begin{trivlist}
\item[ 1.] We have not displayed the $J^P$ quantum numbers of the baryonic
  $\psi_{ijk}$ or the mesonic $\psi_{i\bar{\jmath}}$ states. As
  indicated by the above construction of ``trial'' mesonic functions
  from the baryon wave function, $\psi_{i\bar{\jmath}}$ must be in the
  $J^P$ state (or in general, in a mixture of $J^P$ states) prescribed
  by the original baryonic wave function $\psi_{ijk}^{(0)}$.

Consider first the $\Delta^{++}$ baryon. In the approximation where $L
\ne 0$ components in the ground state wave function are ignored, the
state with $J_z(\Delta^{++}) = 3/2$ consists of three up quarks with
parallel spins: $u \uparrow u \uparrow u \uparrow$, with the arrow
indicating spin direction.  Each of the $\psi_{i\bar{\jmath}}^{(0)}$
will be, in this case, $u \uparrow \bar{u} \uparrow$, in a spin
triplet state $\rho$ or $\omega$. The inequality reads
\begin{equation}
   m_{\Delta^{++}} \geq \frac{3}{2} m_\rho \, \,  
      (\mbox{or} \, \, m_\omega) \qquad
      (ijk = u \uparrow u \uparrow u \uparrow)
\>.
\label{eq:mdelta}
\end{equation}
Likewise,
\begin{mathletters}
   \begin{equation}
      m_{\Omega^{-}} \geq \frac{3}{2} m_\phi \qquad
      (ijk = s \uparrow s \uparrow s \uparrow)
   \end{equation}
   \begin{equation}
      2 m_{\Xi^0} \geq 2  m_{K^{\ast}} + m_\phi \qquad
      (ijk = s \uparrow s \uparrow u \uparrow)
   \end{equation}
   \begin{equation}
      2 m_{\Sigma^{+}} \geq 2 m_{K^{\ast}} + m_\rho \qquad
      (ijk = u \uparrow u \uparrow s \uparrow).
   \end{equation}
\end{mathletters}
The situation is different for the $J = 1/2$ ({\em i.e.} $S = 1/2$ in
this $L = 0$ approximation) baryons such as the nucleon (see
App.~\ref{app:bmineq}). In this case the diquark systems are, on
average, with equal probability in the $S=0$ and $S=1/2$ states, as
can be readily verified from $\langle (s_1 + s_2 + s_3)^2 \rangle =
\langle s_N^2 \rangle = 3/4$. [Strictly speaking, the $uu$ diquark is
pure triplet and the $ud$ diquarks are in a triplet (singlet) state with
probabilities 1/4 (3/4), respectively.]

To see the effect of having a mixed rather than a pure trial mesonic
state, we revert back to Eq.~(\ref{eq:mepsiij}). Instead of a single
matrix element of $H_{i\bar{\jmath}}$ in a specific
$\tilde{\psi}_{(i\bar{\jmath})}^{(k)}$ state, we have, in general, a
weighted sum of matrix elements corresponding to the mixture of
two-body states generated from the three-body baryonic wave function:
\begin{equation}
   \langle \psi_{ijk}^{(0)} | H_{i\bar{\jmath}} | \psi_{ijk}^{(0)} \rangle
      = \sum_n c_n^2 \langle \psi_{i\bar{\jmath}(n)}^{(k)} | 
      H_{i\bar{\jmath}} | \psi_{i\bar{\jmath}(n)}^{(k)} \rangle
\>.
\end{equation}
Here $n$ labels the different ($J^P$) states and $c_n^2$ are the
normalized weights ($\sum_n c_n^2 = 1$). 

The variational argument can be applied separately for each of the (trial)
$\tilde{\psi}_{(n)}$ states and to their matrix elements $\langle
\tilde{\psi}_{(n)} | H_{i\bar{\jmath}} | \tilde{\psi}_{(n)} \rangle$ to
obtain
\begin{equation}
   \langle \psi_{ijk}^{(0)} | H_{i\bar{\jmath}} | \psi_{ijk}^{(0)} \rangle
      \geq \sum_n c_n^2 m_{i\bar{\jmath}(n)}
\>.
\label{eq:spinhalf}
\end{equation}
For the specific case of the nucleon the weights of the $qq$ singlet
(triplet) configuration are $c_{s(qq) = 1}^2 = c_{s(qq) = 1/2}^2 =
1/2$.  The corresponding $\qqb$ states are the $\rho (\omega)$ and the
$\pi$, and the inequality (\ref{eq:spinhalf}) (plus similar ones for
the matrix elements of $H_{j\bar{k}}$ and $H_{k\bar{\imath}}$) yields:
\begin{equation}
   2 m_N \geq \frac{3}{2} (m_\pi + m_\rho)
\>.
\end{equation}
Likewise by considering the $\Lambda$ hyperon we find
\begin{equation}
   2 m_\Lambda \geq \frac{1}{2} \left[ m_\pi + m_\rho 
      + 3 m_K + m_{K^{\ast}} \right]
\>.
\end{equation}

\item[ 2.] A simplified version of the inequalities applied just to
  the spin-averaged (``center of mass'') multiplets was suggested by
  Richard \cite{ref:richplb139}.  Note that unlike for the case of the
  meson-meson mass relation in Sec.~\ref{sec:mmineq}, the baryon-meson
  mass inequalities do apply even when we have mass dependent $q_i
  q_j$ (or $q_i q_{\bar{\jmath}}$) interactions such as the hyperfine
  and/or spin orbit interactions generated by one gluon exchange. The
  point is that the same flavor (mass) combinations $(ij), (jk), (ki)$
  appear in both the diquark subsystems and in the mesons
  $M_{i\bar{\jmath}}, M_{j\bar{k}},$ and $M_{k\bar{\imath}}$. Hence
  unlike the case of the meson-meson relation there is no motivation
  to consider only the above simplified version.
  
\item[ 3.] The above discussion utilized the nonrelativistic quark model
  with a one gluon exchange potential. We show next (and in
  Sec.~\ref{sec:nonpertmbineq}) that the inequalities are valid in a
  vastly larger domain \cite{ref:nussprl51}.
  
  First we note that any vertex or propagator insertions into the one
  gluon exchange diagram leave the $\lambda_1 \cdot \lambda_2$ color
  structure intact (see Fig.~\ref{fig:onegluon}). Such insertions
  generate a running coupling constant $\alpha(q^2)$ or $\alpha(r)$.
  Even if the non-perturbative effect of the propagator insertions can
  generate a confining potential \cite{ref:man83,ref:bbz} $V \sim
  \sigma r$, this potential still has the $\lambda_1 \cdot \lambda_2$
  color structure, and the derivation of the inequalities still
  applies.
  
\item[ 4.] The most general two-body interaction [due to any number of
  gluon and $\qqb$ exchanges; see Fig.~\ref{fig:twoquark}] in the $qq$
  or $\qqb$ system, has, from the $t$-channel point of view, a
  $\bar{3} \otimes 3 = 8 \oplus 1$ color structure.  The octet part
  corresponds to the $\lambda_1 \cdot \lambda_2$ structure; the
  singlet to $1 \cdot 1$. Thus we need not assume that the important
  interactions are due to one gluon exchange. Rather, we have to
  assume that the octet, $\lambda_1 \cdot \lambda_2$ part of the full
  two-body interaction is dominant. We note that in the large $N_c$
  limit we expect the $\lambda_1 \cdot \lambda_2$ part to dominate
  over the $1 \cdot 1$ part.
  
\item[ 5.] The trilinear gluon couplings are a potential obstacle to
  having the operator relation (\ref{eq:mbham}) in a general setting.
  These couplings could yield genuinely nonseparable interactions in
  the nucleon of the type shown in Fig.~\ref{fig:threegluon}. However,
  it turns out that the trigluon diagram vanishes in the baryon state.
  The relevant color factor is
\begin{displaymath}
   f^{rst} \lambda_r^{aa'} \lambda_s^{bb'} \lambda_t^{cc'}
      \epsilon_{abc} \epsilon_{a'b'c'}
\>,
\end{displaymath}
with $rst = 1, \ldots, 8; \, abc \, (a{\, '}b{\, '}c{\, '}) = 1,
\ldots, 3$ the adjoint and fundamental color indices, respectively.
This color factor vanishes since an exchange $r \rightarrow s, a
\rightarrow b, a{\, '} \rightarrow b{\, '}$ changes the sign of the
summand.  This cancellation is not modified by any further dressing of
this diagram with more gluon exchanges.

\item[ 6.] The above discussion did not require any specific form of the
  one-body kinetic terms $T_i$. We could have the nonrelativistic
  $T_i = m_i + \frac{p_i^2}{2m_i}$, a relativistic $T_i = \sqrt{p_i^2
    + m_i^2}$ form, or, by considering the operator relation
  (\ref{eq:mbham}) as a matrix in spinor space, also a Dirac $T_i =
  \not \! p_i + m_i$ form.
  
\item[ 7.] Just as for the meson-meson relation, we could use the
  variational principle in the space of the lowest $n$ states to
  obtain a relation between masses of radially excited baryons and
  mesons analagous to Eq.~(\ref{eq:sumeng}).

\item[ 8.] A diquark in a baryon cannot annihilate into gluons. This is not
  the case, however, for the $u\bar{u}$ or $d\bar{d}$ in a meson.
  These annihilations are avoided by choosing $I=1$ [{\em i.e.} $\rho$
  rather than $\omega$ in Eq.~(\ref{eq:mdelta})] combinations.
  
\item[ 9.] A simple intuitive feeling for the deviation from equality
  expected in the baryon-meson relation has been offered by Cohen and
  Lipkin \cite{ref:cohen} (in a paper which predated the QCD
  inequalities). The point is that the diquark systems in the baryon
  are not at rest but recoil against the third quark with a typical
  momentum $\vp$ of a few hundred MeV/$c$. Thus the trial meson wave
  functions correspond to a meson moving with $\vp \neq 0$.
  Consequently
\begin{eqnarray*}
   2 m_{ijk} &\geq& E_{i\bar{\jmath}} + E_{j\bar{k}} + E_{k\bar{\imath}}
   \nonumber \\
   &\approx& \sqrt{m_{ij}^2 + \vec{p}_k^{\, 2}} 
      + \sqrt{m_{jk}^2 + \vec{p}_i^{\, 2}}
      + \sqrt{m_{ki}^2 + \vec{p}_j^{\, 2}} \geq m_{ij} + m_{jk} + m_{ki}
\>.
\end{eqnarray*}

A more comprehensive investigation of the pattern of deviations from
equality as a function of the quark masses, and in particular for
logarithmic interquark potentials was recently performed by Imbo
\cite{ref:imbo}.

\item[ 10.] The baryon-meson relation can be easily extended via
  Eq.~(\ref{eq:lblm}) to any number of colors $N$:
\begin{equation}
   m_{B_{i_1 \ldots i_N}}^{(0)} \geq \frac{1}{N-1} \sum_{a \neq b}
      m^{(0)}_{i_a \bar{\imath}_b}
\>.
\label{eq:sunmbineq}
\end{equation}
\end{trivlist}

%
\section{Baryon-baryon mass inequalities}
\label{sec:bbineq}

The meson-meson mass inequalities above (see Sec.~\ref{sec:mmineq})
follow in potential models only if the various two-body
quark-antiquark potentials are independent of the quark masses.
Baryon-baryon mass inequalities motivated by similar convexity
arguments were suggested \cite{ref:nussprl52} originally to also hold
when all two-body quark-quark interactions are flavor-independent. It
turns out however, that inequalities of the form
\begin{equation}
   E^{(0)}(m,m,m) + E^{(0)}(m,M,M) \leq 2 E^{(0)}(m,m,M)
\label{eq:emmm}
\end{equation}
do not hold in general \cite{ref:lieb,ref:richtaxil,ref:martin}. As
will be clearly indicated in Sec.~\ref{sec:interflavor}, the key
requirement for proving either meson-meson or baryon-baryon mass
inequalities is the flavor symmetry of the ground state wave function.
In the two-body, pure potential case, the relevant, relative
coordinate wave function $\psi^{(0)}(\vecr), \vecr = \vecr_1 -
\vecr_2$ is guaranteed to have the $\vecr_1 \leftrightarrow \vecr_2$
``flavor'' symmetry.  This however is not the case for three-body
potential systems
\begin{equation}
   H = T_1(\vec{p}_1) + T_2(\vec{p}_2) + T_3(\vec{p}_3) 
      + V_{12}(\vecr_1 - \vecr_2) + V_{23}(\vecr_2 - \vecr_3)
      + V_{31}(\vecr_3 - \vecr_1)
\>.
\end{equation}
Indeed even for flavor-independent potentials $V_{12} = V_{23} =
V_{31}$, and simple nonrelativistic kinetic terms $T_i = \vec{p}_i^{\,
  2} / 2m_i$, the ground state wave function
$\psi^{(0)}(\vecr_1,\vecr_2,\vecr_3)$ is {\em not} (flavor) symmetric
under interchange of $1 \leftrightarrow 2$, {\em etc.} As various
counter examples show \cite{ref:lieb,ref:martin} (particularly the
simplest one due to Lieb \cite{ref:lieb} which we present in
App.~\ref{app:lieb1}), for certain, rather ``extreme'' potentials
$V_{ij}(r) = V(r)$, the kinematic assymetry due to the different quark
masses $(m_1 = m_3 = M, m_2 = m)$ is strongly enhanced so that
Eq.~(\ref{eq:emmm}) fails.  This notwithstanding, the elegant work of
Lieb \cite{ref:lieb} has shown that the conjectured equation
(\ref{eq:emmm}) does hold for a wide class of two-body potentials
$V(\vecr_i,\vecr_j)$ for which the operator
\begin{equation}
   L_\beta(\vec{x},\vec{y}) = {\rm e}^{-\beta V(\vec{x},\vec{y})}
\label{eq:lbeta}
\end{equation}
is positive semidefinite. The latter is equivalent, for $V = V(\vec{x}
- \vec{y})$ and $L_\beta = L_\beta (\vec{x} - \vec{y})$, to a positive
semidefinite Fourier transform of $L_\beta$. A sufficient condition
for this is that $V(\vecr_i, \vecr_j) = V(\vecr_i - \vecr_j) = V(r)$
satisfies 
\begin{mathletters}
\label{eq:vcond}
   \begin{equation}
      V'(r) \geq 0 \qquad \mbox{(monotonically increasing)}
   \label{eq:vcond1}
   \end{equation}
   \begin{equation}
      V''(r) \leq 0  \qquad \mbox{(convex)}
   \label{eq:vcond2}
   \end{equation}
and
   \begin{equation}
      V'''(r) \geq 0.
   \end{equation}
\end{mathletters}
Interestingly for the case of heavy quarks (where potentials can be
derived via the Wilson loop construction), Eqs.~(\ref{eq:vcond1}) and
(\ref{eq:vcond2}) can indeed be proven, as we will show in
Sec.~\ref{sec:qqbar}.  Indeed all potentials used in quark model
phenomenology to date satisfy all of Eqs.~(\ref{eq:vcond}) and hence
(\ref{eq:lbeta}).

We present in App.~\ref{app:lieb2} Lieb's proof of Eq.~(\ref{eq:emmm})
for positive semidefinite $\exp\left[-\beta V_{ij}(\vec{x},
  \vec{y})\right]$. We do this in some detail, even elaborating
somewhat beyond the original concise paper, since the baryon-baryon
inequalities are indeed born out by data.  Also, Lieb's line of
argument invoking the full three-particle Green's function serves as a
``bridge'' between the Hamiltonian, variational, largely potential
model motivated, first part of our review; and the more formal
Lagrangian correlator inequalities proved via the path integral
representation which is the approach of the second part.

%
\section{Relating masses in different flavor sectors}
\label{sec:interflavor}

In the following we investigate the interflavor mass relations
(\ref{eq:mijineq}) in a lattice Hamiltonian formulation of QCD.

The conservation of quark flavors allows us to break the QCD Hilbert
space into flavor sectors. Each flavor sector $(U,D,S,C,\ldots)$
consists of a net number $U$ of up $(u)$ quarks ($U$ is negative for
$\bar{u}$ excess), $D$ of down $(d)$ quarks {\em etc.} with $U + D
+ \ldots = 0$ (mod 3). We will be interested in the following ``low''
sectors:
\begin{enumerate}
\item $M^{(0)}$: The flavor vacuum $U = D = \cdots = 0$. It consists
  of states with an arbitrary number of gluons and ($q_l \bar{q}_l$)
  pairs.
\item $M_{(i\bar{\jmath})}, \, i \neq j$: The meson sector, with an
  excess of one quark flavor of type $i$ and a different antiquark
  flavor $\bar{\jmath}$ together with any number of gluons and ($q_l
  \bar{q}_l$) pairs.
\item $B_{(ijk)}$: The baryon sector with a net excess of three
  quarks, $q_i, q_j,$ and $q_k$.
\item $M_{(i\bar{\jmath}k\bar{l})} (i \neq j, i \neq l, k \neq j, k
  \neq l)$: The exotic meson sector with a net excess of two quark
  flavors and two (different) antiquark flavors.
\end{enumerate}
We will also discuss in Sec.~\ref{sec:exotic} and
App.~\ref{app:fourfive} other sectors such as the pentaquark and
hybrid sectors.

We can now define $H_{i\bar{\jmath}}$ to be the QCD Hamiltonian
restricted to the sector $M_{i\bar{\jmath}}$; $H_{ijk}$ the
Hamiltonian restricted to $B_{ijk}$ {\em etc.} The meson spectrum
will be given by
\begin{equation}
   H_{i\bar{\jmath}} | \Psi_{i\bar{\jmath}} \rangle =
      m_{i\bar{\jmath}} | \Psi_{i\bar{\jmath}} \rangle 
\>,
\end{equation}
with $\Psi$ a wave functional belonging to the $M_{i\bar{\jmath}}$ sector.
Similar equations hold in other sectors.

We show in App.~\ref{app:eq42} that \cite{ref:nussprl52}
\begin{equation}
   H_{i\bar{\jmath}} + H_{k\bar{l}} = H_{i\bar{l}} + H_{k\bar{\jmath}}
\>.
\label{eq:hflavors}
\end{equation}
Particle masses are given by the Schr\"{o}dinger equation, {\em e.g.}
\begin{equation}
   H_{ij} | \Psi_{ij} \rangle = (m_{ij} + \Delta_0)
      | \Psi_{ij} \rangle  
\>,
\label{eq:schmass}
\end{equation}
with $\Delta_0$ an additive, common constant representing the vacuum
energy, {\em i.e.} the lowest energy obtainable for functionals in
$M^{(0)}$.  To ensure that $\Delta_0$ is finite we restrict ourselves
to finite lattices so that $\Delta_0 = V \epsilon_{\text{vac}}^{(0)}$,
with $V$ the volume and $\epsilon_{\text{vac}}^{(0)}$ the vacuum
energy density. In writing Eq.~(\ref{eq:schmass}) we assumed that the
systems have no net translational motion.\footnote{To ensure $\vec{P}
  = 0$, {\em i.e.}  translational invariance, we need to sum over all
  locations of the centroid of the wave functional. For finite
  lattices of size $L$, only $P \leq 1/L$ seems achievable, yet for
  periodic boundary conditions a discrete version of translational
  invariance persists.}  Additional symmetries (rotations, parity, and
for certain states, charge conjugations) could be used to project any
desired state of given $J^{PC}$ quantum numbers. To simplify notation,
the $J^{PC}$ labels are omitted. The lattice formulation reduces the
symmetry from the full rotation to the cubic subgroup.  However, the
operator relations most likely also hold in the continuum limit where
the full rotation symmetry is regained.

The $|\Psi_{ij} \rangle$ are wave functionals:
\begin{equation}
   |\Psi_{ij} \rangle = \sum_{\text{configurations}} A^{ij}_
      {(\text{conf})} | (\text{conf}) \rangle
\>.
\end{equation}
Even for a finite lattice the (discrete) summation includes infinitely
many configurations characterized by the locations and spinors of all
quarks and/or antiquarks and by $E^2_{\vn, \vn + \nhat}$ at all links
where the latter are constrained by Gauss' law [Eq.~(\ref{eq:gauss})]
(recall that arbitrarily high SU(3)$_C$ representations are {\em a
  priori} allowed).  The specific flavor $(ij)$ dependence enters only
into the probability amplitudes $A^{ij}_{({\rm conf})}$ for finding a
given configuration in $| \Psi_{ij} \rangle$. The generic
configurations used in App.~\ref{app:eq42} have been defined in such a
way so as to make operator relations such as Eq.~(\ref{eq:hflavors})
manifestly true. $| \Psi_{ij} \rangle$ is normalized:
\begin{equation}
   \langle \Psi_{ij} | \Psi_{ij} \rangle = \sum_{\text{configurations}} 
      |A^{ij}_{(\text{conf})}|^2 = 1
\>.
\end{equation}

A key observation is that the variational principle applies to wave
functional solutions of the Schr\"{o}dinger equation
(\ref{eq:schmass}) much in the same way as it does to wave functions.
In particular, the ground state $| \Psi_{ij}^{(0)} \rangle$ in any
specific channel (say a mesonic $ij$ channel with given $J^{PC}$) is
given by the requirement that:
\begin{equation}
   \Delta_0 + m_{ij} =  \langle \Psi_{ij}^{(0)} |
      H_{ij} | \Psi_{ij}^{(0)} \rangle = {\rm min}
      \langle \Psi | H_{ij} | \Psi \rangle
\>,
\end{equation}
with the minimum sought in the space of all (normalized) $| \Psi_{ij}
\rangle$ functionals with the given quantum numbers:
\begin{equation}
   | \Psi \rangle = \sum_{\text{conf}} A_{\text{conf}} 
      | (\text{conf (mesonic)}) \rangle
\>.
\end{equation}

We would like next to argue that we can set $i = l, j = k$ in
Eq.~(\ref{eq:hflavors}) and still obtain a meaningful operator relation
\begin{equation}
   H_{i\bar{\jmath}} + H_{j\bar{\imath}} 
      = H_{i\bar{\imath}} + H_{j\bar{\jmath}}   
\>.
\label{eq:hij}
\end{equation}

In the light $(ud)$ quark sector $|m_u - m_d| \ll
\Lambda_{\text{QCD}}$, and since $\alpha_{\text{EM}} \ll
\alpha_{\text{QCD}}$, isospin is a good flavor symmetry.  The $I = 1
\, \, |u\bar{u} - d\bar{d} + \mbox{gluons} + \mbox{pairs} \rangle$
states do not then mix with the $I = 0, M^{(0)}$ sector, so these
states should then be used in deriving the inequalities.  For heavier
states $s\bar{s}, c\bar{c}, b\bar{b}$ there is a strong ``Zweig rule''
\cite{ref:lip73,ref:zweig,ref:lipalex} suppression of $s\bar{s},
c\bar{c}, b\bar{b}$ annihilation. Hence mesonic sectors
$M_{i\bar{\imath}}$ distinct from $M^{(0)}$ can be defined for which
the $i = l, j = k$ version of Eq.~(\ref{eq:hflavors}), namely
Eq.~(\ref{eq:hij}), holds.

Unfortunately, we {\em cannot} use this relation to obtain mass
inequalities without further assumptions. In general,
$\Psi_{i\bar{\jmath}}^{(0)}$, the ground state wave functional for
(for example) a heavy quark $q_i$ and a light antiquark $\bar{q}_j$,
may be {\em different} from $\Psi_{j\bar{\imath}}^{(0)}$, even though
$\Psi_{i\bar{\jmath}}$ and $\Psi_{j\bar{\imath}}$ are related by
charge conjugation.\footnote{This important point (missed in
  Ref.~\cite{ref:nussprl52}) will be further elaborated in the Summary
  section.} Indeed one of the inequalities, namely $2m_{K^{\ast}}\leq
m_\phi + m_\rho$, is not manifest. Symmetry is automatically
guaranteed in the simple potential model where
$\Psi_{i\bar{\jmath}}^{(0)}$ depends only on $\vecr = \vecr_i -
\vecr_j$, the relative coordinate of $q_i \bar{q}_j$.  It is also
plausible in a large $N_c$ limit where the important degrees of
freedom are the gluonic ones. In the following we will assume $i
\leftrightarrow j$ symmetry.

By taking the expectation value of the left hand side of
Eq.~(\ref{eq:hij}) in $ \Psi_{i\bar{\jmath}}^{(0)}$, we have $\langle
\Psi_{i\bar{\jmath}}^{(0)} | H_{i\bar{\jmath}} + H_{j\bar{\imath}} |
\Psi_{i\bar{\jmath}}^{(0)} \rangle = 2 m_{i\bar{\jmath}}^{(0)} + 2
\Delta_0$, and the variational argument for deriving the inequalities
proceeds exactly as in Sec.~\ref{sec:mmineq}. After cancelling a
common $2 \Delta_0$ term representing vacuum energy we obtain
\begin{equation}
   2 m_{i\bar{\jmath}}^{(0)} \geq m_{i\bar{\imath}}^{(0)} 
      + m_{j\bar{\jmath}}^{(0)} 
\>.
\end{equation}
We also have the relations for the sum of the first $n$ excited states
in any $M^{ij} (J^{PC})$ channel [Eq.~(\ref{eq:mijineq})].

For the baryonic sector we readily find a relation
\begin{equation}
   H_{ijr} + H_{klr} = H_{ilr} + H_{jkr}
\>.
\label{eq:hbaryonflavors}
\end{equation}
Eq.~(\ref{eq:hbaryonflavors}) is analagous to relation
(\ref{eq:hflavors}) and is obtained by adding a common ``spectator''
quark $r \neq i, j, k, l$.

Assuming a flavor symmetric baryon ground state, we obtain from
variants of Eq.~(\ref{eq:hbaryonflavors}) with some of the flavor
indices set equal to each other, analog convexity relations for
baryonic states
\begin{mathletters}
\label{eq:baryonmasses}
   \begin{equation}
      m_{iij}^{(0)} \geq \frac{1}{2} \left[ m_{iii}^{(0)} +  
      m_{ijj}^{(0)}\right]
   \label{eq:miij}
   \end{equation}
   \begin{equation}
      m_{ijk}^{(0)} \geq \frac{1}{6} \left[ m_{iij}^{(0)} +  
      m_{iik}^{(0)} + m_{jji}^{(0)} + m_{jjk}^{(0)} + 
      m_{kki}^{(0)} + m_{kkj}^{(0)} \right].
   \label{eq:mijk}
   \end{equation}
\end{mathletters}
It should be emphasized \cite{ref:richprl} that for the baryonic case
we have two relative CMS coordinates, and flavor symmetry of the wave
function is not guaranteed even in the framework of a potential model.
Indeed, as discussed in Sec.~\ref{sec:bbineq} and
App.~\ref{app:lieb1}, for certain extreme type of potentials one can
show that the inequalities such as Eq.~(\ref{eq:hij}) are violated --
though as elaborated in App.~\ref{app:lieb2}, they do hold
\cite{ref:lieb} for the class of potentials which are of interest in
QCD.

%
\section{Baryon-meson inequalities in a non-perturbative approach}
\label{sec:nonpertmbineq}

Inequalities relating masses of baryons and mesons were motivated in
Sec.~\ref{sec:mbineq} via a potential model. It was suggested there
that such inequalities may persist beyond the (dressed) one gluon
exchange approximation. In this section we will employ a
nonperturbative framework which still allows the application of
variational arguments and the derivation of the baryon-meson mass
inequalities \cite{ref:nusssath}.

It has been argued that confinement in QCD is the electric analog of
the Meissner effect in a superconductor. The nonperturbative QCD
vacuum develops a condensate of color monopole pairs and/or of large
loops of magnetic flux \cite{ref:thooftnpb,ref:man79,ref:aharonov}.
More recently, lattice and other approaches have made this much more
concrete, particularly in the context of ``center vortices''
\cite{ref:tomb,ref:cornwall}.  In this vacuum, the chromoelectric flux
emitted from a quark and ending on an antiquark is localized along a
thin tube \cite{ref:greensite} -- the analog of the magnetic vortex in
the ordinary superconductor.  In the $N \rightarrow \infty$ limit the
chromoelectric flux tube may become infinitely thin
\cite{ref:polyakov} and the original dual string model
\cite{ref:nambu,ref:sus70,ref:suss77} for hadrons could emerge.  Also
in the strong coupling limit of Hamiltonian lattice QCD a single set
of minimally excited links of total minimum length connects the quarks
and antiquarks in a single hadron.

A more general approximation for the meson wave functional is a sum
over configurations of strings [or chains of lattice links, see
Fig.~\ref{fig:color}(a)] connecting $q$ and $\bar{q}$ (indicated by the
symbol $\leadsto$ below), with a probability amplitude for each
configuration:
\begin{equation}
   | \psi_{M12} \rangle = \sum_{1 \leadsto 2} 
      A(\leadsto) | \leadsto \rangle
\>.
\end{equation}
The end points 1,2 can also vary. The kinetic ($B^2$ and $\psibar D
\psi$) parts of the QCD Hamiltonian move the string and the fermions
at the end points, respectively; while the ``potential'' $E^2$ term
gives a diagonal contribution proportional to the length (number of
links) of the string multiplied by the string tension (weighted by the
values of the second Casimir operator for the representation residing
on each link).

In the same approximation, the baryon's wave functional is a sum over
configurations where the three $E$ vortex lines (symbolized for
convenience by $Y$ in the following equation, though the vortices need
not be straight) emanate from quarks 1, 2, and 3, and join together at
a common ``junction point'' $x$ [see Fig.~\ref{fig:color}(b)]:
\begin{equation}
   | \psi_{B123} \rangle = \sum_{Y} A(Y) | Y \rangle   
\>.
\end{equation}

Just as in the case of the nonrelativistic potential model, we would
like to extract trial wave functionals for the ground states of the
mesons $q_1 \bar{q}_2$, $q_2 \bar{q}_3$, and $q_3 \bar{q}_1$ from the
ground state baryon wave functional. This is achieved in the following
way (see Fig.~\ref{fig:bmtrial}). We consider quarks 1 and 2, together
with the string connecting them in the baryon as a possible
configuration in the wave functional of the meson $M_{1\bar{2}}$. To
this end we need to color conjugate the quark $\bar{q}_2$ and reverse
the chromoelectric flux in the section $x-2$.  A detailed analysis
using a lattice formulation indicates this can be done for the case of
minimal flux strings, without effecting the matrix elements of the
Hamiltonian. Likewise quarks 2 and 3 with the string $2-x-3$ are
viewed as a possible configuration for the meson $M_{2\bar{3}}$ and
similarly for $M_{3\bar{1}}$.

This suggests the operator relation
\begin{equation}
   2 H_{ijk} = H_{i\bar{\jmath}} + H_{j\bar{k}} + H_{k\bar{\imath}}
\>,
\label{eq:hbaryonic}
\end{equation}
with $ijk$ the flavors of $q_1 q_2 q_3$. Indeed by applying $
H_{i\bar{\jmath}}, H_{j\bar{k}}$ and $H_{k\bar{\imath}}$ to each of
the above mesonic configurations we see that each part of the baryonic
Hamiltonian $H_{ijk}$ is encountered twice. Thus the kinetic term and
mass term $(\dslasha_1 + m_1)$ associated with the motion of $q_1 (=
q_i)$ occurs in both $H_{i\bar{\jmath}}$ and $H_{k\bar{\imath}}$, and
so does the kinetic and potential energy associated with the motion,
and total length, of the string bit $\leadsto$ connecting quark $q_i$
with the junction point $x$. A similar argument can be applied
to the $2-x$ and $3-x$ string bits.

Let us next take the expectation value of Eq.~(\ref{eq:hbaryonic}) in the
ground state wave functional of the baryon at rest. As in
Sec.~\ref{sec:mbineq}, we obtain one one hand simply $2m_{ijk}^{(0)}$ and
on the other hand the sum of the expectation values of $ H_{i\bar{\jmath}},
H_{j\bar{k}}$ and $H_{k\bar{\imath}}$ in the mesonic wave functionals
extracted, in the manner described above, from the baryon's ground state
wave functional. In analogy to fixing the coordinates of the ``third
quark'' (say $q_k$) in the potential model discussion of
Sec.~\ref{sec:mbineq}, we fix in each case the string bit ($x-q_k$ for
$H_{i\bar{\jmath}}$) which should be ignored in order to form a mesonic
$(M_{i\bar{\jmath}})$ trial wave functional. The expectation values of
$H_{i\bar{\jmath}}$ in these trial wave functionals
\begin{equation}
   | \psi^t_{ijk} \rangle = \sum_{Y} A(Y) | Y \rangle   
\label{eq:psitrial}
\end{equation}
are then integrated with the weight implied by the original baryonic
amplitude $A(Y)$ over the ``ignored'' section as well.

Note that because of the special character of the baryonic state, the
configurations in the trial wave functional (\ref{eq:psitrial}), used
here for the three mesonic $q_i \bar{q}_j, q_j \bar{q}_k, q_k
\bar{q}_i$ ground states, are in fact correlated. Specifically, the
strings (or flux lines) associated with all three of these meson
states are forced to have a common junction point $x$ (which again is
integrated over at the end).

This extra constraint only reinforces the general result of the
variational principle, namely that
\begin{mathletters}
   \begin{equation}
      \langle \psi_{ijk}^{(0)} | H_{i\bar{\jmath}} | \psi_{ijk}^{(0)}
      \rangle \geq m_{i\bar{\jmath}}^{(0)}
   \end{equation} 
   \begin{equation}
      \langle \psi_{ijk}^{(0)} | H_{j\bar{k}} | \psi_{ijk}^{(0)}
      \rangle \geq m_{j\bar{k}}^{(0)}
   \end{equation} 
   \begin{equation}
      \langle \psi_{ijk}^{(0)} | H_{k\bar{\imath}} | \psi_{ijk}^{(0)}
      \rangle \geq m_{k\bar{\imath}}^{(0)}.
   \end{equation} 
\end{mathletters}
Specifically these inequalities state that the mesonic trial wave
functionals extracted from the subsystems of the baryon's ground state
wave functional are not optimized so as to minimize the energy of the
mesonic subsystems. Rather $\psi_{ijk}^{(0)}$ is constructed so as to
minimize the expectation value of $H_{ijk}$ and
\begin{equation}
      \langle \psi_{ijk}^{(0)} | H_{ijk} | \psi_{ijk}^{(0)}
      \rangle = m_{ijk}^{(0)}
\>.
\end{equation} 
Thus we have from Eq.~(\ref{eq:hbaryonic}) the required inequality
\begin{equation}
   2 m_{ijk}^{(0)} \geq m_{i\bar{\jmath}}^{(0)} + 
      m_{j\bar{k}}^{(0)} + m_{k\bar{\imath}}^{(0)}
\>.
\end{equation}

It is amusing to see how this inequality is realized for the strong
coupling limit. The flux then proceeds in straight lines in both the
mesons and the baryons so as to minimize the total string length. This
yields the potentials $V_{12} = \sigma |\vecr_1 - \vecr_2 |, V_{23} =
\sigma |\vecr_2 - \vecr_3 |, V_{31} = \sigma |\vecr_3 - \vecr_1 |$ for
the meson and the genuine three-body interaction \cite{ref:horgan}
\begin{displaymath}
   V_{123} = {\rm min}_x \sigma ( | \vecr_1 - \vec{x} | + 
      | \vecr_2 - \vec{x} | +  | \vecr_3 - \vec{x} | )
\end{displaymath}
(with the total length of the Y-shaped string configuration minimized
over the choice of function point $\vec{x}$) for the baryon. The
inequality reduces in this case simply to $2 V_{123} \geq V_{12} +
V_{23} + V_{31}$ which is just a sum of three triangular inequalities:
$ | \vecr_1 - \vec{x} | + | \vecr_2 - \vec{x} | \geq | \vecr_1 -
\vecr_2 | $ {\em etc.} (see Fig.~\ref{fig:triangle}). Since the
potentials are linear for both systems, the virial theorem implies
that the total energy is given by $2 \langle V \rangle$, with $\langle
V \rangle$ the expectation value of the potential energy, and $2
E_{123} \geq E_{12} + E_{23} + E_{31}$.\footnote{The virial theorem
  (with massless quarks and linear potentials) was used in
  \cite{ref:nussplb180} to motivate the equipartition of the light
  cone momenta between quarks and gluons, and more recently in
  connection with the mass dependence of Bose-Einstein correlations in
  multi-particle final states \cite{ref:acl}.}

We have so far considered the approximation in which the quarks are
connected by a single set of minimally excited links on the lattice or
by a single Y-type configuration for the baryons.  We still have many
paths of different lengths. All of these configurations can be
generated by repeated application of the kinetic $(B^2)$ term in the
Hamiltonian which generates a closed flux line around a plaquette and
shifts the initial flux line as indicated in Fig.~\ref{fig:lattice}.
The problem of finding the mesonic or baryonic ground state wave
functionals, even in this approximation, is intractable, and at best
we could hope for some numerical results. We would like to point out,
however, that the baryon-meson inequalities can be proven in an even
broader context when $\qqb$ pair creation, bifurcation of flux lines,
and excitation of links to higher SU(3)$_C$ representations are all
allowed.

If we consider the procedure for extracting mesonic trial wave
functionals from the baryonic wave functional, we can pinpoint the
crucial ingredient for deriving Eq.~(\ref{eq:hbaryonic}). It is that
the network of links in any configuration in the baryonic wave
functional includes only one junction point $x$, where we can separate
the network into three ``patches'' $P_1, P_2,$ and $P_3$ connected to
the external quarks $q_i, q_j,$ and $q_k$ respectively (see
Fig.~\ref{fig:bmsep}). We consider $P_1 \bar{P}_2$, where $\bar{P}_2$
is the patch with all flux flows reversed and $q_j \rightarrow
\bar{q}_j$, as a configuration in the trial meson functional
$|\psi_{Mi\bar{\jmath}} \rangle$ with an amplitude $A(P_1, P_2, P_3),
(P_3$ fixed) inferred from the baryonic wave functional. Likewise
$P_2$, $\bar{P}_3$ and $P_3$, $\bar{P}_1$ offer trial wave functionals
for the $q_2 \bar{q}_3 (q_j \bar{q}_k)$ and $q_3 \bar{q}_1 (q_k
\bar{q}_i)$ mesonic systems. We can easily verify that the operator
relation $2 H_{123} = H_{1\bar{2}} + H_{2\bar{3}} + H_{3\bar{1}}$
still holds for this class of baryonic and corresponding mesonic
functionals. The kinetic and potential parts of the full QCD
Hamiltonian operate on each patch $P_1, P_2, P_3$ separately and those
contributions are counted twice in $H_{1\bar{2}} + H_{2\bar{3}} +
H_{3\bar{1}}$.

The variational argument is applicable in the larger class of baryonic
and mesonic trial wave functionals and it yields the required
baryon-meson mass inequalities $2 m_{ijk} \geq m_{i\bar{\jmath}} +
m_{j\bar{k}} + m_{k\bar{\imath}}$.

The fact that the same inequality, Eq.~(\ref{eq:mbmassineq}),
motivated by the color factor for the perturbative one-gluon exchange,
can be rederived in a strong coupling, string-like limit (and some
generalizations thereof), suggests that these inequalities may indeed
be a true consequence of the full-fledged QCD theory. Indeed a
particular relation involving the nucleon and pion masses has been
rigorously proved by Weingarten \cite{ref:weingarten}, who utilized
the correlation function techniques, and is reproduced in the second
half of this report.\footnote{Despite many efforts
  \cite{ref:nussunpub,ref:iwao} to extend the correlation function
  technique to more detailed, flavor-dependent baryon-meson and
  baryon-baryon inequalities, no fully convincing results have been
  obtained.}

Finally, we note that a construction of mesonic trial wave functionals
from the baryon's ground state functional $\psi_{(i_1 \ldots i_N)}$
can be carried out for SU(N) with general N as well. The configuration
contributing to $\psi_{(i_1 \ldots i_N)}$ consists of N strings joined
at a common junction point. It can be formally separated into N(N-1)/2
mesonic subsystems with each part of the baryonic Hamiltonian counted
(N-1) times (since a segment or patch $P_{ia}$ connecting the quark
$q_{1_a}$ and the junction $x$ appear in the (N-1) $q_{i_a}
\bar{q}_{i_b}, a=b$ systems), and we have:
\begin{displaymath}
   (N-1) H_{i_1 \ldots i_N} = \sum_{i_a = i_b} H_{i_a i_{\bar{b}}}
\>,
\end{displaymath}
leading to the baryon-meson inequality (\ref{eq:sunmbineq}) for
general SU(N).

The case of $N_c = 2$ is rather special. The lightest diquark baryon
is in fact a $0^{++}$ mesonic state. The inequality $m_B \geq m_\pi$
is in this case most likely an equality:
\begin{equation}
   m_{\qqb}^{(0^+)} = m_{\qqb}^{(0^-)}
\>.
\label{eq:mqqb}
\end{equation}

Indeed the gluon couplings inside the meson and ``baryon'' are the
same here, so that to all orders in perturbation theory we expect the
S-wave $qq$ (or $\qqb$) states to be degenerate. To obey the
generalized Pauli principle, the S-wave spin and color singlet $\qqb$
state ought to be a ``flavor'' antisymmetric $ud$ combination which is
to be compared with the $u\gamma_5 \bar{d}$ pion. The pseudoreality of
the SU(2) group implies that in the nonperturbative string or flux
tube picture, the flux emitted by one quark in the bosonic diquark can
readily end on the other quark. There is no junction point in this
case, the wave functionals and Hamiltonians for the $0^-$
$q\gamma_5 \bar{q}$ and $0^+$ $qq$ systems are identical, and
Eq.~(\ref{eq:mqqb}) follows.

%
\section{Comparison of the inequalities with hadronic masses}
\label{sec:compdata}

We proceed next to list the various inequalities and compare them with
available particle data \cite{ref:pdg}.

First consider the meson-meson relations. As emphasized above, these
inequalities do rely on an {\em additional} assumption of a
flavor-symmetric wave function. In all testable cases (with one
possible exception) these inequalities are satisfied.  The
inequalities could then be used as a rule of thumb to restrict the
masses of as yet undiscovered new particles.

There are six relevant $m_{i\bar{\jmath}}, m_i^{0} \neq m_j^{0}$
flavor combinations: $u\bar{s}, u\bar{c}, s\bar{c}, s\bar{b},
u\bar{b}$, and $c\bar{b}$.\footnote{Since the decay width of the top
  quark exceeds a few GeV, it decays before $t\bar{u}$, {\em etc.}
  states can form.} Because of the small violation of $I$-spin
symmetry (via radiative electromagnetic corrections and the effect of
$|m_u - m_d| \simeq $ 4-5 MeV) we have not separately considered
members of an $I$-spin multiplet (obtained by $u \rightarrow d$
substitutions), but rather averaged over these states. Also in order
to minimize the effects of the annihilation channels we consistently
choose the I = 1 $u\bar{d}$ rather than the I = 0 $u\bar{u}$ states
({\em e.g.} $\rho$ and not $\omega$, $a_2$ and not $f_0$ {\em etc.}).

The comparison with measured masses is summarized in Table
\ref{table:one}, which indicates the specific particles and masses
relevant for each inequality. In some cases all three masses in
Eq.~(\ref{eq:mijineq}) are fairly well known or reliably estimated. In
other cases lower bounds are predicted for certain
$M_{i\bar{\jmath}}^{(0)J^{PC}}$. All masses are listed in MeV.

For the pseudoscalars the inequalities have been derived by Witten
\cite{ref:witt83} using the euclidean Green's function method to be
discussed at length in the second part of this review, and indeed hold
with a fairly wide margin.  The light pseudoscalar mesons $\pi, K,
\eta$ can also be viewed, to a good approximation, as pseudo-Goldstone
bosons. Current algebra methods \cite{ref:adler} then lead to
$m^2_{ps_{ij}} \propto (m^{(0)}_{q_i} + m^{(0)}_{q_j})$. Hence in this
approximation $2 m^2_{ps_{ij}} = m^2_{ps_{ii}} + m^2_{ps_{jj}}$ and $2
m_{ps_{ij}} > m_{ps_{ii}} + m_{ps_{jj}}$ is guaranteed.\footnote{The
  last argument is reminiscent of that made earlier for harmonic
  potentials. It was noticed \cite{ref:rosner} that the ``dual''
  Lovelace-Shapiro-Veneziano formula \cite{ref:lsv} for $\pi-\pi$
  scattering has a remarkable tendency to conform to soft pion
  theorems. Considering the harmonic string origin of ``dual''
  amplitudes, these may perhaps be related features.}

For the $I = 0$ pseudoscalars the strong coupling to the gluonic $M^0$
channel (which in particular accounts for the massive $\eta'$)
\cite{ref:thooftprep,ref:wittven} suggests strong mixing between the
two $I = 0$ states that are made of light quarks, $\frac{1}{\sqrt{2}}
(u \bar{u} + d \bar{d})$, and the $s\bar{s}$ state.  Hence $|s\bar{s}
\rangle = \alpha | \eta \rangle + \beta | \eta' \rangle$. The
reasonable expectation \cite{ref:gellmann} that the eigenstates $\eta,
\eta'$ are the SU(3) flavor octet $\frac{1}{\sqrt{6}} (u\bar{u} + d
\bar{d} - 2 s\bar{s})$ and singlet $\frac{1}{\sqrt{3}} (u \bar{u} + d
\bar{d} + s\bar{s})$ implies $\alpha = \sqrt{2}/3, \beta = \sqrt{1}/3$
and
\begin{displaymath}
   m_{s\bar{s}}^{(0^-)} = \left( \alpha \langle \eta |
      + \beta \langle \eta' | \right) H \left( \alpha | \eta \rangle 
      + \beta | \eta' \rangle \right) = \alpha^2 m_\eta 
      + \beta^2 m_{\eta'} \ldots
\end{displaymath}
and yields $ m_{s\bar{s}}^{(0^-)} \approx$ 706 MeV. This indeed
satisfies $m_{s\bar{s}}^{(0^-)} + m_{u\bar{d}}^{(0^-)} \leq 2
m_{s\bar{d}}^{(0^-)}$ [{\em i.e.}  $m_{s\bar{s}}^{(0^-)} + m_\pi \leq 2
m_K$] with a large margin. In general $2 m_K > m_\pi +
m_{s\bar{s}}^{(0^-)}$ holds as long as $\beta^2 / \alpha^2 < 1.5$, a
constraint satisfied by all mixing schemes suggested for the $0^-$
mesons \cite{ref:gilman}.

In the relation $ 2m_{u\bar{b}}^{(0^-)} \geq m_{u\bar{u}} +
m_{b\bar{b}}$ we use the mass of the vectorial state, $m_\Upsilon$,
instead of the as yet unknown $m_{\eta_b}$. Following Weingarten we
show in Sec.~\ref{sec:quarkbi} that, when mixings with $M^0$ are
neglected (which seems justified by asymptotic freedom in the case of
heavy quarks), the lowest state in any $M_{i\bar{\jmath}}$ sector is
indeed pseudoscalar and $m_\Upsilon \geq m_{\eta_b}$, so that the
original inequality is {\em a fortiori} satisfied.

The vector meson mass inequalities hold with smaller margins than the
corresponding inequalities for the pseudoscalars. This reflects the
hyperfine splittings (as indicated in the conclusion of
Sec.~\ref{sec:mmineq}), which tend to weaken (enhance) the inequality
for the vector (pseudoscalar) mesons. We find it impressive that
despite the very large spin splittings, {\em e.g.} $m_\rho - m_\pi
\simeq 650 {\rm MeV} \simeq 5 m_\pi$, the inequality may still
marginally hold for the vector mesons. For the specific case of
$K^{\ast}$, $\rho$, and $\phi$, the inequality seems to fail, although
only within ${\cal O}(1\%)$ of the widths of the states considered.

The inequalities for the tensor mesons can be tested in the sectors
not involving the $b$ quark. In the $b$ quark sectors we can use the
inequalities to predict lower bounds for the masses of the $B_{{\rm
    tensor}}$ particles. The $D_{{s2^+}}$ is not known for certain to
be a $2^{++}$ state, but this is the quark model prediction. The
inequalities for the axial $1^{++}$ mesons all hold at the level of a
few percent. Finally, the inequalities for scalars can be tested (and
verified) only for the case of the $u\bar{c}$ and $u\bar{b}$ flavor
combinations.\footnote{A. Falk \cite{ref:falk} tried using heavy quark
  symmetries to fix some of the unknown $J^P$. The QCD inequalities
  nicely complement this program.}

The deviations from equality
\begin{equation}
   \delta = \frac{\delta m}{m} = \frac{m_{ij} - (1/2) (m_{ii} + m_{jj})}
      {m_{ij}}
\label{eq:deviation}
\end{equation}
are often small [${\cal O}(1\% - 2\%)$]. This may be traced to the
variational principle. Since $\langle \psi| H_{ii} | \psi \rangle,
\langle \psi| H_{jj} | \psi \rangle$ are extremized at $\psi =
\psi_{ii}^{(0)}, \psi = \psi_{jj}^{(0)}$, respectively, the deviations
$\Delta_i = \langle \psi_{ij}^{(0)}| H_{ii} | \psi_{ij}^{(0)} \rangle
- \langle \psi_{ii}^{(0)}| H_{ii} | \psi_{ii}^{(0)} \rangle$ and
$\Delta_j = \langle \psi_{ij}^{(0)}| H_{jj} | \psi_{ij}^{(0)} \rangle
- \langle \psi_{jj}^{(0)}| H_{jj} | \psi_{jj}^{(0)} \rangle$ are
quadratic in the wave function shifts, {\em e.g.}
\begin{equation}
   \delta m \simeq (\delta\psi)^2 \approx 
      ( \psi_{ij}^{(0)} - \psi_{ii}^{(0)})^2
\>.
\end{equation}
The $\delta\psi$ in turn are expected, on the basis of first order
perturbative estimates, to be
\begin{equation} 
   \delta \psi \simeq \frac{\delta H \psi^{(0)}}{\Delta m}
      \simeq \frac{(H_{ij} - H_{ii})}{\Delta m} \psi^{(0)}
      = \frac{\delta H}{\Delta m} \psi^{(0)} 
\>,
\label{eq:deltapsi}
\end{equation}
with $\Delta m \simeq$ 1 GeV the typical splitting $m_{ij}^{(0)} -
m_{ij}^{(1)}$ between the ground and first excited states in the specific
channel considered. $\delta H$ is solely due to the quark mass differences.
Taking $i = u, j = s$ we have for the simple model of
Sec.~\ref{sec:mmineq}: 
\begin{equation}
   \delta H \simeq \sqrt{m_s^2 + \langle p^2 \rangle}
      -  \sqrt{m_u^2 + \langle p^2 \rangle}
\>,
\label{eq:deltah}
\end{equation}
with $\langle p^2 \rangle$ the average (momentum)$^2$ in the wave
functions. If we then use the bare quark masses $m_u^0 \simeq 0, m_s^0
\simeq 150$ MeV and $\langle p^2 \rangle \simeq (300 {\rm MeV})^2$
then $|\delta H| \simeq \frac{m_s^2}{2\sqrt{\langle p^2 \rangle}}
\simeq 50$ MeV, and Eqs.~(\ref{eq:deviation}) -- (\ref{eq:deltah})
yield $\delta \simeq 1 \%$.

For heavy-light quark combinations the difference in the kinetic parts
of the Hamiltonian are larger. However, most of the masses are given
in this case by the heavy quark mass itself, so that fractional
deviations again remain small.

Approximate mass equalities $2 m_{ii} = m_{ii} + m_{ij}$ or
$2 m_{ij}^2 = m_{ii}^2 + m_{ij}^2$ were suggested quite a while
back on the basis of the SU(3) (Gell-Mann -- Ne'eman) flavor symmetry
\cite{ref:gellmann}.  The underlying assumption was that the flavor
symmetry breaking part of the Hamiltonian behaves like $H^8$, the
neutral isoscalar member of an octet (a feature which is manifestly
true for the QCD Hamiltonian), and that mass splittings can be
obtained by using $\langle \phi^{(0)} | H^8 | \phi^{(0)} \rangle$ with
$\phi^{(0)}$ the SU(3) symmetric wave functions. The present
discussion suggests that these relations can be transformed into
inequalities for the linear mass combinations.

When the particles in question are broad resonances, the mass
inequalities need not apply. This is not just due to the technical
difficulty of precisely defining, for example, $m_\rho$ (different
methods based on the $\pi\pi$ mass distributions or the Argand
diagrams for phase shifts yield somewhat different values for the
masses of resonances \cite{ref:pdg}). Strictly speaking, the
inequality $2 m_{u\bar{s}}^{(0)1^{-}} \geq m_{u\bar{u}}^{(0)1^{-}} +
m_{s\bar{s}}^{(0)1^{-}}$ holds only for the {\em lowest} $1^{-}$
states in the $M_{u\bar{s}}, M_{u\bar{u}}, M_{s\bar{s}}$ sectors which
are $K\pi, \pi\pi,$ and (neglecting Zweig rule violations, {\em i.e.}
mixing with the gluonic $M^{0}$ sector) $K\bar{K}$ P-wave states at
threshold. The inequality would then appear to degenerate into the
trivial statement on the kinematical threshold: $m_K + m_\pi \leq
(1/2) (2m_\pi + 2 m_K)$. As we will show in Sec.~\ref{sec:exotic}, the
operator relation enables us to go beyond this and deduce relations
for the phase shifts. To the extent that the threshold physics is
completely dominated by narrow $K^{\ast}, \rho,$ and $\phi$ resonances
the mass inequality $2m_{K^{\ast}} \geq m_\rho + m_\phi$ can be
regained from the phase shift inequalities.

In the large $N_c$ limit, resonance widths and Zweig rule violations
vanish like 1/$N_c$ and $1/N_c^2$ respectively \cite{ref:coleman} and
the distinction between the relation (\ref{eq:hflavors}) and the
approximate version (\ref{eq:hij}) is lost. Also the flavor
symmetry assumption may be on better footing as the gluonic degrees of
freedom dominate.

We proceed next to the baryon-meson inequalities, with the results
listed in Table \ref{table:two}. The specific flavor-spin combinations
appearing there are fixed in the manner explained in some detail in
App.~\ref{app:bmineq}.\footnote{The specification of the combinations
  of spins and flavors of the mesons appearing in these inequalities
  has used the notation and level assignments of the nonrelativistic
  quark model. This is, however, mainly done for convenience and does
  {\em not} detract from the degree of rigor of the derivation. Thus
  let us consider the comparison between the polarized $\Delta^{++}$
  and $\rho$ masses. Assume that the $\Delta^{++}$ ground state has
  important $L \neq 0$ components. Then we could have a $uu$ $S = 1$
  diquark with $L = 2$, or $S = 0$ with $L = 1, \ldots,$ and we would
  then simply need to construct, from the $\Delta^{++}$ wave
  functional, the corresponding trial wave functional for the lowest
  $J = 1$ $u\bar{d}$ state. The latter is still the $\rho$ -- no
  matter what orbital or other components it may have.}  At the
present the $J^P$ of $\Lambda_c^+$ and $\Xi_c$ are not established and
the choice made here ($J^P = 1/2^{+}$, rather than $3/2^{+}$) is
essential for the inequalities to be satisfied. In general the
baryon-meson inequalities are satisfied with a higher margin than the
meson-meson or baryon-baryon inequalities.  This could be attributed
to the weak basis for the interflavor inequalities whose derivation
required, beyond the rigorous operator relations, also an assumption
of predominantly flavor symmetric ground state wave functions (or
functionals). Also the mesonic wave functions for the different
flavors may indeed be very similar to each other [see
Eqs.~(\ref{eq:deviation}) -- (\ref{eq:deltah})], whereas the baryonic
wave functions are intrinsically different from the mesonic ones. In
the simple potential model picture the diquark subsystems are not at
rest. Also the Y-configuration in the baryon's functionals are
different from the $\leadsto$ configurations in the corresponding
mesonic functionals.  The baryonic ground state wave functions are
therefore a poorer approximation for the mesonic wave functions than
the wave functions of mesons with different flavors.

Baryon-baryon inequalities are listed in Table \ref{table:three}.
From Eqs.~(\ref{eq:baryonmasses}) we have also an inequality version
of the equal spacing for baryons in the decuplet (\ref{eq:miij}) and
of the Gell-Mann -- Okubo (GMO) mass formula for the baryon octet
\cite{ref:gellmann}. For the latter we use the inequality
(\ref{eq:mijk}): $m_{uds} \geq (1/6) (m_{\Sigma^{+}} + m_{\Sigma^{-}}
+ m_p + m_n + m_\Xi + m_{\Xi^0})$.  Identifying $uds$ with
$(1/\sqrt{2}) (\Lambda^0 + \Sigma^0)$, this becomes $3 m_{\Lambda^0} +
3 m_{\Sigma^0} \geq (m_{\Sigma^{+}} + m_{\Sigma^{-}} + m_p + m_n +
m_\Xi + m_{\Xi^0})$, which in the limit where $I$-spin splittings are
neglected becomes the GMO relation. While both the linearity in the
decuplet and the GMO formula are very accurate it is gratifying that
the small deviations are consistent with the inequalities. We should
emphasize however that the flavor symmetry of the three quark baryonic
ground state wave functions which underlies these inequalities is
strictly an additional assumption. In spin independent quark-quark
potentials $V(|\vecr|)$, the inequalities hold for the fairly large
class of potentials discussed in App.~\ref{app:lieb2}.

We have not listed most of the many mass relations involving radially
excited states, Eq.~(\ref{eq:sumeng}).  While we have several known
radially excited states in the heavy quark $Q\bar{Q} = c\bar{c},
b\bar{b}$ systems (particularly in the $1^{--}$ channel), there are
very few known radial excitations in the $Q\bar{q}$ or $\qqb$ systems
to allow useful comparisons. An exception is the case of radial
excitations in the $D \, (c\bar{q})$ system, for which the continuous
experimental effort at Fermilab \cite{ref:fermilab} keeps providing a
relatively elaborate charmed meson spectrum. We have therefore
indicated the relevant charm radial excitations in
Table~\ref{table:four}.

\begin{table}
\caption{Meson-meson Inequalities}
   \begin{tabular}{ccrcl} 
   \multicolumn{5}{c}{$u\bar{s}$ sector: $m_{u\bar{s}} \geq 
      \frac{1}{2} \left( m_{u\bar{u}} + m_{s\bar{s}}\right)$}
      \\ \hline
   pseudoscalar & $m_K \geq \frac{1}{2} \left(m_\pi + \frac{2}{3}m_\eta
      + \frac{1}{3}m_{\eta'} \right)$ & 495.009 & $\geq$ & 411.08 \\
   vector & $m_{K^{\ast}} \geq \frac{1}{2} \left( m_\rho + m_\phi \right)$ 
      & 893.14 & $\geq$ & 894.7 \\
   tensor & $m_{K_2^{\ast}} \geq \frac{1}{2}
      \left( m_{a_2} + m_{f_2} \right)$ & 1427.7 & $\geq$ & 1296.55 \\ 
   axial & $m_{K_1} \geq \frac{1}{2}
      \left( m_{a_1} + m_{f_1} \right)$ & 1273 & $\geq$ & 1256.0 \\
   scalar & $m_{K_0^{\ast}} \geq \frac{1}{2}
      \left( m_{a_0} + m_{f_0} \right)$ & 1429 & $\geq$ & 982 
      \\ \hline \hline
   \multicolumn{5}{c}{$u\bar{c}$ sector: $m_{u\bar{c}} \geq 
      \frac{1}{2} \left( m_{u\bar{u}} + m_{c\bar{c}}\right)$}
      \\ \hline
   pseudoscalar & $m_D \geq \frac{1}{2} \left(m_\pi + m_{\eta_c}
      \right)$ & 1867.7 & $\geq$ & 1558.9 \\ 
   vector & $m_{D^{\ast}} \geq \frac{1}{2} \left( m_\rho + m_{J/\psi} \right)$ 
      & 2008.9 & $\geq$ & 1933.4 \\ 
   tensor & $m_{D_2^{\ast}} \geq \frac{1}{2}
      \left( m_{a_2} + m_{\chi_{c2}} \right)$ & 2458.9 & $\geq$ & 2437.1 \\ 
   axial & $m_{D_1^0} \geq \frac{1}{2}
      \left( m_{a_1} + m_{\chi_{c_1}} \right)$ & 2422.2 & $\geq$ & 2370 \\ 
   scalar & $m_{D_{0^+}} \geq \frac{1}{2}
      \left( m_{a_1} + m_{\chi_{c_0}} \right)$ & $m_{D_0^+}$ &
      $\geq$ & 2200.4 \\ \hline \hline
   \multicolumn{5}{c}{$u\bar{b}$ sector: $m_{u\bar{b}} \geq 
      \frac{1}{2} \left( m_{u\bar{u}} + m_{b\bar{b}}\right)$}
      \\ \hline
   pseudoscalar & $m_B \geq \frac{1}{2} \left(m_\pi + m_\Upsilon
      \right)$ & 5279.0 & $\geq$ & 4799.2 \\
   vector & $m_{B^{\ast}} \geq \frac{1}{2} \left( m_\rho + m_\Upsilon \right)$ 
      & 5324.9 & $\geq$ & 5115.2 \\
   tensor & $m_{B_2^{\ast}} \geq \frac{1}{2}
      \left( m_{a_2} + m_{\chi_{b_2}} \right)$ & 5739 & $\geq$ 
      & 5615.65 \\
   axial & $m_{B_1^0} \geq \frac{1}{2}
      \left( m_{a_1} + m_{\chi_{b_1}} \right)$ & $m_{B_1^0}$ & $\geq$ 
      & 5560.95 \\
   scalar & $m_{B_{0^+}} \geq \frac{1}{2}
      \left( m_{a_0} + m_{\chi_{b_0}} \right)$ 
      & $m_{B_{0^+}}$ & $\geq$ & 5421.6 \\ \hline \hline
   \multicolumn{5}{c}{$s\bar{c}$ sector: $m_{s\bar{c}} \geq 
      \frac{1}{2} \left( m_{s\bar{s}} + m_{c\bar{c}}\right)$}
      \\ \hline
   pseudoscalar & $m_{D_s} \geq \frac{1}{2} \left(
      \frac{2}{3}m_\eta + \frac{1}{3}m_{\eta'} + m_{\eta_c}\right)$ 
      & 1968.5 & $\geq$ & 1832.0 \\ 
   vector & $m_{D_s^{\ast}} \geq \frac{1}{2} \left( m_\phi 
      + m_{J/\psi} \right)$ 
      & 2112.4 & $\geq$ & 2058.15 \\
   tensor & $m_{D_{s2^+}} \geq \frac{1}{2}
      \left( m_{f_2} + m_{\chi{c_2}} \right)$ 
      & 2573.5 & $\geq$ & 2415.6 \\
   axial & $m_{D_{s1^+}} \geq \frac{1}{2}
      \left( m_{f_1} + m_{\chi{c_1}} \right)$ 
      & 2535.35 & $\geq$ & 2396.2 \\
   scalar & $m_{D_{0^+}} \geq \frac{1}{2}
      \left( m_{f_0} + m_{\chi{c_0}} \right)$ & $m_{D_{0^+}}$ & 
      $\geq$ & 1709 \\ \hline \hline
   \multicolumn{5}{c}{$s\bar{b}$ sector: $m_{s\bar{b}} \geq 
      \frac{1}{2} \left( m_{s\bar{s}} + m_{b\bar{b}}\right)$}
      \\ \hline
   pseudoscalar & $m_{B_s^0} \geq \frac{1}{2} \left(
      \frac{2}{3}m_\eta + \frac{1}{3}m_{\eta'} + m_\Upsilon \right)$ 
      & 5369.3 & $\geq$ & 5072.2 \\
   vector & $m_{B_s^{\ast}} \geq \frac{1}{2} \left( m_\phi + m_\Upsilon \right)$ 
      & 5416.3 & $\geq$ & 5239.89 \\
   tensor & $m_{B_{2^+}} \geq \frac{1}{2}
      \left( m_{f_2} + m_{\chi_{b_2}} \right)$ & $m_{B_{2^+}}$ & 
      $\geq$ & 5594.1 \\
   axial & $m_{B_{1^+}} \geq \frac{1}{2}
      \left( m_{f_1} + m_{\chi_{b_1}} \right)$ & $m_{B_{1^+}}$ &
      $\geq$ & 5586.9 \\
   scalar & $m_{B_{0^+}} \geq \frac{1}{2}
      \left( m_{f_0} + m_{\chi_{b_0}} \right)$ & $m_{B_{0^+}}$ &
      $\geq$ & 5421.9 \\ \hline \hline
   \multicolumn{5}{c}{$c\bar{b}$ sector: $m_{c\bar{b}} \geq 
      \frac{1}{2} \left( m_{c\bar{c}} + m_{b\bar{b}}\right)$}
      \\ \hline
   pseudoscalar & $m_{B_c^\pm} \geq \frac{1}{2} \left(
      m_{\eta_c} + m_\Upsilon \right)$ 
      & 6400 & $\geq$ & 6219.6 \\
   vector & $m_{B_{c1^-}} \geq \frac{1}{2} \left( m_{J/\psi} 
      + m_\Upsilon \right)$ 
      & $m_{B_{c 1^-}}$  & $\geq$ & 6278.6 \\
   tensor & $m_{B_{c2^+}} \geq \frac{1}{2}
      \left( m_{\chi_{c_2}} + m_{\chi_{b_2}} \right)$ 
      & $m_{B_{c2^+}}$ & $\geq$ & 6734.7 \\
   axial & $m_{B_{c1^+}} \geq \frac{1}{2}
      \left( m_{\chi_{c_1}} + m_{\chi_{b_1}} \right)$ 
      & $m_{B_{c1^+}}$ & $\geq$ & 6701.2 \\
   scalar & $m_{B_{c0^+}} \geq \frac{1}{2}
      \left( m_{\chi_{c_0}} + m_{\chi_{b_0}} \right)$ 
      & $m_{B_{c0^+}}$ & $\geq$ & 6638.6 \\
   \end{tabular}
\label{table:one}
\end{table}

\begin{table}
\caption{Baryon-meson Inequalities}
\begin{tabular}{crcl}
   \multicolumn{4}{c}{$J^P = 3/2^{+}$} \\ \hline
   $m_{\Delta} \, (uuu) \geq (3/2) m_\rho$ 
      & 1232.0 & $\geq$ & 1155 \\
   $m_{\Sigma} \, (suu) \geq (1/2) (m_\rho + 2 m_{K^{\ast}})$ 
      & 1384.6 & $\geq$ & 1278 \\ 
   $m_{\Xi} \, (ssu) \geq (1/2) (m_\phi + 2 m_{K^{\ast}})$ 
      & 1533.4 & $\geq$ & 1402.85 \\ 
   $m_{\Omega^{-}} \, (sss) \geq (3/2) m_\phi$ 
      & 1672.45 & $\geq$ & 1529.12 \\ 
   $m_{\Sigma_c^{++}} (cuu) \geq (1/2) (m_\rho + 2 m_{D^{\ast}})$ 
      & 2519.4 & $\geq$ & 2394 \\ 
   $m_{\Xi_c} \, (csu) \geq (1/2) (m_{K^{\ast}} + m_{D^{\ast}} + 
      m_{D_s^{\ast}})$ & $m_{\Xi_c}$  & $\geq$ & 2507.2 \\ 
   $m_{\Omega_c^0} \, (css) \geq (1/2) (m_\phi + 2 m_{D^{\ast}})$ 
      & $m_{\Omega_c^0}$ & $\geq$ & 2518.6\\ 
   $m_{\Xi_{cc}} \, (ccu) \geq (1/2) (m_{J/\psi} + 2 m_{D^{\ast}})$ 
      & $m_{\Xi_{cc}}$ & $\geq$ & 3557.3 \\ 
   $m_{\Omega^+_{cc}} \, (ccs) \geq (1/2) (m_{J/\psi} + 2 m_{D_s^{\ast}})$ 
      & $m_{\Omega^+_{cc}}$ & $\geq$ & 3660.8 \\
   $m_{\Omega_{ccc}^{++}} (ccc) \geq (3/2) (m_{J/\psi})$ 
      & $m_{\Omega_{ccc}^{++}}$ & $\geq$ & 4645.32 \\ 
   $m_{\Sigma_b} \, (buu) \geq (1/2) (m_{\rho} + 2 m_{B^{\ast}})$ 
      & $m_{\Sigma_b}$ & $\geq$ & 5710 \\ 
   $m_{\Xi_b} \, (bsu) \geq (1/2) (m_{K^{\ast}} + m_{B^{\ast}}
      + m_{B_s^{\ast}})$ & $m_{\Xi_b}$ & $\geq$ & 5793.7 \\
   $m_{\Omega_b} \, (bss) \geq (1/2) (m_{\phi} + 2 m_{B_s^{\ast}})$ 
      & $m_{\Omega_b}$ & $\geq$ & 5879.0 \\ 
   $m_{\Xi_{bb}} \, (bbu) \geq (1/2) (m_{\Upsilon} + 2 m_{B^{\ast}})$ 
      & $m_{\Xi_{bb}}$ & $\geq$ & 10055.1\\ 
   $m_{\Omega_{bb}} \, (bbs) \geq (1/2) (m_{\Upsilon} + 2 m_{B_s^{\ast}})$ 
      & $m_{\Omega_{bb}}$ & $\geq$ & 10099.5\\ 
   $m_{\Omega_{bbb}} \, (bbb) \geq (3/2) (m_{\Upsilon})$ 
      & $m_{\Omega_{bbb}}$ & $\geq$ & 14190.56 \\ \hline
   \multicolumn{4}{c}{$J^P = 1/2^{+}$} \\ \hline
   $m_N \, (uud) \geq (3/4) ( m_\pi + m_\rho)$ 
      & 938.919 & $\geq$ & 681 \\
   $m_{\Sigma} \, (suu) \geq (1/4)( 2 m_\rho
      + 3 m_K + m_{K^{\ast}})$ & 1193.15 & $\geq$ & 980 \\
   $m_{\Xi} \, (uss) \geq (1/4)( 2 m_\phi
      + 3 m_K + m_{K^{\ast}})$ & 1318.1 & $\geq$ & 1104.25 \\ 
   $m_{\Sigma_c} \, (cuu) \geq (1/4)( 2 m_\rho
      + 3 m_D + m_{D^{\ast}})$ & 2452.9 & $\geq$ & 2288 \\ 
   $m_{\Omega_c} \, (ssc) \geq (1/4)( 2 m_\phi
      + 3 m_{D_s} + m_{D_s^{\ast}})$ & 2704 & $\geq$ & 2514.1 \\ 
   $m_{\Xi_c} \, (ccu) \geq (1/4)( 2 m_{J/\psi}
      + 3 m_D + m_{D^{\ast}})$ & $m_{\Xi_c}$ & $\geq$ & 3451.5 \\
   $m_{\Omega^+_{cc}} \, (ccs) \geq (1/4)( 2 m_{J/\psi}
      + 3 m_{D_s} + m_{D_s^{\ast}})$ & $m_{\Omega^+_{cc}}$ & $\geq$ & 3552.9 \\
   $m_{\Sigma_b} \, (buu) \geq (1/4)( 2 m_\rho
      + 3 m_B + m_{B^{\ast}})$ & $m_{\Sigma_b}$ & $\geq$ & 5675.5 \\ 
   $m_{\Lambda} \, (uds) \geq (1/4)( 2 m_\pi
      + 3 m_K + m_{K^{\ast}})$ & 1115.683 & $\geq$ & 663.56 \\
   $m_{\Lambda_c^+} \, (udc) \geq (1/4)( 2 m_\pi 
      + 3 m_D + m_{D^{\ast}})$ & 2284.9 & $\geq$ & 1973.0 \\
   $m_{\Lambda_b^0} \, (udb) \geq (1/4)( 2 m_\pi
      + 3 m_B + m_{B^{\ast}})$ & 5624 & $\geq$ & 5359.5 \\ 
   $m_{\Xi_c} \, (cus) \geq (1/4) \left[ (m_K + m_{K^{\ast}}) 
      + (m_{D_s} + m_D) + (m_{D_s^{\ast}} + m_{D^{\ast}}) \right]$ 
      & 2469.0 & $\geq$ & 2336.4 \\
   $m_{\Xi_b} \, (bus) \geq (1/4) \left[ (m_K + m_{K^{\ast}}) 
      + (m_{B_s} + m_B) + (m_{B_s^{\ast}} + m_{B^{\ast}}) \right]$ 
      & $m_{\Xi_b}$ & $\geq$ & 5682.7 \\
   $m_{bcs} \, (bcs) \geq (1/4) \left[ (m_{D_s} + m_{D_s^{\ast}}) 
      + (m_{B_s} + m_B) + (m_{B_{s^{\ast}}} + m_{B^{\ast}}) \right]$ 
      & $m_{bcs}$ & $\geq$ & 6355.9 \\
\end{tabular}
\label{table:two}
\end{table}

\begin{table}
\caption{Baryon-baryon Inequalities}
\begin{tabular}{crcl}
   \multicolumn{4}{c}{for the octet (Gell-Mann -- Okubo)} \\ \hline
   $3m_\Lambda + \frac{1}{3} \left( m_{\Sigma^-} + m_{\Sigma^+}
      + m_{\Sigma^0} \right) \geq m_p + m_n + m_{\Xi^-} + m_{\Xi^0}$
      & 4731.6 & $\geq$ & 4514.1 \\ \hline\hline
   \multicolumn{4}{c}{for the decuplet} \\ \hline
   $m_{\Sigma} \geq (1/2) \left( m_\Delta + m_\Xi \right)$
      & 1384.6 & $\geq$ & 1382.7 \\
   $m_{\Xi} \geq (1/2) \left( m_\Omega + m_\Sigma \right)$
      & 1533.4 & $\geq$ & 1528.5 \\
\end{tabular}
\label{table:three}
\end{table}

\begin{table}
\caption{Inequalities involving charm meson radial excitations}
\begin{tabular}{ccrcl}
   pseudoscalar & $m_D^{(0)} + m_D^{(1)} \geq
      \frac{1}{2}\left[m_{\eta_c}^{(0)} + m_{\eta_c}^{(1)} +
      m_{\pi}^{(0)} + m_{\pi}^{(1)} \right]$
      & $m_D^{(1)}$ & $\geq$ 
      & 2138.2\footnote{We can predict a lower bound on the mass of the as yet
      undiscovered $D^{(1)}$, since the other masses in the inequality
      are known.} \\ \hline
   vector & $m_{D^\ast}^{(0)} + m_{D^\ast}^{(1)} \geq
      \frac{1}{2}\left[m_{J/\psi}^{(0)} + m_{J/\psi}^{(1)} +
      m_{\rho}^{(0)} + m_{\rho}^{(1)} \right]$
      & 4645.9 & $\geq$ & 4508.95 \\
\end{tabular}
\label{table:four}
\end{table}

%
\section{QCD inequalities for correlation functions of quark bilinears}
\label{sec:quarkbi}

The set of all euclidean correlation functions, {\em i.e.} ordinary
Green's functions continued to the euclidean domain, contains the
complete information on any field theory \cite{ref:schwinger}. In
particular, two-point euclidean correlation functions are closely
related to the spectrum of the theory, and indeed play a key role in
most attempts to compute the spectrum of QCD. Following Weingarten
\cite{ref:weingarten}, we will prove in this section inequalities for
mesonic correlation functions.  Let
\begin{equation}
   F_a(x,y) = \langle 0 | J_a(x) J^{\dag}_a(y) | 0 \rangle
\>,
\end{equation}
with $J_a(x)$ a general, local, gauge invariant ({\em i.e.} color
singlet) operator with the index $a$ indicating Lorentz and/or flavor
indices, be such a two-point function. Note that in the euclidean case
all $x-y$ intervals are spacelike. The usual time ordering stating
that the creation operator $J^{\dag}_a(x)$ should act prior to the
annihilation $J_a(x)$ is redundant.

With an eye to the original Minkowski configuration, we will still assume
that $x^0 - y^0 \geq 0$. In particular, we could use rotational (Lorentz)
invariance to make $x^0 - y^0 = |x - y| = t$ by choosing $y = (0, \vec{0}),
x = (t, \vec{0})$. Inserting a complete set of physical energy momentum
eigenstates and using the (euclidean) time translation operator ${\rm
e}^{-Ht} J_a(0) {\rm e}^{Ht} = J_a(t)$, we obtain a spectral function
representation \cite{ref:luscher}
\begin{equation}
   F_a(x-y) = \int_{\mu_0}^\infty \, {\rm d}\mu^2 \sigma_a(\mu^2)
      {\rm e}^{-\mu |x-y|}
\label{eq:specrep}
\end{equation}
for the correlation function. The spectral function is given by
\begin{equation}
   \sigma_a(\mu^2) = \sum_n | \langle 0 | J_a | n \rangle |^2
      \delta(p_n^2 - \mu^2)
\>,
\label{eq:spectral}
\end{equation}
with $p_n$ the four momentum of the state $n$.

The asymptotic behavior of $F_a(x-y)$ (as $|x-y| \rightarrow \infty$)
is controlled by the state of lowest mass contributing in
Eq.~(\ref{eq:spectral}):
\begin{equation}
   F_a(x-y) \sim {\rm e}^{-\mu_0 |x-y|}
\>,
\label{eq:faxy}
\end{equation}
where $\sim$ means equality up to a residue factor $\gamma_a$ and
powers of $|x-y|^{-1}$. Here $\mu_0$ could correspond to a physical
two particle threshold. It could also be an isolated contribution
$p^2(n_a) = \left[m_a^{(0)}\right]^2$ of a particle with the quantum
numbers of the current $J_a$, {\em i.e.} a state for which $\langle
n_a | J_a^{\dag} | 0 \rangle \neq 0$. In this case,
\begin{equation}
   F_a(x-y) \sim {\rm e}^{-m_a^{(0)} |x-y|} \gamma_a
\>.
\label{eq:fa}
\end{equation}

We note that due to color confinement in QCD the finite energy
physical spectrum consists of color singlet states and therefore it is
sufficient to consider only correlation functions of color singlet
operators.\footnote{We will not address the interesting suggestion
  \cite{ref:domokos} that the inequalities mandate specific patterns
  of color symmetry breaking.}

Eq.~(\ref{eq:fa}) is used in QCD lattice Monte Carlo calculations of
the spectrum \cite{ref:creutzetal,ref:creutz}. An exponential form is
fitted to the large (many lattice spacings) distance behavior of the
appropriate correlation function. The latter is estimated by numerical
sampling via the functional path integral expression for the
correlation functions:
\begin{equation}
   \langle 0 | J_a(x) J^{\dag}_a(y) | 0 \rangle =
      \frac{\intD{A_\mu(x)} \intD{\psi(x)} \intD{\psibar(x)}
      J_a(x) J_a^{\dag}(x) {\rm e}^{-S(x)}}
      {\intD{A_\mu(x)} \intD{\psi(x)} \intD{\psibar(x)} {\rm e}^{-S(x)}}
\>,
\label{eq:pathint}
\end{equation}
with the action
\begin{equation}
   S_{\text{QCD}} = \intd{x} 
      {\cal L}_{\text{QCD}} = \intd{x} \left[ \sum_i \psibar_i
      (\dslasha_A + m_i) \psi_i + {\rm tr} (F_{\mu\nu}^a \lambda^a)^2
      \right]
\end{equation}
and $\dslasha_A = \gamma^\mu (\partial_\mu + i \lambda^a A_{a\mu})$.
For the purpose of lattice calculations, and also for the derivation
of the QCD inequalities, we should eliminate the Grassman variables
$\psi(x)$ and $\psibar(x)$ (corresponding to the anti-commuting
fields) in the functional integral. Since $\psi_i$ and $\psibar_i$
appear only in the bilinear kinetic and mass term in the action, this
integration can be done in closed form \cite{ref:creutz}.  In the
partition function in the denominator of Eq.~(\ref{eq:pathint}) this
results simply in the additional factor $\prod_{i=1}^{N_f} {\rm Det}
(\dslash_A + m_i)$, which modifies the integration weight from
\begin{displaymath}
   d\mu(A) = {\rm e}^{-S_{YM}(A)} {\rm d}[A_\mu(x)] \ldots
\>,
\end{displaymath}
where $S_{YM}(A) = {\rm tr} (F_{\mu\nu}^a \lambda^a)^2$ is the pure
Yang-Mills part of the QCD action, to 
\begin{equation}
   {\rm d}\mu(A) =  {\rm e}^{-S_{YM}(A)} \prod {\rm Det}
      (\dslasha_A + m_i) {\rm d}[A_\mu(x)]
\>.
\label{eq:intmeasure}
\end{equation}
The same change occurs in the numerator if $J_a$ are functions of gluonic
$(F_{\mu\nu})$ degrees of freedom only.

The currents of interest -- for the non-glueball part of the QCD
spectrum -- are, however, bilinears of quark fields, of the form
$J_{\Gamma}^{i\bar{\jmath}} = \psibar_i^a \Gamma \psi_{ja}$. We use the
scalar, pseudoscalar, vector, and pseudovector currents and bilinear
expressions with extra $D_\mu$ derivatives:
\begin{equation}
   \begin{array}{ll}
      J_s^{i\bar{\jmath}} = \psibar_i^a \psi_{ja}, &
      J_{ps}^{i\bar{\jmath}} = \psibar_i^a \gamma_5 \psi_{ja}, 
   \\
      J_v^{i\bar{\jmath}} = \psibar_i^a \gamma_\mu \psi_{ja}, &
      J_{pv}^{i\bar{\jmath}} = \psibar_i^a \gamma_5 \gamma_\mu \psi_{ja}, 
   \\
      J_T^{i\bar{\jmath}} = \psibar_i^a \dslasha_\nu \gamma_\mu
      \psi_{ja}, & {}
   \end{array}
\label{eq:currents}
\end{equation}
in order to extract the spectrum of the $0^+, 0^-, 1^+, 1^-,$ and $2^+$
mesons made up of quark flavors $q_i$ and $\bar{q}_j$.

In all of these cases we need to contract the extra $(\psibar) \psi$
at $(x)$ and $(\psibar) \psi$ at $(y)$. One chain of consecutive
contractions can involve both $\psi(x)$ and $\psibar(y)$. This gives a
``connected'' flavor structure with the quark $q_i$ propagating from
$y$ to $x$ and antiquarks $\bar{q}_j$ from $x$ to $y$ [see
Fig.~\ref{fig:2ptcorr}(a)]; or, if $i = j$, we could also have the
$q_i(x) \bar{q}_i(x)$ and $q_i(y) \bar{q}_i(y)$ in two separate
contractions, yielding a (flavor) disconnected structure [see
Fig.~\ref{fig:2ptcorr}(b)]. The expression for the two-point
correlation function then becomes
\begin{mathletters}
\label{eq:tp}
   \begin{equation} 
      \intoned{\mu(A)} {\rm tr} \left[ \Gamma S^i_A(x,y) \Gamma S^j_A(y,x) 
      \right] \left/ \intoned{\mu(A)} \right.
   \label{eq:corr1}
   \end{equation}
   \mbox{for the flavor connected case, or}
   \begin{equation}
      \intoned{\mu(A)} {\rm tr} \left[ \Gamma S^i_A(x,x) \right] 
      {\rm tr} \left[ \Gamma S^i_A(y,y) \right] \left/ 
      \intoned{\mu(A)} \right.
   \label{eq:corr2}
   \end{equation}
   \mbox{for the flavor disconnected case.}
\end{mathletters}

In Eqs.~(\ref{eq:tp}) $S^i_A(x,y)$ is the full euclidean fermionic
propagator (for flavor $q_i $) in the background field $A_\mu(x)$;
$\Gamma$ is the Dirac matrix in the expression for the current
(\ref{eq:currents}); and the trace refers to both spinor and color
indices (which are suppressed).

The quark propagator solves the equation
\begin{displaymath}
   (\dslasha_A + m_i) S^i_A(x,y) = \delta(x-y)
\end{displaymath}
and formally is given by
\begin{equation}
   S^i_A(x,y) = \langle x | (\dslasha_A + m_i)^{-1} | y \rangle
\label{eq:sfermion}
\end{equation}
with $\dslash = \gamma_\mu D_\mu(A)$ involving the covariant derivatives
$D_\mu(A) = \partial_\mu + i \lambda_a A^a_\mu$ in the given background
field configuration $A_\mu(x)$. The euclidean $\dslasha$ is purely
antihermitian and the $\gamma_\mu$ are all hermitian and satisfy $ \{
\gamma_\mu, \gamma_\nu \} = 2 \delta_{\mu\nu}$.

It is well known that the singular behavior of tr[$\gamma_5 S_A(x,y)$]
when $x \rightarrow y$ \cite{ref:schwinger,ref:adler69,ref:bell}, or,
alternatively some subtleties in the fermionic integration
\cite{ref:fujikawa}, may induce an anomaly $\sim F \tilde{F}$ in the
flavor disconnected parts for the pseudoscalar case. Thus for much of
the following discussion we will take $i \neq j$ and avoid altogether
the flavor disconnected contribution.

An important element for the derivation of the inequalities is the
positivity of the determinantal factor $\prod {\rm Det} (\dslasha_A +
m_i)$, and hence also of the integration measure ${\rm d} \mu(A)$ in
Eq.~(\ref{eq:intmeasure}), for any given $A_\mu(x)$ configuration
\cite{ref:weingarten,ref:vafanp}. The euclidean QCD action ${\cal
  L}_{E} = E^2 + B^2$ is real so long as we do not have an $i \theta E
\cdot B$ term, and hence $\exp \left[ {-S_{YM}(A)} \right]$ is
positive. To show the positivity of the determinant factor, let us
consider the eigenmodes $\psi_A (\lambda_A)$ of the hermitian operator
$i \dslasha_A$ satisfying
\begin{equation}
   i \dslasha_A \psi_\lambda^A = \lambda(A) \psi_\lambda^A 
\label{eq:idslash}
\end{equation}
with real eigenvalues $\lambda(A)$. The $\gamma_5$ anticommutation
$\gamma_5 \gamma_\mu \gamma_5 = - \gamma_\mu$ implies the relation
\begin{equation}
   \dslasha_A = - \gamma_5 \dslasha_A \gamma_5
\>.
\label{eq:dslash}
\end{equation}
This in turn forces all nonzero eigenvalues of $\dslash_A$ to appear
in complex conjugate pairs. Indeed from (\ref{eq:idslash}) and
(\ref{eq:dslash})
\begin{equation}
   i\dslasha_A (\gamma_5 \psi_\lambda) = \gamma_5^2 i\dslasha_A \gamma_5
      \psi_\lambda = - \gamma_5 (i\dslasha_A) \psi_\lambda 
      = - \lambda(A) \gamma_5 \psi_\lambda
\>,
\end{equation}
so that $\gamma_5 \psi_\lambda$ is the eigenfunction corresponding to
the eigenvalues $- \lambda(A)$. Thus we have explicitly positive
determinants
\begin{equation}
   {\rm Det} (\dslasha_A + m_i) = \prod_\lambda \left[ 
      i \lambda(A) + m_i \right] \left[ -i \lambda(A) + m_i \right]
      = \prod_\lambda \left[ \lambda^2(A) + m_i^2 \right]
\end{equation}
and hence
\begin{equation}
   {\rm d}\mu(A) \geq 0
\end{equation}
for all $A_\mu(x)$. 

It is important to notice that the above argument applies only for
vectorial, non-chiral models with the quarks being Dirac fermions. If
$\gamma_5 \psi_\lambda$ is not a distinct new spinor (as is the case for
chiral fermions) then the positivity argument breaks down.

The measure positivity allows us to prove inequalities between correlation
functions if we can show that the corresponding integrands in
Eq.~(\ref{eq:corr1}) satisfy, for any $A_\mu(x)$, the same inequality.
Thus, let us assume that by simple algebra one can prove
\begin{equation}
   {\rm tr} \left[ \Gamma_a S^i(x,y) \Gamma_a S^j(y,x) \right]
      \geq  {\rm tr} \left[ \Gamma_b S^i(x,y) \Gamma_b S^j(y,x) \right]
\>.
\label{eq:trace}
\end{equation}
The same inequality continues to hold after we integrate over the
normalized, positive ${\rm d}\mu(A)$, yielding
\begin{equation}
   \langle 0 | J_a(x) J^{\dag}_a(y) | 0 \rangle \geq   
      \langle 0 | J_b(x) J^{\dag}_b(y) | 0 \rangle
\label{eq:jab}
\end{equation}
with
\begin{equation}
   J_{a(b)}(x) = \psibar_i(x) \Gamma_{a(b)} \psi_i(x)
\>.
\end{equation}

We will next show that a relation like Eq.~(\ref{eq:trace}) holds with
$\Gamma_a = \gamma_5$. To this end we use Eq.~(\ref{eq:sfermion}) defining
the propagator $S_A(x,y)$. We then have 
\begin{eqnarray}
   \gamma_5 S_A(x,y) \gamma_5 &=& \langle x | \gamma_5 
      (\dslasha_A + m)^{-1} \gamma_5 | y \rangle
      = \langle x | - (\dslasha_A + m)^{-1} | y \rangle 
   \nonumber \\
   &=&  \langle x | \left[(\dslasha_A + m)^{-1}\right]^{\dag} | y \rangle
      = S_A^{\dag}(y,x)
\>,
\label{eq:tracegamma5}
\end{eqnarray}
where the last dagger refers to the conjugate matrix in color and spinor
space.

It is instructive to see how the same result is obtained in various
more explicit expressions for the propagator.  Thus consider the
contribution of any fermionic path on the lattice to the propagator
$S^i_A(x,y)$ in the hopping parameter expansion \cite{ref:hopping}. It is
\cite{ref:creutz}
\begin{equation}
   \prod_{\vn, \hat{\mu}} U(\vn, \vn + \hat{\mu}) (a + b \gamma_\mu)
\>,
\label{eq:lattprop}
\end{equation}
where the product extends over any set of links connecting the initial
point $y$ to the final point $x$. Multiplying by $\gamma_5$ on the
left and right and commuting the $\gamma_5$ factor through, we have
$\prod_{\vn, \vn + \hat{\mu}} U(\vn, \vn + \hat{\mu}) (a - b
\gamma_\mu) = \prod_{\vn + \hat{\mu}, \vn} \left[ U(\vn + \hat{\mu},
  \vn) (a - b \gamma_\mu)\right]^{\dag}$ (the euclidean $\gamma_\mu$
are hermitian), which is the same as the contribution of the reversed
path of links (connecting $x$ to $y$) to $\left[ S^i_A(y,x)
\right]^{\dag}$. Summing over all fermionic paths, we reconstruct
$S_A^i(x,y)$ and $S_A^i(y,x)$ and Eq.~(\ref{eq:tracegamma5}) is
satisfied.

The key observation now is that if
\begin{equation}
   S_A^i(x,y) = S_A^j(x,y)
\end{equation}
then Eq.~(\ref{eq:tracegamma5}) implies that the integrand in the
pseudoscalar correlation function is a positive definite sum of squares:
\begin{equation}
   {\rm tr} \left[ \gamma_5 S^i_A(x,y) \gamma_5 S^j_A(x,y) \right] =
       {\rm tr} \left[(S^i_A)^{\dag}(y,x) S^j_A(y,x) 
       \right] = \sum_{\stackrel{a,a'}{\tau,\tau'}}
      |(S^i_A)_{aa',\tau\tau'}|^2 \geq 0
\>,
\end{equation}
with the sum extending over color $(a)$ and spinor $(\tau)$ indices.

The integrand for any correlation functions of the other,
non-pseudoscalar currents in Eq.~(\ref{eq:currents}) involves again
sums of bilinear products of the same matrix elements
$(S^i_A)_{aa',\tau\tau'}$, but in general with alternating signs. Thus
the integrand for the pseudoscalar correlation functions is larger
than all the other integrands. This then yields the desired result:
\begin{equation}
    \langle 0 | J_{ps}^{i\bar{\jmath}}(x) (J_{ps}^{i\bar{\jmath}})^{\dag}(y) 
      | 0 \rangle \geq \langle 0 | J_{\Gamma}^{i\bar{\jmath}}(x) 
      J_{\Gamma}^{i\bar{\jmath}}(y) | 0 \rangle
\>,
\label{eq:pscurrent}
\end{equation}
with $J_\Gamma$ any non-pseudoscalar current. 

The requirement $S^i_A(x,y) = S^j_A(x,y)$, with $i \neq j$, is made in
order to avoid the flavor disconnected contribution (\ref{eq:corr2})
which could invalidate the derivation. It can be satisfied in the
$I$-spin limit -- with $m_u - m_d \ll \Lambda_{\text{QCD}}$ and EM
effects neglected -- by taking $i = u, j = d$. Since $S^u_A(x,y)$ and
$S^d_A(x,y)$ then satisfy identical equations, $S^u_A(x,y) =
S^d_A(x,y)$ as required.  The lowest mass particle in this $\psibar_u
\gamma_5 \psi_d$ channel is then the pion. The inequality
(\ref{eq:jab}) between the correlation functions implies, by going to
the asymptotic $|x-y| \rightarrow \infty$ region and utilizing the
asymptotic behavior (\ref{eq:faxy}), the reverse inequality between
the lowest mass particles in the corresponding channels:
\begin{equation}
   m_a^{(0)} \leq m_b^{(0)}
\>.
\end{equation}
Thus we obtain the result that the pion is the lightest meson in the
$u\bar{d}$ channels
\begin{mathletters}
   \begin{equation}
      m_\pi \leq m_\rho \, (1^-)
   \end{equation}
   \begin{equation}
      m_\pi \leq m_{a_1} \, (1^+), \, \, m_{a_2} \, (2^+)
   \end{equation}
   \begin{equation}
      m_\pi \leq m_\chi \, (0^+)
   \label{eq:mpimchi}
   \end{equation}
\end{mathletters}
We note that these level orderings are indeed expected in any $\qqb$
potential model \cite{ref:rujula} where the singlet S-wave state --
{\em i.e.} the $\pi$ meson -- is the lowest lying state.

Could the various renormalizations which are required in order to
render the correlation functions and the local products $\psibar^i(x)
\Gamma \psi^j(x)$ finite invalidate the derivation? The regularization
required in order to make the path integral $\intD{A_\mu(x)}$
well-defined will not in general interfere with the derivation of the
inequalities. Thus we could use a lattice discretization with any
spacing $a$ and with any action -- the minimal Wilson action or with
terms in the adjoint or higher representation. The inequalities will
continue to be satisfied at every step of the procedure of letting $a
\rightarrow 0$ and $V$, the volume of the lattice, to infinity. Thus,
to the extent that any sensible regularization exists, the
inequalities will be satisfied.

Instead of the local (non-pseudoscalar) current $J^\Gamma(x) = \bar{q}(x)
\Gamma q(x)$, we could use a non-local gauge invariant version
\cite{ref:weingarten} 
\begin{eqnarray}
   J^\Gamma_c(x,x') &=& \bar{q}_i(x) \Gamma \exp
      {i \oint_x^{x'} A_\mu dx^\mu} q_j(x')
   \nonumber \\
   &=& \bar{q}_i(x) \Gamma U_c q_j(x')
\end{eqnarray}
which creates an extended state -- a quark at $x$, antiquark at $x'$,
and a connecting flux string along a curve $c$. The path-ordered
Wilson line factor $U_c$ is a unitary matrix. The corresponding
correlation function
\begin{equation}
   \left\langle 0 \left| J^\Gamma(x,x') \left[J^\Gamma(y,y')
      \right]^{\dag} \right| 0 \right\rangle
      = \intD{\mu(A)} {\rm tr} \left[ \Gamma U_c S_A(x,y) U_c^{\dag} \Gamma
      S_A(y',x') \right]
\end{equation}
can readily be shown, using a Schwartz inequality, to satisfy:
\begin{equation}
   \left|\left\langle 0 \left| J^\Gamma(x,x') 
   \left[J^\Gamma(y,y')\right]^{\dag} 
   \right| 0 \right\rangle\right| \leq \langle 0 | J_p(x) J_p^{\dag}(y) 
   | 0 \rangle \langle 0 | J_p(x') J^{\dag}_p(y') | 0 \rangle
\>,
\end{equation}
from which it follows that the mass of any particle $\chi$ created by
$J_c^\Gamma(x,x')$ still satisfies $m_\chi \geq m_\pi$.\footnote{The
  advantage of the more general construction using a nonlocal
  $J_c^\Gamma$ is that $\chi$ states of any spin can be created.}

The derivation of the inequalities still applies regardless of the
choice of ``gauge fixing'' which limits the allowed $A_\mu(x)$ (or
$U_{\vn, \vn + \hat{\mu}}$ in the lattice version) which should be
summed over, but does not affect the positivity of the measure. Also,
the derivation is not sensitive to the issue of how we put the
fermions on the lattice. We can use Kogut-Susskind fermions
\cite{ref:susprd} [which would correspond to $a=0, b=1$ in
Eq.~(\ref{eq:lattprop})], Wilson fermions \cite{ref:wilson}
($a=1,b=1$), or any intermediate procedure.\footnote{Since the
  derivation of the inequalities may formally apply to any dimension
  and any shape of lattice, it should also hold for the new domain
  wall fermions \cite{ref:kaplan}.}

The inequalities are independent of the issue of quark and color
confinement.\footnote{In the limit of unbound massive quarks we could
  have additional relations between the correlation functions. To
  leading order, the asymptotic behavior of all correlation functions
  is now the same: $\exp \left[-(m_i + m_j)|x-y| \right] $. However,
  the P-wave intermediate states in the scalar or axial current
  correlation function cause a stronger (power) vanishing of
  $\sigma(\mu^2)$ at $\mu = m_i + m_j$; hence the pseudovector
  correlation functions have an inferior power behavior
  \cite{ref:feinberg} when $|x-y| \rightarrow \infty$, as compared
  with the vectorial correlation functions.  The inequalities $\langle
  J_V^{u\bar{d}}(x)J_V^{u\bar{d}}(y) \rangle \geq \langle
  J_A^{u\bar{d}}(x)J_A^{u\bar{d}}(y) \rangle$ and are true to all
  orders of perturbation theory and thus may be true even in the
  confining phase.  In this case we expect additional mass ordering
  relations such as $m_\rho \, (1^-) \leq m_{a_1}$.}  In fact the
inequalities also apply in non-confining vectorial theories like QED
and in the short distance $|x-y| \rightarrow 0$ limit where
perturbative QCD is adequate. They would amount to the statement that
the pseudoscalar spin singlet ``para-positronium'' type state is the
lightest (this happens in QED even with $i = j = $ electron since the
effects of the $F \tilde{F}$ anomaly are weak for $m_e \neq 0$).
Likewise, the euclidean pseudoscalar correlation function dominates,
as $x \rightarrow 0$ (or $q \rightarrow \infty$ in momentum space),
all the other correlation functions in the zero and one gluon
approximation respectively. These inequalities first derived by
Weingarten are the simplest of the exact QCD inequalities and yet, as
will be made clear in the next section, are extremely useful.  The
positivity of the contribution of each $A_\mu(x)$ (or $U_{\vn, \vn +
  \hat{\mu}}$) configuration to the path integral defining the
pseudoscalar propagator $\langle 0 | J_{ps}(x) J_{ps}^{\dag}(y) | 0
\rangle$ is a novel feature specific to QCD-type theories. It is quite
distinct from the positivity (by unitarity) of the contribution of
each physical intermediate state $|\langle 0 | J_a | n \rangle |^2
\delta(p_n^2 - \mu^2)$ to the spectral function which holds for an
arbitrary current $J_a$ in any theory, vectorial or otherwise. We
cannot in general appeal to {\em both} types of positivity at the {\em
  same time}, since, in order to have the correct physical states, we
have to sum over all the background $A_\mu(x)$ configurations first.
Otherwise, we do not respect even the translational invariance used in
order to derive Eq.~(\ref{eq:spectral}).

Anishetty and Wyler \cite{ref:wyler} and Hsu \cite{ref:hsu} noticed
that the measure positivity and ensuing inequalities hold also for
SU(2) chiral gauge theories with an even number of flavors (required
in order to avoid global anomalies \cite{ref:witt82}. Indeed in this
case, we can define
\begin{displaymath}
   \psi^i = \psi^i + (\psi^{i+N_f/2})^C, \qquad i = 1, \ldots, N_f/2
\end{displaymath}
to be $N_f/2$ vectorial, interacting Dirac fields.

%
\section{QCD inequalities and the non-breaking of global vectorial
  symmetries}
\label{sec:vecsym}

The symmetry structure of a field theory, {\em e.g.} whether it
spontaneously breaks vectorial and/or axial global symmetries present
in the original Lagrangian, is closely tied in with the zero mass
sector. Thus spontaneous breaking of global symmetries implies, via
the Goldstone theorem \cite{ref:goldstone}, the existence of massless,
scalar Goldstone bosons.  Likewise spontaneous breaking of axial
global symmetries implies Nambu-Goldstone massless pseudoscalars
\cite{ref:njl}. Finally, an unbroken axial symmetry may require for
its realization massless fermions in the physical spectrum (or a
parity doubled spectrum). Evidently mass relations such as
Eqs.~(\ref{eq:pionineqs}) between baryons (fermions) and scalar and/or
pseudoscalar masses could restrict some of these possibilities and
dictate the patterns of global symmetry realization in QCD and other
vectorial theories.

As a first illustration we will use the inequality (\ref{eq:mpimchi}) --
due to Weingarten -- to motivate the Vafa-Witten theorem. The theorem
states that in QCD (and in vectorial theories in general) global vectorial
symmetries do not break down spontaneously. This will then be supplemented
by some of the more rigorous discussion in the original Vafa-Witten paper.

To be specific let us first consider a nonabelian global vectorial
symmetry such as the SU(2) isospin symmetry for QCD. This symmetry is
generated by
\begin{eqnarray}
   I^+ &=& \int {\rm d}^3 x \psi_u^{\dag}(x) \psi_d(x)
   \nonumber \\
   I^- &=& \int {\rm d}^3 x \psi_d^{\dag}(x) \psi_u(x)
   \nonumber \\
   I^3 &=& \frac{1}{2}\int {\rm d}^3 x \left[ 
      \psi_u^{\dag}(x) \psi_u(x) - \psi_d^{\dag}(x) \psi_d(x) \right]
\>.
\label{eq:isospin}
\end{eqnarray}
In the limit of $m_u^{(0)} = m_d^{(0)}$ and no electromagnetic
interactions it is an exact symmetry of $H_{\text{QCD}}$.  If the
symmetry breaks down spontaneously (either completely or to a U(1)
subgroup generated by $I_3$) we expect massless Goldstone scalars
$0^+_{u\bar{d}}$ and $0^+_{u\bar{d}}$.  The essence of the argument
against spontaneous breaking is that such massless scalars can be
ruled out if $m_u^{(0)} = m_d^{(0)} \neq 0$.  Alternatively an
exponential falloff as $|x-y| \rightarrow \infty$ of all correlation
functions can be proven for $m_u^{(0)} = m_d^{(0)} > 0$. This
conflicts with the power falloff expected due to the intermediate zero
mass $0^+_{u\bar{d}}$ state contributing via Eq.~(\ref{eq:fa}) to the
spectral representation of the scalar correlation function.

Let us present first the more heuristic argument which is analogous to
the first argument of Vafa and Witten. The inequality
(\ref{eq:mpimchi}) $m^{(0^+)}_{u\bar{d}} \geq m^{(0^-)}_{u\bar{d}}$
states that the lowest scalar state in the $u\bar{d}$ channel must
have a mass equal to or larger than the mass of the lowest $u\bar{d}$
pseudoscalar. If the $I^+$ symmetry is spontaneously broken, then
$m^{(0^+)}_{u\bar{d}} = 0$ and hence $m^{(0^-)}_{u\bar{d}}$ must
vanish as well. If the axial global symmetry is not spontaneously
broken, such a vanishing could only be accidental and hence most
implausible. We next make a more precise and specific argument why
$m^{(0^-)}_{u\bar{d}}$ should {\em not} vanish.

In the limit $m_u^{(0)} = m_d^{(0)} = 0$ the QCD Lagrangian possesses an
extra global SU(2) symmetry generated by the axial analogs of
Eq.~(\ref{eq:isospin}): 
\begin{equation}
   I^+_5 = \int {\rm d}^3 x \, \psi_u^{\dag}(x) \gamma_5 \psi_d(x), \, \,
   \mbox{{\em etc.}}
\end{equation}
The spontaneous breaking of $I_5^+$ could then yield a massless
pseudoscalar $u\bar{d}$ Nambu-Goldstone boson and
$m^{(0^+)}_{u\bar{d}} \geq m^{(0^-)}_{u\bar{d}}$ would then be
trivially satisfied as $0 \geq 0$.  However, following Vafa and Witten
we keep $m_u^{(0)} = m_d^{(0)} \neq 0$.  No exact chiral symmetry then
exists and $m^{(0^-)}_{u\bar{d}}$ should not vanish. The Weingarten
inequality implies $m^{(0^+)}_{u\bar{d}} \geq m^{(0^-)}_{u\bar{d}} >
0$, which as we argue next, negates the spontaneous breaking of $I^+$
symmetry.

As we approch the  $m_u^{(0)} = m_d^{(0)} = 0$ limit, the lowest lying
pseudoscalar particle becomes a pseudo-Goldstone particle whose (mass)$^2$
is given by \cite{ref:adler,ref:gor}
\begin{equation}
   m_\pi^2 f_\pi^2 = \left( m_u^{(0)} + m_d^{(0)} \right) 
      \langle \psibar \psi \rangle
\label{eq:mpi2}
\end{equation}
and $m_\pi^2$ vanishes linearly with the explicit chiral symmetry breaking
$(m_u^{(0)} + m_d^{(0)})$ in the Lagrangian. Similar manipulations
utilizing the analog of the soft pion theorem and current algebra yielding
Eq.~(\ref{eq:mpi2}) suggest an analogous relation for the mass of the
tentative $0^+_{u\bar{d}}$ Goldstone boson $\chi$:
\begin{equation}
   m_\chi^2 f_\chi^2 = \left( m_d^{(0)} - m_u^{(0)} \right) 
      \left( \langle \psibar_u \psi_u \rangle 
      - \langle \psibar_d \psi_d \rangle \right)
      \leq  m_\pi^2 f_\pi^2
\>.
\label{eq:mchi2}
\end{equation}
If $f_\chi \neq 0$, {\em i.e.} the Goldstone $0^+$ boson does not
decouple, then Eqs.~(\ref{eq:mpi2}) and (\ref{eq:mchi2}) are
inconsistent with the QCD inequality $m^{(0^+)}_{u\bar{d}} \geq
m^{(0^-)}_{u\bar{d}}$ so long as $m_u^{(0)} - m_d^{(0)} \ll m_u^{(0)}
+ m_d^{(0)}$, which we can maintain even as we approach $m_u^{(0)} =
m_d^{(0)} = 0$.

The inequality (\ref{eq:mpimchi}) was derived in
Sec.~\ref{sec:quarkbi} by using $S_A^u(x,y) = S_A^d(x,y)$. If,
however, $I$-spin ({\em i.e.} $u \leftrightarrow d$) symmetry breaks
spontaneously, this last equality could be violated as well! Is our
argument then circular, and we proved $I$-spin symmetry only after
assuming it at an intermediate stage?\footnote{A similar argument was
  made recently \cite{ref:azcoiti} in a more forceful manner in
  connection with the second Vafa-Witten theorem, regarding the
  non-breaking of the discrete parity symmetry in QCD.} We do not
think that this is the case. The rationale behind our argument was
alluded to in Sec.~\ref{sec:quarkbi} above. Spontaneous symmetry
breaking is a collective long-range phenomenon manifest in the $V
\rightarrow \infty$ limit.

The inequalities are proved first for finite lattices. The proof is
valid for any volume $V$ (and also for any UV momentum cutoff), no
matter how large. We thus expect the inequalities to hold in the
$V \rightarrow \infty$ limit.

Vafa and Witten present also a more sophisticated and compelling argument.
They show that the propagator $S_\Delta^A(x,y)$ of a massive quark between
two regions of size $\Delta$ and around $x$ and $y$ is bound for any
$A_\mu(x)$ by
\begin{equation}  
   S^A_\Delta(x,y,m_0) \leq \frac{{\rm e}^{-m_0|x-y|}{\rm e}^{2m_0\Delta}}
      {m_0 \Delta^4} \qquad \left(m_u^{(0)} = m_d^{(0)} = m_0\right)
\>.
\label{eq:sA}
\end{equation}
The square of this bound applies to ${\rm tr} \left[ \Gamma
  S^A_\Delta(x,y,m_0) \Gamma S^A_\Delta(y,x,m_0) \right]$. Using the
weighted normalized averaging with $\intoned{\mu(A)}, {\rm d}\mu(A)
\geq 0$ in Eq.~(\ref{eq:corr1}), the same bound is also derived for
the ``smeared'' correlation functions for any current:
\begin{equation}
   \langle 0 | J^a_\Delta(x) [J^a_\Delta(y)]^{\dag} | 0 \rangle
      \leq \frac{{\rm e}^{-2m_0|x-y|} {\rm e}^{4m_0^2\Delta}}
      {m_0^2 \Delta^8}
\>.
\label{eq:smear}
\end{equation}
The finite smearing around $x$ and $y$ does not affect the asymptotic $|x-y|
\gg \Delta$ behavior, which is still given by Eq.~(\ref{eq:fa}).
Eq.~(\ref{eq:smear}) then implies that the lowest state in any mesonic
channel $q_i \bar{q}_j$ has a mass satisfying
\begin{equation}
   m_{ij}^{(0)} \geq 2m_0 \qquad ({\rm or} \,\, 
      m_{ij}^{(0)} \geq m_i^{(0)} + m_j^{(0)}
      \, \, {\rm for} \, \, m_i^{(0)} \neq m_j^{(0)})
\>.
\label{eq:msmear}
\end{equation}
Thus for non-vanishing bare quark masses there are no massless bosons
(and by a simple extension also no massless fermions) in the physical
spectrum.  In particular for $m_i^{(0)} = m_j^{(0)} \neq 0$ we could
have no scalar Goldstone bosons.

We now review the argument in detail. It is easy to show, first,
that the propagator for a colored scalar in a background field $A_\mu$ is
always maximized by the free propagator $\langle x | K_0^{-1} | y \rangle,
K_0 = - \partial_\mu \partial^\mu + m_0^2$ \cite{ref:kato}. The latter can
be cast in the form of a path integral expression \cite{ref:feynman} with a
positive definite integrand
\begin{eqnarray}
   D_0(x,y,m_0) &=& \langle x | K_0^{-1} | y \rangle = \frac{1}{2}
      \int_0^\infty {\rm d}T \, \langle x | {\rm e}^{-(1/2)TK_0}
      | y \rangle 
   \nonumber \\
   &=& \frac{1}{2} \int_0^\infty {\rm d}T \, \int_{x(0)=x}^{x(T)=y}
      \, {\rm d}x_\mu(t) \exp \left\{ {-\frac{1}{2} \left[ \int_0^T \, 
      \left(\frac{dX}{dt} \right)^2 {\rm d}t + m^2T \right]} \right\}
\>.
\label{eq:freeprop}
\end{eqnarray}
The effect of any external gauge field is to introduce just the additional
path-ordered ``phase factor'' for each path connecting $x(0) = x$ and $x(T)
= y$:
\begin{equation}
   P \left\{ \exp\left[i \int_{x(0)}^{x(T)}  \, {\rm d}x_\mu(t)
      A^a_\mu \lambda_a \right] \right\}
\>.
\end{equation}
This extra unitary matrix, when integrated over the positive measure
in Eq.~(\ref{eq:freeprop}), decreases the functional integral $\int
{\rm d}\mu(A)$ . Consequently $D_A(x,y,m_0) \leq D_0(x,y,m_0)$ (and in
particular $D_A(x,y,m_0) \leq \exp \left[ {-m_0|x-y|}\right]$).

The motion in Eq.~(\ref{eq:freeprop}) is formally that of a particle in
four space dimensions and a proper time. Vafa and Witten utilize an
analogous expression for the fermionic propagator 
\begin{equation}
   S^A_\Delta = \langle \alpha | (D+m)^{-1} | \beta \rangle =
      \int_0^\infty \, {\rm d}\tau \, {\rm e}^{-m\tau}
      \langle \alpha | {\rm e}^{-i\tau(-iD)} | \beta \rangle
\>,
\label{eq:sdelta}
\end{equation}
where $D$ is to be interpreted as the Dirac Hamiltonian in a (4+1)
formulation, and $\alpha$ and $\beta$ are smeared states of spatial
extent $\Delta$ around $x$ and $y$ respectively. The norms
$\langle\alpha|\alpha\rangle = \langle\beta|\beta\rangle =
\Delta^{-4}$ (norm $f$ = max$f$ is required for the purpose of the
subsequent discussion) are finite, and unitarity of
$\exp\left[{i\tau(iD)}\right]$ implies
\begin{equation}
   \langle\alpha | \exp\left[{i\tau(iD)}\right] 
      | \beta\rangle \leq \Delta^{-4}
\>.
\label{eq:eitau}
\end{equation}

The minimal distance between the supports of $\alpha$ and $\beta$ is $R =
|x-y| - 2\Delta$. Since the motion of the Dirac particle in 4+1
Minkowski space is causal, the minimal ``time'' required for propagation
between $\alpha$ and $\beta$ is $\tau_{{\rm min}} = R$. Using
Eq.~(\ref{eq:eitau}) and the lower limit $\tau_{{\rm min}} = R$ in
Eq.~(\ref{eq:sdelta}) we obtain the desired bound (\ref{eq:smear}) for the
smeared propagator
\begin{equation}
   S_\Delta^A \leq \left( \int_R^\infty \, {\rm d}\tau \, {\rm e}
      ^{-m_0 \tau} \right) \Delta^{-4} = \frac{{\rm e}^{-m_0 |x-y|}
      {\rm e}^{2m_0\Delta}}{m_0 \Delta^4}
\>.
\end{equation}
The smearing of the states $\alpha$ and $\beta$ can be made gauge invariant
by a refined procedure (analogous to that used by Schwinger
\cite{ref:schwinger}) without affecting the basic argument.

For the abelian case the current, {\em e.g.} $J_B^0 = \sum_i \psibar_i
\gamma_\mu \psi_i$ carries no net baryon number. However, Vafa and
Witten argue that if spontaneous baryon number violation is to occur
we will have a vacuum expectation value of some baryon number carrying
operator such as $T(x) = B_{ijk}(x) \simeq \psi^3(x)$.  This, in turn,
would imply that $\langle T_\Delta(x) T_\Delta(y) \rangle$ cannot have
an exponential falloff. Such an exponential falloff is implied by
Eq.~(\ref{eq:sA}) and $\langle T(x) T(y) \rangle \simeq
\intoned{\mu(A)} \left[S_A(x,y)\right]^6$.

The need for smearing the fermionic propagator arises from the
existence of zero modes of the Dirac operator $D_A$
\cite{ref:aharonov79} in any external $B$ field $B_z(x,y)$, with $\int
B_z(x,y) \, {\rm d}x \, {\rm d}y \geq \phi_0 \simeq (h/e)$.  This lowest
Landau level corresponds to the classical spiraling motion of
particles along $B$ with a Larmor radius $r \simeq B^{-1/2}$. If $B$
is aligned with ($x - y$) the particles originating in $x$
do not diverge geometrically with a resultant $|x-y|^{-2}$ factor in
the four-dimensional propagator. Rather, particles can be funneled
from $x$ to a spot of size $r^2 \simeq B^{-1}$ at $y$ (see
Fig.\ref{fig:bfield}) yielding an enhanced propagator $S_A \simeq B {\rm
  e}^{-m_0|x-y|}$ (instead of $S_A \simeq {\rm
  e}^{-m_0|x-y|}/|x-y|^2$).  The explicit $B$ dependence prevents the
derivation of the ${\rm e}^{-m_0|x-y|}$ bound directly for $S_A(x,y)$
and requires the spreading over a region of geometric size $\Delta^2$.
Once $B \gg \Delta^{-2}$, further focusing of the motion to regions
smaller than $\Delta^2$ will not enhance the $\alpha\rightarrow\beta$
propagation and the uniform ($B$ independent) bound
Eq.~(\ref{eq:smear}) is obtained.

The same type of spiraling motion also occurs for a charged scalar
particle in a magnetic field. However, in this case the motion has no
corresponding zero modes. The contribution of the spiraling
trajectories to the path integral expression for the bosonic
propagator $D_A(x,y)$ will decay as $\sim {\rm e}^{\Delta ET}$ at
large proper times and is therefore unimportant, explaining the fact
that $D_A(x,y)$ can be bound without utilizing the smearing procedure.

At first sight Eq.~(\ref{eq:msmear}) seems somewhat surprising. In QED
we have positronium states with nonzero binding and thus we might
expect Eq.~(\ref{eq:msmear}) to be violated in a weak coupling limit.
However, recall that $m_0$ refers to the {\em bare} mass. If we have
an electron and positron (or quark and antiquark), then besides the
attractive interaction between $e^+$ and $e^-$ there are the
self-energies of $e^+$ and $e^-$ Due to the vector nature of the gauge
interaction this self-interaction of a smeared (or otherwise
regularized) charge distribution is repulsive. Since
\begin{equation}
   \int \, {\rm d} \vecr \, {\rm d} \vecr{\, '} \rho_1(\vecr)
      K(\vecr - \vecr{\, '}) \rho_2(\vecr{\, '}) \leq \frac{1}{2} 
      \left[ \int \rho_1(\vecr) K(\vecr - \vecr{\, '}) 
      \rho_1(\vecr{\,'}) + \int \rho_2(\vecr) 
      K(\vecr - \vecr{\, '}) \rho_2(\vecr{\, '}) \right]
\end{equation}
for any positive kernel $K$, the positive contribution exceeds the
coulombic binding and Eq.~(\ref{eq:msmear}) does hold. Scalar
self-interactions can be attractive and hence this reasoning is
specific to gauge theories \cite{ref:vafanp}. Indeed for the scalar
case $m_0 \rightarrow g\phi + m_0$, and by choosing $\phi \simeq
-m_0/g$, we can generate propagators $D^\phi$ which do not have the
characteristic ${\rm e}^{-m_0|x-y|}$ behavior and the above proof of
the Vafa-Witten theorem fails. For a $\gamma_5$ Yukawa coupling, $m_0
\rightarrow m_0 + i g \gamma_5$, and we even lose the positivity of
the determinantal factor and of the measure.

%
\section{Baryon-meson mass inequalities from correlation functions}
\label{sec:mbineqcorr}

We next extend the discussion of Sec.~\ref{sec:quarkbi} to baryonic
correlation functions of currents trilinear in quark fields:
\begin{mathletters}
   \begin{equation}
      F^B_{ijk} = \langle 0 | B_{ijk}(x) B_{ijk}^{\dag}(y) | 0 \rangle
   \end{equation}
   \begin{equation}
      B_{ijk} = \psi_a^i(x) \psi_b^j(x) \psi_c^k(x) \Gamma_{abc},
   \end{equation}
\end{mathletters}
with $\Gamma_{abc}$ indicating some matrix in spinor space plus the
$\epsilon_{abc}$ color factor. Eq.~(\ref{eq:fa}) implies that the
asymptotic behavior of $F^B_{ijk}(x,y)$ is prescribed by the lowest-lying
baryon in this channel (say the nucleon for $i = j = u, k = d$ and a
coupling to overall spin 1/2):
\begin{equation}
   F^B_{uud}(x,y) \approx {\rm e}^{-m_N|x-y|}, \qquad \qquad
      |x-y| \rightarrow \infty
\>.
\label{eq:fbasymp}
\end{equation}
Also in analogy with Eq.~(\ref{eq:corr1}) we have a path integral
representation of $F^B(x,y)$
\begin{equation}
   F^B_{ijk}(x,y) = \intoned{\mu(A)} S^i_A(x,y)_{aa'} S^j_A(x,y)_{bb'}
      S^k_A(x,y)_{cc'} \Gamma_{abc} \Gamma_{a'b'c'}
\>.
\label{eq:fbpathint}
\end{equation}
The strategy for deriving the inequalities is to compare this trilinear
expression with the positive definite quadratic expression for the
correlation function of the pseudoscalar in the $m_i = m_j$ limit
\begin{equation}
   F_{ij}^\pi(x,y) = \intoned{\mu(A)} \sum_{aa'} |S^i_A(x,y)_{aa'}|^2
\>.
\label{eq:fbcorr}
\end{equation}
If an inequality of the form
\begin{equation}
   F^B_{ijk}(x,y) \leq \left[ F_{ij}^\pi(x,y) \right]^{1/p}
\label{eq:fbfpi}
\end{equation}
can be derived, with $(1/p) > 1/2$, then Eqs.~(\ref{eq:fbasymp}) and
(\ref{eq:fa}) imply the inequality
\begin{equation}
   m_N > (1/p) m_\pi
\>.
\label{eq:mnpmpi}
\end{equation}
[If $(1/p) < 1/2$ the last relation still allows for $m_\pi \geq 2m_N$, in
which case the two nucleon threshold is the lowest physical state in the
$J^{(0^-)}_{u\bar{d}}$ channel and the inequality derived from
Eq.~(\ref{eq:fbfpi}) becomes the trivial $m_N \geq (2/p) m_N$.]

The baryonic correlation function in Eq.~(\ref{eq:fbfpi}) is readily
bound by
\begin{equation}
   F_{ijk}^B(x,y) \leq \intoned{\mu(A)} \left[ \sum_{aa'} 
      |S^i_A(x,y)_{aa'}|^2 \right]^{3/2}
\>.
\label{eq:fbbound}
\end{equation}
In order to bound the right hand side by $\left[ F_{ij}^\pi(x,y)
\right]^p$, Weingarten \cite{ref:weingarten} appeals to the H\"{o}lder
inequality \cite{ref:titch}
\begin{equation}
   \left| \intoned{\mu} fg\right| \leq \left(\intoned{\mu} |f|^p \right)^{1/p}
      \left(\intoned{\mu} |g|^{p/(p-1)} \right)^{[(p-1)/p]}
\>,
\label{eq:genholder}
\end{equation}
valid for $p>1$, and chooses
\begin{eqnarray}
   f &=& \left( \sum_{aa'}  |S^i_A(x,y)_{aa'}|^2 \right)^
      {\frac{n-3}{n-2}}
   \nonumber \\
   g &=& \left( \sum_{aa'}  |S^i_A(x,y)_{aa'}|^2 \right)^
      {\frac{3}{2} - \frac{n-3}{n-2}}
\end{eqnarray}
so that $\intoned{\mu}fg$ is the right hand side in Eq.~(\ref{eq:fbbound}).
For $p = (n-2) / (n-3)$ the general H\"{o}lder inequality (\ref{eq:genholder})
implies 
\begin{equation}
   F^B_{ijk} \leq \left\{  \intoned{\mu(A)} \sum_{aa'}
      |S^i_A(x,y)_{aa'}|^2 \right\}^{\frac{n-3}{n-2}}
      \left\{ \intoned{\mu(A)} \left[ \sum_{aa'} 
      |S^i_A(x,y)_{aa'}|^2 \right]^{\frac{n}{2}} \right \}
      ^{\frac{1}{n-2}}
\>.
\label{eq:holder}
\end{equation}
Note that the first term in curly brackets is simply $F_{ij}^\pi(x,y)$
which is then raised to a $1/p = (n-3)/(n-2)$ power. Let the number of
light degenerate flavors be larger than $n$ which we take to be even.
The second term in the square brackets in Eq.~(\ref{eq:holder}), in
which the integrand is raised to the $n/2$ power, then has a direct
physical interpretation. It is the two-point correlation function of
products of $n/2$ pseudoscalar currents with all flavors $i_l,j_m$
different:
\begin{equation}
   K(x,y) = \left\langle 0 \left| \left[ J^{ps}_{i_1 \bar{\jmath}_1}(x) 
      J^{ps}_{i_2 \bar{\jmath}_2}(x) \cdots J^{ps}_{i_{n/2} 
      \bar{\jmath}_{n/2}}(x) \right] \left[ J^{ps}_{i_1 \bar{\jmath}_1}(y) 
      J^{ps}_{i_2 \bar{\jmath}_2}(y) \cdots J^{ps}_{i_{n/2} 
      \bar{\jmath}_{n/2}}(y) \right]^{\dag} \right| 0 \right\rangle
\>.
\label{eq:kcorr}
\end{equation}
There is no possibility of contracting a quark and an antiquark
emanating from $x$ (or from $y$), and we also have a unique pattern of
contracting quark lines from $x$ with those from $y$ [see
Fig.~\ref{fig:flavor}(a)] so as to form $n/2$ loops with separate
color and spinor traces.

Each of these traces then yields a factor $ \sum_{aa'}
|S^i_A(x,y)_{aa'}|^2$ and altogether we have $ \left( \sum_{aa'} 
|S^i_A(x,y)_{aa'}|^2 \right)^{n/2}$ as required. If we now make the rather
weak assumption that the lattice regularized correlation function $K(x,y)$
is finite, we deduce from Eq.~(\ref{eq:holder}), that, up to a numerical
constant, Eq.~(\ref{eq:fbfpi}) is indeed valid. Finally, we conclude from
Eq.~(\ref{eq:mnpmpi}) that
\begin{equation}
   m_N \geq \frac{n-3}{n-2} \, m_\pi
\>,
\label{eq:mnbound}
\end{equation}
with $n$ the number of light degenerate quark flavors, which should be
even. Thus we can obtain $m_N \geq (3/4) m_\pi$ if $n = N_f = 6$. This
large number of light degenerate flavors required is associated with
the particular derivation. Thus, if we assume only two light ($u$ and
$d$) flavors we can still construct Eq.~(\ref{eq:kcorr}) by taking all
$i_1 \ldots i_{n/2} = u$ and $\bar{\jmath}_1 \ldots \bar{\jmath}_{n/2}
= d$ and avoiding any flavor ``disconnected'' $xx$ or $yy$
contractions.  We have, however, several contractions between $x$ and
$y$, yielding not only $\left\{ {\rm tr} \left[ S_A^{\dag}(x,y)
    S_A(x,y) \right] \right\}^{n/2}$ but also terms like ${\rm
  tr}(S^{\dag}_A S_A S^{\dag}_A S_A)$ [see Fig.~\ref{fig:flavor}].

We could improve the bound (\ref{eq:mnbound}) if, instead of $K(x,y) \leq$
constant, we appeal to the asymptotic behavior
\begin{equation}
   K(x,y) \simeq {\rm e}^{-m^{(0)}_{n/2}|x-y|}
\>,
\end{equation}
with $m^{(0)}_{n/2}$ the lowest mass state created by the action of
$(J^{ps}_{u\bar{d}})^{n/2}$ on the vacuum. The bound derived from
Eq.~(\ref{eq:holder}) would then read
\begin{equation}
   m_N \geq \left(\frac{n-3}{n-2}\right) m_\pi 
      + \left(\frac{1}{n-2}\right) m^{(0)}_{n/2}
\>.
\label{eq:mnbound2}
\end{equation}

If there were no bound states in the exotic channel with $\pi_1^+ \ldots
\pi_{n/2}^+$ quantum numbers then the lowest state is at threshold and
$m^{(0)}_{n/2} = (n/2) m_\pi$. In this case Eq.~(\ref{eq:mnbound2}) becomes
\begin{equation}
   m_N \geq (3/2) \, m_\pi
\>,
\end{equation}
the result suggested by the more heuristic discussion in Secs.
\ref{sec:mbineq} and \ref{sec:interflavor}. To our knowledge the
nonexistence of exotic $\pi^+ \pi^+$-type bound states has not been
proved.  This can be done in the large $N_c$ limit (as will be
discussed in Sec.~\ref{sec:largenc}), in which case
\begin{equation}
   m_N \geq \frac{N_c}{2} m_\pi
\>.
\label{eq:mnbound3}
\end{equation}
Weingarten \cite{ref:weingarten} makes the simple observation that if
$S_A(x,y)$ can be uniformly bound by an $A_\mu(x)$ independent constant,
\begin{equation}
   S_A(x,y) \leq {\rm constant}
\>,
\label{eq:sabound}
\end{equation}
then from Eqs.~(\ref{eq:fbpathint}) and (\ref{eq:fbcorr}) one can directly
show that
\begin{equation}
   F^B(x,y) \leq F^\pi(x,y)
\>,
\end{equation}
implying
\begin{equation}
   m_N \geq m_\pi
\>.
\end{equation}
We can utilize for Eq.~(\ref{eq:sabound}) the Vafa-Witten bound on
$S_A(x,y)$ [Eq.~(\ref{eq:sA})]. The requisite smearing of the points $x$
and $y$ into two regions of size $\Delta$ around $x$ and $y$ respectively
was shown not to effect the asymptotic $|x-y| \rightarrow \infty$ behavior
and the ensuing mass relations.

Weingarten \cite{ref:weingarten} has independently motivated the bound
(\ref{eq:mnbound3}) by considerations of lattice QCD:
$(\dslasha)_{{\rm lattice}} + m_0 = m_0 + R + iI$ (with $R,I$ being
Hermitian, and $R$ having a non-negative spectrum
\cite{ref:weinchal}).  Thus for $m_0 > 0$, $\dslasha + m_0$ has no
non-vanishing eigenvalues which implies that $S_A(x,y) = \langle x |
(\dslasha_A + m_0)^{-1} | y \rangle$ is regular and bounded for all
$A_\mu$ configurations. For $m_0 = 0$, $\dslash_A$ does in general
have zero modes which could be important for the issue of spontaneous
breaking of chiral symmetry.

%
\section{Mass inequalities and S$\chi$SB in QCD and vectorial theories}
\label{sec:sxsb}

The question of whether global axial symmetries are spontaneously
broken in QCD (or in other field theories) is of utmost importance. In
the $m_u^{(0)} = m_d^{(0)} = 0$ limit, QCD has the global axial SU(2)
symmetry generated by $I_5^+, I_5^-$, and $I_5^3$. If we can show that
this symmetry is spontaneously broken ({\em i.e.} S$\chi$SB occurs),
then QCD is quaranteed to reproduce Goldstone pions and the successful
rich phenomenology of soft pion theorems and current algebra
\cite{ref:adler} developed in the late sixties.

Conventionally, S$\chi$SB is achieved via formation of a $\qqb$
condensate in the QCD vacuum, $\langle \psi\psibar \rangle \neq 0$.
$\langle \psi\psibar \rangle$ is the order parameter for a phase
transition from an axial SU(2) symmetric phase at high temperature to
the broken phase at low temperature \cite{ref:pisarski}. $\langle
\psi\psibar \rangle$ (at zero temperature) serves, along with $\langle
F^2 \rangle$, as one of the nonperturbative inputs in the QCD sum
rules \cite{ref:shifman,ref:reinders}. Many attempts have been made to
prove $\langle \psi\psibar \rangle \neq 0$ in lattice QCD, to estimate
its magnitude, and its finite temperature behavior.  Evidently
\begin{equation}
   \langle \psi\psibar \rangle = \intoned{\mu(A)}
      \langle \psi\psibar \rangle_A
\>,
\end{equation}
and $\langle \psi\psibar \rangle_A$ was shown by Banks and Casher
\cite{ref:banks} to equal $\pi \rho(0)$, with $\rho(\lambda)$ the
spectral (eigenvalue) density of the Dirac operator $i \dslasha_A
\psi_A^\lambda = \lambda \psi_A^\lambda$. These authors also argued
that the same set of ``fluxon'' configurations of $A_\mu(x)$ are
responsible for both confinement and the requisite dense set of zeroes
of $\dslasha_A$. This supports earlier, more heuristic arguments
\cite{ref:casher} suggesting that confinement, or binding, of a
massless fermion in a vectorial theory necessarily leads to S$\chi$SB.

The issue of S$\chi$SB is also crucial for the case of composite
models for quarks and leptons \cite{ref:harari,ref:peskin,ref:bars}.
In such theories one assumes that the axial symmetry is {\em not}
broken. The existence of practically massless fermions ($m_e, m_u, m_d
\simeq$ a few MeV and even $m_b$ = 5 GeV are very small when compared
with the compositeness scale $\Lambda_p \geq$ TeV
\cite{ref:bars,ref:lane}) in the physical spectrum is then believed to
be a manifestation of these axial symmetries. However, the
baryon-meson mass inequalities such as $m_N \geq m_\pi$ provide an
additional strong argument for S$\chi$SB in QCD and similar vectorial
theories.

Let us assume that the axial isospin symmetry (which holds in the
limit $m_u^{(0)} = m_d^{(0)} = 0$) is not broken spontaneously, {\em
  i.e.}  $I_5^\alpha | 0 \rangle = 0$. This symmetry can then be
realized linearly in the massive spectrum by having degenerate parity
doublets.  However, one can show via the 't Hooft anomaly matching
condition \cite{ref:thooftcarg} that we must also have at least one
$I=1/2$ massless physical spin-1/2 fermion (the nucleon).

Anomaly constraints have been discussed extensively
\cite{ref:thooftcarg,ref:frishman,ref:colegros}. Following
Ref.~\cite{ref:colegros}, we consider the vertex
\begin{equation}
   \Gamma_{\mu\nu\lambda}(q_1 q_2 q_3) = \intd{x_1} \intd{x_2}
      {\rm e}^{i(q_1 x_1 + q_2 x_2)} \langle 0 | T \left[ J_\mu(x_1)
      J_\nu(x_2) J_\lambda(x_3) \right] | 0 \rangle
\end{equation}
of three identical flavor currents, which, in terms of chiral right and
left combinations are
\begin{equation}
   J^\mu = \psibar_i^a \gamma^\mu \left[ A_{+(ij)}(1 + \gamma_5)
      + A_{-(ij)}(1 - \gamma_5) \right] \psi_{aj}
\>,
\end{equation}
with $A_{\pm}$ matrices in flavor space. We take ${\rm tr}A_+ = {\rm tr}
A_- = 0$ to avoid the $\gamma_5$ anomaly involving SU(N$_c$) gluons
\begin{equation}
   \partial_\mu J^\mu = \alpha_c ({\rm tr} A_+ - {\rm tr} A_-)
      {\rm tr} F\bar{F} = 0
\>.
\end{equation}
It was shown \cite{ref:adler69,ref:bell,ref:georgi} that
$\Gamma_{\mu\nu\lambda}$ satisfies the anomalous Ward identity
\begin{equation}
   q^\lambda_3 \Gamma_{\mu\nu\lambda} = \frac{N_c}{\pi^2}
      {\rm tr}(A_+^3 - A_-^3)
\>.
\label{eq:wardid}
\end{equation}
The right hand side is given by the lowest triangular graph involving the
massless fundamental fermions of the theory.

Equation (\ref{eq:wardid}) is true to all orders
\cite{ref:colegros,ref:adbard} and most likely is a genuine
nonperturbative result. Equation (\ref{eq:wardid}) implies that some
invariant amplitudes in the decomposition of $\Gamma_{\mu\nu\lambda}$
have a pole at $(q_3)^2 = 0$ of prescribed residue, given by ``the
anomaly'', {\em i.e.} the right hand side of Eq.~(\ref{eq:wardid}).
Such a singularity can arise either by 
\newcounter{bean}
\begin{list}
{(\alph{bean})}{\usecounter{bean}\setlength{\rightmargin}{\leftmargin}}
\item having zero mass (pseudo-) scalar boson states with appropriate
  nonvanishing $\langle n^0 | J^+_\mu | 0 \rangle$ matrix elements; or
  by
\item having a multiplet of massless spin 1/2 physical fermions
  (``nucleons'') such that in this basis
\begin{equation}
   J^\mu = \psibar^i_N \gamma^\mu \left[ B_{+(ij)}(1 + \gamma_5)
      + B_{-(ij)}(1 - \gamma_5) \right] \psi^j_N
\>,
\end{equation}
with a ``matched anomaly''
\begin{equation}
   {\rm tr}(B_+^3 - B_-^3) = {\rm tr}(A_+^3 - A_-^3)
\>.
\label{eq:match}
\end{equation}
\end{list}   
Possibility (a) implies $Q|0\rangle \neq 0$ with $Q$ the charge $\int
{\rm d}^3x J_0(x)$, {\em i.e.} a spontaneous breaking, which we
assumed does not occur.  This leaves us then with possibility (b),
{\em i.e.} the need to have massless fermions satisfying
Eq.~(\ref{eq:match}). The last algebraic anomaly condition constrains
the various composite models of quarks and leptons where the latter
are the massless physical states $N_i$ \cite{ref:thooftcarg,ref:bars}.

Let us now appeal to a new element, namely to the nucleon-pion (or
fermion-boson) mass inequalities
\begin{equation}
   m_N \geq m_\pi \qquad (m_F \geq m_B )
\>.
\label{eq:fbineq}
\end{equation}
Possibility (b) of massless fermions can then be ruled out.
Eq.~(\ref{eq:fbineq}) implies that if the nucleons are massless so are
the pions. However, unless the matrix element $\langle 0 | J^5_\mu |
\pi \rangle \simeq f_\pi q_\mu$ accidentally vanishes and the pion
``decouples'', we then have spontaneous breaking of the axial symmetry
which is precisely the possibility that we were trying to exclude.

It has been shown \cite{ref:colewitt} that in the $N_c \rightarrow
\infty$ limit, that anomaly matching via possibility (b) is ruled out.
Amusingly in this limit we have, as indicated in
Sec.~\ref{sec:largenc} below, the inequality (\ref{eq:mnbound3}): $M_N
\geq (N_c/2)m_\pi$. It excludes, for infinite $N_c$ (and some very
small non-vanishing $m_u^{(0)}$ and $m_d^{(0)}$ and consequently
finite $m_\pi$) even finite baryon masses, so that alternative (b)
cannot {\em a fortiori} be realized.

It is interesting to see how the inequality (\ref{eq:fbineq}) is
satisfied if we introduce a small explicit breaking of the axial
symmetry via $m_u^{(0)} = m_d^{(0)} \neq 0$. In this case we have no
strict arguments for the vanishing of either $m_N$ and/or $m_\pi$.

From Eq.~(\ref{eq:mpi2}), we have $m_\pi \sim \sqrt{m^0}$. If the
nucleon mass is generated exclusively via the explicit chiral symmetry
breaking term $m^0 \psibar_q \psi_q$, then this term and $m_N
\psibar_N \psi_N$ both have to flip sign under the discrete
$\exp[{i\pi Q_5}]$ transformation. This suggests that $m_N \simeq a
m^0 < m_\pi \sim \sqrt{m^o}$ for small $m^0$, violating the inequality
$m_N \geq m_\pi$. This in turn rules out the possibility of explicit
chiral symmetry breaking only and S$\chi$SB is again indicated.

An alternate approach to proving S$\chi$SB \cite{ref:nussspiegel} would
be to start from the Weingarten inequality (\ref{eq:pscurrent}) for
the pseudoscalar and scalar two-point correlation functions:
\begin{equation}
   J_{ps}^{ij} = \psibar_i \gamma_5 \psi_j, \qquad
      J_{s}^{ij} = \psibar_i \psi_j; \qquad
      i = u, j = d
\>,
\end{equation}
and show that the inequality is strict     
\begin{equation}
    \langle 0 | J_{ps}^{i\bar{\jmath}}(x) J_{ps}^{i\bar{\jmath}}(y) 
      | 0 \rangle > \langle 0 | J_{s}^{i\bar{\jmath}}(x) 
      J_{s}^{i\bar{\jmath}}(y) | 0 \rangle
\>.
\label{eq:pss}
\end{equation}
If chiral symmetry remains unbroken ($Q_5 | 0 \rangle = 0$), we can show,
by utilizing
\begin{displaymath}
   {\rm e}^{-i\pi Q_5} J_s^{ij} {\rm e}^{i\pi Q_5} = J_{ps}^{ij}  
\end{displaymath}
that
\begin{equation}
    \langle 0 | J_{ps}^{i\bar{\jmath}}(x) J_{ps}^{i\bar{\jmath}}(y) 
      | 0 \rangle = \langle 0 | J_{s}^{i\bar{\jmath}}(x) 
      J_{s}^{i\bar{\jmath}}(y) | 0 \rangle
\>.
\end{equation}
Thus, proving Eq.~(\ref{eq:pss}) would amount to proving S$\chi$SB.

The fermionic propagator in a background field can in general be decomposed
in terms of the Dirac $\gamma$ matrices:
\begin{eqnarray}
   S_F(x,y;A) &=& S(x,y;A) I + \gamma_\mu v^\mu(x,y;A) 
      + \sigma_{\mu\nu} t^{\mu\nu}(x,y;A) \nonumber \\
      &+& \gamma_5 \gamma_\mu a^\mu(x,y;A)
      + \gamma_5 p(x,y;A)
\>.
\label{eq:fprop}
\end{eqnarray}
Substituting this last expression into Eq.~(\ref{eq:corr1}) with $\Gamma =
\gamma_5$ and $\Gamma = I$, and utilizing Eq.~(\ref{eq:tracegamma5}), we
find that $v^\mu$ and $a^\mu$ contribute equally to $(J_{ps}J_{ps})$ and
$(J_s J_s)$. However, $s, t^{\mu\nu},$ and $p$ contribute $|s|^2 +
|t^{\mu\nu}|^2 + |p|^2$ to the first correlation and $- |s|^2 -
|t^{\mu\nu}|^2 - |p|^2$ to the second. Thus, if we can pinpoint any set of
gauge field configurations for which
\begin{equation}
   \intoned{\mu(A)} ( |s|^2 + |t^{\mu\nu}|^2 + |p|^2 ) > 0
\label{eq:stp}
\end{equation}
then those configurations could generate S$\chi$SB by making the strict
inequality hold. Note that the positivity of expression (\ref{eq:stp})
and of ${\rm d}\mu(A)$ insures that the effect of such configurations
cannot be cancelled by some other configurations.

The topologically invariant relation
\begin{equation}  
 \intd{x} p(x,x;A) = \intd{x} {\rm tr}\left[S(x,x;A)\gamma_5 \right]
      = \frac{1}{m_0} \intd{x} F_{\mu\nu}(x) \tilde{F}_{\mu\nu}(x)
\end{equation}
suggests that if we have an instanton \cite{ref:belavin} -- anti-instanton
\cite{ref:callan} gas, the contribution of the region near any
pseudoparticle is
\begin{equation}
   \intd{x} p(x,x;A) \simeq \pm 1, \, \, \mbox{and} \,\,
      \intoned{\mu(A)} |p(x,x;A)|^2 > 0
\>.
\end{equation}
Indeed, it has been speculated by several authors
\cite{ref:callan,ref:caldi} that instantons may be the source of S$\chi$SB.
The detailed quark propagator structure in Eq.~(\ref{eq:fprop}) was
utilized in \cite{ref:rafecas,ref:ferrando}.

One may try showing directly the existence of a zero mass state in the
QCD spectrum in the $m_u^{(0)} = m_d^{(0)} = 0$ limit. Thus, if the
representation for the pseudoscalar propagator
\begin{displaymath}
    \langle J_{ps}^{u\bar{d}}(x) J_{ps}^{u\bar{d}}(y) \rangle
      = \intoned{\mu(A)} {\rm tr} \left[ S_A^{\dag}(x,y) S_A(x,y) \right]
\end{displaymath}
as a sum of positive definite contributions could be used to show a power
falloff of the correlation function, then the existence of massless
hadronic states with $0^-$ quantum numbers (massless pions or massless
nucleons) would follow.

The above program attempts to achieve the opposite goal as compared with
the original work of Vafa and Witten \cite{ref:vafanp} described in
Sec.~\ref{sec:vecsym}, where upper bounds on the euclidean correlation
function were found. It has not been realized in the form described above.
Vafa and Witten have, however, succeeded by combining the measure
positivity and various index theorems to prove a closely related result
\cite{ref:vafacmp} which we will briefly describe next. They considered
the $k$-point function
\begin{equation}
   S(x_1, \ldots, x_k) = \langle 0 | \psibar_1 \psi_2(x_1)
      \psibar_2 \psi_3(x_2) \ldots \psibar_k \psi_1(x_k) | 0 \rangle
\>.
\end{equation}
With all fermionic flavors distinct, there is a unique, cyclic pattern
of contractions. The $k$-point function then has the path integral
representation
\begin{equation}
   S(x_1, \ldots, x_k) = \intoned{\mu(A)} {\rm tr} \left[
      S_A(x_1,x_2) \ldots S_A(x_k,x_1) \right]
\>.
\label{eq:spathint}
\end{equation}
Integrating over all $x_i$ and using translational invariance defines
\begin{eqnarray}
   S(k) &=& \intd{x_1} \ldots \intd{x_{k-1}} S(x_1,x_2,\ldots,x_{k-1},
      x_k = 0)
   \nonumber \\
   &=& \frac{1}{V} \intd{x_1} \ldots \intd{x_k} 
      S(x_1,x_2) \ldots S(x_{k-1},x_k)
\>.
\label{eq:sk}
\end{eqnarray}
The idea is to prove that (after appropriate regularization)
\begin{equation}
   S(k) \geq C_k S_0(k)
\>,
\label{eq:sks0}
\end{equation}
with
\begin{equation} 
   S_0(k) = 4 d(R) \int \frac{{\rm d}^4p}{(2\pi)^4}
      \left( \frac{1}{p^2} \right)^{(k/2)}
\end{equation}
[here $d(R)$ is the dimension of the (color) representation of the
fermions], the corresponding expression for the free massless
fermions.  This implies that for $k \geq 4$, $S(k)$ has, like
$S_0(k)$, infrared divergences which in terms of the physical states
can be understood only if we had zero mass hadronic states. $S(k)$ can
be written, using Eqs.~(\ref{eq:spathint}) and (\ref{eq:sk}), as
\begin{mathletters}
   \begin{equation}
      S(k) = \intoned{\mu(A)} S_A(k)
   \end{equation}
   \begin{equation}
      S_A(k) = \frac{1}{V} \intd{x_1} \ldots \intd{x_k} {\rm tr} \left[ 
         S_A(x_1,x_2) \ldots S_A(x_k,x_1) \right] = \frac{1}{V}
         {\rm tr} \left[ \left(\frac{1}{D_A} \right)^k \right] \>,
   \end{equation}
\end{mathletters}
where the last trace and operator multiplication refer not only to
color and spinor space but $x$ space as well. To prove
Eq.~(\ref{eq:sks0}), it is clearly sufficient to show, due to the
measure positivity, that
\begin{equation}
   S_A(k) \geq C_k S_0(k)
\end{equation}
holds for each $A_\mu$ configuration separately.  $S_A(k)$ can be
expressed in terms of a Dirichlet sum over the eigenvalues of $D_A$:
\begin{equation}
   i D_A \psi = \lambda \psi
\end{equation}
\begin{equation}
   S_A(k) = \frac{1}{V} \sum_{i=1}^{\infty} \lambda_i^{-k}
\>,
\end{equation}
with $\lambda_1, \lambda_2, \ldots, \lambda_N$ the eigenvalues in ascending
order. Vafa and Witten then prove that the lowest eigenvalue $\lambda_1$
scales with the size of the box considered as
\begin{equation}
   \lambda_1 \leq \frac{C}{L} = C V ^{-1/4}
\>,
\label{eq:lambda1}
\end{equation}
with $C$ independent of the gauge field. This implies, for even $k$, that
$S(k) \geq C^{-k}V^{\frac{k-4}{4}}$ and the required infrared divergence
ensues when $k > 4$ (the limiting case of $k = 4$ can also be handled by a
more refined discussion).  We will not present here the proof of
Eq.~(\ref{eq:lambda1}) which involves topological considerations and
tracing out the flow of eigenvalues under gauge deformations and refer the
reader to the original paper \cite{ref:vafacmp}. 

All the above discussion, while strongly suggesting S$\chi$SB in QCD,
fails to demonstrate that $\langle\psibar\psi\rangle$ is indeed
non-zero. Recently Stern \cite{ref:stern} suggested a novel pattern of
S$\chi$SB in QCD with the pseudoscalars still being the
Nambu-Goldstone bosons associated with this spontaneous breaking, but
where $\langle\psibar\psi\rangle = 0$. Indeed, as Kogan, Kovner, and
Shifman \cite{ref:kovner} noted, there could be some residual
``custodial'' discrete $Z_N$ axial symmetry which allows only higher
$(\psibar\psi)^N$ order parameters to have non-vanishing VEVs. The new
scheme is phenomenologically interesting. In particular, since now
$m_\pi^2 = {\cal O}(m_q^2)$, larger values of $m^{(0)}_u$,
$m^{(0)}_d$, and a more symmetric $m^{(0)}_u$, $m^{(0)}_d$,
$m^{(0)}_s$ mass pattern would be implied.

However, as pointed out by Kogan, Kovner, and Shifman
\cite{ref:kovner}, the inequality
\begin{equation}
   C_A(x) \equiv \langle 0 | J^{\dag}_{\mu, A}(x) \, J_{\mu, A}(0)
      | 0 \rangle \leq \langle 0 | J^{\dag}_{ps}(x) 
      \, J_{ps}(0) | \rangle \equiv C_{ps}(x)
\>,
\label{eq:cacps}
\end{equation}
can be judiciously utilized to rule out this pattern. Both correlators
have their asymptotic behavior controlled by the lightest $0^-$ pion
states, {\em i.e.} $C_A(x), C_{ps}(x) \simeq {\rm e}^{-m_\pi|x|}$ for
$x \rightarrow\infty$. However, the vacuum-to-pion axial matrix
element{ has the conventional form
\begin{displaymath}
   \langle 0 | J_\mu^a | \pi^b \rangle = i q_\mu F_\pi \delta^{ab}
\>,
\end{displaymath}
with $F_\pi$ scaling like $\Lambda_{\text{QCD}}$ and not vanishing in
the $m_q \rightarrow 0$ limit; whereas the vacuum-to-pion pseudoscalar
matrix element behaves in this scheme as
\begin{displaymath}
   \langle 0 | J_{ps}^a | \pi^b \rangle = \delta_{ab} {\cal O}(m_q)
\>.
\end{displaymath}
If $M \simeq {\cal O}(\Lambda_{\text{QCD}})$ is the mass of the first
massive (non-pion) state in the $0^-$ channel, then for $m_\pi^{-1}
\gg x \gg M^{-1}$, we have [recalling that the pion intermediate state
still dominates $C_{ps}(x)$]
\begin{displaymath}
   C_{ps}(x) \simeq m_q {\rm e}^{-m_\pi|x|} \simeq
      m_\pi^2  {\rm e}^{-m_\pi|x|}
\>,
\end{displaymath}
and
\begin{displaymath}
   C_A(x) \simeq \frac{ {\rm e}^{-m_\pi|x|} }{x^2}
\>,
\end{displaymath}
and the inequality (\ref{eq:cacps}) is violated.

The analog of Eq.~(\ref{eq:fbineq}) holds also for composite models based
on an underlying vectorial gauge interaction \cite{ref:lieb}. This
interaction should confine the massless hyperons at a scale
$(\Lambda_p)^{-1}$, leaving us with the quarks, leptons, Higgs particles,
and conceivably also $W^{\pm}, Z^0$ (and even the photon in some models!)
as the low lying physical states.

Eq.~(\ref{eq:fbineq}) suggests that global axial symmetry, assumed to
protect the (almost!) massless fermions from acquiring large masses
$\sim \Lambda_p$, does spontaneously break down. The analog of
(\ref{eq:fbineq}) here is $m_F \geq m_H$ and the fact that we have no
Higgs particles lighter than the lightest quarks and leptons is
bothersome. (A scenario with only three Higgs states, which
disappeared in the process of SU(2) $\times$ U(1) breaking, appears to
conflict the existence of several fermionic generations.) Finally, we
note that the $m^{(1^-)} \geq m^{(0^-)}$ inequality suggests that any
composite vector particle should be massive, and gauge symmetries, and
in particular the exact U(1)$_{\text{EM}}$ (with $m_\gamma \leq
10^{-20}$ eV!  \cite{ref:goldhaber}) should not arise ``accidentally''
due to the existence of almost zero mass vectorial bound states.

All the constraints stemming from mass inequalities do not apply to
chiral preonic theories \cite{ref:gipson,ref:raby,ref:georgi86},
and/or preonic theories with fermions and bosons with Yukawa couplings
\cite{ref:pati}. In this case the measure positivity ${\rm d}\mu(A)$
is lost (inequalities cannot be proven in supersymmetric models). Also
in chiral models the pseudoscalar current $\psibar \gamma_5 \psi$
which was used extensively above vanishes identically. The mass
inequalities, together with the study of the anomaly constraints,
shifted the focus of research for composite models from the early work
on vectorial models \cite{ref:lieb,ref:zhou} to chiral and/or
fermion-boson composite models
\cite{ref:gipson,ref:georgi86,ref:gerard,ref:king} suggested
earlier.\footnote{Aharony {\em et al.} \cite{ref:aspy} found that
  QCD-type inequalities were still useful in SUSY theories, though in
  the $m_{\text{Higgsino}} \rightarrow\infty$ limit (see
  App.~\ref{app:fourfive}). Nishino \cite{ref:nishino} has shown,
  using the Vafa-Witten inequalities, that in SUSY theories parity is
  conserved.}  Indeed it was observed \cite{ref:cvetic} that in gauge
and scalar (not pseudoscalar!) field theories, one can then prove the
inequality
\begin{equation}
   \langle 0 | \phi^{\dag}(x) \phi(x) \phi^{\dag}(0)\phi(0) | 0
      \rangle
   \langle 0 | \psibar(x) \gamma_5 \psi(x) \psibar(0) \gamma_5 
      \psi(0) | 0 \rangle \geq
   \left| \langle 0 | \phi^{\dag}(x) \psi(x) \phi^{\dag}(0)\psi(0) | 0
      \rangle \right|^2
\>,
\end{equation}
by integrating over $\psi$, $\psibar$. Using the positivity of the
${\rm d}[A_\mu] {\rm d}[\phi] {\rm e}^{-S[A_\mu,\phi]}$ integration
and of the pseudoscalar correlator, the desired relation follows
readily as a Schwartz-type inequality. However, we {\em cannot} now
infer the mass inequality $m_F \geq (1/2)(m_\pi + m_\chi)$ between the
lightest particles in the $\phi^{\dag}\psi$, $\psibar\gamma_5\psi$,
and $\phi^{\dag}\phi$ channels, simply because the $\langle
\phi^{\dag} \phi(x)\phi^{\dag} \phi(0)\rangle$ correlator always has
the constant $\left(\langle 0 | \phi^{\dag}\phi|0\rangle\right)^2$
contribution due to the vacuum state, which cannot be avoided without
spoiling the positivity. This negates the claim made by Nussinov in
\cite{ref:nussplb139}.

Hsu \cite{ref:hsu} suggested that QCD, and in particular the
Vafa-Witten inequalities along with the measure positivity for SU(2)
chiral with an even number of flavors, can be used to exclude the
possibility that such strongly interacting chiral theories underlie
the standard EW model \cite{ref:abbott,ref:claudson}. Since his
analysis relies on Vafa-Witten upper bounds for {\em both} fermionic
and scalar propagators, one needs to choose a regularization point
where the $\lambda\phi^4$ coupling vanishes (otherwise the quadratic
integral defining $S_\phi[A_\mu(x,0)]$ cannot be performed).

For a long period it was not clear, in view of difficulties
encountered in putting the theory on the lattice \cite{ref:nielsen},
if consistent chiral gauge theories could be defined. However, the
recent domain wall fermions \cite{ref:kaplan} and overlap formalism
\cite{ref:neunara} put the (lattice) regularization of fermionic
theories with gauge interactions on a much more solid foundation.
These developments also allowed for lattice calculations incorporating
chiral symmetries in a more explicit manner
\cite{ref:shamir,ref:soni}. Most likely proper regularization of
chiral gauge theories \cite{ref:luschnpb549} will also soon be
feasible. Our preceding discussion then suggests that chiral composite
models for quarks and leptons may revive.

We will not pursue here the possibility that this novel approach could
``rigorize'' the derivation of (some) of the QCD inequalities. 

%
\section{Inequalities between masses of pseudoscalar mesons}
\label{sec:pseumeson}

In this section we conclude the discussion of mass inequalities in the
non-exotic $qqq$ (baryonic) and $\qqb$ (mesonic) channels, by proving
that \cite{ref:witt83}:
\begin{mathletters}
   \begin{equation}
      m_{i\bar{\jmath}}^{(0^-)} \geq \frac{1}{2} \left[ 
         m_{i\bar{\imath}}^{(0^-)} + m_{j\bar{\jmath}}^{(0^-)} \right],
   \label{eq:psineq}
   \end{equation}
with $m_{i\bar{\jmath}}^{(0^-)}$ the mass of the lowest lying
pseudoscalar meson in the $q_i \bar{q}_j$ channel; and
\cite{ref:witt83,ref:nussplb139}
   \begin{equation}
     m_{\pi^+} \geq m_{\pi^0}.
   \label{eq:pipiineq}
   \end{equation}
\end{mathletters}
The derivation of both inequalities relies on the positivity of the
integrand in the functional integral expressing correlation functions
of pseudoscalar currents. We have, however, to make the additional
assumption (a) that the $\qqb$ annihilation channels generating
the flavor disconnected part [Eq.~(\ref{eq:corr2})] make negligible
contributions. For the derivation of Eq.~(\ref{eq:pipiineq}) we could
assume instead (b) the validity of the soft pion expression
\cite{ref:das} for $(m_{\pi^+}^2 - m_{\pi^-}^2)$ in terms of the (axial)
vector spectral functions \cite{ref:witt83}. In order to derive
Eq.~(\ref{eq:psineq}) we compare the three correlation functions
\begin{mathletters}
   \begin{equation}
      \langle 0 | J_{i\bar{\jmath}}^p(x) \left[ J_{i\bar{\jmath}}^p(y) 
      \right]^{\dag} | 0 \rangle
      = \intoned{\mu(A)} {\rm tr} \left\{ \left[ S_A^i(x,y) \right]^{\dag}
      S_A^j(x,y) \right\}
   \label{eq:corrij} 
   \end{equation}
   \begin{equation}
      \langle 0 | J_{i\bar{\imath}}^p(x) \left[ J_{i\bar{\imath}}^p(y) 
      \right]^{\dag} | 0 \rangle
      = \intoned{\mu(A)} {\rm tr} \left\{ \left[ S_A^i(x,y) \right]^{\dag}
      S_A^i(x,y) \right\}
   \label{eq:corrii} 
   \end{equation}
   \begin{equation}
      \langle 0 | J_{j\bar{\jmath}}^p(x) \left[ J_{j\bar{\jmath}}^p(y) 
      \right]^{\dag} | 0 \rangle
      = \intoned{\mu(A)} {\rm tr} \left\{ \left[ S_A^j(x,y) \right]^{\dag}
      S_A^j(x,y) \right\}
   \label{eq:corrjj} 
   \end{equation}
\end{mathletters}
where we have left out, following our assumption (a), the flavor
disconnected contribution, {\em e.g.} $\intoned{\mu(A)} {\rm tr} \left[
  \gamma_5 S^i_A(x,x) \right] {\rm tr} \left[ \gamma_5 S^i_A(y,y)
\right]$ in Eq.~(\ref{eq:corrii}). The expressions in
Eqs.~(\ref{eq:corrii}) and (\ref{eq:corrjj}) have the form of perfect
squares of vectors $S^i_{A\alpha a}$ (or $S^j_{A\alpha a}$) with
$A_\mu(x)$ and $\alpha, a$ the spinor, color indices viewed as a joint
index. Equation (\ref{eq:corrij}) is the corresponding scalar product
$S^i_{A\alpha a} \cdot S^j_{A\alpha a}$. The Schwartz inequality
$(S^i)^2 (S^j)^2 \geq |S^i \cdot S^j|^2$ therefore implies that
\begin{equation}
   \left| \langle 0 | J_{i\bar{\jmath}}^p(x) \left[ J_{i\bar{\jmath}}^p(y) 
      \right]^{\dag} | 0 \rangle \right|^2 \leq \langle 0 | 
      J_{i\bar{\imath}}^p(x) 
      \left[ J_{i\bar{\imath}}^p(y) \right]^{\dag} | 0 \rangle 
      \langle 0 | J_{j\bar{\jmath}}^p(x) \left[ J_{j\bar{\jmath}}^p(y) 
      \right]^{\dag} | 0 \rangle
\label{eq:corrineq}
\end{equation}
and Eq.~(\ref{eq:psineq}) immediately follows from the last inequality by
using Eq.~(\ref{eq:fa}) and going to the $|x-y| \rightarrow \infty$ limit,
where
\begin{equation}
    \langle 0 | J_{i\bar{\jmath}}^p(x) \left[ J_{i\bar{\jmath}}^p(y) 
      \right]^{\dag} | 0 \rangle \simeq {\rm e}^{-m_{i\bar{\jmath}}^{(0^-)}
      |x-y|}
\>.
\label{eq:schwartz}
\end{equation}
Eq.~(\ref{eq:corrineq}) is the lowest in a hierarchy of inequalities
stating that all the principle minors of the matrix $M_{ij} = S^i
\cdot S^j$ are positive. However, as can be readily verified by
considering the 3 $\times$ 3 minors, no useful new mass inequalities
follow.

We have noted in our earlier more heuristic discussion of the
inequalities the need for making a ``Zweig rule'' assumption (a). It
amounts to suppressing the annihilation of $q_i \bar{q}_i$ into gluons
and allows us to consider the mesonic sector $m_{i\bar{\imath}} = q_i
\bar{q}_i$ to be distinct from $m^{(0)}$, the flavor vacuum sector.
This is valid particularly in the heavy flavor $Q\bar{Q}$ sector.

The inequalities are also automatically satisfied in a soft pion limit
with $\left[m_{ij}^{(0^-)} \right]^2 \simeq (f_\pi)^{-2} \langle
\psibar\psi \rangle \left[m_i^{(0)} + m_j^{(0)}\right]$
\cite{ref:gor}. We recall that in the actual comparison with particle
data in Sec.~\ref{sec:compdata}, the pseudoscalar inequalities were
indeed satisfied with a larger margin than the other inequalities.

We next prove Eq.~(\ref{eq:pipiineq}) along similar lines
\cite{ref:nussplb139}. We compare the propagators
\begin{mathletters}
   \begin{equation}
      \langle J^p_{\pi^+}(x) J^p_{\pi^-}(y) \rangle = \intoned{\mu(A)}
      {\rm tr} \left\{ \left[S_A^u(x,y)\right]^{\dag} S_A^d(x,y) \right\}
   \label{eq:propchgpi}
   \end{equation} 
and
   \begin{equation}
      \langle J^p_{\pi^0}(x) J^p_{\pi^0}(y) \rangle = \frac{1}{2}
      \intoned{\mu(A)} {\rm tr} \left\{ \left[S_A^u(x,y)\right]^{\dag} 
      S_A^u(x,y) + \left[S_A^d(x,y)\right]^{\dag} S_A^d(x,y) \right\},
   \label{eq:propneutpi}
   \end{equation} 
\end{mathletters}
with
\begin{mathletters}
   \begin{equation}
      J^p_{\pi^+} = \psibar_u \gamma_5 \psi_d
   \end{equation}
   \begin{equation}
      J^p_{\pi^0} = \frac{1}{\sqrt{2}} (\psibar_u \gamma_5 \psi_u
      - \psibar_d \gamma_5 \psi_d),
   \end{equation}
\end{mathletters}
{\em i.e.} the pseudoscalar currents with the quantum numbers of the
$\pi^+,\pi^0$ mesons respectively.  We can take into account
electromagnetic effects, which are the source of the $\pi^+ - \pi^0$
mass difference\footnote{It was realized early on that the Cottingham
  formula expressing the EM contributions to the mass differences,
  converges to the correct $\pi^+ - \pi^0$ mass difference
  \cite{ref:cotting,ref:harari67}, and fails for the $K^+ - K^0$ and $n
  -p$ difference -- for which $m^{(0)}_d - m^{(0)}_u$ makes the
  dominant contribution.}, by considering the gauge group to be
SU(3)$_C \times$ U(1)$_{\text{EM}}$. The path integral measure then
becomes
\begin{equation}
   {\rm d}\mu(A) = {\rm d}[A_\mu^C] \, {\rm d}[A_\mu^{{\rm EM}}] 
      {\rm e}^{-S_{{\rm YM}}(A_\mu^C)}{\rm e}^{-S_{{\rm EM}}
      (A_\mu^{{\rm EM}})} \prod_{j=1}^{N_f} {\rm Det} \left[
      \dslasha_{(A_C,A_{{\rm EM}})} + m_j \right]
\>,
\end{equation}
and the propagator in the joint external fields is
\begin{equation}
   S_A^j(x,y) = \left\langle x \left| 
      \left[ \dslasha_{(A_C,A_{{\rm EM}})} + m_j \right]^{-1}
      \right| y \right\rangle
\>.
\end{equation}
Since the EM interaction is vectorial, the arguments leading to the measure
positivity, ${\rm d}\mu(A) \geq 0$, and the positive definite form of the
integrand for the pseudoscalar correlation functions remains intact.
Comparing Eqs.~(\ref{eq:propchgpi}) and (\ref{eq:propneutpi}) we can then
show, from $a^2 + b^2 \geq 2 a \cdot b$, that
\begin{equation}
   \langle J^p_{\pi^0}(x) J^p_{\pi^0}(y) \rangle \geq 
      \langle J^p_{\pi^+}(x) J^p_{\pi^-}(y) \rangle 
\>,
\label{eq:jpipi}
\end{equation}
from which $m_{\pi^+} \geq m_{\pi^0}$ readily follows. 

The starting point for the original derivation of $m_{\pi^+} \geq
m_{\pi^0}$ by Witten \cite{ref:witt83} is the soft pion -- current algebra
relation \cite{ref:das}
\begin{equation}
   m_{\pi^+}^2 -  m_{\pi^0}^2 = \frac{e^2}{f_\pi^2} \int \frac{{\rm d}^4k}
      {k^2} \left[ \langle V^3_\mu(k) V^3_\mu(-k) \rangle -
      \langle A^3_\mu(k) A^3_\mu(-k) \rangle \right]
\>,
\label{eq:pimassdiff}
\end{equation}
expressing the $\pi^+ - \pi^0$ mass difference as an integral over the
momentum space vector and axial vector (isovector) correlation functions.
The latter are Fourier transforms of the euclidean space correlation
functions which can be expressed via Eq.~(\ref{eq:corr1}), so that we
have
\begin{eqnarray}
\lefteqn{\langle V^3_\mu(k) V^3_\mu(-k) \rangle -
      \langle A^3_\mu(k) A^3_\mu(-k) \rangle} \qquad \nonumber \\
   & & = \frac{2}{V} \intd{x}
      {\rm e}^{ikx} \intd{y} {\rm e}^{-iky}
   \nonumber \\
   & & \times \intoned{\mu(A)} {\rm tr} \left[ \gamma_\mu S_A(x,y) 
      \gamma_\mu S_A(y,x)
      - \gamma_\mu \gamma_5 S_A(x,y) 
      \gamma_\mu \gamma_5 S_A(y,x) \right]
   \nonumber \\
   & & = \frac{2}{V} \intd{x}
      {\rm e}^{ikx} \intd{y} {\rm e}^{-iky}
   \nonumber \\
   & &\times \intoned{\mu(A)} {\rm tr} \left[ \gamma_\mu S_A(x,y) 
      \gamma_\mu
      - \gamma_\mu \gamma_5 S_A(x,y) 
      \gamma_\mu \gamma_5 \right] S_A(y,x)
\>.
\label{eq:vavcorr}
\end{eqnarray}
To this lowest, $\alpha_{{\rm EM}}$ order, there are no flavor disconnected
contributions to the isovector correlation function. The factor of 2 comes
from $V_\mu = \psibar \gamma_\mu T^3 \psi$ with ${\rm tr}
\left[(T^3)^2\right] = 2$.

We note that in the present case -- unlike in all previous inequalities --
the actual sign of the expression in (\ref{eq:vavcorr}) is of crucial
importance. A minus sign coming from the fermion loop has been cancelled by
the fact that the euclidean currents are $V_\mu = i \psibar \gamma_\mu
\psi$ (and $A_\mu \simeq i \psibar \gamma_\mu \gamma_5 \psi$) and the $i^2$
supplies the extra minus sign. [$\gamma_\mu^M \rightarrow i \gamma_\mu^E$
is essential since the euclidean $\gamma_\mu$ are hermitian and satisfy $\{
\gamma_\mu^E,\gamma_\nu^E \} = \delta_{\mu\nu}$, whereas $\{
\gamma_\mu^M,\gamma_\nu^M \} = g_{\mu\nu}$.]

The difference $\left[ \gamma_\mu S_A(x,y) \gamma_\mu - \gamma_\mu
  \gamma_5 S_A(x,y) \gamma_\mu \gamma_5 \right]$ occurring in
Eq.~(\ref{eq:vavcorr}) is simply $\gamma_\mu E_A(x,y) \gamma_\mu$,
with $E_A(x,y) = \left\langle x \left| m_0(-D^2_A + m_0^2)^{-1}
  \right| y \right\rangle$, the $\gamma_5$ even part of $S_A(x,y) =
\left\langle x \left|( D_A + m_0 )^{-1}\right| y \right\rangle =
E_A(x,y) + O_A(x,y)$.  The odd $O_A(x,y)$ behaves like a product of an
odd number of $\gamma$ matrices. Thus ${\rm tr}(\gamma_\mu E_A
\gamma_\mu O_A) = 0$, and Eq.~(\ref{eq:vavcorr}) can be rewritten as
\begin{eqnarray}
   \langle V^3_\mu(k) V^3_\mu(-k) \rangle -
      \langle A^3_\mu(k) A^3_\mu(-k) \rangle &=& \frac{2}{V} \intd{x}
      {\rm e}^{ikx} \intd{y} {\rm e}^{-iky}
   \nonumber \\
   &\times& \intoned{\mu(A)} {\rm tr} \left[ \gamma_\mu E_A(x,y) 
      \gamma_\mu E_A(y,x) \right]
\>.
\end{eqnarray}
Viewing $M_\mu \equiv {\rm e}^{ikx} \gamma_\mu = {\rm e}^{ikx}
(\gamma_\mu)_{\alpha\alpha'}$ as a matrix in spinor $(\alpha\alpha')$ and
coordinate space jointly, we can rewrite this last expression as
\begin{equation}
   \langle V^3_\mu(k) V^3_\mu(-k) \rangle -
      \langle A^3_\mu(k) A^3_\mu(-k) \rangle = \frac{2}{V} \intoned{\mu(A)}
      {\rm tr} \left[ M_\mu E_A M_\mu^{\ast} E_A \right]
\>,
\end{equation}
where the last trace and matrix multiplications refer also to the
coordinates $x,y$ as indices.

The operator $E_A$ is positive definite, as can be seen by going to the basis
defined by $i D_A | \lambda_A \rangle = \lambda_A | \lambda_A \rangle$. In
this basis
\begin{displaymath}
   {\rm tr} \left[ M_\mu E_A M_\mu^{\ast} E_A \right] = 
      \sum_{\lambda_A, \lambda_{A'}} \left| (M_\mu)\lambda_A \lambda_{A'}
      \right|^2 \frac{m_0^2}{(\lambda_A^2 + m_0^2)(\lambda_{A'}^2 + m_0^2)}
\end{displaymath}
is manifestly positive. The measure positivity then implies that
$\intoned{\mu(A)}{\rm tr} \left[ M_\mu E_A M_\mu^{\ast} E_A \right]$ is
also positive (non-negative). We finally arrive at
\begin{equation}
   \langle V^3_\mu(k) V^3_\mu(-k) \rangle -
      \langle A^3_\mu(k) A^3_\mu(-k) \rangle \geq 0
\>,
\label{eq:vav}
\end{equation}
and $m_{\pi^+} \geq m_{\pi^0}$ follows from Eq.~(\ref{eq:pimassdiff}).

It is amusing to note that the Eq.~(\ref{eq:vav}) complements the
asymptotic chiral symmetry \cite{ref:das67} $\langle V^3_\mu(k)
V^3_\mu(-k) \rangle \simeq \langle A^3_\mu(k) A^3_\mu(-k) \rangle$, as
$k\rightarrow\infty$.  Such an asymptotic equality is indeed required
for the ${\rm d}^4k$ integration in Eq.~(\ref{eq:pimassdiff}) to
converge. It motivated the spectral function sum rules of Weinberg
\cite{ref:weinberg} some time ago.  Using unsubtracted dispersion
relations for $\langle V^3_\mu(k) V^3_\mu(-k) \rangle$ and $\langle
A^3_\mu(k) A^3_\mu(-k) \rangle$, the inequalities (\ref{eq:vav}) imply
\begin{equation}
   \intoned{\mu^2} \sigma_V(\mu^2)(\mu^2 + k^2)^{-1} \geq
      \intoned{\mu^2} \sigma_A(\mu^2)(\mu^2 + k^2)^{-1}
\>,
\end{equation}
with $\sigma_V (\sigma_A)$ the vector (axial vector) spectral functions.

The configuration space analog of the inequality 
\begin{equation}
   \langle A_\mu(x) A_\mu(0) \rangle \leq 
      \langle V_\mu(x) V_\mu(0) \rangle 
\label{eq:av}
\end{equation}
has been discussed at length \cite{ref:shuryak} but no convincing
proof exists at present. One difficulty is that the one pion
contribution to the ``longitudinal'' part of the axial correlator
$C_A(x) \simeq {\rm e}^{-m_\pi |x|}$ dominates at large distances and
an appropriate transverse projection of $\langle A_\mu A_\nu\rangle$
is required. (This is not the case in momentum space since the large
pion pole contribution at small $k$ is suppressed by $k^\mu k^\nu$
``derivative coupling'' factors.) 

Efforts to prove such an inequality are motivated not only by the
experimental fact that the lightest hadrons in the $1^-$, $1^+$
sectors satisfy $m_{a_1} > m_\rho$.  Precision tests of the weak
interactions indicate that the ``$S$ parameter''
\cite{ref:altarelli,ref:pesktake}
\begin{equation}
   S \equiv \frac{{\rm d}}{{\rm d}k^2} \left[F_V(k^2) - F_A(k^2)
      \right]|_{k^2=0}
\>,
\end{equation}
with $F_V$, $F_A$ the covariant (transverse) parts of $\langle
A_\mu(k) A_\mu(-k) \rangle$ and $\langle V_\mu(k) V_\mu(-k) \rangle$,
is negative. The second moment of the conjectured inequality
(\ref{eq:av}) will imply that $S > 0$ in all vectorial theories such
as the original technicolor models \cite{ref:eichten,ref:appelquist},
thereby ruling those out as viable mechanisms for dynamical breaking
of the EW symmetry of the Standard Model.

The previous derivation of $m_{\pi^+} \geq m_{\pi^0}$, based on
Eq.~(\ref{eq:jpipi}), did not utilize the soft pion limit (tantamount
to letting $m_u^{(0)} + m_d^{(0)} \rightarrow 0$), but only the weaker
assumption of isospin $|m_u^{(0)} - m_d^{(0)}| \rightarrow 0$. [In
particular isospin was utilized to argue that there are no
disconnected purely gluonic intermediate state contributions to
$\langle 0 | J^{p(\pi^0)}(y) | n \rangle.$]

At first sight the previous derivation appears to lead to a stronger
result. This however is not the case \cite{ref:cash}. The point is
that there are intermediate states with one photon which make
\begin{equation}
   \langle 0 | (\psibar_u \gamma_5 \psi_u - 
      \psibar_d \gamma_5 \psi_d) | \mbox{multigluon state}
      + \gamma \rangle \neq 0
\>,
\label{eq:mgs}
\end{equation}
and which contribute to the same order ${\cal O}(\alpha_{{\rm EM}})$.
(Because of charge conjugation parity and color neutrality, the lowest
perturbative state of this type consists of at least three gluons and
a photon.) This contribution could be neglected if we appeal to the
analog of the Zweig rule hypothesis (a) used above. Alternatively, we
could go to the soft $\pi^0$ limit in which case all the couplings of
the Goldstone particle to the photon and any neutral hadronic system
[the ``multigluon state'' in Eq.~(\ref{eq:mgs})] must vanish.

Since the EM interaction conserves $J^3_{\mu A}$, soft pion theorems
imply that the EM contributions to a massless neutral Goldstone
particle vanish to all orders. Thus, if $m_{\pi^0} (\alpha = 0) = 0$
then also $m_{\pi^0} (\alpha \ne 0) = 0$.  The result $m_{\pi^+}^2
\geq m_{\pi^0}^2$, and its analog in other vectorial gauge theories,
is therefore of great importance.  Otherwise $m_{\pi^+}^2 \leq
m_{\pi^0}^2 = 0$ and the $\pi^+$ becomes tachyonic.  A condensate of
charged pions could then form, leading potentially to spontaneous
violations of EM charge conservation.  Similar results like
$m_{\pi^+}^{(T)} \geq m_{\pi^0}^{(T)}$ could help fix the pattern of
symmetry breaking or ``vacuum alignment''
\cite{ref:peskin80,ref:preskil} in other vectorial theories such as
technicolor \cite{ref:weinberg2,ref:sus79}.

The inequality $m_{\pi^+} \geq m_{\pi^0}$ is one aspect of the general
property mentioned in Sec.~\ref{sec:vecsym} [see in particular
Eq.~(\ref{eq:msmear})] that vectorial interactions make positive
contributions to the masses of physical particles. We can motivate
this result (see Sec.~\ref{sec:vecsym}) by considering the EM
self-interactions and mutual interactions between arbitrary charge
distributions (representing {\em e.g.} the extended constituent quarks and
including effects of $\qqb$ pairs as well). This leads to the
conjecture that the EM contributions to hadronic masses are always
positive.

In order to extract genuine EM contributions to hadron masses we need to
form $\Delta I = 2$ mass combinations such as
\begin{eqnarray}
   \delta_2[\Sigma] &=& m_{\Sigma^+} + m_{\Sigma^-} - 2 m_{\Sigma^0}
   \nonumber \\
   \delta_2[\rho] &=& 2 ( m_{\rho^+} - m_{\rho^0})
\>,
\end{eqnarray}
in which the effects of $(m_u^{(0)} - m_d^{(0)}) \neq 0$ have been
cancelled to first order. (Next order QCD effects are smaller than the
observed splitting.) In terms of a naive quark model the $\Sigma$
isotriplets are $xuu$, $xud$, and $xdd$ states with $x = s,c,b
\ldots$, some heavy quark. The EM contribution to $\delta_2(\Sigma)$
comes from mutual $q_i,q_j, i \neq j$ interactions and we find
\begin{equation}
   \delta_2 \simeq \alpha | Q_u - Q_d|^2 \int \rho_n(\vecr) 
      \rho_n(\vec{r}{\, '}) / |\vecr - \vec{r}{\, '}| \geq 0
\>,
\end{equation}
with $\rho_n$ referring to the density of the light $u$ or $d$ quark,
which in the $I$-spin symmetric limit are equal. Our conjecture is then
that this positivity is not an artifact of the simple model but would
persist in the full-fledged theory. Experimentally \cite{ref:pdg}, we
find
\begin{mathletters}
   \begin{equation}
      \delta_2[\Sigma(1190)] = 1.5 \pm 0.18 \, \mbox{MeV}
   \end{equation}
   \begin{equation}
      \delta_2[\Sigma(1380)] = 2.6 \pm 2.1 \, \mbox{MeV}
   \end{equation}
   \begin{equation}
      \delta_2[\Sigma_c(2455)] = 2.0 \pm 1.6 \, \mbox{MeV}
   \end{equation}
   \begin{equation}
      \delta_2[\rho] = - 0.2 \pm 1.8 \, \mbox{MeV}.
   \end{equation}
\end{mathletters}
The positivity of $\delta_2[\Sigma(1190)]$ and
$\delta_2[\Sigma(1380)]$ are statistically very significant (in view
of the fairly small width). Estimating $m_{\rho^+} - m_{\rho^0}$ is
difficult because of the large widths $\Gamma_\rho \simeq 150$
MeV.\footnote{Also for the special case of the vectorial $\rho^+
  \rho^0 \rho^-$ multiplet there is a relatively important one photon
  annihilation contribution, which the argument presented above
  misses.} We also note that the positivity of the $\Delta I = 2$
energy shift should apply not only to the ground state, but also to
any excited states.

%
\section{The absence of spontaneous parity violation in QCD}
\label{sec:parity}

A beautiful result also proved by Vafa and Witten \cite{ref:vafaprl}
via the QCD inequalities technique is that there is no spontaneous
breaking of parity symmetry in vectorial theories.  Such a breaking
should manifest via a nonzero vacuum expectation value of some parity
odd operator $A$. The simplest candidates for such operators are the
local quantities constructed from the gauge fields only
\begin{equation}
   A = \tilde{F}F = \epsilon^{\mu\nu\lambda\sigma} F_{\mu\nu}
      F_{\lambda\sigma}, \epsilon^{\mu\nu\alpha\beta} D_\sigma
      F_{\mu\nu} D^\sigma F_{\alpha\beta}, \ldots
\label{eq:aforms}
\end{equation}
Alternatively we could use quantities involving fermions such as $A =
\psibar \gamma_5 \psi$, with $\gamma_5 =
\epsilon^{\mu\nu\lambda\sigma} \gamma_\mu \gamma_\nu \gamma_\lambda
\gamma_\sigma$. The distinguishing feature of any parity odd operator
is that it includes an odd number of $\epsilon^{\mu\nu\lambda\sigma}$
tensors. While the $\gamma^\mu$ matrices, fields $A^\mu$, and metric
$g_{\mu\nu}$ are hermitian and real in the euclidean continuation,
$\epsilon^{\mu\nu\lambda\sigma}$, which transforms like ${\rm d}x^1
{\rm d}x^2 {\rm d}x^3 {\rm d}x^4$, picks up a factor of $i$ and
becomes purely imaginary. Consequently $A_{\text{euc}}$ corresponding
to all parity odd operators is imaginary.

If $\langle A \rangle \neq 0$, then from ordinary first order
perturbation theory it follows that a (hypothetical) theory with a
modified Lagrangian
\begin{equation}
   {\cal L}_\lambda = {\cal L}_{\text{QCD}} (\lambda = 0)
      + \lambda \int {\rm d}^3 x \, A
\end{equation}
has, for small $\lambda$, a shifted vacuum energy density
\begin{equation}
   E(\lambda) = E(0) + \lambda \langle A \rangle
\>,
\end{equation}
with $\langle A \rangle = \langle A \rangle_0$ still computed for the
$\lambda = 0$ vacuum. For a $\lambda$ with an appropriate sign,
$\lambda\langle A \rangle < 0$ and $E(\lambda) < E(0)$ -- just as the
energy of a spontaneously magnetized ferromagnet is lowered by adding an
external $\vec{B}$ field anti-parallel to the magnetization $\vec{\mu}$.

The euclidean path integral representation of the vacuum energy density
for ${\cal L}_\lambda$ is
\begin{equation}
   E(\lambda) = -\frac{1}{V} \ln \intD{A_\mu}
      \intD{\psi} \intD{\psibar} {\rm e}^{-\intd{x} {\cal L}_\lambda}
\>.
\end{equation}
We will next show that $E(\lambda) > E(0)$, negating the possibility that
$E(\lambda) < E(0)$ and forbidding $\langle A \rangle \neq 0$.

If $A$ is of the type indicated in Eq.~(\ref{eq:aforms}), the $\intD{\psi}
\intD{\psibar}$ integration can be carried out and
\begin{equation}
   E(\lambda) = -\frac{1}{V} \ln \intD{A_\mu}
      {\rm e}^{i\lambda A} \geq -\frac{1}{V} \ln \intD{A_\mu}
      = E(0)
\>,
\label{eq:elambda}
\end{equation}
since the oscillatory factor ${\rm e}^{i\lambda A}$ always decreases
the integral. The same is true if $\langle A \rangle = \langle \psibar
\gamma_5 \psi\rangle \neq 0$. Adding $\lambda A$ to the Lagrangian is
equivalent to introducing complex masses in the determinantal factor
$\prod_i {\rm Det}( \dslash + m_i )$ ({\em e.g.} $m_i \rightarrow m_i
+ i\lambda$) which again destroys the positivity of the determinantal
factor, reduces $Z(\lambda)$, and increases $E(\lambda)$, so that
Eq.~(\ref{eq:elambda}) holds.

The fact that $Z(\lambda)$ is minimal for $\lambda = 0$ allows also
the proof that of all the different QCD $\theta$ vacua (which can be
generated by adding the topological density $\theta \tilde{F}F$ term
to the Lagrangian), the one with the lowest vacuum energy is that with
$\theta = 0$. (Also $\theta = \pi$ was excluded by similar arguments
\cite{ref:sikivie}, a result suggested by dilute instanton
calculations \cite{ref:callan} and other considerations
\cite{ref:wittven}).

%
\section{QCD inequalities and the large $N_c$ limit} 
\label{sec:largenc}

We have commented on several occasions on the inequalities in the $N_c
\rightarrow\infty$ limit. We will next show \cite{ref:nusssath} that in
this limit we have indeed the stronger version of the baryon-meson
inequality
\begin{equation}
   m_B \geq \frac{N_c}{2} \, m_\pi
\>.
\label{eq:mB}
\end{equation}
We will also indicate that the interflavor mesonic mass inequalities
could be extended to non-pseudoscalars.

The Schwartz inequality implies that the baryonic correlation function
$\langle B(x) B(y) \rangle = \intoned{\mu(A)} \left[ \Gamma S_A(x,y)^n
  \Gamma \right]$ with $S_A(x,y)$ the common fermion propagator, is
smaller than $\left[ \intoned{\mu(A)} {\rm tr}^n
  S_A^{\dag}(x,y)S_A(x,y) \right]^{1/2}$. This last expression
represents the correlation function of products of $n$ pseudoscalar
currents, $\left\langle 0 \left| \left[J^{ps}(x) \right]^n
    \left[J^{ps}(y) \right]^n \right| 0 \right\rangle$. We can show
that in the large $N_c$ limit this joint correlation function
effectively factors into a product of $[\langle J^{ps}(x) J^{ps}(y)
\rangle]^n$.  Specifically, gluon exchanges between different $\qqb$
bubbles in Fig.~\ref{fig:fplanar} do not modify the energy $N_c
m_\pi$ by more than ${\cal O}(1)$, of the system viewed as separately
propagating $N_c$ $\qqb$ pairs. Indeed, the color trace counting
argument \cite{ref:coleman,ref:thooftnpb75} indicates that the interaction
energy between any $(\qqb) (\qqb)$ pair of bubbles is ${\cal
  O}(1/N_c)$.  Since we have $N_c/2$ pairs, which can interact via
planar diagrams of gluon exchanges, the total effect is ${\cal O}(1)$.
The interaction of triplets (or a higher number) of bubbles is ${\cal
  O}(1/N_c^2)$, or ${\cal O}(1/N_c^k)$, $k > 2$, and is less
important. Thus in the large $|x-y|$ limit $\intoned{\mu(A)}{\rm tr}^n
S_A^{\dag}(x,y)S_A(x,y) \simeq {\rm e}^{-N m_\pi}$, and since $\langle
B(x) B(y) \rangle \simeq {\rm e}^{-m_B |x-y|}$, Eq.~(\ref{eq:mB})
follows.

To extend the interflavor mass relations consider the following three
correlation functions of four flavor currents (see
Fig.~\ref{fig:fplanar}): 
\begin{eqnarray*}
   F_1 &=& \left\langle 0 \left| J_{\pi^+}(x_1) J_{K^+}(x_2)
      J_{\pi^-}(y_1) J_{K^-}(y_2) \right| 0 \right\rangle
   \\
   F_2 &=& \left\langle 0 \left| J_{K^-}(x_1) J_{K^+}(x_2)
      J_{K^-}(y_1) J_{K^+}(y_2) \right| 0 \right\rangle
   \\
   F_3 &=& \left\langle 0 \left| J_{\pi^-}(x_1) J_{\pi^+}(x_2)
      J_{\pi^-}(y_1) J_{\pi^+}(y_2) \right| 0 \right\rangle
\>,
\end{eqnarray*}
with the euclidean points $x_1,x_2,y_1,y_2$ forming a rectangle of height
$T$ and width $\tau$ in the, for example, $3-4$ plane. In the large
$N_c$ limit, quark annihilation is suppressed and we have the planar quark
diagrams indicating the possible contractions. In the case of $F_2$ we
specifically choose the contraction with $s\bar{s}$ exchanged in the
vertical ($T$) direction, so that in all three cases we have $u\bar{u}$
exchanged in the horizontal ($\tau$) direction. We can use the euclidean
Hamiltonian to evolve the system in $T$, {\em e.g.}
\begin{eqnarray}
   F_1 &=& \sum_n \left\langle 0 \left| J_{\pi^+}(x_1) J_{K^+}(x_2)
      \right| n \right\rangle \left\langle n \left|
      J_{\pi^-}(x_1) J_{K^-}(x_2) \right| 0 \right\rangle 
      {\rm e}^{-E_n T}
   \nonumber \\
   &=& \sum_n \left| \left\langle 0 \left| J_{\pi^-}(x_1) J_{K^+}(x_2)
      \right| n \right\rangle \right|^2  {\rm e}^{-E_n T}
\>,
\label{eq:f1}
\end{eqnarray}
with $ \left\langle 0 \left| J_{\pi^-}(x_1) J_{K^+}(x_2) \right| n
\right\rangle$ independent, by translational invariance, of the specific
location in $T$ of $x_1,x_2$.

We can also evolve the system in $\tau$ to find analogous expressions:
\begin{mathletters}
   \begin{equation}
      F_1 = \sum_{n'} \left\langle 0 \left| J_{K^+}(x_2) J_{K^-}(y_2)
      \right| n' \right\rangle \left\langle n' \left| J_{\pi^+}(x_2) 
      J_{\pi^-}(y_2) \right| 0 \right\rangle {\rm e}^{-E_{n'} \tau}
   \end{equation}
   \begin{equation}
      F_2 = \sum_{n'} \left| \left\langle 0 \left| J_{K^+}(x_2) 
      J_{K^-}(y_2) \right| n' \right\rangle \right|^2  
      {\rm e}^{-E_{n'} \tau}
   \label{eq:f2}
   \end{equation}
   \begin{equation}
      F_3 = \sum_{n'} \left| \left\langle 0 \left| J_{\pi^+}(x_2) 
      J_{\pi^-}(y_2) \right| n' \right\rangle \right|^2  
      {\rm e}^{-E_{n'} \tau}
   \label{eq:f3}
   \end{equation}
\end{mathletters}
where, due to the choice of $F_2$, we have the same states $|n'\rangle$
(with $\bar{u}u$ flavor) in all three cases. From the Schwartz inequality
we have
\begin{equation}
   |F_1|^2 \leq F_2 F_3
\end{equation}
which, by going back to the representation like Eq.~(\ref{eq:f1}) for $F_1,
F_2,$ and $F_3$, yields inequalities between the masses of the lowest
intermediate states for the three cases
\begin{equation}
   m_{u\bar{s}}^{(0)} \geq \frac{1}{2} \left[ m_{u\bar{u}}^{(0)} +
       m_{s\bar{s}}^{(0)} \right]
\>,
\label{eq:mus}
\end{equation}
since by construction the intermediate ``$T$-channel'' states for $F_2$ are
$s\bar{s}$ type states (and $u\bar{u}$ for $F_3$). As emphasized earlier in
the large $N_c$ limit the $m_{u\bar{u}}, m_{s\bar{s}}$ sectors are distinct
from $m^{(0)}$, the flavor vacuum, and from each other. To get perfect
squares in Eqs.~(\ref{eq:f2}) and (\ref{eq:f3}), we need to have identical
currents at $x_1,x_2$ and $y_1,y_2$. Thus the quantum numbers of these
$u\bar{u}, s\bar{s}$ (and $u\bar{s}$) states are those of $J_\pi J_\pi$, $C
= +,$ even G-parity, {\em e.g.} $0^{++}$ states.

Indeed, similar unitarity motivated inequalities were suggested
earlier in the context of dual models and were applied to intercepts
of Regge trajectories on which $0^{++}$ states lie \cite{ref:geffen},
and it has been conjectured that in the large $N_c$ limit QCD
approaches a dual, string-like model. The present derivation
pinpoints, however, the crucial element for deriving
Eq.~(\ref{eq:mus}). It is that in the large $N_c$ limit we can
meaningfully separate the various planar contributions to the
correlation function and each evolves separately under the appropriate
$H_{i\bar{\imath}}$ (or $H_{i\bar{\jmath}}$) Hamiltonian.

We note that the last argument did not utilize the measure positivity
in the euclidean path integral, but rather the unitarity based, more
general, spectral positivity. QCD inequalities have also been applied
in the large $N_c$ limit in (1+1) dimensions \cite{ref:aoki}.

%
\section{QCD inequalities for glueballs} 
\label{sec:glue}

QCD also predicts ``glueball'' states consisting of gluonic degrees of
freedom and no quark flavors. Experimental evidence for such states
exists at present \cite{ref:crystal} -- but mixing with the many
$\qqb$ and/or $\qqb \, \qqb$ states in the 1 -- 1.5 GeV region (where
the glueball spectrum is estimated to start \cite{ref:meshkov})
complicates the analysis.

The local density $F^2 = F_{\mu\nu}^a F_{\mu\nu}^a$ becomes, in the
euclidean domain, $\vec{E}^2 + \vec{B}^2$ and maximizes all other $F$
bilinears such as $\tilde{F} F \rightarrow i \vec{E} \cdot \vec{B}$. Thus,
as pointed out by Muzinich and Nair \cite{ref:muzinich,ref:muzinich2}, we
have $F^2(x) F^2(y) \geq \tilde{F}F(x) \tilde{F}F(y)$. Due to measure
positivity this translates into an inequality between the corresponding
correlation functions
\begin{eqnarray}
   \langle 0 | F^2(x) F^2(y) | 0 \rangle &=& \intoned{\mu(A)}
      F^2(x) F^2(y) 
   \nonumber \\ 
      &\geq& \intoned{\mu(A)} \tilde{F}F(x) \tilde{F}F(y)
      = \langle 0 | \tilde{F}F(x) \tilde{F}F(y) | 0 \rangle
\>.
\label{eq:ffff}
\end{eqnarray}
$F^2(x), \tilde{F}F(x)$ create from the vacuum $0^{++}$ and $0^{-+}$ states
respectively. Thus, following the standard argument, we derive the
inequality
\begin{equation}
   m_{{\rm gb}}^{(0^{++})} \leq  m_{{\rm gb}}^{(0^{-+})}
\label{eq:mgb}
\end{equation}
between the masses of the lowest glueball states in the respective
channels. Confinement and the existence of a mass gap in the gluonic sector
prevents interpretation of Eq.~(\ref{eq:mgb}) as a statement about gluon
thresholds. 

Any polynomial in the component of $F_{\mu\nu}$, used to create a glueball
state of arbitrary $J^{PC}$, is bound by an appropriate power of $F^2$.
Thus Eq.~(\ref{eq:mgb}) can be generalized to
\begin{equation}
   m_{{\rm gb}}^{(0^{++})} \leq \mbox{(mass of any glueball)}
\>.
\label{eq:manygb}
\end{equation}

All these correlation function inequalities are automatically
satisfied if $F^2$ has a nonvanishing vacuum expectation value:
$\langle 0 | F^2 |0 \rangle \neq 0$. In this case, correlation
functions involving $F^2$ [or $(F^2)^n$] have constant terms as $|x-y|
\rightarrow \infty$, corresponding to the obvious zero mass $0^{++}$
``state'': the vacuum, and Eq.~(\ref{eq:manygb}) need not hold.

In Ref.~\cite{ref:west}, West suggested that the inequalities
(\ref{eq:mgb}) and (\ref{eq:manygb}) can be recovered if we use the
time derivative of the correlator:
\begin{eqnarray}
   \frac{{\rm d}}{{\rm d}t} \langle 0 | F^2(0) F^2(t,\vec{0})| 0
      \rangle &=& \langle 0 | F^2(0) [H,F^2(t,\vec{0})]| 0 \rangle
   \nonumber \\
      &=& \langle 0 | F^2(0) H F^2(t,\vec{0})| 0 \rangle 
\>,
\end{eqnarray}
and the corresponding expression for $\tilde{F}F$. The offending
constant term in the intermediate summation would be absent in the new
inequalities. The idea then is to prove that
\begin{equation}
   \langle 0 | FF(0) H FF(t,\vec{0})| 0 \rangle >
      \langle 0 | \tilde{F}F(0) H \tilde{F}F(t,\vec{0})| 0 \rangle
\>.
\label{eq:fhf}
\end{equation}
This could be accomplished by using the positivity of the Hamiltonian
for each $A_\mu(x)$ background configuration separately in a path
integral representation of the last two correlators. There are however
subtleties in this argument which West tried to address.

In general, while the Hamiltonian is positive when operating on
physical states, the latter states are obtained only after the path
integral has been carried out, and the pointwise positivity of $H$ for
every $A_\mu(x)$ configuration is not {\em a priori} guaranteed. It is
the case for the pure Yang-Mills theory if $H = \int \, {\rm d}x (
\vec{E}^2 + \vec{B}^2)$;\footnote{Actually it is the euclidean
  Lagrangian which has this positive definite $\vec{E}^2 + \vec{B}^2$
  form, and the euclidean Hamiltonian then has the form $\int
  (\vec{B}^2 - \vec{E}^2)$. A non-vanishing $\langle 0 | H | 0 \rangle$
  VEV then amounts to the nontrivial dynamical assumption of the dual
  Meissner effect in the QCD vacuum -- namely the preponderance of
  large {\em magnetic} fluctuations.}  however, in this case
$H|0\rangle = 0$, which has been assumed in the above argument, holds
only after an appropriate subtraction has been made. Such a common
subtraction will then modify Eq.~(\ref{eq:fhf}) to
\begin{eqnarray}
   &{}& \langle 0 | FF(0) H FF(t,\vec{0})| 0 \rangle
      -c \langle 0 | FF(0) FF(t,\vec{0})| 0 \rangle >
   \nonumber \\
   &{}& \langle 0 | \tilde{F}F(0) H \tilde{F}F(t,\vec{0})| 0 \rangle
      -c \langle 0 | \tilde{F}F(0)\tilde{F}F(t,\vec{0})| 0 \rangle
\>,
\end{eqnarray}
with $c$ positive. Due to Eq.~(\ref{eq:ffff}) this may invalidate the
desired inequality.

In passing we remark that $\langle 0 | F^2 |0 \rangle \neq 0$ has been
conjectured to be the driving mechanism for confinement. Its value
controls the bag constant \cite{ref:hasenfrantz}, the string tension
in extended hadronic states \cite{ref:johnson}, and enters QCD sum
rules.

The glueball sector can be divided into ``even'' and ``odd'' parts
consisting of states with quantum numbers of two or three gluons with
appropriate orbital and spin angular momentum. Thus the $J^{PC}$
states $0^{++}, 2^{++}, 0^{-+}, 2^{-+}$ belong to the even part and
$1^{--}$ to the odd. If we adopt a simple ``constituent'' description
of the lowest even (odd) states in terms of two (three) gluons bound
by one gluon exchange potentials, then the arguments of
Sec.~\ref{sec:mbineq} leading to the baryon-meson mass inequality $m_B
\geq (3/2) m_M$ can be repeated here to prove the inequality
\begin{equation}
   m_{3g}^{(0)} \geq \frac{3}{2} m_{2g}^{(0)} \label{eq:m3gm2g}
\end{equation} 
between the masses of the lowest odd and even states. Just as in the
meson (baryon) $3 \otimes \bar{3} \rightarrow 1 (3 \otimes \bar{3}
\rightarrow \bar{3}$), we have in the 2g (3g) state $8 \otimes \bar{8}
\rightarrow 1 (8 \otimes \bar{8} \rightarrow \bar{8}$) and we can
repeat the argument by replacing $\lambda_1 \cdot \lambda_2$ in the
fundamental representation by $\Lambda_1 \cdot \Lambda_2$ in the
adjoint representation of SU(3)$_C$.

Evidently, as the previous discussion suggests, a constituent valence
model for glueballs is on much weaker footing than that for mesons and
baryons. In particular, the latter (but not the former!) can be
justified in the large $N_c$ limit. If the even glueball spectrum
starts at 1.5 -- 2 GeV, then Eq.~(\ref{eq:m3gm2g}) suggests that the
lightest $1^{--}$ glueball may be near the J/$\psi$, $\psi'$ states.
Its putative mixing with the latter could then modify the perturbative
approach to J/$\psi\rightarrow 3g$ decays and/or other matrix elements
\cite{ref:brodsky}.

%
\section{QCD inequalities in the exotic sector}
\label{sec:exotic}

So far we have focused on correlation functions of $q_a \bar{q}_a$
bilinear and $\epsilon^{abc}q_a q_b q_c$ trilinear currents. In the
following we utilize $\langle J^{\text{ex}}(x)
J^{\text{ex}}(y)\rangle$ with quartic ``currents'':
\begin{equation}
   J^{\text{ex}} \sim q_i(x) 
      \bar{q}_j(x) q_k(x) \bar{q}_l(x)
\>,
\end{equation}
to probe the ``exotic'' $M_{i\bar{\jmath}k\bar{l}}$ sector.  We also
address the possibility that the lowest lying intermediate state in
$\langle J(x) J(y) \rangle$ is not a stable one particle state, but
rather a two particle threshold at $m_1 + m_2$.

This last issue has been encountered already for non-exotic currents. Thus,
the lowest lying state in the vector $1^- \, u\bar{d}$ channel is {\em
not} the $\rho(760)$ but the $\pi\pi$ threshold at $\approx 280$ MeV. Since
this happens because $m_\rho > 2m_\pi$, the Weingarten inequality $m_\rho
\geq m_\pi$ is not invalidated.

Let us next analyze the general case $\langle 0 | J_a(x) J^{\dag}_a(y)
| 0 \rangle \geq \langle 0 | J_b(x) J^{\dag}_b(y) | 0 \rangle$, when
the lowest, non-exotic mesons in {\em both} channels are resonances of
widths $\Gamma_a (\Gamma_b)$. If the states are narrow, {\em i.e.}
$\Gamma_a < m_a^{(0)}, \Gamma_b < m_b^{(0)}$, then a mass inequality $
m_a^{(0)} \leq m_b^{(0)}$ can be derived. The point is that the
correlation function inequalities are derived for {\em all} $|x-y|$.
By varying $|x-y|$ we effectively scan the spectrum since we keep
changing the relative weight of different $(\mu^2)$ regions in
Eqs.~(\ref{eq:specrep}) and (\ref{eq:spectral}). In the $|x-y|
\rightarrow 0$ limit the high $\mu^2$ region dominates. The
perturbative expressions for the correlation functions are adequate
and indeed conform to the inequalities. In the other extreme, $|x-y|
\rightarrow \infty$, the threshold region dominates and we have, {\em
  e.g.}, $\langle VV \rangle \sim \exp \left\{-2m_\pi|x-y|\right\}$.

However, following lattice calculations of hadronic masses, let us
consider ``intermediate times'' during which the local
$u(x)\gamma_\mu \bar{d}(x)$ state, for example, evolves into a true
$\rho$, $\qqb$ bound state, but has {\em not yet} evolved into the
final decayed form of two pions (see Fig.~\ref{fig:pion}).  In the
range $\Gamma_\rho^{-1} \gg |x-y| \gg m_\rho^{-1}$, from which
meaningful mass values (and mass inequalities) could be extracted for
the $\rho$ resonance, $\langle 0 | J_\mu(x) J^{\dag}_\mu(0) | 0
\rangle \sim \exp \left\{-m_\rho|x-y|\right\}$ is a valid
approximation.

In this section we focus on the threshold regions and the
$|x-y|\rightarrow\infty$ domain in the correlator inequalities. Since
near threshold the kinetic energies are much smaller than the masses
of the stable particles, nonrelativistic kinematics, and some aspects
of a potential model (in the $\pi\pi, \pi K$, and $KK$ channels) may
be applicable.

The inequalities $m_{i\bar{\j}}^{(0)} \geq \frac{1}{2}
\left(m_{i\bar{\i}}^{(0)} + m_{j\bar{\j}}^{(0)} \right)$ often become
in the continuum limit trivial equalities $m_i + m_j = \frac{1}{2}
(2m_i + 2m_j)$ [or $m_K + m_\pi = \frac{1}{2} (2m_K + 2m_\pi)$ when we
have confined $i = u$, $j = \bar{s}$ quarks and a meson-meson
continuum].  This is analogous to the trivialization of $m_\delta \leq
2 m_\pi$ into $m_{2\pi} \leq 2 m_\pi$ mentioned above.

If we consider systems confined to a volume of diameter $R$, we expect
the various inequalities to be satisfied with a margin of order
$\Delta E \simeq 1/2mR^2$, the level splitting for a system of
size $R$.  For the continuum, $R \rightarrow\infty$, and $\Delta E$
vanishes. However, the relevant quantities become the phase shifts. We
wish to interpret the inequalities as nontrivial statements about the
latter \cite{ref:nusssath,ref:gupta}.

To this end, let use consider the ``${\rm tr}_{v_n}h_{12}$''
inequalities (\ref{eq:sumeng}):
\begin{equation}
   \sum_{n=0}^N E_{ij}^{(n)} \geq \frac{1}{2}  \sum_{n=0}^N 
      \left( E_{ii}^{(n)} + E_{jj}^{(n)} \right)
\>,
\label{eq:nrsumeng}
\end{equation}
with the sum extending over the first $N$ excited states. If we put
our system in a large box, then each time a ``bound state'' is
generated (by increasing the strength of the attractive potential),
the phase shift $\delta$ in the relevant channel changes by $\pi$.
Levinson's theorem \cite{ref:gottfried} suggests replacing the
discrete sum $\sum_{n=0}^N$ by $\frac{1}{\pi} \int^{\Delta} {\rm
  d}\delta$, and Eq.~(\ref{eq:nrsumeng}) then becomes
\begin{equation}
   \frac{1}{\pi} \int^{\Delta} {\rm d}\delta  \, E_{ij}(\delta) \geq
      \frac{1}{2} \frac{1}{\pi} \int^{\Delta} {\rm d}\delta \,
      \left[  E_{ii}(\delta) +  E_{jj}(\delta) \right]
\>.
\label{eq:nrinteng}
\end{equation}
Evidently (\ref{eq:nrinteng}) reverts back to (\ref{eq:nrsumeng}) in
the narrow resonance approximation with $\frac{{\rm d}\delta}{{\rm
    d}E}$ localized near the resonances $E^{(n)}$.

Near threshold $\delta = ka$, with $k$ the center of mass momentum and $a$
the scattering length. Using NR kinematics for $k_{ij}(E)$ we can obtain
the relation \cite{ref:nusssath}
\begin{equation}
   \frac{1}{a_{ij}^2} \left( \frac{1}{m_i^2} + \frac{1}{m_j^2} \right)
      \geq \left( \frac{1}{a_{ii}^2} \frac{1}{m_i^2}
      +  \frac{1}{a_{jj}^2} \frac{1}{m_j^2} \right)
\>.
\label{eq:aij}
\end{equation}
Quarks are confined and for $i=s, j=\bar{u}$ in $0^{++}, 1^{--},
2^{++}$, {\em etc.} channels, the lowest states are $\pi^+ \pi^-, K
\pi^-$, and $K\bar{K}$, and one might use this inequality with $i,j
\rightarrow K,\pi$. Scattering data for $\pi\pi$ and $\pi K$ can be
analyzed by considering $\pi N \rightarrow 2 \pi N$ and $KN
\rightarrow K\pi N$ scattering and extrapolating to the one pion
exchange pole \cite{ref:chew,ref:pennington}. This extrapolation is
much more difficult fo the $K$ exchange case, making a direct test of
(\ref{eq:aij}) difficult.  However, the existence of the $0^{++}$
state in the $K\bar{K}$ system slightly below threshold is expected to
enhance $a_{K\bar{K}}$ so that (\ref{eq:aij}) would most likely
hold.\footnote{The inequality
\begin{displaymath}
   \left\langle J_{K^+ K^{'-}}(x) J_{K^+ K^{'-}}(y) \right\rangle
   \left\langle J_{\pi^+ \pi^{'-}}(x) J_{\pi^+ \pi^{'-}}(y) \right\rangle
   \geq \left| \left\langle J_{K^+ \pi^{'+}}(x) 
   J_{K^- \pi^{'-}}(y) \right\rangle \right|^2 
\>, 
\end{displaymath} 
with $ J_{K^+ K^{'-}} = J_{K^+}J_{ K^{'-}}$, $J_{K^+} = \psibar_s(x)
\gamma_5 \psi_u(x)$, {\em etc.} is readily derived, if we introduce
additional degenerate flavors $u',s',m_{u'} = m_u, m_{s'} = m_s$. As
$|x-y| \rightarrow \infty$ the various correlation functions are
dominated by the respective meson-meson thresholds. If we neglect the
interaction between the propagating mesons, the above inequality
becomes a trivial equality since each of the two-point functions
factorizes, {\em e.g.} $\langle 0 | J_{\pi^+}(x) J_{K^{-'}}(x) J_{\pi^+}(y)
J_{K^{-'}}(y)| 0 \rangle = f_\pi f_K D_\pi(x,y) D_K(x,y)$ with $f_\pi
= \langle 0 | J_\pi|\pi\rangle$ and $D_\pi$ the pion propagator. In
this long distance, low energy limit, we can effectively treat the
pions as elementary with interactions $\lambda_{\pi\pi} \pi^4,
\lambda_{KK} K^4,$ and $\lambda_{\pi K} \pi^2 K^2$. To first order in
the $\lambda$s these interactions modify the two-point correlation
function as follows: $\langle J_1(x) J_2(x) J_1^{\dag}(y)
J_2^{\dag}(y) \rangle = F_1^2 F_2^2 D_{m_1}(x,y) D_{m_2}(x,y) -
\lambda_{12} \intd{x} D_{m_1}(x,z) D_{m_2}(x,z)D_{m_1}(z,y)
D_{m_2}(z,y)$.  The second term represents the effect of one $m_1 m_2$
collision. By going to momentum space the $z$ integration can be done
and an inequality between the $\lambda$s is obtained
\cite{ref:nusssath}. Unfortunately, it contains, besides the masses,
also a subtraction point $\mu_0$ and will not be reproduced here. Note
that the minus sign in front of the first order euclidean perturbative
contribution reflects the expansion of e$^{-Ht}$.  Thus a negative
$\lambda_{12}$ -- corresponding to an attractive $\pi\pi$ interaction
-- enhances the joint propagation.}

Consider next the exotic current with $0^{++}$ quantum numbers:
\begin{equation}
   J_\delta = J^{ps}_{i\bar{\jmath}}(x) J^{ps}_{k\bar{l}}(x) =
      \psibar_i(x) \gamma_5 \psi_j(x) \psibar_k(x) \gamma_5 \psi_l(x) 
\>.
\end{equation}
If all flavors are distinct, we have only one possible contraction,
illustrated in Fig.~\ref{fig:flavor}(a), contributing to $J_\delta
J_\delta$ which, using Eq.~(\ref{eq:tracegamma5}), can be written as
\begin{equation}
   \langle 0 | J_\delta(x) J^{\dag}_\delta(y) | 0 \rangle =
      \intoned{\mu(A)} {\rm tr}\left\{ \left[ S_A^i(x,y)\right]^{\dag}
      S_A^j(x,y) \right\} {\rm tr} \left\{ \left[ S_A^k(x,y)\right]^{\dag}
      S_A^l(x,y) \right\} 
\>.
\label{eq:jdelta}
\end{equation}
If $m_i^{(0)} = m_j^{(0)}, m_k^{(0)} = m_l^{(0)}$, then the integrand in
Eq.~(\ref{eq:jdelta}) becomes the product of two perfect squares and
$\langle J_\delta J_\delta \rangle$ maximizes all other exotic correlation
functions. Espriu {\em et al.} \cite{ref:espriu} speculate that this is
associated with the fact that the lowest exotic $\qqb \, \qqb$ states
found in bag model calculations \cite{ref:jaffe77} are $0^{++}$ states.

Even more dramatic results follow \cite{ref:espriu} if we take all
four, distinct, flavors, $i,j,k,$ and $l$ to be degenerate. In this
case we have $\langle J_\delta J_\delta \rangle = \intoned{\mu(A)}
\left\{ {\rm tr} \left[ S_A(x,y)^{\dag} S_A(x,y)\right] \right\}^2$,
which by the Schwartz inequality is larger than the square of the
pseudoscalar propagator:
\begin{equation}
   \langle J_\delta(x) J_\delta(y) \rangle \geq
      \left| \langle J^{ps} J^{ps} \rangle \right|^2
\>.
\label{eq:jdeltajps}
\end{equation}
While the last inequality is consistent with having a bound $\delta$ state
in the $\pi^+ \pi^{+'} (\pi = \bar{u}\gamma_5 d, \pi' = \bar{u}' \gamma_5
d')$ channel, it does {\em not require} such a state. The ``$\delta$'' state
could simply be a $\pi \pi'$ threshold state and the inequality would be
trivially satisfied as
\begin{equation}
   m(\pi\pi) \,\, \mbox{threshold} \,\, \leq m_\pi + m_{\pi'} 
\>.
\end{equation}
We would like to interpret this, in analogy with our above
discussion, as a statement that the low energy (threshold) interacts
attractively. This in turn enhances the density of threshold states
relative to the non-interacting, free case, and enhances the long
distance euclidean correlators.

The connection between an attractive potential in the $i\bar{\jmath}$ channel
and the propagator $\langle 0 | J^{\dag}_{ij}(x) J_{ij}(y)| 0 \rangle$ can
be directly seen in a potential model limit. With one of the particles
infinitely heavy, the $ij$ propagator is essentially that of the other
particle moving in the attractive potential. The path integral expression
for such a propagation is
\begin{equation}
   \sum_{{\rm paths}} {\rm e}^{-\int L {\rm d}\tau} =
      \sum_{{\rm paths}} \exp \left\{ - \int \left[ \frac{1}{2}
      \left( \frac{{\rm d}x^\nu}{{\rm d}\tau} \right)^2
      - V(x^\nu) \right] {\rm d}\tau \right\}
\>,
\end{equation}
with $\tau$ a ``proper time'' and $x^\nu(0) = x, x^\nu(\tau) = y$.
Evidently an attractive potential ($V < 0$) will enhance the positive
contributions of the individual paths. This is indeed expected from
the interpretation of the free euclidean propagator as a diffusion
kernel: the probability of the diffusing particle to return to the
origin is clearly enhanced by an attractive potential. A more
rigorous and systematic approach directly relating euclidean lattice
correlators to phase shifts and scattering lengths was suggested by
L\"uscher \cite{ref:lusch91}. It was utilized by Gupta {\em et al.}
\cite{ref:gupta} to prove that the $\pi-\pi$ scattering length is
positive.

The idea in L\"uscher's approach (see also Neuberger's \cite{ref:neu}
suggestion of a calculational lattice method for determining $f_\pi$)
is to use the finite size corrections to the correlators. The latter
are the configuration space analog of the $1/R$ energy shifts
discussed here. We refer the reader to the original L\"uscher work
\cite{ref:lusch91} for further details.

We have argued that $m(\pi\pi') \leq (m_\pi + m_{\pi'})$ can be
fulfilled by an interacting $\pi^+ \pi^{+'}$ threshold state if the
scattering length is attractive (positive). The $\pi$ and $\pi'$ can
only interact via gluon exchanges since they are composed of different
quarks. The interesting fact that this interaction is attractive is in
accord with an old result \cite{ref:feinsuch} that the retarded Van
der Waals forces between systems of identical polarizabilities are
always attractive.

Indeed the Casimir-Polder two-photon exchange
interaction:
\begin{displaymath}
   V_{CP}(\vecr) = \frac{1}{4\pi r^7} \left[ -23 \left( \alpha_E^{(1)}
      \alpha_E^{(2)} + \alpha_M^{(1)} \alpha_M^{(2)} \right)
      + 7 \left( \alpha_E^{(1)} \alpha_M^{(2)} + 
      \alpha_E^{(2)} \alpha_M^{(1)} \right) \right]
\end{displaymath}
remains attractive so long as the ratios of electric and magnetic
polarizabilities $\alpha_E / \alpha_M$ for the two neutral systems (1)
and (2) in question are similar. Our above result generalizes this
attractive nature of the photon exchange to the full non-perturbative
case, and applies also for nonabelian gauge interactions. (If, like in
QCD, the theory is confining, then there is a mass gap in the pure
glue sector, and we expect an exponential rather than power law
falloff of the interaction. The attractive nature does however
persist.)

We note that two independent lines of argument can suggest that the
two-photon exchange interaction is attractive. The first
nonrelativistic, second order perturbation theory argument applies if
the systems (1) and (2) considered are in their respective ground
states.  The second, more general, relativistic field argument uses
$t$-channel dispersion \cite{ref:feinberg} and the positivity of the
corresponding spectral functions when (1) = (2). Amusingly the
conditions required for proving the inequality (\ref{eq:jdeltajps})
simultaneously conform to both arguments, since the $i\bar{\j},
k\bar{l}$ pseudoscalars are indeed the lowest states in their
channels, and taking $m_i = m_k, m_j = m_l$ makes them (dynamically)
identical. Finally we would like to mention that the second argument
suggests \cite{ref:kenneth} that contrary to some lore, two halves of
a conducting spherical shell attract rather than repel.

In the real world the $\pi^+ \pi^+$ scattering length is repulsive
\cite{ref:weinpion}. Indeed, with $u = u', d = d'$ we have the
alternate contraction of Fig.~\ref{fig:flavor}(b). It has one fermion
loop and makes hence a contribution of the opposite sign:
\begin{equation}
   \langle J_\delta J_\delta \rangle = \intoned{\mu(A)} \left\{ \left[ {\rm tr}
       (S_A^{\dag} S_A) \right]^2 - {\rm tr} \left[ (S_A^{\dag} S_A)^{\dag}
       (S_A^{\dag} S_A) \right] \right\}
\>,
\label{eq:joneloop}
\end{equation}
so that Eq.~(\ref{eq:jdeltajps}) can no longer be proven. Indeed at
short distance, when the mesons' wave functions overlap, we expect a
repulsive Pauli effect which is reflected in the minus sign in
Eq.~(\ref{eq:joneloop}). We have not been able to show that this extra
negative term reverses the sign in Eq.~(\ref{eq:jdeltajps}), so that
no $\pi^+ \pi^+$ exotic bound state exists and $a_{\pi^+ \pi^+} =
a_{\pi\pi}(I=2)$ is repulsive.

%
\section{QCD inequalities for finite temperature and finite chemical
  potential}
\label{sec:finite}

Recently there has been much interest in QCD at finite temperature and
finite baryon density, {\em i.e.} finite chemical potential. This
interest is partially motivated by the desire to better understand
compact neutron / (strange) quark matter stars, and by the prospect
that heavy ion collisions at the Relativistic Heavy Ion Collider
(RHIC) can indicate the expected finite temperature phase transition.

In the following we would like to comment on the possible relevance of
QCD inequalities in these cases. First, we note that all the
correlator inequalities are maintained for $T > 0$. In the euclidean
formulation, introduction of finite temperature is simply equivalent
to the imposing periodicity $\beta = 1/T$ in the time direction
\cite{ref:kapusta}. The restriction of the gauge field configurations
in the euclidean path integral to such periodic configurations clearly
does not spoil the positivity of the measure. Also the relation
$\gamma_5 S^A_F(x,0) \gamma_5 = S_F^{\dag A}(x,0)$ and the ensuing
positivity of the fermionic determinant and the integrand in the
pseudoscalar correlators are maintained at finite temperature -- and
therefore so are all the correlator inequalities.

Precisely because of the periodicity in time, we cannot use the
Hamiltonian and its $t \rightarrow\infty$ limiting ${\rm e}^{-mt}$
behavior in order to infer bounds on masses smaller than $T =
1/\beta$. Therefore we cannot attempt, when $T \ne 0$, to prove that the
axial global symmetry spontaneously breaks down, as this feature is
closely tied to the massless pseudoscalar Nambu-Goldstone bosons.
Indeed, numerous thoeretical arguments \cite{ref:yaffe,ref:weiss} and
lattice simulations have virtually proven that in QCD there is in fact
a phase transition corresponding to axial symmetry restoration (and
quark deconfinement) at a temperature $T_c \simeq
\Lambda_{\text{QCD}}$. The exact character of this phase transition
and its dependence on $N_f$, the number of quark flavors, was for a
long time unclear.\footnote{It is generally believed to be weakly
  first order \cite{ref:lattice}, and could be $N_f$ dependent.  We
  will later make a conjecture that $T_c$ monotonically decreases with
  $N_f$.}  However, we believe that the Vafa-Witten results regarding
the non-breaking of vectorial global symmetry [which rely on $S_F^A(x)
\leq S^0_F(x)$] and even more so the one concerning the nonbreaking of
parity (which depends only on bulk, {\em i.e.} free energy,
properties) continue to hold for finite $T$ QCD.\footnote{T.  Cohen
  \cite{ref:cohenhit} utilized the QCD inequalities to suggest that
  not only is SU(3) axial symmetry restored above $T_c$, but also
  U(1)$_A$. Since the latter is broken via the QCD anomaly and not
  spontaneously, this result is rather surprising. Indeed as noted by
  Cohen the proof involves a technical assumption which may fail.}

The QCD inequalities technique may be even less useful in discussing
transient varying $T$ phenomena, such as disoriented chiral
condensates \cite{ref:dcc}, suggested to occur in domains of cooling
quark gluon plasma.

The introduction of a finite chemical potential, {\em i.e.} consideration of
QCD in a background of uniform baryon density, has a much more drastic
effect. It amounts to changing the fermionic part of the euclidean
Lagrangian to
\begin{equation}
   {\cal L}(\mu) = \sum_{i = 1}^{N_f} \left[ \psibar_i \gamma_\mu
      D_\mu \psi_i + \mu \psibar_i \gamma_0 \psi_i + m_i \psibar_i
      \psi_i \right]
\>,
\end{equation}
so that the new $\mu \ne 0$ propagator no longer satisfies $\gamma_5
S_F \gamma_5 = S_F^{\dag}$, but rather
\begin{equation}
   \gamma_5 S_F \gamma_5 = \gamma_5 \frac{1}
      {\dslasha + \mu\gamma_0 + m}\gamma_5 = \frac{1}
      {- \dslasha - \mu\gamma_0 + m} \ne S_F^{\dag}
\>,
\end{equation}
and hence the positivity of the fermionic determinant can no longer
be inferred. This not only prevents the proof of the QCD inequalities,
but also excludes the utilization of lattice numerical simulations in
which the statistics of occurance of lattice gauge configurations
prescribes their (positive!) weights. The various dramatic
speculations concerning the high $\mu$ phase (involving parity
breaking and even ``color superconductivity'' \cite{ref:csc}) are
therefore not excluded. 

The special case of $N_c = 2$, {\em i.e.} SU(2) gauge theory, is a notable
exception, and one can show measure positivity in this case even for
$\mu \ne 0$. This result, which has been utilized for some time in
lattice simulations \cite{ref:kogut} is readily proven by using
\begin{equation}
   \gamma_5 C I_2 S_F \gamma_5 C I_2 = D^{\ast}
\>.
\label{eq:gcgc}
\end{equation}
In Eq.~(\ref{eq:gcgc}) $C$ is $i \gamma_0 \gamma_2$ and $I_2$ is the
generator of the SU(2) color isospin, and it holds for arbitrary
$\mu$.  This feature, which is due to the pseudoreality of SU(2), is
essentially the same one used by Anishetty and Wyler \cite{ref:wyler}
and Hsu \cite{ref:hsu} to extend the inequalities to chiral SU(2). It
has been used by Kogut, Stephanov, and Toublan \cite{ref:kogut} in
order to prove that the correlator of the $0^+$, $I = 0$ ({\em i.e.}
antisymmetric in flavor) $\psi\psi$ combination
\begin{displaymath}
   M_{\psi\psi} = \psi^{\dag} C I_2 \gamma_5 \psi
\end{displaymath}
can serve, for $\mu \ne 0$, as an upper bound for any other
correlator. This implies that for $\mu \ne 0$ the $0^+$ $\psi\psi$
diquark is the lightest boson. This complements the claim that for
$\mu = 0$ and $N_c = 2$, the $0^+$ $\psi\psi$ diquark is degenerate
with the $0^-$ $\psibar\gamma_5\psi$ pion (see the end of
Sec.~\ref{sec:nonpertmbineq}), and nicely fits with the unique
patterns of symmetry breaking suspected in this case \cite{ref:kogut}.

External electric and magnetic fields modify the hadronic spectrum.
Indeed lattice calculations utilized external magnetic fields to
prove, via the $\vec{\mu}\cdot\vec{B}$ interactions, the hadronic
magnetic moments. Also very strong fields (the analogs of
supercritical fields in superconductors) can even modify the phase
structure of spontaneously broken gauge theories. In the context of
QCD and other vectorial theories the measure postivity is clearly
maintained in the presence of the external $\vec{B}$ field. However,
QCD inequalities such as BE (parapositronium) $\geq$ BE
(orthopositronium) or $m_\rho \geq m_\pi$ may be modified since the
vector (triplet) state has a magnetic moment, and for a sufficiently
strong external $\vec{B}$ field ($|\vec{B}| \gtrsim
\Lambda^2_{\text{QCD}}$), the state with $\vec{\mu}$ antiparallel to
$\vec{B}$ may become lower than the singlet pion. In this connection
we recall the amusing suggestion \cite{ref:rubinstein} that in strong
enough magnetic fields, existing in appropriate astrophysical
environments, that this effect can reverse $m_n > m_p$ to $m_p > m_n$
with an inverted $\beta$ decay!

Unlike the effect of high temperature and/or chemical potentials, we
believe that strong magnetic fields do not cause deconfinement or
chiral symmetry restoration. Indeed in the limit $B\gg \Lambda^2$ we
expect, in analogy with the case of atomic physics
\cite{ref:salpeter}, that in strong magnetic field stars the motion of
the quarks becomes one dimensional, along the $\vec{B}$ field lines.
It is well known that in such cases confinement only gets stronger
(even U(1) theories confine!), and so should S$\chi$SB. It is an
interesting, open conjecture which we would like to make here, that
even moderate $\vec{B}$ fields tend to enhance S$\chi$SB, {\em e.g.}
by enhancing the density near $\lambda=0$ eigenvalues of the Dirac
operator and thus, according to the Banks-Casher criterion, enhance
the $\langle\psibar\psi\rangle$ condensate.

%
\section{QCD inequalities for $\bar{Q}Q$ potentials, quark masses, and weak
transitions}
\label{sec:qqbar}

The main application of the techniques developed above is to obtain
inequalities between directly measurable quantities such as hadron
masses. However, there are several calculational approaches to QCD
such as the potential model for heavy quarkonia
\cite{ref:quigg,ref:richnpb,ref:kwong,ref:roncaglia}, chiral
perturbation theory for the low-lying mesonic sector
\cite{ref:gasser}, and QCD sum rules \cite{ref:shifman,ref:reinders}.
Each of these schemes depends on a few input parameters and makes many
predictions.  Applying the techniques of QCD inequalities to these
input parameters could therefore yield a very large body of suggestive
results.

The area in which most research along these lines has been done (in a
large measure prior to, and independent from, the introduction of QCD
inequalities in 1983) is that of potential models. The point is that
given some general properties of the potential such as convexity of
$V(r)$ or other features, we have a wealth of information concerning
the level ordering, thanks to the work of Martin and collaborators
\cite{ref:bertlmann,ref:common,ref:baum}; Lieb \cite{ref:lieb}; Fulton
and Feldman \cite{ref:feldman}; and others.

An early important observation was that the P-wave excitation in
$c\bar{c}$ (or $b\bar{b}$) systems is lower than the first radial
excitation: $E_{n_r+1,l} \geq E_{n_r,l+1}$. This deviation from the
famous degeneracy in the pure Coulombic case is related to the fact
that the QCD potential $V_{{\rm QCD}}(r)$ does not satisfy $\nabla^2
V_{{\rm QCD}} = 0$ but rather $\nabla^2 V_{{\rm QCD}} \geq 0$
\cite{ref:feldman}. Also the assumption of a monotonically increasing,
$\frac{{\rm d}V}{{\rm d}r} > 0$, convex potential, $\frac{{\rm
    d}^2V}{{\rm d}r^2} < 0$ allowed Baumgartner, Grosse and Martin
\cite{ref:baum} (BGM) to prove $E_{n,l} \leq E_{n-1,l+2}$.

In addition if $\frac{{\rm d}^3V}{{\rm d}r^3} > 0$ (or if e$^{-\lambda
  V}$ has a positive Fourier transform), then the baryon mass
relations Eqs.~(\ref{eq:miij}) and (\ref{eq:mijk}) can be proved (see
\cite{ref:lieb}). These relations, which prescribe the sign of
deviation from linearity of masses in the decuplet or from the
Gell-Mann -- Okubo relation in the octet, were proven above only under
the explicit additional assymption of flavor symmetric wave functions,
a point emphasized by Richard and Taxil \cite{ref:richtaxil}. [Indeed
as shown by Lieb \cite{ref:lieb} and by Martin, Richard, and Taxil
\cite{ref:martin86}, Eqs.~(\ref{eq:miij}) and (\ref{eq:mijk}) are
violated for potentials $V = r^a$ with sufficiently large $a$.]

It is not clear at the present what are all of the model independent
statements about $V_{\bar{Q}Q}(R)$ that can be proven. We would like to
mention, however, the very elegant result of Bachas \cite{ref:bachas}
on the convexity
\begin{equation}
   V(R) \geq \frac{1}{2} \left[ V\left(\frac{R-r}{2}\right) 
      + V\left(\frac{R+r}{2}\right) \right]
\label{eq:VRineq}
\end{equation}
of $V(R)$. Together with earlier work by Simon and Yaffe \cite{ref:simon}
which shows a monotonically increasing $V(R)$, this promotes
the BGM level ordering into a QCD theorem, which is confirmed in the
$c\bar{c}$ system (and partially tested in the $b\bar{b}$ system). 

Let us next reproduce Bachas' proof. The static potential is definted
in terms of a rectangular Wilson loop $W$ \cite{ref:wilson} in the
$(t,\vecr)$ plane of height $T \rightarrow\infty$ and width $R$:
\begin{equation}
   V(R) = \lim_{T \rightarrow\infty} \left[ -\frac{1}{T} 
      \ln \langle {\rm tr} U(W) \rangle \right]
\>,
\label{eq:VR}
\end{equation}
with $U(W)$ the ordered product of $U_{{\rm links}}$ around the loop $W$.
We have the path integral representation
\begin{equation}
   \langle {\rm tr} U(W) \rangle = \frac{ \int \prod_{{\rm links}}
      {\rm d}\mu(U) \, {\rm tr} U(W)}{ \int \prod_{{\rm links}}
      {\rm d}\mu(U)}
\>,
\label{eq:traceu}
\end{equation}
with d$\mu(U) = \prod {\rm d}[U] \exp \left( -\frac{1}{g^2} \right)
\sum_{\Box} {\rm tr}U_p$ the positive lattice measure. We can write $W
= W_1(\tilde{\phi}\tilde{W}_2)$, where $\tilde{\phi} \tilde{W}_2$, the
line reversed version of the reflected path, denotes the reflection of
the portion of $W$ to the left of the hyperplane indicated by the
dotted line in Fig.~\ref{fig:wilson}. Obviously $U(\tilde{x}) =
U^{\dag}(x)$ and thus ${\rm tr} U(W) = {\rm tr} \left[ U(W_1)
  U^{\dag}(W_2) \right]$. We can compare $\langle {\rm tr} U(W)
\rangle$ with the corresponding expressions for the symmetric Wilson
loops of size $R + r, R - r$ obtained by joining $W_1$ and $\phi W_1$
or $W_2$ and $\phi W_2$. By dividing the d$\mu(U)$ integration into
variables to the left of, right of, and on the reflection plane, the
expectation values of $\langle {\rm tr} (W_1 \phi W_1) \rangle$ and
$\langle {\rm tr} (W_2 \phi W_2) \rangle$ can be written as perfect
squares, {\em e.g.} $\langle {\rm tr} (W_1 \phi W_1) \rangle =
\intoned{\mu_{{\rm on}}} {\rm tr} \left[ \left( \intoned{\mu_R} W_1
  \right) \left( \intoned{\mu_R} W_1 \right)^{\dag} \right] =
\intoned{\mu_{{\rm on}}} {\rm tr} \left[ \left( \intoned{\mu_L} W_1
  \right) \left( \intoned{\mu_L} W_1 \right)^{\dag} \right]$ and the
Schwartz inequality implies $\langle {\rm tr} U(W_2) U(\phi W_2)
\rangle \cdot \langle {\rm tr} U(W_2) U(\phi W_2) \rangle \geq
|\langle {\rm tr} U(W_2) U(\phi W_2) \rangle |^2$.  The definition of
$V(R)$, Eq.~(\ref{eq:VR}), then readily yields the desired convexity
(\ref{eq:VRineq}).

Similar arguments were applied \cite{ref:gromes} to the
Eichten-Feinberg \cite{ref:eichten} representation of the tensor and
spin-spin potentials. The inequalities obtained are basically in
accord with a scalar long range confining potential \cite{ref:buch}.

The masses of the different quark flavors in the QCD Lagrangian (or
Hamiltonian) $H = H_0 + \sum_{i=1}^{N_f}m_i^{(0)} \psibar_i \psi_i$
are the only explicit dimensional parameters. ``Dimensional
transmutation'' generates, however, an additional scale
$\Lambda_{\text{QCD}}$ ($\approx$ a few hundred MeV).  Thus, unlike
QED with $r_{\text{Bohr}} \simeq (m_e)^{-1}$ and Coulombic binding
$\simeq m_e$, we expect that scaling all quark masses $m_i^{(0)}
\rightarrow c m_i^{(0)}$ changes physical masses (lengths) by less
than a factor $c$ (1/$c$).  As we next argue, the simple linear
dependence of $H_{\text{QCD}}$ on $m_i^{(0)}$ further restricts the
variation of mass parameters as a function of $m_i^{(0)}$.

In quantum mechanics the ground state energy $E^{(0)}(\lambda)$ is a
convex function of any set of parameters that the Hamiltonian depends
upon linearly \cite{ref:thirring}. If $H(\vec{\lambda}) = H_0 + \sum
\lambda_i H_i = H_0 + \vec{\lambda} \cdot \vec{H}$, and $\vec{\lambda}
= \alpha \vec{\mu} + (1 - \alpha) \vec{\nu}$, then
$E^{(0)}(\vec{\lambda}) \leq \alpha E^{(0)}(\vec{\mu}) +
(1-\alpha)E^{(0)}(\vec{\nu})$. [This result follows immediately from
$H(\vec{\lambda}) = \alpha H(\vec{\mu}) + (1-\alpha) H(\vec{\nu})$ by
taking expectation values in $\psi^{(0)}(\vec{\lambda})$, the ground
state of $H(\vec{\lambda})$, and using the variational principle.]

We wish to apply this result to the masses $m_{ij}^{(0)} (J^{PC})$ of
ground state mesons with different flavor and Lorentz quantum numbers.
We are hindered by the fact that we are not free to vary $m_i^{(0)}$
and by the need to subtract the vacuum energy.  However, in a large
$N_c$ approximation where $q_i \bar{q}_i$ do not annihilate and
effects of closed quark loops are neglected, it is meaningful to
describe each sector $m_{ij}$ by a separate Hamiltonian
$H_{i\bar{\jmath}} = H_{\text{QCD}}^{(0)} + m_i \psibar_i \psi_i +
m_{\bar{\jmath}} \psibar_j^c \psi_j^c$.  The knowledge of the masses
of the $0^{--}, 1^{--}, 2^{++}, \ldots$ flavor multiplets can then be
used to constrain ratios of quark masses like $R = (m_c^{(0)} -
m_s^{(0)}) / (m_s^{(0)} - m_u^{(0)})$ \cite{ref:nussplb164}. The
result obtained is somewhat large but still consistent with the
estimates of Gasser and Leutwyler \cite{ref:gasser}, and correspond to
a lower bound $m_s^{(0)} \ge $ 80-100 MeV.\footnote{It is amusing to
  note that the relatively ``large'' $\epsilon^{\prime}/\epsilon$
  ratio (as compared with previous theoretical estimates
  \cite{ref:derafael}) recently found in high precision experiments on
  CP violating kaon decays \cite{ref:ktev} do indeed suggest a
  relatively low $m^{(0)}_s \simeq 80$ MeV. The fact that we have so
  far been able to obtain only the above modest lower bound, rather
  than, say, $m_s^{(0)} \geq$ 190 MeV, makes it easier to accomodate
  the measured $\epsilon^{\prime}/\epsilon$ in the standard model. In
  passing we note that the smaller $m_s^{(0)}$ value would also favor
  the stability of strange quark matter \cite{ref:derafael,ref:ktev}.
  Note however that if the new lattice calculation of
  $\epsilon^{\prime}/\epsilon$ (see next Footnote) is adopted, the
  above remark need not apply.}

Over the last twenty years there has been an ongoing effort to address
in a systematic way the low energy sector of hadronic physics, and in
particular the sector containing light quarks only via chiral
perturbation theory ($\chi$PT). The idea is to incorporate S$\chi$SB,
the Goldstone pions and current algebra, and the ensuing low energy
theorems via effective Lagrangians \cite{ref:cjt,ref:ert} which will
manifest the desired symmetries and which are written in terms of the
pionic fields only. Then one uses these chiral Lagrangians to perform
a systematic expansion in the external momenta of the problem, and/or
the pion mass divided by some ``hadronic scale'' usually taken to be
$4\pi f_\pi$.  These effective Lagrangians contain a series of terms
${\cal L}_1, {\cal L}_2, \ldots$ which in general are ranked according
to the number of derivatives appearing in each term. One then can also
systematically compute higher loop processes. A particularly simple
and elegant form of such an effective Lagrangian is the Skyrme model
\cite{ref:skyrme}, which even incorporates the nucleon as a soliton
state.

In principle all the coefficients of the terms in the effective
Lagrangian are computable from QCD. In practice they are often fixed
by fitting some low energy data. In any event, QCD inequalities can
often be applied to constrain the range of these parameters. While no
systematic program of this kind has been completed, some steps have
been taken by Comellas {\em et al.}  \cite{ref:comellas}. By
expressing the currents in the inequality relating vector and
pseudoscalar currents as a $\chi$PT expansion in the pionic field, and
also by using the momentum space version of these inequalities, bounds
on parameters in the effective Lagrangian were obtained, and are well
satisfied.

Over the last decade the heavy quark approximation was often used in
connection with $Q\bar{q}$ or $Qqq$ systems. This is based on the
realization that in the infinite $m_Q$ limit the heavy quark simply
becomes a static color source leading to some universality relations,
heavy quark symmetry, and a systematic expansion in inverse powers of
$m_Q$ \cite{ref:wetzel,ref:voloshin,ref:wise}.  The Witten
pseudoscalar mass inequalities such as $2 m(q,\bar{Q}) > m(q,\bar{q})
+ m(Q,\bar{Q})$ become trivial in this limit where the $Q\bar{Q}$
system is essentially Coulombic, with infinite binding $\alpha_s^2 m_Q
/ r$.  Despite an interesting effort on the part of Guralnik and
Manohar \cite{ref:guralnik}, it is not clear \cite{ref:bigi} how this
can be amended to yield useful inequalities.

Precise information on nonperturbative QCD parameters of weak decay is
of particular importance and may decide the fate of the Standard Model
with the three generation KM scheme \cite{ref:3km}. It is worth
pointing out that some inequalities between such matrix elements can
be motivated \cite{ref:alfaro}. Let us then consider the $K
\rightarrow \pi\pi$ decay. In addition to the standard left-handed
four fermion operators [say $A = \bar{s}(x) \gamma_\mu L u(x)
\bar{u}(x) \gamma_\mu L d(x)$, with $L = 1 - \gamma_5$ the left
projection operator], we can extract from ``penguin diagrams''
additional mixed terms of the form $B = \bar{s}(x) \gamma_\mu L u(x)
\bar{u}(x) \gamma_\mu R d(x)$, with $R = 1 + \gamma_5$. After using
soft pion and current algebra techniques to reduce one pion we need to
evaluate $\langle K | A ({\rm or}\,\, B)| \pi \rangle$ matrix
elements. The latter can be related to the asymptotic limit when
$|x-y|$ and $|y-z| \rightarrow \infty$ of the three-point function $
\langle 0 | \bar{s}(x) \gamma_5 u(x) A(y) \bar{u}(z) \gamma_5 d(z) | 0
\rangle$. The latter has the path integral form
\begin{displaymath}
   \langle K | A | \pi\rangle = \intoned{\mu(A)} {\rm tr} \left[ \gamma_5 
      S_A(x,y) \gamma_\mu (1 - \gamma_5) S_A(y,z) \gamma_5 
      S_A(z,y) \gamma_\mu (1 - \gamma_5) S_A(y,x) \right]
\>,
\end{displaymath}
corresponding to the unique contraction in Fig.~\ref{fig:8pattern}. A
similar expression with the second $1 - \gamma_5 \rightarrow 1 + \gamma_5$
applies when $A \rightarrow B$. We have assumed a flavor symmetric limit
and used the same propagator $S_A$ for all the quarks. Using $\gamma_5
S_A(x,y) \gamma_5 = S^{\dag}(y,x)$, the hermiticity of the euclidean
$\gamma_\mu$, and $(\gamma_5 \gamma_\mu)^{\dag} = - \gamma_5 \gamma_\mu$,
we can show that the mixed $L - R$ expression (case $B$) has a path
integral with an integrand which is an absolute square. It is therefore
larger than the expression for the ``pure'' $L - L$ case ($A$) and
allows the conclusion that
\begin{displaymath}
   \langle K | B | \pi\rangle \geq  \langle K | A | \pi\rangle
\>.
\end{displaymath}
While the soft pion limit {\em and} flavor symmetry assumed in the
derivation are considerably weaker than this result we do still find
it interesting and suggestive.\footnote{It is worth noting that the
  $B$ matrix element with the ``8'' construction is one of the
  ingredients fixing the positive sign of
  $\epsilon^{\prime}/\epsilon$. The QCD inequalities are generally not
  operative when we have disconnected flavor contractions as in the
  ``eye'' contraction of Fig.~\ref{fig:eye}.  Since the latter appears
  to dominate the lattice calculated matrix element \cite{ref:blum},
  we unfortunately cannot fix the sign of this very important quantity
  by QCD inequalities alone.}

%
\section{QCD inequalities beyond the two-point functions} 
\label{sec:beyond2pt}

Most of the preceding sections, and of the work on QCD inequalities to
date, has focused on euclidean two-point functions. These correlation
functions are sufficient for obtaining hadronic masses via the
spectral representation. An obvious advantage of two-point functions
is that the path integrals expressing them [Eqs.~(\ref{eq:corr1}) and
(\ref{eq:corr2})] contain products of quark propagators $S_A(x,y)$
between the {\em same} points $x$ and $y$. Thus it is easy to prove
positivity of certain combinations, {\em e.g.}
tr[$S_A^{\dag}(x,y)S_A(x,y)$] and the ensuing inequalities between the
integrands of the path integrals for various two-point correlation
functions. These algebraic inequalities which are true ``pointwise''
for each external $A_\mu(x)$ configuration do survive the path
integration with the positive measure ${\rm d}\mu(A)$.

Can we make general statements in the form of inequalities also for
three-, four-, and higher point euclidean correlation functions?
Consider a generic four-point correlation function $\langle J_a(x)
J_b(u) J_c(y) J_d(v) \rangle$, with $a,b,c,d$ referring to Lorentz and
flavor quantum numbers. By appropriate contraction we obtain an
expression for correlation functions of the following general type,
where for simplicity we assumed degenerate quark flavors and hence the
same propagator $S_A$:
\begin{equation}
   \langle J_a(x) J_b(u) J_c(y) J_d(v) \rangle = \intoned{\mu(A)} {\rm tr}
      \left[ \Gamma_a S_A(x,u) \Gamma_b S_A(u,y) \Gamma_c S_A(y,v)
      \Gamma_d S_A(v,x) \right]
\>.
\label{eq:gencorr}
\end{equation}
All propagators $S_A$ in the path integral refer to {\em different}
pairs of points and it is not obvious how to construct positive
definite combinations for each external $A_\mu(x)$ configuration
separately.

However, if we wish to study hadronic properties beyond the mass
spectrum such as couplings, scattering amplitudes, weak interaction
matrix elements, or wave functions, we need to consider more than
two-point correlation functions \cite{ref:nussplb202}. In particular
correlation functions of the form given in Eq.~(\ref{eq:gencorr}) have
been used in order to study the charge distribution of mesonic states.
Let the current
\begin{equation}
   J^{\dag}_a = \psibar_i(x) \Gamma_a \psi_j(x)
\>,
\end{equation}
with $x = (-T, \vec{0})$, create at the origin, at some remote past
instant, a quark and antiquark of flavors $i,\bar{\jmath}$ in a
specific Lorentz state. As the system evolves under the QCD
Hamiltonian it settles into the wave function of the ground state
meson in this channel.  Specifically, we can infer from the spectral
expression Eq.~(\ref{eq:spectral}) that after time $T$ the components
of excited states in the wave function of the system are suppressed
relative to the ground state amplitude by e$^{-T \Delta m}$, with
$\Delta m$ the mass gap between the ground state and the first excited
state. If we wish to find the relative separation of the quarks in the
ground state, we can probe the system again with two external currents
$J_b(u) J_d(v)$ at $u = (0, \vecr_1)$, $v = (0, \vecr_2)$, with
$\vecr_{1,2} = (\vec{R} \pm \vecr)/2$ and $\vecr$ the relative
separation. The currents $J_b, J_d$ should refer to the flavors $i$
and $j$ of the quarks respectively
\begin{equation}
   J_b = \psibar_i \Gamma_b \psi_i, \qquad
      J_d = \psibar_j \Gamma_d \psi_j
\>,
\end{equation}
and we will take $\Gamma_b = \Gamma_d = \Gamma$. Finally, in order to
obtain the complete gauge invariant correlation function the two
quarks are propagated back to the origin where they are annihilated at
$v = (T, \vec{0})$ by the current $J_c = J_a^{\dag}$. Various attempts
have been made to measure, via lattice Monte-Carlo calculations,
charge distributions (or form factors) for the ground state mesons
\cite{ref:barad,ref:wilcox}. In the following we will focus on the
particular case of the pion, {\em i.e.} using
\begin{eqnarray*}
   J_a^{\dag} &=& \psibar_u \gamma_5 \psi_d(x) = J_c \\
   J_b &=& \psibar_u \Gamma \psi_u, J_d = \psibar_d \Gamma \psi_d
\end{eqnarray*}
we define
\begin{equation}
   F(t,\vecr) = \int {\rm d}^3 \vec{R} \, \left\langle 0 \left|
      J_a^{\dag}(-T,\vec{0}) J_b\left(0,\frac{\vecr + \vec{R}}{2}\right) 
      J_a(T,\vec{0}) J_d\left(0,\frac{-\vecr + \vec{R}}{2}\right) 
      \right| 0 \right\rangle
\>.
\label{eq:fvr}
\end{equation}
This function has the following general properties \cite{ref:nussspiegel}:
\begin{enumerate}
\item $F(T,\vecr) \leq F(T,\vec{0})$, {\em i.e.} the ``charge
  density'' is maximal at the origin.
   \item The Fourier transform ({\em i.e.} the ``form factor'')
\begin{equation}
   L(T,\vp) = \int {\rm e}^{i\vp \cdot \vecr}F(T,\vecr)
\end{equation}
is positive for all $T$ and $\vp$ values
\begin{equation}
   L(T,\vp) \geq 0
\>.
\end{equation}
\end{enumerate}
Evidently (2) implies (1): $F(t,\vecr)$ is also the Fourier transform
of the positive $L(t,\vp)$, and hence has its maximum value $\int
L(t,\vp){\rm d}^3 \vp$ at the origin ($\vecr = 0$).

In order to prove (2) let us consider the four-point correlation function
\begin{eqnarray}
   F&=& \intoned{\mu(A)} {\rm tr} \left[ \gamma_5 S_A^i(x,u) \Gamma
      S_A^i(u,y) \gamma_5 S_A^i(y,u) \Gamma S_A^i(u,x) \right]
   \nonumber \\
   &=& \intoned{\mu(A)} {\rm tr} \left\{ \left[S_A^i(u,x)\right]^{\dag} 
      \tilde{\Gamma} \left[S_A^i(y,u)\right]^{\dag}
      S_A^i(y,u) \Gamma S_A^i(u,x) \right\}
\>,
\end{eqnarray}
where we have used the ``charge conjugation'' property $\gamma_5 S_A(x,y)
\gamma_5 = S_A^{\dag}(y,x)$. $\tilde{\Gamma}$ is defined as
\begin{equation}
   \tilde{\Gamma} = \gamma_5 \Gamma \gamma_5 = \pm \Gamma 
      = \pm \Gamma^{\dag}
\>.
\end{equation}
Using $u,v = (0, \vecr_{1,2})$ we take the Fourier transform with respect
to $\vecr = \vecr_1 - \vecr_2$, assigning momentum $\vp \, (-\vp)$ to the
quark (antiquark) at $\vecr_1 (\vecr_2)$. Exchanging the order of the
d$\mu(A)$ and d$\vecr_1$d$\vecr_2$ integration the Fourier transform can be
written as
\begin{eqnarray}
   L(T,\vp) &=& \intoned{\mu(A)} \intoned{\vecr_1} \intoned{\vecr_2}
      {\rm e}^{i\vp \cdot(\vecr_1 - \vecr_2)} {\rm tr} \left\{
      S(y,u) \Gamma S(u,x) \left[S(y,v) \Gamma S(v,x)\right]^{\dag}
      \right\}
   \nonumber \\
   &=& \intoned{\mu(A)} {\rm tr} \left| \intoned{\vecr_1} {\rm e}
      ^{ i\vp \cdot \vecr_1}  S(y,u) \Gamma S(u,x) \right|^2
\>,
\end{eqnarray}
and the manifest positivity of the integrand persists after the d$\mu(A)$
($\geq$ 0) integration, yielding the desired result $L(T,\vp) \geq 0$.

While $F(t,\vecr)$ and $L(t,\vp)$ are well-defined, gauge invariant,
and, in principle, measurable quantities, the suggestive
interpretations as ``charge density'' and ``form factor'' are more
heuristic. In particular it is not evident from Eq.~(\ref{eq:fvr}) why
$F(T,\vecr)$ is positive definite as $T \rightarrow\infty$. Here we
interpret $F$ as $|\psi_\pi(\vecr)|^2$, with $\psi_\pi$ the ``pion wave
function''. Recall, however, that, as emphasized in the conclusion of
Sec.~\ref{sec:quarkbi}, the positivity of norms [and spectral weight
functions in Eq.~(\ref{eq:spectral})] emerges only after the d$\mu(A)$
integration has been performed and need not be manifest for each
external $A_\mu(x)$ configuration separately.

As $\vecr_2$ approaches $\vecr_1$, $S(y,v)$ and $S(y,u)$ [and likewise
$S(x,u),S(x,v)$] in Eq.~(\ref{eq:gencorr}) connect one vertex to two
nearby points. For a typical smooth $A_\mu(x)$ configuration we expect
$S(y,v) \rightarrow S(y,u)$ and the mixed products will gradually
become squares (see Fig.~\ref{fig:diamond}).  This suggests that
$F(T,\vecr)$ is not only maximum at $\vecr = 0$ but is also
monotonically increasing towards $\vecr = 0$. This property does not
hold for all gauge configurations, but only after averaging, since we
expect smooth $A_\mu(x)$ configurations to dominate in
$\intoned{\mu(A)}$. Thus a proof of the conjectured monotonicity of
$F(T,\vecr)$ requires understanding the correlation between $A_\mu(x)$
along neighboring paths and would depend on the specific form of the
Yang-Mills action $S_{YM}$.  This action prefers small variations of
the gauge fields over a small region [gradients of $A_\mu$ yield
$\vec{E}$ and $\vec{B}$ and $S_{YM} = \intd{x} (\vec{E}^2 +
\vec{B}^2)$].

If we identify $F(T = \infty,\vecr)$ with the pion's charge density,
then it is amusing to note that the monotonicity of the ground state
wave function can be proven in nonrelativistic quark models
\cite{ref:herbst} when the potential is purely attractive. The fairly
simple argument employs the spherical rearrangement technique
\cite{ref:brascamp}, showing that given any trial wave function will
always lower the energy by ``rearranging'' its values into a radially
monotonically decreasing set.

The present discussion of four-point correlation functions applies
only to pseudoscalar currents $J_5(x)J_5(y)$. Indeed it is precisely
for the spin singlet S-wave ``pion'' state that {\em all} components
of the nonrelativistic quark model potential -- the confining linear
part, the Coulomb force at shorter range, {\em and} the very short
range, hyperfine color magnetic interactions -- are attractive.

\section{Summary and suggested future developments}
\label{sec:conc}

We have presented above many inequalities for hadronic masses and
analyzed the possible theoretical and phenomenological aspects. Many
results follow essentially from the positivity of the measure in the
functional path integral for euclidean correlation functions in QCD
(or other vectorial theories). This may be on occasion complemented by
fairly mild assumptions on the number of light degenerate flavors or
the Zweig rule (large $N_c$) suppression of $\qqb$ annihilation. These
results include the Weingarten mass relations, the two Vafa-Witten
theorems, the convexity of the $\bar{Q}Q$ potential, $m_{\pi^+} >
m_{\pi^0}$, and Witten's interflavor relation for pseudoscalars.

The fact that so many results, which have far-reaching implications,
can be proven with such minimal input, is truly fascinating. We
believe, however (and will try to make slightly more concrete
conjectures later), that many more results would follow if we appeal
to the specific form of the QCD action and in particular to its
``ferromagnetic'' character.

A large class of meson-meson and baryon-baryon mass relations follow
from the flavor independence (apart from the explicit mass terms) of
$H_{\text{QCD}}$. This allows us to prove operator relations for
$H_{\text{QCD}}$ restricted to different flavor sectors, from the mass
relations following -- though {\em only} for flavor symmetric wave
functions. Under fairly general assumptions we can also use
Hamiltonian variational techniques to prove detailed baryon-meson
inequalities. All of the above together with some general level
ordering rules for the different $J^{PC}$ states (again motivated by
QCD inequalities applied to $\qqb$ potentials) can serve as extremely
useful ``Hund like'' rules of thumb in the hadronic domain.  This in
turn can restrict the masses (or $J^{PC}$ quantum numbers) of new
flavor combinations or radial excitations.

In many cases the inequalities are more sophisticated versions of
relations suggested by a naive quark model. Eventually we hope that
much of the vast information available on hadronic parameters
(including scattering, wave functions, {\em etc.}) will be
constrained by such inequalities.

On the more theoretical side the generic properties of QCD and all
vector QCD-like theories should be analyzed. The constraints imposed
by the inequalities on composite models of quarks and leptons,
together with the anomaly matching conditions, are particularly
interesting.  The fermion-boson inequalities, $m_F^{(0)} \geq
m_B^{(0)}$, exclude the protection of small masses for composite
quarks and leptons via an unbroken global axial symmetry, in all cases
when we have vectorial, QCD-like, underlying dynamics. Such
constraints are avoided by going to chiral gauge models or
(supersymmetric) models with scalars where the measure positivity
d$\mu(A)$ is lost. (In passing we note that large classes of scalar
and scalar plus fermion theories are ruled out by ``triviality''
difficulties \cite{ref:aizen}.)

We have often referred to the euclidean correlation function approach
to deriving the QCD inequalities as ``rigorous'' and to that based on
the Hamiltonian variational approach as more ``heuristic''. It should
be emphasized that this does not reflect any true {\em objective}
distinction. Clearly the Hamiltonian and Lagrangian approaches to
classical mechanics, and to quantum field theory, are equivalent and
equally rigorous (or not) depending on the practitioner.

The Hamiltonian variational approach allows inputting various
approximations and/or physics intuition much more readily, however.
These inputs can motivate certain restrictions on the wave
function(al)s allowed. An example is our restriction in the
nonperturbative derivation of the baryon-meson inequalities
(Sec.~\ref{sec:nonpertmbineq}) to baryonic configurations with only
{\em one} junction point (see Fig.~\ref{fig:color}).

Clearly any restrictions on the class of wave function(al)s allowed
potentially pushes the energy of the ground state higher. Hence, in
particular, our restricted baryon is {\em not} the true ground state
baryon. If, however, we can assume that any network containing some
extra junction (and in this case, also extra anti-junction) points
corresponds to ``massive'' components in the wave function(al) (with
essentially extra virtual baryon-antibaryon pairs), then such
components are likely to be small in the true ground state, and thus
fairly safe to neglect. Similar comments apply to the neglect of
possible sextet internal fluxons in our ansatz functionals for the
quadri- and pentaquarks (see App.~\ref{app:fourfive}).

Much of the intuition for hadronic physics stems from naive quark
models. In these, the baryon-meson wave functionals are approximated
by $q_{\text{con}} q_{\text{con}} q_{\text{con}}$ and
$\bar{q}_{\text{con}} q_{\text{con}}$ wave functions. Here
$q_{\text{con}}$ refers to ``constituent quarks'', namely some
effective quasiparticles obtained when some short distance modes are
integrated. Clearly this concept would be on firmer ground if we could
indeed show that the mass of the constituent $u$ and $d$ quarks --
which in the $m_u^{(0)} = m_d^{(0)} = 0$ axial SU(2) symmetry limit is
purely dynamical -- is indeed generated by physics (instaneous, or
other) operating on distance scales smaller than the radius of the
baryon or meson.\footnote{The celebrated example of the extended yet
  most useful Cooper pair quasiparticles indicates that this may not
  be a necessary condition.}

Because the constituent quarks are extended, complex entitities, the
``potentials'' acting betwen them are in general rather complex --
with spin-, flavor-, and possibly energy-dependence and non-locality.
Still, the naive expectations for potentials generated via gluon
exchange seem to lead to very successful predictions. Overall the
naive quark model, which predated QCD by almost a decade, still
provides more insight and results than any of the more sophisticated
effective Lagrangian approaches. Justifying its usage from first
principles would therefore constitute a major triumph and is certainly
a worthwhile effort. In our discussion of the QCD inequalities we have
encountered on several occasions [{\em e.g.} in discussing the pion
mass and wave function, the interflavor relations, and the positivity
of EM and more general vectorial interaction energies (see
Secs.~\ref{sec:vecsym} and \ref{sec:pseumeson} and
App.~\ref{app:scatt})] a remarkable coincidence between the ``naive''
quark model and rigorous QCD inequalities.  Thus the efforts to
extract constraints on the interquark potentials from QCD, and then to
utilize this rather broad class of potentials allowed to derive level
ordering theorems and/or baryon-baryon mass relations (see
App.~\ref{app:lieb2}), is quite worthwhile.

Even when we have mass dependent $q_{\text{con}}\bar{q}_{\text{con}}$
interactions, the system does retain its symmetry under
$q_i^{\text{con}} \leftrightarrow \bar{q}_j^{\,
  \mbox{\scriptsize{con}}}$, and one may thus wonder if just this
single feature, abstracted from the constituent quark model, is
sufficient to supply the flavor symmetry assumption which was the
missing link in our proof of meson the interflavor relations in
Sec.~\ref{sec:interflavor}. Unfortunately this is {\em not} the case.
Essentially the operation $\psi_i^{(0)}(d) \leftrightarrow
\psi_j^{(0)}(\bar{u})$ in Sec.~\ref{sec:interflavor} corresponds to
exchanging the ``fundamental'', pointlike, flavor carrying, entities
inside the extended constitutent quarks. For sufficiently heavy quarks
the ``cloud'' of light degrees of freedom ($u\bar{u},d\bar{d},$ and
possibly $s\bar{s}$ and gluons) are universal and flavor-independent.
However, the clouds around $s$ and $u$ quarks may differ
significantly. Hence the above symmetry assumption cannot be justified
in this case when {\em one} quark is heavy, and indeed $2m_{K^\ast}
\leq m_\rho + m_\phi$, which relies on an $s^{(0)} \leftrightarrow
u^{(0)}$ exchange, does marginally fail.

Clearly the $c\leftrightarrow u$ and $b\rightarrow u$ exchanges are
even less justified, yet since the hyperfine mass splittings are very
small ($\approx 1/m_c$ or $1/m_b$), the naive, spin-dependent,
potential model derivation of the inequalities holds.

The rigor -- and corresponding paucity of the euclidean correlation
function QCD inequalities -- reflects, in our view, the lack of
intuition as to which $A_\mu^a(x)$ field configurations are more
important in the functional path integral.\footnote{This is clearly a
  subjective statement, as such intuition might have been developed by
  lattice gauge calculations. Indeed the positivity of contributions
  of field configurations to the pseudoscalar propagators was known to
  practitioners in this field prior to the advent of QCD
  inequalities.} This makes justifying the approximation of keeping
certain field configurations more difficult, and is more of a handicap
than a virtue. By developing such an intuition akin to that inspired
by the quark model or strong coupling approximation (which suggested
the more important components of wave function(al)s in the variational
approach), we could obtain many more heuristic, yet very important and
useful inequalities. In delineating directions for future research on
the subject of this review, an obvious goal is to try and find more
rigorous proofs for likely correct inequalities, whose present
derivations are lacking. These inequalities would include the
detailed, flavor-dependent baryon-meson mass inequalities; relations
between masses of different mesons such as $m_{a_1}(1^+) \geq m_\rho
(1^-)$; and mass relations for glueballs such as $m_{{\rm gb}}^{0++} \leq$
(mass of any glueball).

We believe, however, that there are additional, promising, richer
avenues for further research. These involve two general (although
unfortunately not completely well defined) conjectures that we would
like to make next, and the application of the QCD inequality
techniques to observables other than hadron masses.


\subsection{A conjecture on the $N_f$ dependence of the QCD
  inequalities}

We have seen that many of the QCD inequalities such as $m_N \geq
m_\pi$ and $m^{(0^+)}_{u\bar{d}} \geq m^{(0^-)}_{u\bar{d}}$ are
related to observed symmetry patterns, namely the spontaneously broken
global axial and the conserved vectorial symmetries. The correlator
inequalities hold for arbitrary $N_f$. Indeed for $N_f$ degenerate
quarks, $N_f$ only appears in the positive determinantal factor via
$\left[\mbox{Det}(\dslasha + m)\right]^{N_f}$. In particular, all the
inequalities hold in the ``quenched approximation'' in which $N_f = 0$
and the determinantal factor disappears altogether. It is well known
that in the quenched limit, lattice calculations of QCD, such as
hadronic masses, couplings, {\em etc.} simplify enormously and one
has (within this approximation!) completely reliable results.

The QCD inequalities, for example $m_N \geq m_\pi$ and $m_\rho \geq
m_\pi$, are often satisfied by a large margin.  If we had, however, a
systematic trend for the inequalities to become weaker in the sense,
say, that $m_N/m_\pi$, $m_\rho/m_\pi$ decrease as $N_f$ increases,
that could furnish very useful information such as
\begin{equation}
   \left. \frac{m_N}{m_\pi} \right|_{\text{expt}} \leq
      \left. \frac{m_N}{m_\pi} \right|_{\text{quenched}}
\>.
\label{eq:quench}
\end{equation}
We would like to conjecture that this is indeed the case. Our
motivation is the likely decrease of the ``strength'' of the S$\chi$SB
as manifested via the magnitude of $\langle\qqb\rangle$ (measured in an
appropriate way) with $N_f$. Indeed increasing $N_f$ weakens -- via
enhanced screening -- the $\qqb$ gluon exchange interactions and
consequently the expected value of $\langle\qqb\rangle$.

Consider $\langle\psibar\psi\rangle_A = \langle
\mbox{tr}S_F\rangle_A$, with $\langle\psibar\psi\rangle_A$ evaluated
for a particular gauge background $A$ [here $\langle {\cal O}\rangle_A
= \int \, {\rm d}\mu(A) \langle{\cal O}\rangle_A / {\cal Z}$]. We can
write
\begin{equation}
   \langle \mbox{tr}S_F\rangle_A = \sum_j \frac{1}{(i\lambda_j(A)+m)}
      = \sum_j \frac{m}{\lambda^2_j(A) + m^2}
\>,
\end{equation}
where in the last expression we paired together the $\gamma_5$
conjugated, non-vanishing eigenvalues $\pm i\lambda_j$ of $\dslasha$.
For the $m\rightarrow 0$ limit of interest, the Lorentzians become
$\delta$-functions and hence we have
\begin{displaymath}
   \langle\qqb\rangle \neq 0 \leftrightarrow \lim_{\lambda \rightarrow
      0} \rho(\lambda) \neq 0
\>,
\end{displaymath}
with $\rho(\lambda)$ the density obtained after the $\frac{1}{{\cal
Z}} \int\, {\rm d}\mu(A)$ averaging of the eigenvalues of $\dslasha$. 

If ${\rm d}\mu_0(A) = \intD{A_\mu(x)}\exp\left[-S_{YM}(A_\mu)\right]$
is the $N_f = 0$ measure, the conjectured decrease of
$\langle\psibar\psi\rangle$ when $N_f = 0 \rightarrow N_f \neq 0$
amounts to:
\begin{equation}
   \left\langle \prod_l \left[ \lambda_l^2(A) + m^2 \right] \sum_j \left[
      \lambda_j^2(A) + m^2 \right]^{-1} \right\rangle_0 \leq 
      \left\langle \prod_l \left[ \lambda_l^2(A) + m^2 \right] \right\rangle_0
      \left\langle \sum_j \left[\lambda_j^2(A) + m^2 \right]^{-1} \right\rangle_0
\>,
\label{eq:nfzero}
\end{equation}
with $\langle f\rangle_0$ indicating averaging with the ${\rm
  d}\mu_0(A)$ measure. Since the factors $\prod_l (\lambda^2_l + m^2)$
and $\sum_j (\lambda_j^2 + m^2)^{-1}$ appearing in the conjectured
inequality (\ref{eq:nfzero}) are, respectively, monotonically
increasing (decreasing) with each $\lambda$, the conjecture is very
suggestive. Indeed for {\em any two} positive functions of one
variable $f(x)$ and $g(x)$ which are monotonically increasing
(decreasing), $\langle f \rangle \langle g \rangle \geq \langle fg
\rangle$. However, as pointed out to us by O. Kenneth
\cite{ref:kenneth2}, this inequality does not generally hold for
functions of many variables. In the present context it implies that
measures ${\rm d}\mu(\lambda_1,\ldots,\lambda_N)$ can be constructed
for which the conjectured inequality (\ref{eq:nfzero}) can be
reversed. This inequality is therefore heuristic and depends on
additional assumptions. Recalling our comment on functions of one
variable, the inequality would apply if there were one dominant
variable in the measure ${\rm d}\mu\left[A(x)\right]$. Color
confinement and asymptotic freedom suggest that the overall scale $R$
of the $A_\mu(x)$ fluctuations could serve in such a role. The
enhanced weight of larger fluctuations coupled with the expected
smallness of the corresponding $\lambda_i(A)$ indicates that this may
indeed be the case.

Changing $N_f$ changes the $\beta$ function and hence, for fixed
$g^2$, also $\Lambda_{\text{QCD}}$. This in turn implies that the
mass/length scales used in two lattice calculations with different
$N_f$ values should be appropriately changed to allow for meaningful
comparisons.

In general hadron masses [and also $\sqrt{\sigma}$, with $\sigma$ the
string constant ($\approx$ the coefficient of the linear potential in
heavy quarkonia, and $\langle\qqb\rangle^{1/3}$)] scale linearly with
$\Lambda_{\text{QCD}}$ and comparing ratios of such quantities for
different $N_f$ values is straightforward. However, $m_{\pi+} \approx
\sqrt{f_\pi m_\pi} \approx \sqrt{\Lambda_{\text{QCD}} m_0}$, with
$m_0$ the $(u+d)/2$ bare quark mass. Hence instead of
Eq.~(\ref{eq:quench}) we should use
\begin{equation}
   \frac{m_N f_\pi}{m_\pi^2} \leq
      \left. \frac{m_N f_\pi}{m_\pi^2} 
      \right|_{\text{quenched}}
\>.
\end{equation}

In conclusion we recall yet another heuristic supporting argument for
the conjectured decrease of $\langle\qqb\rangle /
(\Lambda_{\text{QCD}})^3$, {\em i.e.} of the quark condensate, with
$N_f$. It is the restoration of $Q_5$ symmetries in exactly SUSY QCD
theories \cite{ref:seiberg} at moderate $N_f / N_c$ ratios, for which
asymptotic freedom (and even confinement) may still hold. Since real
QCD is {\em not} supersymmetric, and further, since QCD inequalities
may not apply for SUSY theories, the significance of this fact in the
present context is not clear.

\subsection{Conjectured inequalities related to the ``ferromagnetic''
  nature of the QCD action}

The monotonic pion (configuration space) wave function was motivated
in the conclusion of Sec.~\ref{sec:beyond2pt} above by the 
ferromagnetic nature of the Yang-Mills
action. In some ferromagnetic spin systems this feature is embodied in a
rigorous set of Griffith's inequalities \cite{ref:griffiths} for the spin
correlations. The analogous $A_\mu$ correlations are not gauge invariant
but some gauge invariant versions can be conjectured.

The ferromagnetic character of QCD can motivate many other conjectured
inequalities, of which we will mention just a few.  An $E_1^a
E_2^a(x)$ excitation at $x$ (with 1 and 2 spatial indices of
$\vec{E}^a$, the color electric field) and an $E_1^a E_2^a(y)$
excitation at $y$\footnote{{\em i.e.} the state of the system obtained
  immediately after $E_1^a(x)E_2^a(x)$ operates on the vacuum.  Here
  $a = 1 \ldots 8$ is a color index, and 1 and 2 indicate the $x_1$
  and $x_2$ components ($E_1 = F_1^0$ {\em etc.}).}  correspond
intuitively to parallel ``spins'' (the corresponding plaquettes at $x$
and $y$ in the euclidean lattice are indeed parallel).  Likewise
$B_1^a B_2^a(x)$ and $B_1^a B_2^a(y)$ are ``parallel''. However,
$E_1^a(x) B_2^a(x)$ and $E_2^a(y) B_1^a(y)$ are not ``parallel''. The
preference of parallel configurations and the stronger positive
correlation between such configurations suggests therefore that
\begin{eqnarray*}
   \langle E_1^a(y) E_2^a(y) E_1^{a'}(x) E_2^{a'}(x) \rangle &\geq&
      \langle E_1^a(y) B_2^a(y) E_2^{a'}(x) B_1^{a'}(x) \rangle \\
   \langle B_1^a(y) B_2^a(y) B_1^{a'}(x) B_2^{a'}(x) \rangle &\geq&
      \langle E_1^a(y) B_2^a(y) E_2^{a'}(x) B_1^{a'}(x) \rangle
\>,
\end{eqnarray*}
and in particular
\begin{equation}
   \langle E_1E_2(x) E_1E_2(y) \rangle \langle B_1B_2(x) B_1B_2(y) \rangle
      \geq | \langle E_1B_2(x) E_2B_1(y) \rangle |^2
\>.
\label{eq:ebineq}
\end{equation}
which has been suggested by Muzinich and Nair
\cite{ref:muzinich,ref:muzinich2} along with many other inequalities. Since
$E_1E_2$ (or $B_1B_2$) and $E_1B_2$ acting on the vacuum creates $2^{++}$
($2^{-+})$ states, the last inequalities suggest that the lowest lying
$2^{++}$ glueball is lighter than the lowest $2^{-+}$ state:
\begin{equation}
   m_{2^{++}}^{(0)} \leq m_{2^{-+}}^{(0)}
\>.
\end{equation}
We believe that Eq.~(\ref{eq:ebineq}), many other Muzinich-Nair
relations (some of which involve $J$ dependence of masses as well),
and other yet to be discovered relations are true, but also furnish
tests not just of the measure positivity in QCD, but rather of the
ferromagnetic nature of its action.

\subsection{Inequalities for quantities other than hadronic masses}

The inequalities between two-point correlation functions and the
variational Hamiltonian techniques naturally tend to yield
inequalities between masses, rather than between other hadronic
observables. We have seen, however, in Secs.~\ref{sec:exotic} and
\ref{sec:beyond2pt}, and at the end of Sec.~\ref{sec:qqbar}, other
applications of the QCD inequalities techniques involving scattering
lengths, form factors, and weak matrix elements. In this last
subsection we would like to consider yet another quantity, namely high
energy hadronic cross sections. These attracted much attention in the
sixties and seventies, in connection with ``Regge pole'' exchanges in
the crossed $t$-channel, and the approximate ``quark counting''
suggested relation $\sigma_{\pi N} \approx \frac{2}{3}\sigma_{NN}$ was
one of the early indicators for the relevance of the quark model.

In retrospect asymptotic cross sections may reflect physics which is
quite different from that controlling meson and baryon masses. Hence
while the quark counting rule also suggested $m_B \gtrsim \frac{3}{2}
m_M$, which has been ``transformed'' here to the baryon-meson mass
inequalities, it is not clear that we have analogous inequalities for
hadronic cross sections:
\begin{equation}
   \sigma_{NN} \gtrsim \frac{3}{2} \sigma_{MN}
\>.
\end{equation}
However, this is precisely what we would like to conjecture here
(along with the more detailed flavor dependent variants). We have
attempted to motivate such a conjecture by using intersecting
chromoelectric fluxons as a model for high energy cross sections
\cite{ref:nussprd94} and the stringy trial wave functionals of
Sec.~\ref{sec:nonpertmbineq}. While we have not succeeded (mainly
because of ``shadowing'' -- multiple intersections which are more
prominent in baryon-baryon collisions), this model is highly
inadequate as it does not account for the rising cross sections.

We note that if $\int_{\Omega_1} E^2_1 \, {\rm d}^3x$ and
$\int_{\Omega_2} E^2_2 \, {\rm d}^3x$ roughly reflect the masses of
hadrons 1 and 2 extending over $\Omega_1,\Omega_2$, and $\int
\vec{E}_1 \cdot \vec{E}_2 \, {\rm d}^3x$ generates the Born scattering
amplitude, we expect some relation of the form\footnote{$\sigma_{12}
  \sim m_1 m_2$ is expected if the total cross section is dominated
  by a tensor particle exchange, which in turn dominates (in the sense
  of vector meson dominance) the graviton couplings
  \cite{ref:freund}.}
\begin{displaymath}
   \sigma_{12} \lesssim \frac{C}{\Lambda^4_{\text{QCD}}}
      m_1 m_2
\end{displaymath}
Clearly the case of the Goldstone, almost massless pion is again very
special.\footnote{The smallness of the pion mass relative to, for
  example, the $\rho$, stems in quark models from large short range
  hyperfine attraction. It is quite surprising that this hardly
  results in a smaller sized pion and in $\sigma_{\pi N} \leq
  \sigma_{\rho N}$ (the ``rho-pi puzzle'' \cite{ref:pagels}).}
The arguments of Sec.~\ref{sec:largenc} could be formally extended to
the (non-planar) exchanges pertinent to high energy cross
sections. This would then suggest relations like
$(\sigma_{\pi\pi})(\sigma_{NN}) \geq (\sigma_{\pi N})^2$, which
unfortunately would be difficult to test.

Physics similar to that motivating the QCD inequalities could be used
at many other length scales, and not only for underlying composite
models. Thus some variant of the Schwartz propagator inequality
(\ref{eq:schwartz}) may have implications for the frequency of
occurence of like-sex non-identical twins, and the more heuristic
interflavor inequalities of Sec.~\ref{sec:interflavor}, can suggest
many inequalities between bindings of polar molecules. We discuss
these applications in App.~\ref{app:bio}.

We believe that much more can and will be done on the subject of
inequalities, both in and out of QCD.

%
\section*{Acknowledgments}

It is often difficult to trace the precise inception of any project.
The motivation of S. Nussinov to prove a baryon-meson inequality,
using lattice techniques, stemmed from a well defined origin. It was
the summary talk in a lattice workshop (held in the summer of 1982 at
Saclay) by the late Claude Itzykson.  Itzykson, a pioneer of lattice
QCD and one of the finest mathematical physicists of our generation,
commented on the lack of rigorous results in QCD. Indeed even now,
seventeen years later, there are preciously few such results.

The work of S.N. on QCD inequalities, and this review in particular,
has been carried out during the last sixteen years at Tel Aviv
University; Brookhaven National Laboratory; Los Alamos National
Laboratory; Universities of Maryland, Pennsylvania, and Minnesota;
MIT; and at Boston University, SUNY Stonybrook, and the University of
South Carolina. The material in Sec.~\ref{sec:exotic} is largely based
on joint work with B.  Sathiapalan, and Sec.~\ref{sec:beyond2pt} and a
portion of Sec.~\ref{sec:sxsb} on work with M. Spiegelglas.

S.N. would like to acknowledge the crucial help of E. Lieb at a very
early stage, and his contribution to Sec.~\ref{sec:beyond2pt}; the
friendly and helpful correspondance with D.  Weingarten and
discussions with E. Witten; the encouragement of J.  Sucher and S.L.
Glashow; and the interest of the late Y. Dothan, W.  Greenberg, R.
Mohapatra, and J.  Pati. The critical comments of A. Casher, E. Lieb,
A. Martin, J.-M.  Richard, P.  Taxil, and E. Witten made S.N.  realize
some shortcomings of his earlier work. During the writing of the
review S.N. enjoyed discussions with M. Cornwall, S. Elitzur, Y.
Hosotani, O. Kenneth, H.J. Lipkin, A. Martin, I.J.  Muzinich, and V.P.
Nair.  We are particularly grateful to B. Svetitsky for many helpful
comments and advice, and to him and to B. Sathiapalan for reading
earlier versions of the manuscript.  S.N.  would like to acknowledge
the very warm hospitality of the Institute for Theoretical Physics at
the University of Minnesota, and in recent years the University of
South Carolina, which helped him very much in writing the review. M.L.
would like to thank S.N. for introducing her to the subject QCD
inequalities and for asking her to help with this review.  The work of
M.L. was supported by the Israel Science Foundation under Grant No.
255/96-1.  Finally, S.N. would like to acknowledge a grant from the
Israel Academy of Sciences.

%
\appendix
\section{Lieb's counterexample to Eq. (4.1)}
\label{app:lieb1}

Following Lieb \cite{ref:lieb} we prove that for the case of $m =
\infty$ {\em and} two-body potentials which are infinite square wells:
\begin{equation}
   V_{ij}(\vecr_i - \vecr_j) = V_{i.s.w.}(r) = \left\{
      \begin{array}{ll}
         \infty , & r \geq r_0 \\
         0 , & r \leq r_0
      \end{array}
   \right.
\end{equation}
the conjectured ``convexity'' relation
\begin{equation}
   E^{(0)}(m,m,m) + E^{(0)}(m,M,M) \leq 2 E^{(0)}(m,m,M)
\label{eq:em}
\end{equation}
is violated.

To achieve minimal (and indeed even finite!) ground state energies,
the wave functions have to vanish when $|\vecr_i - \vecr_j| > r_0$ for
any quark pair. For heavy quarks this can be done with no kinetic
energy penalty. Thus let us fix $\vecr_1 = 0$ in all three wave
functions $\psi^{(0)}(m,m,m), \psi^{(0)}(m,m,M),$ and
$\psi^{(0)}(m,M,M)$, thereby disposing also of the overall
translational degrees of freedom. For $\psi^{(0)}(m,m,m)$ we could
take all $\vecr_i = 0$, achieving the minimal possible energy:
$E^{(0)}(m,m,m) = 0$. In $\psi^{(0)}(m,m,M)$ it is clearly
advantageous to put also $\vecr_2 = \vecr_1 = 0$ so that the condition
$|\vecr_3 - \vecr_2| \leq r_0$ is satisfied automatically once
$|\vecr_3 - \vecr_1| \leq r_0$. Thus finding $E^{(0)}(m,m,M)$ reduces
to finding the minimal energy of the one-body Hamiltonian $\vec{p}^{\,
  2} / 2M + 2 V(\vecr \,)$, which for $V = V_{i.s.w.}$ is $h \equiv
\vec{p}^{\, 2} / 2M + V(\vecr \,)$.

On the other hand, $E^{(0)}(m,M,M)$ is the minimal energy of
\begin{equation}
   \hat{h} = \left[ \vec{p}_2^{\, 2} / 2M + V(\vecr_2) \right]
      + \left[ \vec{p}_3^{\, 2} / 2M + V(\vecr_3) \right]
      + V(\vecr_2 - \vecr_3)
\>.
\end{equation}
If $V(\vecr_2 - \vecr_3)$ were absent, $\hat{h}$ separates into two
one-body infinite square well problems and
\begin{displaymath}
   E^{(0)}(\hat{h}) = 2 E^{(0)}(h)
\>.
\end{displaymath}
However, since $\langle V_{23}\rangle \geq 0$ we have
\begin{equation}
   E^{(0)}(\hat{h}) \equiv E^{(0)}(m,M,M)
      \geq 2 E^{(0)}(h) = 2 E^{(0)}(m,m,M)
\>,
\end{equation}
and the desired inequality (\ref{eq:em}) is violated. Indeed
$\psi^{(0)}_{mMM}(\vecr_1 = 0, \vecr_2, \vecr_3)$ has to vanish now in
all of the region $|\vecr_2 - \vecr_3| \geq r_0$ in addition to the
vanishing for $r_2 \geq r_0, r_3 \geq r_0$. Incorporating this extra
constraint reduces the allowed six-dimensional volume in $(\vecr_2,
\vecr_3)$ space from $(4\pi r_0^3 / 3)^2$ to a fraction thereof.
Obviously this will increase the kinetic energy from $2E^{(0)}(h)$ to
$(2+\alpha)E^{(0)}(h)$, with $\alpha \approx 1$.

By continuity we therefore expect Eq.~(\ref{eq:em}) to fail already
when we approach the limit $m \rightarrow \infty$ and $V(r)
\rightarrow V_{i.s.w.}(r)$, {\em e.g.} via 
\begin{equation}
   V(r) = c_n (r/r_0)^n, \qquad n \rightarrow \infty
\>.
\end{equation}

In passing we note that $\exp\left[-\beta V_{i.s.w.}(r)\right] =
\Theta(r - r_0)$ is indeed not a postive semidefinite operator, since
its Fourier transform is not positive. This should be the case since
we have, as indicated in the following Appendix, Lieb's theorem that
the desired inequality holds when $\exp\left[-\beta V \right]$ is
positive semidefinite. Also for $V(r) = r^n, n \geq 4$, a case for
which another counter-example was produced first \cite{ref:martin},
$\exp\left[-\beta V(r)\right]$ is not positive semidefinite --
consistent with the fact that it fails the convexity conditions.

\section{Discussion of Lieb's results for three-body Hamiltonians}
\label{app:lieb2}

Following Lieb \cite{ref:lieb}, we prove the inequality (\ref{eq:em})
when the potentials are flavor independent:
\begin{displaymath}
   V(m,m) = V(M,M) = V(m,M) \equiv V
\>,
\end{displaymath}
and $\exp\left[-\beta V(\vec{x} - \vec{y})\right]$ positive
semidefinite. 

The propagation over (imaginary) euclidean time $\beta$ of the
three-body system is given by $\langle X | {\rm e}^{-\beta H} | X'
\rangle$, with $X = (x_1, x_2, x_3)_{initial}, X' = (x_1, x_2,
x_3)_{final}$. We compare three systems with
\begin{eqnarray}
   H_a &=& T_1(x_1) + T_2(x_2) + T_2(x_3) 
      + V_2(x_1,x_2) + V_2(x_1,x_3) + V_1(x_2,x_3)
   \nonumber \\
   H_b &=& T_1(x_1) + T_2(x_2) + T_3(x_3) 
      + V_3(x_1,x_2) + V_2(x_1,x_3) + V_1(x_2,x_3)
   \nonumber \\
   H_c &=& T_1(x_1) + T_3(x_2) + T_3(x_3) 
      + V_3(x_1,x_2) + V_3(x_1,x_3) + V_1(x_2,x_3)
\>.
\label{eq:habc}
\end{eqnarray}
For $X = X'$, ${\cal Z}_\beta(x) = \langle X | {\rm e}^{-\beta H} | X
\rangle$ is dominated, for $\beta \rightarrow \infty$, by the lowest
energy state
\begin{equation}
   \langle X | {\rm e}^{-\beta H} | X \rangle = \sum_n \left| \langle
      X | n \rangle \right|^2 {\rm e}^{-\beta E_n} \stackrel 
      {{\textstyle \longrightarrow}}{{\scriptstyle \beta\rightarrow\infty}}
      \left| \langle X | 0 \rangle \right|^2 {\rm e}^{-\beta E_0}
\>,
\end{equation}
where we use the completeness sum over energy eigenstates. 

Thus to prove the desired inequality
\begin{displaymath}
   E^{(0)}(a) + E^{(0)}(c) \leq 2 E^{(0)}(b)
\end{displaymath}
it suffices to show that
\begin{equation}
   {\cal Z}_\beta(a) {\cal Z}_\beta(c) \geq {\cal Z}_\beta^2(b)
\>.
\label{eq:zschwartz}
\end{equation}

In general the path integral, and ${\cal Z}_\beta$ in particular, is
obtained  by dividing the total evolution into many consecutive
evolutions over small time steps, so that
\begin{displaymath}
   {\cal Z}_\beta = \lim_{N \rightarrow\infty} {\cal Z}_\beta(N)
      = \left[ \left( {\rm e}^{-\beta T/N} \, {\rm e}^{-\beta V/N}
      \right)^N \right]
\>,
\end{displaymath}
with $T$ and $V$ the total one-body kinetic and two-body potential
parts of the Hamiltonian $H = T + V$. Clearly it is sufficient to
prove the inequality for each $N$.

We next insert a complete set of $X$-space states between each pair of
${\rm e}^{-\beta T/N} \, {\rm e}^{-\beta V/N}$ factors.  This gives
${\cal Z}_\beta$ as a path integral. Each such path consists of three
(spatially) closed polygonal paths $\tilde{X}_1, \tilde{X}_2,
\tilde{X}_3$, with $\tilde{X}_1$ consisting of the $N+1$ points
\begin{displaymath}
   \tilde{X}_1(0) = X_1, X_1(\beta/N), X_1(2\beta/N), 
      \ldots, X_1(N\beta/N) = X_1(\beta) = X_1
\>,
\end{displaymath}
and likewise for $\tilde{X}_2$ and $\tilde{X}_3$ (see
Fig.~\ref{fig:lieb}). Since each factor
\begin{equation}
   \langle X_1(j) X_2(j) X_3(j) | {\rm e}^{-\beta T/N} \, 
      {\rm e}^{-\beta V/N} | X_1(j+1) X_2(j+1) X_3(j+1) \rangle \\
\end{equation}
involves an evolution over an infinitesimal ``time'' $\beta / N$, the
non-commutativity of the kinetic and potential parts of $H$ is
neglected. The kinetic single particle operators contribute to ${\cal
  Z}_\beta^N(a)$ a factor of
\begin{displaymath}
   F_a = F_1(\tilde{X}_1) F_2(\tilde{X}_2) F_2(\tilde{X}_3)
\>,
\end{displaymath}
and to ${\cal Z}_\beta^N(b)$ and ${\cal Z}_\beta^N(c)$
\begin{eqnarray*}
   F_b &=& F_1(\tilde{X}_1) F_2(\tilde{X}_2) F_3(\tilde{X}_3) \\
   F_c &=& F_1(\tilde{X}_1) F_3(\tilde{X}_2) F_3(\tilde{X}_3)
\end{eqnarray*}
with $F_i(X)$ corresponding to $T_i$ in Eq.~(\ref{eq:habc}) above. 

The potential two-body operators contribute to ${\cal Z}_\beta^N$ a
product of three terms depending on the three pairs of
paths. Specifically, the contributions in cases (a), (b), and (c) are
\begin{eqnarray*}
   {\cal Z}_\beta^N(a) &:& G_2^N(\tilde{X}_1,\tilde{X}_2) 
      G_2^N(\tilde{X}_1,\tilde{X}_3)
      G_1^N(\tilde{X}_2,\tilde{X}_3) \\
   {\cal Z}_\beta^N(b) &:& G_2^N(\tilde{X}_1,\tilde{X}_2) 
      G_3^N(\tilde{X}_1,\tilde{X}_3)
      G_1^N(\tilde{X}_2,\tilde{X}_3) \\
   {\cal Z}_\beta^N(c) &:& G_3^N(\tilde{X}_1,\tilde{X}_2) 
      G_3^N(\tilde{X}_1,\tilde{X}_3)
      G_1^N(\tilde{X}_2,\tilde{X}_3)
\>,
\end{eqnarray*}
where the $G_i(\tilde{X}_k, \tilde{X}_l)$ are generated from $V_i(X_k,
X_l)$ via
\begin{displaymath}
   G_i(\tilde{X}_k,\tilde{X}_l) = \prod_{j=1}^N \exp 
      \left[ - ( \beta/N ) V_i \left(
      X_k(\beta j / N), X_l(\beta j / N) \right) \right]
\>.
\end{displaymath}

For the particular case of $V_1$, the fact that e$^{-\beta V_1/N}$ is
positive semidefinite ensures that the $N$-fold tensor product
defining $G_1(\tilde{X}_2,\tilde{X}_3)$ is also positive semidefinite.
Collecting all terms and separating the d$^{3(N-1)}X_1$ integrations
over the $N-1$ intermediate points $X_1^2, \ldots, X_1^N$ along the
polygonal path $\tilde{X}_1$ we have
\begin{eqnarray}
   {\cal Z}_\beta^N(a) &=& \int \, {\rm d}^{3(N-1)}X_1 F_1(\tilde{X}_1) \, 
      \int \, {\rm d}^{3(N-1)}X_2 \, \int \, {\rm d}^{3(N-1)}X_3 \,
   \nonumber \\
      &\times& G_2^N(\tilde{X}_1,\tilde{X}_2) F_2(\tilde{X}_2) 
      G_2^N(\tilde{X}_1,\tilde{X}_3) 
      F_2(\tilde{X}_3) G_1^N(\tilde{X}_2,\tilde{X}_3)
   \nonumber \\  
   {\cal Z}_\beta^N(b) &=& \int \, {\rm d}^{3(N-1)}X_1 F_1(\tilde{X}_1) \, 
      \int \, {\rm d}^{3(N-1)}X_2 \, \int \, {\rm d}^{3(N-1)}X_3 \,
   \nonumber \\
      &\times& G_2^N(\tilde{X}_1,\tilde{X}_2) F_2(\tilde{X}_2) 
      G_3^N(\tilde{X}_1,\tilde{X}_3) 
      F_3(\tilde{X}_3) G_1^N(\tilde{X}_2,\tilde{X}_3)
   \nonumber \\  
   {\cal Z}_\beta^N(c) &=& \int \, {\rm d}^{3(N-1)}X_1 F_1(\tilde{X}_1) \, 
      \int \, {\rm d}^{3(N-1)}X_2 \, \int \, {\rm d}^{3(N-1)}X_3 \,
   \nonumber \\
      &\times& G_3^N(\tilde{X}_1,\tilde{X}_2) F_3(\tilde{X}_2) 
      G_3^N(\tilde{X}_1,\tilde{X}_3) 
      F_3(\tilde{X}_3) G_1^N(\tilde{X}_2,\tilde{X}_3)
\>.
\label{eq:polypath}
\end{eqnarray}
Defining next the $N \times N$ matrices
\begin{eqnarray}
   V_2^N(\tilde{X}_1,\tilde{X}_2) &=& \sqrt{F_1(\tilde{X}_1)} 
      G_2^N(\tilde{X}_1,\tilde{X}_2) F_2(\tilde{X}_2)
   \nonumber \\
   V_3^N(\tilde{X}_1,\tilde{X}_2) &=& \sqrt{F_1(\tilde{X}_1)} 
      G_3^N(\tilde{X}_1,\tilde{X}_2) F_3(\tilde{X}_2)
\>,
\label{eq:vmat}
\end{eqnarray}
we can write Eq.~(\ref{eq:polypath}) in the concise form
\begin{eqnarray}
   {\cal Z}_\beta^N(a) &=& {\rm tr} \, (V_2^T \cdot G_1 \cdot V_2)
   \nonumber \\
   {\cal Z}_\beta^N(b) &=& {\rm tr} \, (V_2^T \cdot G_1 \cdot V_3)
   \nonumber \\
   {\cal Z}_\beta^N(c) &=& {\rm tr} \, (V_3^T \cdot G_1 \cdot V_3)
\>.
\label{eq:ztr}
\end{eqnarray}
We note that the positivity of $F_1(\tilde{X}_1)$ was implicitly assumed in
taking the square root in Eq.~(\ref{eq:vmat}). For the case of
interest with $T_1 = {\vec{p}}^{\, 2}_1 / (2m_1)$, $\langle X_1 | {\rm
  e}^{-\beta T} | X_1 \rangle$ is the probability for returning to the
initial point in a Gaussian random walk after a time $\beta$.
$F^N(X_1)$ is the probability that this happens for a specific path
$X_1$ with $N-1$ specific intermediate steps, and hence is clearly
positive. This positivity of the ``heat kernel'' can apparently be
generalized to other relativistic forms of the kinetic energy, for
example $T_1 = \sqrt{{\vec{p}}_1^{\, 2} + m_1^2}$. 

To complete the proof we note that Eqs.~(\ref{eq:ztr}) define a
``scalar product'' of $V_2$ with itself, $V_2$ and $V_3$, and $V_3$
with itself. This can be most clearly seen by transforming to the
basis in which $G_1$ is diagonal with all diagonal elements real and
positive. This change of basis leaves tr $(V_2^T G_1 V_3)$ invariant
but casts it in the form
\begin{displaymath}
   \sum_{\alpha,n}(V_2)_{\alpha n} (G_1)_{nn} (V_3)_{n \alpha}
\>,
\end{displaymath}
which is bilinear in $V_2$ and $V_3$ and positive whenever two
identical ``$V$-vectors'' ( of length $[3(N-1)]^2$ ) are used. Hence
these scalar products satisfy the Schwartz inequality which is
precisely the desired result (\ref{eq:zschwartz}).

It has been noted by Lieb that the assumption of having only two-body
potentials is rather restrictive and we can allow also genuine
three-body potentials $V(X_1,X_2,X_3)$ as long as e$^{-\beta V}$ is
positive semidefinite as a function of $X_2,X_3$.

%
\section{Proof of Eq.~(5.2)}
\label{app:eq42}

We prove Eq.~(\ref{eq:hflavors}) using a concrete basis of states in
which it holds for all matrix elements. We will use the Hamiltonian
version of QCD with a spatial cubic lattic (sites $\vn$, unit vectors
$\nhat$ along the positive $x,y,z$ axes).

The QCD Hamiltonian is written as \cite{ref:creutz,ref:kogsuss}
\begin{eqnarray}
   H_{\text{QCD}} &=& \frac{g^2}{2} \sum_{\vn,\nhat > 0} {\rm tr}
      (E_{\vn, \vn + \nhat})^2 + \frac{1}{2g^2} \sum_{{\rm plaquettes}}
      {\rm tr} (U U U^{\dag} U^{\dag} + {\rm h.c.})
   \nonumber \\
   \qquad &+& \sum_{i=1}^{N_f} \sum_{\vn,\nhat > 0} \psi_i^{\dag}(\vn)
      U_{\vn,\vn + \nhat} \psi_i(\vn + \nhat)
      + {\rm h.c.} + \sum_{i=1}^{N_f} \sum_{\vn} m_i 
      \psibar_i(\vn) \psi_i(\vn) + \cdots
\>.
\label{eq:hqcd}
\end{eqnarray} 
It depends on the SU(3) ``connection'' matrices $U_{\vn, \vn + \nhat}$
which live on lattice links between $\vn$ and $\vn + \nhat$, the
canonically conjugate generators $E_{\vn, \vn + \nhat}$, and the fermionic
spinors $\psi_i(\vn)$ [and conjugate $\psibar_i(\vn)$] at each lattice
site. The first three terms are analagous to the $\vec{E}^2, \vec{B}^2$ and
$\psibar(x) \dslasha \psi(x)$ terms in the continuum Hamiltonian.

We adopt the ``strong coupling'' basis in which all tr$(E_{\vn, \vn +
\nhat})^2$ are diagonal.  At each lattice site we specify the spinors
$\psi_i(\vn)$. The mass term is
\begin{equation}
   \sum_{\vn} \sum_{i=1}^{N_f} \psi_i^{\dag}(\vn) \gamma_0
      m_i \psi_i(\vn)
\>.
\end{equation}
Each of the plaquette terms creates a minimal, closed, $\vec{E}$ flux
line going clockwise (or anti-clockwise) around an elementary
plaquette. This will, in general, change tr$(E_{\vn, \vn + \nhat})^2$
on each of the four adjoining links and hence gives rise to
non-diagonal elements in the representation considered here. Also the
third [$\psibar(x)\!\!\dslash\psi(x)$] term changes the lattice
configuration by allowing the quark to ``hop'' to a neighboring site
(dragging along a flux line), and by creating or annihilating
$q_l\bar{q}_l$ pairs at neighboring sites.

We illustrate some of the matrix elements of $H_{\text{QCD}}$ on a
small $3 \times 3$ two-dimensional lattice in Fig.~\ref{fig:lattice}.
Each of the columns and rows is labeled by a complete lattice
configuration. In the matrix we indicate both the actual value of the
matrix elements (above dotted line) and the specific terms of
$H_{\text{QCD}}$ contributing to it (below dotted line). We omit
additional four-fold spinor indices for each of the occupied sites.

Local gauge invariance 
\begin{eqnarray}
   \psi_i(\vn) &\rightarrow& V(\vn) \psi_i(\vn)
   \nonumber \\
   \psibar_i(\vn) &\rightarrow& \psibar_i(\vn) V^{\dag}(\vn)
   \nonumber \\
   U_{\vn, \vn + \nhat} &\rightarrow& V(\vn) U_{\vn, \vn + \nhat}
      V^{\dag}(\vn + \nhat)
   \nonumber \\
   E_{\vn, \vn + \nhat} &\rightarrow& V(\vn) E_{\vn, \vn + \nhat}
      V^{\dag}(\vn + \nhat)
\>,
\end{eqnarray}
is respected by all terms of $H_{\text{QCD}}$. It constrains the
physically allowed states via Gauss' equation
\begin{equation}
   \sum_{\pm \nhat} E_{\vn, \vn + \nhat} = \sum_l
      \delta(\vecr - \vecr_l) Q_l
\>,
\label{eq:gauss}
\end{equation}
stating that the sum of all outgoing $E_{\vn, \vn + \nhat}$ flux lines
[which generate $V(\vn)$] vanishes at all lattice sites $\vn$ except
those $\vecr_l$ with external color charges $Q_l$. (The $Q_l$ are due
to quarks. Up to $2 N_f$ quarks and $2 N_f$ antiquarks are allowed at
each site by Fermi statistics.)

It is precisely via this Gauss condition that the specific (mesonic,
baryonic {\em etc.}) sector becomes relevant.  Figure
\ref{fig:lattice} pertains to the $M_{u\bar{d}}$ sector. The simple
configurations in Fig.~\ref{fig:lattice} have just one $u$ and one
$\bar{d}$ in the lattice.  Eq.~(\ref{eq:gauss}) then simplifies into
\begin{equation}
   \sum_{+ \nhat} E_{\vn, \vn + \nhat} =
      \delta(\vecr - \vecr_u) \lambda + \delta(\vecr - \vecr_d) 
      \lambda^{\dag}
\>,
\label{eq:vgauss}
\end{equation}
where $r_{u(\bar{d})}$ are the locations of $u(\bar{d})$, and
$\lambda$, $\lambda^{\dag}$ are the corresponding color matrices.

Since $H_{\text{QCD}}$ commutes with the gauge transformations, the
evolution of the state on the lattice, allowing for creation of $\qqb$
pairs at one lattice point and their subsequent separation and
possible annihilation with other $\qqb$, maintains the Gauss condition
(\ref{eq:gauss}) [though not, in general, its simple ``valence''
version (\ref{eq:vgauss})].  Since the Gauss law constraints are
common to all the mesonic sectors, they do not interfere with the
suggested operator relation Eq.~(\ref{eq:hflavors}).

We start by showing that (\ref{eq:hflavors}) is satisfied when we switch
off the quark creation and annihilation terms and then work our way
gradually to the general case.
\begin{trivlist}
\item[ 1.] In the simple case where $M_{i\bar{\jmath}}$ consists of
  just the $q_i \bar{q}_j$ quarks and not other pairs, it it easy to
  see that all the non-mass terms in $H_{\text{QCD}}$ have identical
  matrix elements in any $H_{ij}$ sector. Also the Gauss law
  constraints are the same. The matrix elements depend only on the
  $U$'s and $E$'s and the generic quarks or antiquarks but not on the
  specific quark flavors. Taking for concreteness $ijkl = udsc$, we
  have flavor dependence only in the mass terms ($\psi^C$ denotes a
  charge conjugated spinor of an antiquark):
\begin{mathletters}
   \begin{equation}
      H^{\text{mass}}_{u\bar{d}} = m_u \psi^{\dag}(\vecr_q) \gamma_0
         \psi(\vecr_q) + m_d (\psi^C)^{\dag}(\vecr_{\bar{q}}) \gamma_0
         \psi^C(\vecr_{\bar{q}})
   \end{equation} 
   \begin{equation}
      H^{\text{mass}}_{c\bar{s}} = m_c \psi^{\dag}(\vecr_q) \gamma_0
         \psi(\vecr_q) + m_s (\psi^C)^{\dag}(\vecr_{\bar{q}}) \gamma_0
         \psi^C(\vecr_{\bar{q}})
   \end{equation} 
   \begin{equation}
      H^{\text{mass}}_{u\bar{s}} = m_u \psi^{\dag}(\vecr_q) \gamma_0
         \psi(\vecr_q) + m_s (\psi^C)^{\dag}(\vecr_{\bar{q}}) \gamma_0
         \psi^C(\vecr_{\bar{q}})
   \end{equation} 
   \begin{equation}
      H^{\text{mass}}_{c\bar{d}} = m_c \psi^{\dag}(\vecr_q) \gamma_0
         \psi(\vecr_q) + m_d (\psi^C)^{\dag}(\vecr_{\bar{q}}) \gamma_0
         \psi^C(\vecr_{\bar{q}}).
   \end{equation} 
\end{mathletters}
Since the same four terms appear in $H_{u\bar{d}} + H_{c\bar{s}}$ as in
$H_{u\bar{s}} + H_{c\bar{d}}$, Eq.~(\ref{eq:hflavors}) is valid in this
valence approximation:
\begin{equation}
   H_{u\bar{s}} + H_{c\bar{d}} = H_{u\bar{d}} + H_{c\bar{s}}
   \qquad \mbox{(subspace with two quarks)}
\>.
\end{equation}

\item[ 2.] The above argument is not affected by the creation/annihilation
  terms of all pairs $x\bar{x} \neq u\bar{u}, d\bar{d}, s\bar{s},
  c\bar{c}$.  The created quarks or antiquarks cannot be Pauli blocked
  or annihilated by the valence quarks or antiquarks. Their effect,
  just like that of gauge fields, is common to $H_{u\bar{d}},
  H_{c\bar{s}}$, {\em etc}.
  
\item[ 3.] Finally, we turn on also pair creation/annihilation of the
  valence flavors. We would then generate configurations such as
  $q_i^v(u_1\bar{u}_1) \ldots (x_p \bar{x}_p)\bar{q}_j^v$ in which
  $u_1$ is at the same lattice site, and same spin and color state as
  the valence quark $q_i^v$. If $q_i^v = u$, such configurations
  should, by the Pauli principle, be disallowed. Also $\bar{u}_1$
  could subsequently annihilate the valence quark. Both effects occur
  in $M_{u\bar{d}}$ and $M_{u\bar{s}}$ but not in $M_{c\bar{s}}$ and
  $M_{c\bar{d}}$. To avoid a possible difficulty in deriving
  Eq.~(\ref{eq:hflavors}) we will {\em define} the configuration
  corresponding to
\begin{equation}
   u_v(u_1 \bar{u}_1) \ldots (x_p \bar{x}_p) \bar{d}_v \,\,
      {\rm in} \,\, M_{u\bar{d}} \,\, \left[{\rm or} \,\,
      u_v(u_1 \bar{u}_1) \ldots (x_p \bar{x}_p) \bar{s}_v\,\,
      {\rm in} \,\, M_{u\bar{s}}\right]
\end{equation}
to be
\begin{equation}
   c_v(c_1 \bar{c}_1) \ldots (x_p \bar{x}_p) \bar{d}_v \,\,
      {\rm in} \,\,M_{c\bar{d}} \,\, \left[{\rm or}\,\, 
      c_v(c_1 \bar{c}_1) \ldots (x_p \bar{x}_p) \bar{s}_v\,\,
      {\rm in} \,\, M_{c\bar{s}}\right]
\>.
\end{equation}
This, rather than the definition with a common set of pairs, ensures
that configurations excluded in $M_{u\bar{d}}$ or $M_{u\bar{s}}$ will
be excluded in $M_{c\bar{d}}$ or $M_{c\bar{s}}$; and that for any
subsequent $\bar{u}_1 u_v$ annihilation induced by $H_{u\bar{d}}$ or
$H_{u\bar{s}}$ there will be a corresponding annihilation of
$\bar{c}_1 c_v$ in $H_{c\bar{d}}$ or $H_{c\bar{s}}$. The diagonal mass
terms which contributed to $H_{u\bar{d}} + H_{c\bar{s}}$ and
$H_{c\bar{d}} + H_{u\bar{s}}$ in such configurations will still be the
same. In each case we have $3m_u, 3m_c, m_d$ and $m_s$ terms and a
common set of non-$uscd$ quark masses due to other pair creation.

The construction of corresponding configurations is readily generalized to
the case of arbitrary number of pairs with the flavors of the valence
quark. A generic state in $M_{i\bar{\jmath}}$ is
\begin{equation}
   q_i (q_i^{\, (1)} \bar{q}_i^{\, (1)}) \ldots (q_i^{(p_v)}
      \bar{q}_i^{\, (p_v)}) 
      (x\bar{x})^{n_x} (q_j^{\, (1)} \bar{q}_j^{\, (1)}) 
      \ldots (q_j^{(\bar{p}_v)} \bar{q}_j^{\, (\bar{p}_v)}) \bar{q}_j
\>.
\end{equation}
The superscripts $p_v, n_x, \ldots$ simply count the number of pairs
of type $(q_i \bar{q}_i)$, $(x,\bar{x})$, {\em etc}.  Altogether we
have $p_v$ pairs of the valence quark flavor, $\bar{p}_v$ pairs of the
valence antiquark flavor, and additional $x\bar{x}$ pairs. With the
specific relation (\ref{eq:hflavors}) in mind we will explicitly
exhibit among those $x\bar{x}$ pairs the $(q_k \bar{q}_k)$ and $(q_l
\bar{q}_l)$ pairs:
\begin{equation}
   M_{i\bar{\jmath}} = q_i (q_i \bar{q}_i)^{p_v} 
      (q_j \bar{q}_j)^{\bar{p}_v}
      \bar{q}_j (q_k \bar{q}_k)^{p_1} (q_l \bar{q}_l)^{p_2}
      (x\bar{x})^{n_x} \ldots
\end{equation}
with $x \neq q_i, q_j, q_k, q_l$.  The corresponding states in the
other three mesonic sectors will be taken as
\begin{eqnarray}  
   M_{k\bar{l}} &=& q_k (q_k \bar{q}_k)^{p_v} (q_l \bar{q}_l)^{\bar{p}_v}
      \bar{q}_l (q_i \bar{q}_i)^{p_1} (q_j \bar{q}_j)^{p_2}
      (x\bar{x})^{n_x} \ldots
   \nonumber \\
   M_{i\bar{l}} &=& q_i (q_i \bar{q}_i)^{p_v} 
      (q_l \bar{q}_l)^{\bar{p}_v}
      \bar{q}_l (q_k \bar{q}_k)^{p_1} (q_j \bar{q}_j)^{p_2}
      (x\bar{x})^{n_x} \ldots
   \nonumber \\
   M_{k\bar{\jmath}} &=& q_k (q_k \bar{q}_k)^{p_v} 
      (q_j \bar{q}_j)^{\bar{p}_v}
      \bar{q}_j (q_i \bar{q}_i)^{p_1} (q_l \bar{q}_l)^{p_2}
      (x\bar{x})^{n_x} \ldots
\end{eqnarray}
All of these configurations have the same total number of quarks and
antiquarks: $N = 2 + 2p_v + 2 \bar{p}_v + 2 p_1 + 2 p_2 + 2 \sum_{x
  \neq ijkl} n_x$. Also the corresponding quarks (or antiquarks) in
$M_{i\bar{\jmath}}, M_{k\bar{l}}$, {\em etc.} are taken to be at
the same locations with identical spinor and color states. The Gauss
constraints, the antisymmetrization effects, and (except for mass
contributions) also all matrix elements are the same for
$H_{i\bar{\jmath}}, H_{k\bar{l}}, H_{i\bar{l}}$ and
$H_{k\bar{\jmath}}$. Finally, we have in $H_{i\bar{\jmath}} +
H_{k\bar{l}}$ the same mass terms as in $H_{i\bar{l}} +
H_{k\bar{\jmath}}$. [Altogether there are $2p_1 + 2p_v + 1$ terms with
$m_i$ as a coefficient, the same number with coefficient $m_k$; and $2
p_2 + 2 \bar{p}_v + 1$ terms with coeffecients $m_j (m_l)$.] All of
these terms make identical contributions to both sides of
Eq.~(\ref{eq:hflavors}), which is therefore correct.
\end{trivlist}

Viewing states with equal of numbers of (say) $c\bar{c}$ and
$u\bar{u}$ pairs as ``corresponding'' states does {\em not} imply that
heavy and low mass pairs occur with equal probability in physical
states. This issue is determined by solving $H^{(ij)} | \Psi_{ij}
\rangle = E^{(ij)} | \Psi_{ij} \rangle$ and heavy pairs are in fact
strongly suppressed \cite{ref:cashnuss}.  This is analagous to the
relative separation $r_{\qqb}$ in the example of
Sec.~\ref{sec:mmineq}, where $r_{c\bar{c}}$ and $r_{u\bar{u}}$
correspond to the same ``generic'' degree of freedom $r$. While
solving the Schr\"{o}dinger equations with $m_c \gg m_u$ leads to very
different expectation values of $r$
\begin{displaymath}
   \langle r \rangle_{J/\psi} \ll \langle r \rangle_\omega
\>,
\end{displaymath}
the operator relation $H_{c\bar{c}} + H_{u\bar{u}} = 2 H_{c\bar{u}}$
still holds for the nonrelativistic Hamiltonians.

%
\section{Enumeration and specification of the baryon-meson inequalities}
\label{app:bmineq}

Altogether we have forty two different flavor-spin baryonic ``ground
state'' combinations consisting of $udscb$ quarks, when we do not
distinguish between members of $I$-spin multiplets and assume no
orbital, radial, or gluonic excitations.  (Electromagnetic mass
splittings were discussed in Sec.~\ref{sec:pseumeson}).

We consider first the $J^P = (3/2)^+$ ``$\Delta$-like'' baryons. In
baryons made of three identical quarks $bbb$, $ccc$, $sss$
($\Omega^-$), and $uuu = ddd$ ($\Delta^{++}, \Delta^-$), each diquark
combination is in a spin triplet with $J^P = (3/2)^+$. Also, in all
other $(3/2)^+$ states all quark pair subsystems, and hence the
corresponding mesonic systems $q_i q_j (\rightarrow q_i \bar{q}_j),
q_j q_k (\rightarrow q_j \bar{q}_k), q_k q_i (\rightarrow q_k
\bar{q}_i)$, must be in the triplet, $S = 1$ state. Hence only vector
meson masses appear on the right hand side of the inequalities. If we
list all mesons in lexicographic order with $b > c > s > u, d$, we
have the following $(3/2)^+$ states: $uuu \, (\Delta)$, $suu \,
(\Sigma^{\ast})$, $ssu \, (\Xi^{\ast})$, $sss \, (\Omega^-)$, $cuu \,
(\Sigma_c^{\ast})$, $csu \, (\Xi_c^{\ast})$, $css \,
(\Omega_c^{\ast})$; the doubly charmed baryons $ccu$ and $ccs$; the
$ccc$ state; and $buu \, (\Sigma_b^{\ast})$, $bsu \, (\Xi_b^{\ast})$,
and $bss \, (\Omega_b^{\ast})$.  There are also the heavier states
$bcc$, $bbu$, $bbs$, $bbc$, and finally $bbb$, which are unlikely to
be discovered soon.

For the sake of completeness, we list in Table \ref{table:two} all the
relevant inequalities. The first five can already be tested and are
satisfied with a reasonable margin in all cases. The next eight
inequalities constitute lower bounds predicted for as yet undiscovered
baryonic states, which hopefully can be verified in the near future.
The remaining three will presumably be verified only in the distant
future.

We move next to the $J^P = (1/2)^+$ baryons. Let us first focus on the
case where we have two identical quarks, namely $duu \, (p)$, $suu$ (or
$sud$ with $I = 1$) ($\Sigma^+$), $ssu \, (\Xi^-)$, $cuu \,
(\Sigma_c^{++})$, $css \, (\Xi_c)$, $ccu$, $ccs$, $buu$, $bss$, and
$bcc$. In all these ``$y x_1 x_2$ - type'' states the identical flavor
quarks $x_1 x_2$ must be in a triplet state, namely
\begin{equation}
   ( \vec{s}_{x_1} + \vec{s}_{x_2} )^2 = s_{x_1 x_2}^2 
      = 1 \cdot (1+1) = 2
\>.
\label{eq:triplet}
\end{equation}
Writing next the total baryon spin as
\begin{mathletters}
   \begin{equation}
      ( \vec{s}_{x_1} + \vec{s}_{x_2} + \vec{s}_{y} )^2 
         = (s_{x_1} + s_{x_2})^2 + 2 \vec{s}_{x_1} \cdot \vec{s}_{y}
         + 2 \vec{s}_{x_2} \cdot \vec{s}_{y} + (s_y)^2
         = s_B^2
   \end{equation}
   {\rm or, using Eq.~(\ref{eq:triplet}) and $s_B = 1/2$:}
   \begin{equation}
      2 + 2 ( \vec{s}_{x_1} \cdot \vec{s}_{y} 
         + \vec{s}_{x_2} \cdot \vec{s}_{y} ) + 3/4 = 3/4,
   \end{equation}
   {\em i.e.}
   \begin{equation}
      \left\langle \vec{s}_{x_1} \cdot \vec{s}_{y} \right\rangle_B
         + \left\langle \vec{s}_{x_2} \cdot \vec{s}_{y} 
         \right\rangle_B = -1,
   \end{equation}
\end{mathletters}
with $\langle \, \, \rangle_B$ denoting the expectation value in the
baryon state.  Since we have $x_1 \leftrightarrow x_2$ symmetry of the
above expectation values, we finally conclude that
\begin{equation}
   \left\langle \vec{s}_{x_1} \cdot \vec{s}_{y} \right\rangle_B
      = \left\langle \vec{s}_{x_x} \cdot \vec{s}_{y} \right\rangle_B
      = - 1/2
\>.
\label{eq:exphalf}
\end{equation}

Hence each of the $y x_i \, ( \rightarrow y \bar{x}_i)$ subsystems
must be a mixture of 3/4 singlet and 1/4 triplet states for which
$\langle \vec{s} \cdot \vec{s} \rangle$ = - 3/4 and + 1/4,
respectively. This then implies, by collecting all the terms in $(x_1
\bar{x}_2)\, , (y \bar{x}_1) \, , (x_2 \bar{y})$, that the baryon-meson
inequalities read as follows:
\begin{eqnarray}
   m_n &\geq& \frac{m_\rho}{2} + \frac{1}{4} (3 m_\pi + m_\rho)
      = \frac{3}{4}(m_\rho + m_\pi) 
   \nonumber \\
   m_{\Sigma} &\geq& \frac{m_\rho}{2} + \frac{1}{4} (3 m_K 
      + m_{K^{\ast}})
   \nonumber \\
   m_{\Xi} &\geq& \frac{m_\phi}{2} + \frac{1}{4} (3 m_K 
      + m_{K^{\ast}})
   \nonumber \\
   m_{\Sigma_c} &\geq& \frac{m_\rho}{2} + \frac{1}{4} (3 m_D
      + m_{D^{\ast}})
   \nonumber \\
   m_{\Sigma_b} &\geq& \frac{m_\rho}{2} + \frac{1}{4} (3 m_B
      + m_{B^{\ast}})
\>.
\end{eqnarray}
We have similar relations for the doubly strange baryons:
\begin{eqnarray}
   m_{\Xi_c} &=& m_{css}(1/2)^+ \geq \frac{m_\rho}{2} 
      + \frac{1}{4} (3 m_{D_s} + m_{D_s^{\ast}})
   \nonumber \\
   m_{\Xi_b} &=& m_{bss}(1/2)^+ \geq \frac{m_\rho}{2} 
      + \frac{1}{4} (3 m_{B_s} + m_{B_s^{\ast}})
\>,
\end{eqnarray}
and the doubly charmed baryons (hopefully to soon be found in FNAL
experiments): 
\begin{eqnarray}
   m_{ccu} (1/2)^+ &\geq& \frac{m_{J/\psi}}{2} + \frac{1}{4} 
      (3 m_D + m_{D^{\ast}})
   \nonumber \\
   m_{ccs} (1/2)^+ &\geq& \frac{m_{J/\psi}}{2} + \frac{1}{4} 
      (3 m_{D_s} + m_{D_s^{\ast}})
\>.
\end{eqnarray}
The specific numerical values for all of the above and a few other
$x_1 x_2 y$ states are listed in Table \ref{table:two}.

Among the remaining $(1/2)^+$ baryonic states composed of three different
quark flavors, we have three ``$\Lambda$-type'' states in which the
light quark $ud$ subsystem coupled to $I = 0$ (and hence to $s_{ud} =
I_{ud} = 0$). The ``$\Sigma$-type'' states with $I_{ud} = 1$ have
already been discussed above. The ``$\Lambda$-type'' states are
\begin{equation}
   s\underbrace{ud}_{I=0,s=0} = \Lambda \qquad
   c\underbrace{ud}_{I=0,s=0} = \Lambda_c \qquad
   b\underbrace{ud}_{I=0,s=0} = \Lambda_b
\>.
\end{equation}
Using
\begin{displaymath}
   (\vec{s}_{ud}) \equiv (\vec{s}_u + \vec{s}_d)^2 = 0
\end{displaymath}
and
\begin{displaymath}
   (\vec{s}_{\Lambda}) \equiv (\vec{s}_x + \vec{s}_{ud})^2 = 3/4
\>,
\end{displaymath}
we can now deduce, using also $u \leftrightarrow d$ exchange symmetry,
that 
\begin{equation}
   \left\langle \vec{s}_{x} \cdot \vec{s}_{u} \right\rangle
      = \left\langle \vec{s}_{x} \cdot \vec{s}_{d} \right\rangle
      = 0
\>.
\label{eq:exp0}
\end{equation}
Equation (\ref{eq:exp0}) implies that in each $xu \, ( \rightarrow x
\bar{u})$ and $xd \, ( \rightarrow x \bar{d})$ subsystem we have the
opposite triplet-singlet mixture as in the previous case of
Eq.~(\ref{eq:exphalf}), namely 1/4 singlet and 3/4 triplet. Thus
collecting these and the term corresponding to the singlet $ud 
\rightarrow u \bar{d} \, (\approx \pi)$ subsystems, we finally have:
\begin{eqnarray}
   m_{\Lambda} &\geq& \frac{m_\pi}{2} 
      + \frac{1}{4} (3 m_{K} + m_{K^{\ast}})
   \nonumber \\
   m_{\Lambda_c} &\geq& \frac{m_\pi}{2} 
      + \frac{1}{4} (3 m_{D} + m_{D^{\ast}})
   \nonumber \\
   m_{\Lambda_b} &\geq& \frac{m_\pi}{2} 
      + \frac{1}{4} (3 m_{B} + m_{B^{\ast}})
\>.
\end{eqnarray}

The remaining baryonic states which consist of three different flavor
quarks $csu$, $bsu$, $bcu$, and $bcs$ are either of the
``$\Lambda$-type'' in which the lighter two quarks couple to $S =
0$ subsystems, or of the ``$\Sigma$-type'' in which the lighter two
quarks couple to $S = 1$. Because the hyperfine interaction [$
\propto (\vec{s}_i \cdot \vec{s}_j)/(m_i m_j)$] is bigger for
the lighter quark system we expect that the $\Lambda$-type states will
be lighter than the corresponding $\Sigma$-type states. This will be
particularly so for the $su$ subsystem in $csu$ or $bsu$, which
have been observed or are likely to be observed soon, than the
$bcu$ and $bcs$ states. When we specify the spin $q$ of the $su$
subsystem to be $s_{q_1 q_2} = 1$ (or $s_{q_1 q_2} = 0$), and follow
the previous discussion, we conclude that, for $csu$ for example:
\begin{mathletters}
   \begin{equation}
      \left\langle \vec{s}_c \cdot \vec{s}_u \right\rangle +
      \left\langle \vec{s}_c \cdot \vec{s}_s \right\rangle = -1
      \qquad ({\rm for} \, \, s_{ud} = 1)
   \end{equation}
   {\rm or}
   \begin{equation}
      \left\langle \vec{s}_c \cdot \vec{s}_u \right\rangle +
      \left\langle \vec{s}_c \cdot \vec{s}_s \right\rangle = 0
      \qquad ({\rm for} \, \, s_{ud} = 0).
   \end{equation}
\end{mathletters}

However, unlike in the previous cases, we cannot deduce that
$\left\langle \vec{s}_Q \cdot \vec{s}_u \right\rangle_B = \left\langle
\vec{s}_Q \cdot \vec{s}_d \right\rangle_B$ without appealing to the
approximate Gell-Mann -- Ne'eman SU(3) $u \leftrightarrow s$ symmetry.
If we do this nonetheless, as a first approximation, we obtain
\begin{mathletters}
   \begin{equation}
      m_{cus} (\Lambda) \geq \frac{m_K}{2} + \frac{1}{4} \left[
      \frac{(m_{D_s} + m_D)}{2} + \frac{3 (m_{D_s^{\ast}} 
      + m_{D^{\ast}})}{2} \right]
   \label{eq:mlam}
   \end{equation}
   \begin{equation}
      m_{cus} (\Sigma) \geq \frac{m_K^{\ast}}{2} + \frac{1}{4} \left[
      \frac{(m_{D_s^{\ast}} + m_{D^{\ast}})}{2} +
      \frac{3(m_{D_s} + m_D)}{2}  \right].
   \label{eq:msig}
   \end{equation}
\end{mathletters}
In the real case, we expect the spin-averaged variant, consisting of
the symmetric combination of Eqs.~(\ref{eq:mlam}) and (\ref{eq:msig})
(this is the value we list in Table~\ref{table:two}, with the left
hand side called $m_{cus}$):
\begin{equation}
   \frac{1}{2} \left[ m_{cus} (\Lambda) + m_{cus} (\Sigma)
      \right] \geq \frac{(m_K + m_{K^{\ast}})}{4} + 
       \frac{(m_D + m_{D^{\ast}})}{4} +
       \frac{(m_{D_s} + m_{D_s^{\ast}})}{4} 
\>,
\end{equation}
and in general we expect that the right hand sides of (\ref{eq:mlam})
and (\ref{eq:msig}) to be respectively an over- (under-) estimate of
the combined mesonic masses. Specifically, we expect that
Eqs.~(\ref{eq:mlam}) and (\ref{eq:msig}) to be satisfied with a
somewhat smaller (larger) than usual margin. The difference between
the right hand side in Eq.~(\ref{eq:mlam}) and Eq.~(\ref{eq:msig}) is
\begin{equation}
   \frac{(m_{K^{\ast}} - m_K)}{2} + \frac{1}{2} \left[ 
       \frac{(m_D - m_{D^{\ast}})}{2} +
       \frac{(m_{D_s} - m_{D_s^{\ast}})}{2}\right] 
       \approx 130 {\rm MeV}
\>.
\end{equation}
We will not pursue this issue further here.

%
\section{Application to quadronium}
\label{app:quad}

It has been suggested that the puzzling narrow resonance found in
heavy ion collisions is in some sense an $e^+ e^- e^+ e^-$ bound state
\cite{ref:griffin}. The need to achieve very strong bindings (BE
$\simeq m_e \simeq$ 0.5 MeV) in this normally weakly
(electromagnetically) coupled system is a serious problem of the
quadronium hypothesis. While the experimental motivations 
are questioned, we find that there is a nice, simple, application of the
variational technique \cite{ref:nussplb314}.  It has been suggested
that we cannot really have a two-body dominated interaction in this
system, and that new, unusual, four-body interactions are called for.

The following inequality on the binding energy of quadronium in terms
of the binding of positronium serves to sharpen the issue. Let $\psi_Q
= \psi^0(1 \bar{1} 2 \bar{2})$ be the wave function of quadronium in
its ground state. The Hamiltonian is
\begin{equation}
   H_Q = T_1 + T_{\bar{1}} + T_2 + T_{\bar{2}}
      + V_{1\bar{1}} + V_{1\bar{2}} + V_{2\bar{1}}
      + V_{2\bar{2}} + V_{12} + V_{\bar{1}\bar{2}}
\>.
\end{equation}
Clearly we expect the repulsive electron-electron ($V_{12}$) and
positron-positron ($V_{\bar{1}\bar{2}}$) interactions to lower the
quadronium binding. Hence
\begin{equation}
   \epsilon^0_B(Q) \leq \tilde{\epsilon}^{\, 0}_B(Q)
\>,
\label{eq:eps0b}
\end{equation}
with $\tilde{\epsilon}^{\, 0}_B(Q)$ the binding of a fictitious Hamiltonian
$\tilde{H}_Q$, which is the original $H_Q$ with these repulsive
interactions eliminated:
\begin{equation}
   \tilde{H}_Q = T_1 + T_{\bar{1}} + T_2 + T_{\bar{2}}
      + V_{1\bar{1}} + V_{1\bar{2}} + V_{2\bar{1}}
      + V_{2\bar{2}}
\>.
\end{equation}
If we compare $\tilde{H}_Q$ with the two-body Hamiltonians,
\[ \begin{array}{lr}
   H_{1\bar{1}} = T_1 + T_{\bar{1}} + V_{1\bar{1}}, \qquad &
   H_{1\bar{2}} = T_1 + T_{\bar{2}} + V_{1\bar{2}} \\
   H_{2\bar{1}} = T_2 + T_{\bar{1}} + V_{2\bar{1}}, \qquad &
   H_{2\bar{2}} = T_2 + T_{\bar{2}} + V_{2\bar{2}}
\end{array} \]
we see that
\begin{equation}
   2 \tilde{H}_Q = H'_{1\bar{1}} + H'_{1\bar{2}} 
      + H'_{2\bar{1}} + H'_{2\bar{2}}
\>,
\label{eq:htilde}
\end{equation}
where $H'_{1\bar{1}}$ is obtained from $H_{1\bar{1}}$ by doubling the
interaction: $H'_{1\bar{1}} = T_1 + T_{\bar{1}} + 2 V_{1\bar{1}}$,
{\em etc}. This can be achieved for the one-photon exchange by
doubling the effective coupling strength $\alpha \rightarrow \alpha' =
2\alpha$.  Equation~(\ref{eq:htilde}) can next be used in the by now
familiar way to put an upper bound on $\tilde{\epsilon}^{\, 0}_B(Q)$, and
thus, via Eq.~(\ref{eq:eps0b}), on $\epsilon^0_B(Q)$ as well. The
bound is
\begin{displaymath}
   \epsilon^0_B(Q) \leq 2\epsilon'^{\, 0}_B(P)
\>,
\end{displaymath}
with $\epsilon'^{\,0}_B(P)$ the binding of positronium in which the
strength of the interaction has been doubled by $\alpha \rightarrow 2
\alpha$. Thus we have
\begin{equation}
   \epsilon^0_B(Q) \leq \tilde{\epsilon}^{\, 0}_B(Q)
      \leq 2\epsilon'^{\, 0}_B(P)
\>.
\end{equation}
The fact that the real positronium spectrum conforms so nicely to QED
predictions with radiative corrections included strongly suggests that
if $\alpha = 1/137$ is scaled up to $\alpha' = 2/237$ we can still
treat the $e^+ e^- e^+ e^-$ system with the perturbatively calculated
potential. In this case
\begin{displaymath}
   \epsilon'^{\, 0}_B(P) \simeq (\alpha')^2 \frac{m}{4} 
      = 2 \alpha^2 \frac{m}{2} = 2 \, {\rm Ry} = 27 \, {\rm eV}
\>,
\end{displaymath}
and hence $\epsilon^0_B(Q) \leq 2\epsilon'^{\, 0}_B(P) \leq$ 54 eV, and
the binding falls short by about $10^4$ of the required 0.5 MeV
binding. 

%
\section{QCD inequalities for exotic novel hadron states} 
\label{app:fourfive}

Most known hadrons are $q_i \bar{q}_j$ mesons and $q_i q_j q_k$
baryons. We will refer to (non-glueball) states belonging to neither
of the above two categories as ``exotic''. Such states have been
searched for and discussed for more than 30 years. ``Duality''
arguments predating QCD \cite{ref:rosner} suggested``$\qqb \qqb$''
states coupling mainly to baryon-antibaryon as the most likely exotic
states. Within QCD hybrids ($\qqb$ + gluon) arise naturally
\cite{ref:horn}. Also $\qqb s \bar{s}$ ``bag states'' were suggested
as candidates for the $f_0$ and $K\bar{K}$ threshold states
\cite{ref:jaffe,ref:close,ref:achasov}. Considerations of hyperfine
chromomagnetic interactions,
\begin{displaymath}
   V_{HF} \approx \sum_{ij} \frac{(\vec{\sigma}_i \cdot \vec{\sigma}_j)
      (\vec{\lambda}_i \cdot \vec{\lambda}_j)}{m_i m_j} V(r_{ij})
\>,
\end{displaymath}
generated via one gluon exchange between the quarks $q_i$ and $q_j$
suggested a particular ``hexaquark'' ${\cal H} = uuddss$
\cite{ref:jaffe} below the $\Lambda\Lambda$ threshold
which could be strong interaction stable.\footnote{The existence of
  a $\Lambda \Lambda$ bound state could encourage the further
  conjecture \cite{ref:farhi,ref:wittprd84} that higher $\Lambda^N =
  u^N d^N s^N$ ``strangelets'' exist and that at large densities
  strange quark matter is the most stable, which if true has
  fascinating astrophysical ramifications.}  More recently similar
considerations \cite{ref:lip87,ref:gignoux} singled out the specific
``pentaquark'' ${\cal P} = \bar{c}suud$ as a more likely strong
interaction stable, new, exotic, bound state. Both states have been
experimentally searched for with inconclusive results
\cite{ref:moinester,ref:fitch,ref:ashery} in almost all cases.

The exotic states can be separated into disjoint color singlet
hadrons.\footnote{With the exception of genuine ``quantum number
  exotics'', for example $0^{\pm}$ or $1^{\pm}$ states that cannot
  decay into mesons.}  Indeed we have seen in Sec.~\ref{sec:exotic}
that for degenerate quark flavors (and most likely in other cases as
well) there is an attractive interaction in the $q_i \bar{q}_j q_k
\bar{q}_l$ channels.  The question of whether these residual color
forces between color singlet states can lead to weakly bound states,
in analogy to the nuclear forces in the deuteron and other nuclei, is
of some interest. However we believe that it is a detailed, delicate,
issue that QCD inequalities cannot decide (much in the same way that
the deuteron's binding requires detailed calculation).

Here we would like to focus on the question whether genuine multiquark
states -- referred to in the literature qualitatively as ``single bag
states'' -- which have significantly higher than nuclear bindings,
exist. The arguments for bound hexa- and pentaquark states utilized
the {\em ad hoc} assumption that in all multiquark hadronic states the
quarks are in the same ``universal'' bag as the baryons and mesons,
and have the same hyperfine interactions. Here we will utilize the
complementary, strong coupling picture, where the quarks in the exotic
states are connected via electric flux lines into a more complex color
network than the $\qqb$ for mesons or the ``Y'' topology with one
junction point for the baryons.

Let us first focus on quadriquarks. In the limit when two quarks (or
antiquarks) are heavy, one can show directly that a novel pattern of
an overall singlet system which cannot be broken into color singlet
clusters will form \cite{ref:manohar}. In a $q_i q_j \bar{Q}_k
\bar{Q}_l$ state, the two heavy quarks always bind coulombically into
a color triplet $(\bar{Q}_k \bar{Q}_l)_3$ diquark with binding and
size proportional to $m_Q$ and $m_Q^{-1}$, respectively. The small
$(\bar{Q}_k \bar{Q}_l)_3$ subsystem acts effectively as a heavy quark.
Together with the light $q_i q_j$ it will then form, as a $\Lambda_Q$
analog, the $q_i q_j \bar{Q}_k \bar{Q}_l$ quadriquark. Because of the
small numerical coefficient in the coulombic energy, $E_B =
\frac{1}{2} \left( \frac{\alpha_S}{2} \right)^2 \frac{m_Q}{2}$, the
latter does not exceed $\Lambda_{\text{QCD}} \simeq 0.2$ GeV even for
$m_Q = m_B \simeq 5$ GeV. Hence the question of the stability of the
novel pattern of color coupling -- which in a strong coupling
chromoelectric flux tube picture corresponds to the connected color
network illustrated in Fig.~\ref{fig:qquark} -- remains open.

In the following we would like to use QCD variational inequalities to
address this problem. This we do by following the derivation in the
same strong coupling, flux tube approximation as the baryon-meson mass
inequalities in Sec.~\ref{sec:nonpertmbineq} above. Thus, let us take
for the quadriquark wave functional:
\begin{equation}
   \left| \Psi_{q q' \bar{Q} \bar{Q}'} \right\rangle 
      = \sum_{\bowtie} A_{\bowtie} \left| \bowtie \right\rangle
\>,
\end{equation}
with the possible generalizations maintaining the two junction points
(1) and (2) (see the discussion in Sec.~\ref{sec:nonpertmbineq}).  We
have used $\bowtie$ to symbolize the graph of Fig.~\ref{fig:qquark}. As
indicated in Fig.~\ref{fig:qquark2} we can extract from $\left| \Psi_{q
q' \bar{Q} \bar{Q}'} \right\rangle$ -- and the wave functional of an
anti-quadriquark superposed on it with flux lines reversed -- trial
wave functionals for the four mesons $Q\bar{Q}'$, $Q\bar{q}'$,
$Q\bar{q}'$, and $q\bar{q}'$. In the above Hilbert space, we
then have the operator relation
\begin{equation}
   2H_{QQ'\bar{q}\bar{q}'} = H_{Q\bar{Q}'} + H_{Q\bar{q}'} 
      + H_{q\bar{q}'} + H_{q'\bar{Q}'}
\>,
\end{equation}        
from which we obtain, via the variational principle, the mass
inequality
\begin{equation}
   2m_{QQ'\bar{q}\bar{q}'}^{(0)} \geq m_{Q\bar{Q}'}^{(0)} 
      + m_{Q\bar{q}'}^{(0)} + m_{q\bar{q}'}^{(0)} 
      + m_{Q'\bar{q}'}^{(0)}
\>.
\end{equation}
For the case $Q,Q' = c, qq' = u,d$, this binds from below the (hypothetical)
quadronium mass by the known charmonium, charmed meson, and light
meson masses. Specifically for $cc\bar{u}\bar{u}$ (or
$cc\bar{d}\bar{d}$), the inequality reads:
\begin{equation}
   m_{cc\bar{u}\bar{u}}^{(0)} \geq \frac{1}{2} \left[ m_{J/ \psi} 
      + m_{\rho} + (1+\alpha) m_{D^{\ast}} 
      + (1-\alpha) m_{D} \right] 
\>, 
\end{equation} 
where $m_{c\bar{c}} = m_{J/\psi}$ and $m_{u\bar{u}} = m_\rho$ follows
from the generalized Pauli principle as discussed
previously.\footnote{The identical $uu$ or $cc$ quarks are in the
  lowest $L=0$ state, and hence the $u\bar{u}$ or $c\bar{c}$ in the
  corresponding meson should be in the triplet state.}  The weights of
the $D$ and $D^{\ast}$ depend on the total spin of the
$cc\bar{u}\bar{u}$ system (see App.~\ref{app:bmineq}).  For example,
consider $J^{P} = 1^+$ quantum numbers. In this case the transition
quadriquark $\rightarrow DD$ is forbidden since the two $D$ mesons can
couple only to $0^+, 1^-, 2^+$ {\em etc}. The lowest state available
for decay is then $DD^{\ast}$. Performing a calculation similar to
those in App.~\ref{app:bmineq} we find that $\alpha = 0$ is the
correct weight and the inequality reads
\begin{eqnarray}
   m_{cc\bar{u}\bar{u}}^{(0)} (1^+) &\geq& \frac{1}{2} \left(
      m_{J/ \psi} + m_{\rho} 
      + m_{D^{\ast}} + m_{D} \right) 
   \nonumber \\
   m_{cc\bar{u}\bar{u}}^{(0)} (1^+) &\geq& 3870.7 \, {\rm MeV}
      \approx m_{D^{\ast}} + m_{D}
\>.
\label{eq:mccuu}
\end{eqnarray}

Our previous comparisons of many baryon-meson inequalities with data
indicate that these inequalities are satisfied with a margin of
150-300 MeV. Cohen and Lipkin \cite{ref:cohen} and Imbo
\cite{ref:imbo} suggest heuristic arguments for this margin. Such a
margin should {\em a fortiori} be valid for the quadriquark
inequalities.\footnote{Here we have for some of the $Q\bar{q}$ mesons
  {\em worse} trial wave functionals as compared with those extracted
  from baryons, namely mesonic strings consisting of three segments
  going through the two junction points.} Thus we expect that the
$cc\bar{u}\bar{u}$ state actually lies well above the $D^{\ast} \, D$
threshold and therefore is unstable. The situation is very different
for $cc\bar{u}\bar{d}$.  In this case we could choose $u\bar{d}$ to be
in the spin singlet state. This is consistent with the $cc$ spin
triplet diquark and the $\bar{u}\bar{d}$ spin singlet anti-diquark,
with relative zero orbital angular momenta, adding up to an overall
$J^{P} = 1^+$. The analog of Eq.~(\ref{eq:mccuu}) is then
\begin{eqnarray}
   m_{cc\bar{u}\bar{d}}^{(0)} (1^+) &\geq& \frac{1}{2} \left(
      m_{J/\psi} + m_{\pi}
      + \frac{3}{2} m_{D^{\ast}} 
      + \frac{1}{2} m_{D} \right) 
   \nonumber \\
   m_{cc\bar{u}\bar{d}}^{(0)} (1^+) &\geq& 3590.6 \, {\rm MeV}
      \approx m_{D^{\ast}} + m_{D} - 300 \, {\rm MeV}
\>,
\end{eqnarray}
A bound $ccud (1^+)$ state with $m_{cc\bar{u}\bar{d}}^{(0)} \leq m_D +
m_{D^{\ast}}$ would satisfy the inequality with a margin of
approximately 300 MeV, and is therefore quite likely.

The analogs in the $Q=b$ system
\begin{eqnarray}
   m_{bb\bar{u}\bar{u}}^{(0)} (1^+) &\geq& \frac{1}{2} \left(
      m_{\Upsilon} + m_{\rho}
      + m_{B^{\ast}} + m_{B} \right)
   \nonumber \\
   m_{bb\bar{u}\bar{u}}^{(0)} (1^+) &\geq& 10416.3 \, {\rm MeV}
\>,
\end{eqnarray}
and
\begin{eqnarray}
   m_{bb\bar{u}\bar{d}}^{(0)} (1^+) &\geq& \frac{1}{2} \left(
      m_{\Upsilon} + m_{\pi} + \frac{3}{2} m_{B^{\ast}} 
      + \frac{1}{2} m_{B} \right) 
   \nonumber \\
   m_{bb\bar{u}\bar{d}}^{(0)} (1^+) &\geq& 10113.3 \, {\rm MeV}
\end{eqnarray}
allow bindings in both flavor combinations, though again with a
considerably higher safety margin in the $bb\bar{u}\bar{d}$ case.

In baryons each quark pair must couple to a $\bar{3}$. In the
$\bar{Q}\bar{Q}^{\prime} q q^{\prime}$ system the
$\bar{Q}\bar{Q}^{\prime}$ and $qq^{\prime}$ could also couple to a
color sextet. Thus the segment connecting the junction points (1) and (2)
in Fig.~\ref{fig:qquark} would then carry a sextet flux -- a
possibility that our discussion above neglected. This is justified
in the strong coupling limit. The potential energy $g^2 \int {\rm d}^3
x E^2$ then dominates, and the quadratic Casimir operator (to which
$\int E^2$ is proportional) of the sextet is 10/3 times larger than
that of $\bar{3}$. Nonetheless, the need to exclude the sextet flux,
and the {\em related} fact that there is no alternative
derivation in the weak coupling, one gluon exchange limit, puts the
present inequalities on a weaker footing than that of the baryon-meson
mass inequalities. We believe however the enhanced likelihood for a
$cc\bar{u}\bar{d}$ quadriquark state suggested by the inequalities. 

It is amusing to note that an alternative, purely hadronic approach
to the problem of a bound $D D^{\ast}$ state also suggests that $D^0
D^{\ast}$ (of a $cc\bar{u}\bar{d}$ composition) is more likely to be
bound than $D^0 D^{0\ast}$. Thus following Tornqvist
\cite{ref:tornqvist} (who coined the name ``deusons'' for these
deuteron-like extended mesonic states), let us consider the one pion
exchange interaction in the $D^{\ast}D$ channel. The resulting Yukawa
force has, due to the $D \pi$ threshold -- $D^{\ast}$ proximity ({\em i.e.}
$m_{D^{\ast}} - m_D - m_\pi \equiv \epsilon \ll m_\pi$), an anomalously
large range $\simeq 1 / \sqrt{2\epsilon m_\pi} \simeq 3 - 7$
fm\footnote{Because of the P-wave $D^{\ast} \rightarrow D\pi$ vertex
  this is compensated by a reduced effective coupling.}, and could
generate a $D^{\ast}D$ bound state. However, because of simple
$I$-spin Clebsch-Gordan coefficients, the potential generated by the
$\pi^+$ exchange in the $D^{\ast +} D^0 \leftrightarrow D^{\ast 0}D^+$
system is {\em twice} as strong as that due to the $\pi^0$ exchange in
the $D^{\ast 0} D^0 \leftrightarrow D^{\ast 0}D^0$ system, and a
$D^{\ast +}D^0$ bound state is more likely.

Let us next consider pentaquarks in the same strong coupling flux
string approximation. The corresponding novel connected color network
is illustrated in Fig.~\ref{fig:pquark}. This ``network'' contains
three junction points. At point (1), two light quark fluxes are
coupled to a $\bar{3}$ flux, and at point (2) the same holds for the
fluxes emanating from the other two light quarks. Finally at the third
``anti-junction'' point the three $\bar{3}$ fluxes originating from
points (1), (2), and from the heavy antiquark $\bar{Q}$, couple to a
singlet.  As in the case of the quadriquark we could have coupled the
fluxons in {\em either} the first vertex (1) or the second vertex (2)
to a color sextet flux, which again is neglected in the strong
coupling limit. For pentaquarks, we furthermore have an alternative
(similar) network obtained by exchanging one of the $u$ quarks from
the $uu$ in vertex (1) with the $d$ quark in vertex (2) -- which
represents a different, though not necessarily orthogonal, color
coupling configuration.  By reversing the flux lines on the segments
(2)-$d$ and (3)-$\bar{Q}$ indicated by the double arrows in
Fig.~\ref{fig:pquark}, and ignoring the intermediate fluxon between
(2) and (3), we can extract from a ``stringy'' pentaquark wave
functional trial wave functionals for a charmed baryon ($cuu$ in this
specific case) and an $s\bar{d}$ meson.

The full Hamiltonian for the quark and fluxon system is
\begin{equation}
   H = \sum_{{\rm quark}} H_{D_i} + \sum_{{\rm flux \, segment}} \int \,
   \left[ (\vec{E})^2 + (\vec{B})^2 \right]
\>,
\end{equation}
where $H_{D_i}$ are the Dirac Hamiltonians for the quark,
$(\vec{E})^2$ represents the ``potential energy'' density for the
fluxons, and $(\vec{B})^2$ plays the role of kinetic energy for moving
and distorting the string segments and the $\nabla_i$ in $H_{D_i}$ ``moves''
the tip quarks.

Omitting (the generally positive contribution of) the (2)-(3)
flux segment will only reduce the energy. This and the fact that the
trial configurations of the baryons and mesons thus extracted do not
optimize the wave functionals of the meson and baryon at rest,
strongly suggest the inequality
\begin{equation}
   m_{{\cal P}} \geq m_{s\bar{d}} + m_{cuu}
\>.
\label{eq:mpq}
\end{equation}
Because of the Pauli principle the two $u$ quarks in $cuu$ which are
color antisymmetric and flavor symmetric must be spin symmetric and
hence couple to $S_{uu} = 1$. Since these subconfigurations originate
from the specific pentaquark state for which the four light quarks'
spin has to add to a total spin $S = 0$, we must also have
$S_{s\bar{d}} = 1$. This means that we have to identify the $s\bar{d}$
meson with $K^{\ast}$ (980) and the baryon with $\Sigma_c$ (2445) with
a total mass of $m_{\Sigma_c} + m_{K^\ast} = 3334$ MeV, which is way
above the threshold at 2906.8 MeV. In general, however, the pentaquark
state consists of a superposition of various flavor assignments at the
tips of the same color network. The right-hand side of
Eq.~(\ref{eq:mpq}) should therefore contain, in general, some
weighted average so that
\begin{eqnarray}
   m_{{\cal P}} &\geq& \alpha \left[ m_{s\bar{d} \, (S=1)} 
      + m_{ccu \, (S=1)} \right] + \beta \left[ m_{s\bar{u}}
      + m_{cud} \right] \nonumber \\
      &+& \gamma \left[ m_{u\bar{d}}
      + m_{csu} \right] + \delta \left[ m_{u\bar{u} \, (S=1)} 
      + m_{csd \, (S=1)} \right]
\>,
\label{eq:avgmpq}
\end{eqnarray}
with $\alpha + \beta + \gamma + \delta = 1$ and $\alpha, \beta, \gamma,
\delta \geq 0$. The spins in the last configuration are fixed by the
same consideration as those used in the case of $s\bar{d} - cuu$
separation discussed above. The corresponding total mass
\begin{displaymath}
   m_{\rho} + m_{\Xi_c} = 3221.3 \, {\rm MeV}
\end{displaymath}
is 315 MeV above the $D_s - p$ threshold. In the other two
configurations we will have in general admixtures of the lighter
pseudoscalar $\pi$ and $K$ states, and of the lighter $\Lambda_c$
baryon. The lightest combination is
\begin{displaymath}
   m_{\pi} + m_{\Xi_c} = 2605.2 \, {\rm MeV}
\>.
\end{displaymath}
(The other alternative including the kaon,
\begin{displaymath}
   m_{K} + m_{\Lambda_c} = 2778.6 \, {\rm MeV}
\end{displaymath}
is only 128 MeV below threshold.)

We need an unlikely large admixture of the particular $\pi + \Xi_c$
configuration to bring the right-hand side of Eq.~(\ref{eq:avgmpq})
below 2906.8 MeV, the $D_s - p$ threshold. The baryon-meson
inequalities are generally satisfied with a substantial margin of
$\simeq$ 150 MeV. If this is also the case for the relation in
Eq.~(\ref{eq:avgmpq}), then it appears that the present inequalities
cannot allow for a stable pentaquark bound state.

It should be emphasized that the present approach, assuming that
confinement via the specific mechanism of color flux tubes (or
strings) is the dominant aspect, is practically orthogonal to the
one-gluon exchange potential approach. Indeed in the latter,
confinement is assumed from the outset by simply postulating that all
the quarks in penta- and hexaquark states are inside the same
universal bag. Yet confinement forces play a minimal role in the
pentaquark binding. Rather, the overriding consideration is to
maximize the overall hyperfine interaction even if that involves
having large components of the wave function with $qq$ pairs in a
color sextet state. It is therefore logically consistent that the
inequalities derived in the framework of one approach conflict with
a prediction of a bound pentaquark in another framework.

Nonetheless, the inequalities do suggest that if the pentaquark is not
found, then neither QCD nor the naive quark models which so nicely
explain the baryonic and mesonic spectra are at fault, but rather a
lack of proper treatment of the basic confinement mechanism as adapted
for these circumstances of pentaquark and/or hexaquark configurations.

One other class of exotics are hybrid states containing an extra gluon
such as $(\psibar_i \psi_j)_8 G$ and $(\psi_i \psi_j \psi_k)_8 G$. The
$(\cdots)_8$ notation indicates that the quarks in the hadron couple to a
color octet, which then forms the color singlet hadron with an
additional gluon. To probe this sector, we should consider correlators
of the form:
\begin{equation}
   H(x,0) = \left\langle 0 \left| \left[\psi_i^a(x) \psibar_j^b(x) 
      \lambda_{ab}^r G_r(x) \right]^\dag \left[\psi_i^{a'}(0) 
      \psibar_j^{b'}(0) \lambda_{a'b'}^{r'} G_{r'}(0)\right] 
      \right| 0 \right\rangle
\>.
\end{equation}
If we represent the gluons via a field strength, then $H(x,0)$ can be
written schematically as
\begin{equation}
   H(x,0) = \intoned{\mu(A)} F_{\mu\nu}(x) F_{\mu\nu}(0)
      S^i_A(x,0) S^j_A(x,0)
\>,
\end{equation}
and hence a Schwartz-type inequality implies
\begin{equation}
   |H|^2 \leq \intoned{\mu(A)} F^2_{\mu\nu}(x)F^2_{\mu\nu}(0)
      \intoned{\mu(A)} S_A^i(x,0)^\dag S_A^i(x,0)
      \intoned{\mu(A)} S_A^j(x,0)^\dag S_A^j(x,0)
\>.
\end{equation}
This would imply, {\em if} the difficulty with the vacuum
expectation value $\langle 0 | F^2 | 0 \rangle$ could somehow be
surmounted, that
\begin{equation}
   m_{\text{hybrid}} \leq \frac{1}{2} \left(
      m_{\text{glueball}} + m_\pi \right) \simeq
      0.8 \, \mbox{GeV}
\>,
\end{equation}
which is a rather weak bound, from a phenomenological point of view.

A similar inequality in the context of SUSY models (in the
$m_{\text{squark}} \rightarrow \infty$ limit)\footnote{From the
  discussion at the end of Sec.~\ref{sec:sxsb}, a decoupling of the
  squarks and elimination of their Yukawa couplings is indeed required
  if any QCD-type inequalities are to be proven.}  has proven useful
in assessing the symmetry (phase) structure of that theory.
Specifically, the possibility that massless, singlet, fermionic
composites of $q_i \bar{q}_j$ and a gluino may manifest unbroken
chiral symmetries, has been ruled out. This was done by proving a QCD
inequality between the mass of this state (the ``hybridino'') and the
mass of the pion \cite{ref:aspy}:
\begin{equation}
   \left[m_{\text{hybrid}} = m^{(0)}(q_i \bar{q}_j
     \tilde{g}) \right] \geq
      \left[m_{q_i \bar{q}_j}^{(0^-)} = m_\pi\right]
\>.
\label{eq:mhyb}
\end{equation}
Since unlike for gluons with nonlinear self-couplings one can define
here the gluino propagator, the derivation of the last inequalities
along lines paralleling those of Weingarten's in deriving $m_N \geq
m_\pi$ (see Sec.~\ref{sec:mbineqcorr}), is fairly straightforward.

The hybridino-hybridino correlator (written in a concise, index-free
form) can be bound by Schwartz inequalities:
\begin{eqnarray}
   \langle J^\dag_{\text{hyb}}(x) 
      J_{\text{hyb}}(0) \rangle &=& \intoned{\mu}
      S_\psi S_{\psibar} S_\lambda \nonumber \\
   &\leq& \intoned{\mu} \left(|S_\psi|^2 \right)^{1/2}
      \left(|S_{\psibar}|^2 \right)^{1/2}\left(|S_\lambda|^2
      \right)^{1/2} \nonumber \\
   &\leq& \left( \intoned{\mu} |S_\psi|^2 \right)^{1/2}
      \left( \intoned{\mu} |S_{\psibar}|^2 |S_\lambda|^2 
      \right)^{1/2} \nonumber \\
   &\leq& \left( \intoned{\mu} |S_\psi|^2 \right)^{1/2}
      \left( \intoned{\mu} |S_{\psibar}|^2 \right)^{1/2}
      \left( \intoned{\mu} |S_\lambda|^2 \right)^{1/2} 
\>.
\end{eqnarray}
The first and second factors in the last expression are the
square root of a pion propagator. Hence
\begin{equation}
   \langle J^\dag_{\text{hyb}}(x) 
      J_{\text{hyb}}(0) \rangle \leq {\rm e}^{-m_\pi |x|}
\>,
\end{equation}
and Eq.~(\ref{eq:mhyb}) follows. 

%
\section{QCD inequalities between EM corrections to nuclear scattering}
\label{app:scatt}

In this Appendix we combine the results of Sec.~\ref{sec:pseumeson} on
the essentially positive nature of the EM contribution to the
energies, and the approach of Sec.~\ref{sec:exotic}, in order to apply
``electromagnetic QCD inequalities'' to scattering states. To this end
we first elaborate on some of the material of Sec.~\ref{sec:exotic}.

We have seen in Sec.~\ref{sec:exotic} that QCD arguments leading to
relations of the form
\begin{equation}
   B(aa) + B(bb) \leq 2 B(ab)
\end{equation}
between binding in channels $(aa)$, $(bb)$, and $(ab)$ are useful even
if there are no bound states in those channels. Thus, imagine that the
relative coordinate $\vecr = \vecr_1 - \vecr_2$ of the particle pair
is confined to a sphere $|\vecr| \leq R$, with $R$ an infrared cutoff
larger than the relevant scales in the problem. Let $\epsilon^0_{n,l}$
and $\epsilon_{n,l}$ denote the levels in various $l$ channels without
and with (respectively) the interparticle interactions turned on. Both
sets of levels become dense as $R \rightarrow\infty$. [In particular,
the free energies are $\epsilon^0_{n,l} = (k^0_{n,l})^2 / 2m$ with
$k^0_{n,l}R \equiv x^0_{n,l} \simeq (n\pi/2 + l\pi/2 + \cdots)$, the
$n$th zero of $j_l(x)$.] Yet a careful study of the shifts
$\Delta(n,l) \equiv - \epsilon_{n,l} + \epsilon^0_{n,l}$ reveals the
full information on the phase shifts in various channels.

The key observation is that the binding energy inequalities hold in
general even when the $R$ cutoff is placed, since the relevant QCD or
nuclear interactions are short range. These imply then that
$\Delta_{n,l}(aa) + \Delta_{n,l}(bb) \leq 2\Delta_{n,l}(ab)$. [For the
unperturbed part we have trivially, from the definition of reduced
masses, $\epsilon_{n,l}^0(aa) + \epsilon_{n,l}^0(bb) \equiv
2\epsilon_{n,l}^0(ab)$.] Therefore the density of levels $\frac{ {\rm
    d}n_l(ab)}{ {\rm d}k}$ in the $(ab)$ channel is larger than the
average of these densities in the $(aa)$ and $(bb)$ channels.

Levinson's theorem implies that the phase shift $\delta_l^{(ab)}(k)$
serves as a ``level counter'' in the continuum limit. Hence
\begin{equation}
   \frac{{\rm d}n_l(ab)}{{\rm d}k} \simeq
      \frac{{\rm d}\delta_l(ab)}{{\rm d}k}
\>.
\end{equation}
However, in the $k \rightarrow 0$ limit the last quantity is the
scattering length for $l = 0$. The inequality
\begin{equation}
   \frac{{\rm d}\delta(aa)}{{\rm d}k}
      + \frac{{\rm d}\delta(bb)}{{\rm d}k}
      \leq 2 \frac{{\rm d}\delta(ab)}{{\rm d}k}
\>,
\label{eq:deltal}
\end{equation}
yields in the $k \rightarrow 0$ limit the desired inequality between
scattering lengths
\begin{equation}
   a(aa) + a(bb) \leq 2 a(ab)
\>.
\end{equation}

We would next like to argue that the statement on the $\Delta I = 2$
energy shifts of $NN$ continuum states, for example
\begin{equation}
   \Delta BE (pp) + \Delta BE (nn) \leq 2 \Delta BE (pn)_{I=1}
\label{eq:deltabe}
\end{equation}
does imply the corresponding inequality for the respective scattering
lengths
\begin{equation}
   a(pp) + a(nn) \leq 2 a(pn)
\>.
\label{eq:scattlen}
\end{equation}
The notation $\Delta BE (pn)_{I=1}$ in Eq.~(\ref{eq:deltabe}) refers
to the shift from the idealized $\Delta BE (pn)_{I=1}$ of some
continuum level in the $I = 1$ (flavor symmetrized) $np$ state for the
case when $\alpha_{EM} = 0$ and $m_u^{(0)} = m_d^{(0)}$. In this exact
$I$-spin limit, all $(pn)_{I=1}, (pp),$ and $(nn)$ states would be the
same, and hence so would be phase shifts, scattering lengths, {\em
  etc}.

The $\Delta I = 2$ combination $2\Delta BE(pn)_{I=1} - \Delta BE(nn) -
\Delta BE(pp)$ and the similar combination of phase shifts, level
densities, and scattering lengths are effected only by
electromagnetism -- {\em i.e.} by $\alpha_{EM} \neq 0$. The general
feature of positivity of such electromagnetic self energies can
therefore be applied. This is so since the positivity applies not only
for ground states but to any set of corresponding states which have
equal charge density in the $\alpha_{EM} = 0, m_u^{(0)} = m_d^{(0)}$
limit. This then implies the positivity of all these combinations and
in particular the desired relation (\ref{eq:scattlen}).

In applying the suggested scattering length inequality
(\ref{eq:scattlen}), we face a ``technical'' difficulty. The
long-range Coulombic interaction yields a (logarithmically) infinite
Coulombic phase. In order to extract a meaningful result for
$a(pp)$, one must subtract this Coulomb phase shift. Our inequality
is based precisely on the positive nature of the Coulombic
self-interaction. Thus subtraction of even a part of this interaction
could, in principle, invalidate the derivation of the inequality.
Indeed, the very approach to continuum phase shifts by quantization in
a finite sphere of radius $R$ is jeopardized by the $1/r$ infinite
range potential.

We can, however, avoid the Coulombic $pp$ phase shifts and still have
a meaningful relation. This relies on the observation that screened
versions of the Coulomb potential ({\em i.e.} Yukawa potential) still
yield positive EM self-interactions [specifically the momentum space
representation $(\mu^2 + k^2)^{-1}$ is positive]. We can choose the
cutoff $\mu^{-1}$ to be of the order of the sum of the radii of the
two nucleons. Thus the long-range Coulombic phase shift will be
screened away, yet all other manifestations of EM interactions which
occur during the nucleons' overlap will be maintained. These include
the EM interactions between quarks in the two different nucleons or
between their mesonic charge clouds, or more subtle indirect EM
effects such as the $\pi^+ - \pi^0$ mass difference and the resulting
different ranges for $\pi^+, \pi^0$ exchange potentials
\cite{ref:kolck}. 

It is the latter effects which occur also in the $np$ and $nn$
systems, which we wish to consider. We cannot prove that the above
cutoff procedure is indeed equivalent to Coulomb phase subtraction.
However, any cutoff $\mu$ [or even arbitrary positive superpositions
of potentials with different cutoffs, such as
$\intoned{\mu^2}\sigma(\mu) {\rm e}^{\mu r} / r$] can be used. Hence
for scattering lengths computed with this cutoff,
Eq.~(\ref{eq:scattlen}) holds
\begin{equation}
   a^{\sigma(\mu)}(pp) + a^{\sigma(\mu)}(nn) 
      \leq 2 a^{\sigma(\mu)}(pn)
\>.
\end{equation}
We therefore believe that this inequality applies to the quoted
``nuclear parts''.  Recent measurements indicate that the suggested
inequality is indeed satisfied with a wide margin.

In principle, the original general relation (\ref{eq:deltal}) contains
much more information than that used to derive the scattering
length inequality by taking $l = 0$ and going to the $k \rightarrow 0$
limit. Thus many other inequalities for other scattering parameters in
all $l$ waves can be derived for the two nucleon system, although
these are not as useful and cannot be readily compared with data.

Inequalities analagous to (\ref{eq:scattlen}) should hold for any $I =
1/2$ isospin multiplet. This in particular we should have
\begin{equation}
   a(^3{\rm He} ^3{\rm He}) + a(^3{\rm H} ^3{\rm H}) 
      \leq 2a(^3{\rm He} ^3{\rm H})
\>.
\end{equation}
The $I = 2$ part can be extracted not only in scattering of two
particles which are members of the same $I$-spin doublet. This suggests
relations of the form
\begin{eqnarray}
   a(p ^3{\rm H}) + a(n ^3{\rm He}) 
      &\geq& a(p ^3{\rm He}) + a(n ^3{\rm H})
   \nonumber \\
   a(K^+ n) + a(K^0 p) &\geq& a(K^+ p) + a(K^0 n)
   \nonumber \\
   a(K^0 n) + a(K^- p) &\geq& a(K^0 p) + a(K^- n)
\>,
\end{eqnarray}
and many more. Since many of these relations involve unstable nuclear
isotopes, testing them requires radioactive beam facilities.

Finally we note that purely EM $\Delta I = 2$ combinations of masses
or scattering lengths can be found for nuclei of higher $I$-spin.

%
\section{QCD-like inequalities in atomic, chemical, and biological
  contexts} 
\label{app:bio}

In this last Appendix we present various inequalities inspired by
(more or less) justified analogies with the binding energy and
correlator inequalities.

\subsection{Mass relations between compounds of different isotopes}

Different isotopes provide an almost ideal example of ``flavor
independent'' interactions. Thus, let there be $n_1$ stable isotopes of
a specific atom $(Z^{(1)}, A^{(1)}_1) \ldots (Z^{(1)}, A^{(1)}_{n_1})$
and $n_2$ of another $(Z^{(2)}, A^{(2)}_1) \ldots (Z^{(2)},
A^{(2)}_{n_2})$. In the adiabatic (Born-Oppenheimer) approximation
\cite{ref:born}, it is clear that the interatomic interactions are
independent of the specific isotopes $Z^{(1)}A^{(1)}_i$ and
$Z^{(2)}A^{(2)}_j$ chosen. The arguments in Sec.~\ref{sec:beyond2pt}
then imply that the binding energies $B_{ij}$ of the $n_1 n_2$
possible compounds thus formed are a convex function of a reduced mass
$\mu$ sampled at the values
\begin{equation}
   \mu_{ij} = \frac{m(Z^{(1)}A^{(1)}_i) \, m(Z^{(2)}A^{(2)}_j)}
      {m(Z^{(1)}A^{(1)}_i) + m(Z^{(2)}A^{(2)}_j)}
\>,
\end{equation}
with $m(Z,A)$ the nuclear masses.  This includes in particular the
analog of the interflavor mass inequalities\footnote{The $BE$ refers
  to the true molecular ground states or to sums over the first $n$
  states. The fact that Bose/Fermi statistics can imply even/odd $J$s
  in the rotation band of homoisotopic compounds has a negligible
  effect and can be corrected for.}  $BE(x,x) + BE(x,y) \leq 2
BE(x,y)$, {\em e.g.}
\begin{eqnarray}
   BE({\rm H}_2) + BE({\rm D}_2) &\leq& 2 BE({\rm HD}) 
   \nonumber \\
   BE(^{16}{\rm O}_2) + BE(^{18}{\rm O}_2) 
      &\leq& 2 BE(^{16}{\rm O}_2 + ^{18}{\rm O}_2)
\>.
\end{eqnarray}
We have not investigated the availability of data verifying these many
true mass inequalities.

The flavor ({\em i.e.} isotopic) independence arguments can be
extended to more complex atomic compounds like $XXZ, \, XZZ$, {\em
  etc.}  and the ``convexity'' relations ({\em e.g.} the analogs of those
conjectured for baryons in QCD) are likely to hold. This will indeed
be the case if the two- and three-body interactions between the nuclei
satisfy the condition of positive $\exp \{ V \}$ utilized in Lieb's
proof in App.~\ref{app:lieb2}. We are indebted to Phil Allen
\cite{ref:allen}, for pointing out to us this rather nice test of
QCD-like inequalities. 

\subsection{Conjectured inequalities for chemical bindings}

Atoms with closed shells or even a given $(n,l)$ subshell constituting
one Slater determinant of all possible $m_l, m_s$ states are singlets
of any relevant quantum numbers. This suggests using such states, say
$({\rm Ne} | Z=10)$, as vacuum states upon which we can build
``particle'' states $X = ({\rm Na} | Z = 11)$ or ``hole'' states
$\bar{X} = ({\rm F} | Z = 9)$. Likewise another noble gas ``vacuum'',
say $({\rm Ar} | Z = 18)$, yields $Y = ({\rm K} | Z = 19)$, $\bar{Y} =
({\rm Cl} | Z = 17)$. If we use this identification of vacuums,
particles, and antiparticles, we might be tempted to conjecture, in
``analogy'' with Eq.~(\ref{eq:engineq}), that
\begin{equation}
   BE(X\bar{X}) + BE(Y\bar{Y}) \geq BE(X\bar{Y}) + BE(Y\bar{X})
\>.
\label{eq:atombe}
\end{equation}
This yields, for example,
\begin{eqnarray}
   BE({\rm NaF}) + BE({\rm KCl}) &\geq& BE({\rm NaCl}) + BE({\rm KF})
   \nonumber \\  
   BE({\rm MgO}) + BE({\rm CaS}) &\geq& BE({\rm MgS}) + BE({\rm CaO})
\>,
\end{eqnarray}
{\em etc.} We might, however, consider half-filled shells, {\em e.g.} C,
Si, {\em etc.} with an equal number of electrons and holes to be the
correct atomic analog of the vacuum, in which case
Eq.~(\ref{eq:atombe}) translates into
\begin{equation}
   BE({\rm LiF}) + BE({\rm NaCl}) \geq BE({\rm LiCl}) + BE({\rm NaF})
\>,
\end{equation}
{\em etc.} 

\subsection{Mass inequalities in nuclear physics}

For a while we were excited about the prospect \cite{ref:cohenpriv}
that true QCD inequalities could show that even-even $N=Z$ states are
-- barring Coulomb effects -- the most tightly bound.

The sophisticated third order Garvey-Kelson \cite{ref:gk}
difference relations connect masses of many isotopes. However, there is no
obvious pattern of deviations from the relations, and we have not
found a simple motivation for any such pattern.

\subsection{A biological analog for inequalities between correlators}

The pseudoscalar mass inequalities of Sec.~\ref{sec:pseumeson} reflect
simple Schwartz inequalities for correlators, {\em i.e.} for weighted
bilinears in quark propagators. Generically the latter have gauge
interaction induced ``phases''. These cancel in the tr$(S_i S_i^\dag)$
combinations appearing in the particular case of the pseudoscalar
propagators.\footnote{Schwartz inequalities have often been used in
  other contexts of particle physics. See, {\em e.g.}, 
  Ref.~\cite{ref:terazawa}.} 

In an extremely wide variety of circumstances we may encounter joint
propagation of two equal or two different entities. Quantum phases
are typically irrelevant and we can assign a positive probability for
the propagation of ``${\cal A}$'' from ``$P_1$'' to ``$P_2$'' for any
given set of relevant influencing ``factors'' $A_n$ -- the analog of
the background gauge fields in the QCD case:
\begin{equation}
   {\cal P} \left\{ {\cal A}(\mbox{at} \; P_1) \rightarrow 
      {\cal A} (\mbox{at} \; P_2) \right\} |_{\left\{ A_n \right\}}
\>.
\end{equation}
The overall propability of ${\cal A}$ ``propagating'' from $P_1$ to
$P_2$ is then given by a ``functional'' (path integrated) averaging
over the distribution of the $\left\{ A_n \right\}$ factors:
\begin{equation}
   {\cal P} \left\{ {\cal A}(\mbox{at} \; 1) \rightarrow 
      {\cal A} (\mbox{at} \; 2) \right\} = \intoned{\mu 
      \left\{ A_1 \ldots A_n \right\} } {\cal P} 
      \left\{ {\cal A}(\mbox{at} \; 1) \rightarrow 
      {\cal A} (\mbox{at} \; 2) \right\} |_{\left\{ A_n \right\}} 
\>,
\end{equation}
with a normalized $\intoned{\mu\left\{A\right\} } = 1$ positive
measure of $\left\{ A_1 \ldots A_n \right\}$. 

The probability of the joint propagation of ${\cal A}(1), {\cal B}(1)
\rightarrow {\cal A}(2), {\cal B}(2)$ is given by the corresponding
weighted average of the bilinear product of propagators:
\begin{equation}
   {\cal P} \left\{ \begin{array}{c}
              {\cal A}(1) \rightarrow {\cal A}(2) \\
              {\cal B}(1) \rightarrow {\cal B}(2)
              \end{array} \right\} =
   \intoned{\mu\left\{A_i\right\} } {\cal P} 
      \left\{ {\cal A}(1)\rightarrow{\cal A}(2) 
      \right\} |_{ \left\{A_n \right\} } \cdot {\cal P} 
      \left\{ {\cal B}(1)\rightarrow{\cal B}(2) 
      \right\} |_{ \left\{A_n \right\} }
\>.
\end{equation}
Likewise, the probability of joint propagation of {\em two} ${\cal A}$
objects from (1) to (2) (or {\em two} ${\cal B}$ objects) is given by
a similar expression involving squares of propagators:
\begin{equation}
   {\cal P} \left\{ \begin{array}{c}
              {\cal A}(1) \rightarrow {\cal A}(2) \\
              {\cal A}(1) \rightarrow {\cal A}(2)
              \end{array} \right\} =
   \intoned{\mu\left\{A_i\right\} } \left[{\cal P} 
      \left\{ {\cal A}(1)\rightarrow{\cal A}(2) 
      \right\} |_{ \left\{A_n \right\} } \right]^2
\>,
\end{equation}
and
\begin{equation}
   {\cal P} \left\{ \begin{array}{c}
              {\cal B}(1) \rightarrow {\cal B}(2) \\
              {\cal B}(1) \rightarrow {\cal B}(2)
              \end{array} \right\} =
   \intoned{\mu\left\{A_i\right\} } \left[{\cal P} 
      \left\{ {\cal B}(1)\rightarrow{\cal B}(2) 
      \right\} |_{ \left\{A_n \right\} } \right]^2
\>.
\end{equation}
The Schwartz inequality readily implies the final desired relation:
\begin{equation}
   {\cal P} \left\{ \begin{array}{c}
              {\cal A}(1) \rightarrow {\cal A}(2) \\
              {\cal A}(1) \rightarrow {\cal A}(2)
              \end{array} \right\} \cdot
   {\cal P} \left\{ \begin{array}{c}
              {\cal B}(1) \rightarrow {\cal B}(2) \\
              {\cal B}(1) \rightarrow {\cal B}(2)
              \end{array} \right\} \geq
   \left[ {\cal P} \left\{ \begin{array}{c}
              {\cal A}(1) \rightarrow {\cal A}(2) \\
              {\cal B}(1) \rightarrow {\cal B}(2)
              \end{array} \right\} \right]^2 
\>.
\end{equation}

There are certainly innumerable applications of the above relation in
all areas of science technology and life sciences (most of which were
very likely well known for some time). In the last page of this review
we consider one particular, somewhat intriguing, and exotic
application concerning the sex of non-identical twins.

The conception of non-identical twins can be viewed as the joint
propagation in the same (or relatively similar\footnote{If the two
  eggs are fertilized in subsequent encounters.}) ``background'' of
two sperms. Hence we expect that for non-identical
twins,\footnote{${\cal P}(F,M)$ refers to half the probability of
  mixed-sex twins -- say in which the female was born first.}
\begin{equation}
   {\cal P}(\mbox{male, male}) {\cal P}(\mbox{female, female})
      \geq \left[{\cal P}(\mbox{female, male}) \right]^2
\>.
\label{eq:twinprob}
\end{equation}
Clearly the probability of any specific single (or double) conception
need not be accurately reflected in the percentages at birth. We could
however still derive the same inequalities if we consider not merely
the propagation from ``inception'' to ``conception'' but also the
subsequent nine month long ``timelike'' propagation to birth.

The extent to which the inequality (\ref{eq:twinprob}) is satisfied,
{\em i.e.} the deviation of the ratio of the RHS and LHS from unity,
can serve as a measure of the overall degree of non-parallelism of the
male and female birth ``vectors'', {\em i.e.} as a crude measure of
the total importance of ``variables'' known, or as yet unknown, in
determining the sex of the fetus, which is a quantity of considerable
interest.

{\em A priori} \cite{ref:eshel} there could be some unique rather
surprising effect which would disqualify our proof and possibly
reverse the sign in Eq.~(\ref{eq:twinprob}). This would be the case if
there were a {\em direct interaction} between the propagating elements
-- in this case the fetuses of the two twins. Thus following the
biblical story of the rather unrestful pregnancy of Rebekah
\cite{ref:rivka} with Esau and Jacob (a clear case of non-identical
twins!), we can assume that the competition between equal sex brothers
(and possibly sisters) extends to the prenatal stages. Clearly if it
is too strong it could conceivably disrupt (MM) and (FF) births, thus
tending to reverse the sign of the effect considered. Hopefully this
is not the case.

We have not attempted to verify any of the above non-QCD mass
inequalities. In particular, the last twin inequalities may require
rather extensive statistical analysis.\footnote{Construction of a
clean sample of non-identical twins (nonbiased by sex consideration!) 
  is a highly demanding task.} We hope that the inequalities will
eventually be tested.

%

%
\begin{figure}
\begin{center}
   \caption{Vertex and propagator insertion of the one gluon exchange 
   diagram with the $\lambda_1 \cdot \lambda_2$ structure intact.}
\label{fig:onegluon}
\end{center}
\end{figure}
%
\begin{figure}
\begin{center}
   \caption{The general two quark interaction is a sum of $\lambda_1 \cdot
   \lambda_2$ (octet exchange) and $1_1 \cdot 1_2$ (singlet exchange)
   interactions.} 
\label{fig:twoquark}
\end{center}
\end{figure}
%
\begin{figure}
\begin{center}
   \caption{A non-separable three body interaction in the baryon due to the
   three gluon vertex which is shown, however, to vanish.}
\label{fig:threegluon}
\end{center}
\end{figure}
%
\begin{figure}
\begin{center}
   \caption{(a) The color string picture for a meson. 
    (b) The color network for a $qqq$ baryon with one junction.}
\label{fig:color}
\end{center}
\end{figure}
%
\begin{figure}
\begin{center}
   \caption{Illustration of how a given Y configuration in a baryonic wave
   functional yields configurations for three mesonic trial wave functionals.
   The relation $2 H_{123} = H_{1\bar{2}} + H_{2\bar{3}} + H_{3\bar{1}}$ is
   illustrated by the fact that the $B^2$ term depicted by the small string
   distortion, the $\psibar D \psi$ terms depicted by the motion of the end
   quarks, and the $E^2$ (corresponding to the weighted string length) all
   appear twice.}
\label{fig:bmtrial}
\end{center}
\end{figure}
%
\begin{figure}
\begin{center}
   \caption{The triangular inequalities and relevant vectors for the 
   comparison of the potential energies in the strong coupling limit 
   of the baryonic Y configuration and mesonic subsystems.}
\label{fig:triangle}
\end{center}
\end{figure}
%
\begin{figure}
\begin{center}
   \caption{Some matrix elements $\langle$ configuration' $ | 
     H_{\text{QCD}} | $ configuration $\rangle$ on a small $3 \times
     3$ dimensional lattice. The upper entry in each box is the matrix
     element with $\mu \simeq 1/a$ the lattice energy scale. The lower
     entry indicates the terms in the Hamiltonian contributing to this
     particular matrix element.}
\label{fig:lattice}
\end{center}
\end{figure}
%
\begin{figure}
\begin{center}
   \caption{Configurations in the generalized baryonic wave functional which
   still allow separation into three mesonic subsystems.}
\label{fig:bmsep}
\end{center}
\end{figure}
%
\begin{figure}
\begin{center}
   \caption{Flavor connected (a) and flavor disconnected (b) contraction
   contributing to two-point correlation functions of quark bilinears.}
\label{fig:2ptcorr}
\end{center}
\end{figure}
%
\begin{figure}
\begin{center}
   \caption{The focusing effect of a parallel $B$ field on charged particles
   propagating from $x$ to $y$ and its limited effect on propagation between
   regions of size $\Delta$ once $B \geq 1/ \Delta^2$.}
\label{fig:bfield}
\end{center}
\end{figure}
%
\begin{figure}
\begin{center}
   \caption{(a) The unique contraction when all flavor indices are distinct.
   (b) An alternate contraction when we have permutable quarks in the two
   currents.} 
\label{fig:flavor}
\end{center}
\end{figure}
%
\begin{figure}
\begin{center}
   \caption{The planar duality diagrams representing the euclidean
   correlations $F_1, F_2,$ and $F_3$. For $F_2$ we use only the specific
   contraction with $u\bar{u} (s\bar{s})$ exchanged in the $\tau (T)$ channel
   respectively.}  
\label{fig:fplanar}
\end{center}
\end{figure}
%
\begin{figure}
\begin{center}
   \caption{The evolution of an initial $u\bar{d}$ $1^{-}$ state on different
   time scales.}
\label{fig:pion}
\end{center}
\end{figure}
%
\begin{figure}
\begin{center}
   \caption{The original Wilson loop $(W)$, its parts $W_1$ and $W_2$, and the
   reflection paths used in proving Eq.~(\protect\ref{eq:traceu}).}
\label{fig:wilson}
\end{center}
\end{figure}
%
\begin{figure}
\begin{center}
   \caption{The unique ``8'' pattern of contraction relevant to the
   $\langle K|B|\pi\rangle \geq \langle K|A|\pi\rangle$ inequality.}
\label{fig:8pattern}
\end{center}
\end{figure}
%
\begin{figure}
\begin{center}
   \caption{The ``eye'' contraction relevant to the $\langle K^0 | A |
   0 \rangle$ matrix element.}
\label{fig:eye}
\end{center}
\end{figure}
%
\begin{figure}
\begin{center}
   \caption{Illustration of how as $\vecr\rightarrow 0$ and $u
     \rightarrow v$ (and the diamond-like configuration degenerates
     into a vertical line), we tend to get, due to the smoothness of
     the $A_\mu$ configuration, similar propagators, and hence a
     monotonic decrease of $\psi_\pi(|\vecr|)$ with $|\vecr|$.}
\label{fig:diamond}
\end{center}
\end{figure}
%
\begin{figure}
\begin{center}
   \caption{The color string network for a quadriquark with two light
     quarks and two heavy quarks.}
\label{fig:qquark}
\end{center}
\end{figure}
%
\begin{figure}
\begin{center}
   \caption{The color string network for a quadriquark (solid line)
     plus its doubled network with flux lines reversed (dashed line).}
\label{fig:qquark2}
\end{center}
\end{figure}
%
\begin{figure}
\begin{center}
   \caption{(a) The color network for the putative pentaquark state with
   three junction points (1), (2), (3). (b) The string picture of the
   trial baryon + meson states obtained from that of the pentaquark by
   omitting the (2) - (3) string bit, and reversing the flux direction
   in the (2) - $d_e$ and (3) - $\bar{Q}_f$ sections.}
\label{fig:pquark}
\end{center}
\end{figure}
%
\begin{figure}
\begin{center}
   \caption{The three polygonal paths $\tilde{X}_1, \tilde{X}_2,
     \tilde{X}_3$ consisting of the overall periodic propagation of
     the particles at $X_1, X_2, X_3$ from ``$t$'' = $\beta = 0$ to
     ``$t$'' = $\beta$ with $X_i(\beta) = X_i(0)$.}
\label{fig:lieb}
\end{center}
\end{figure}
%
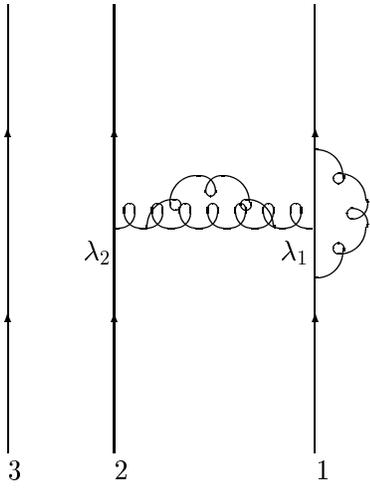
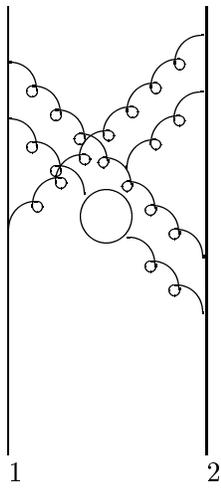
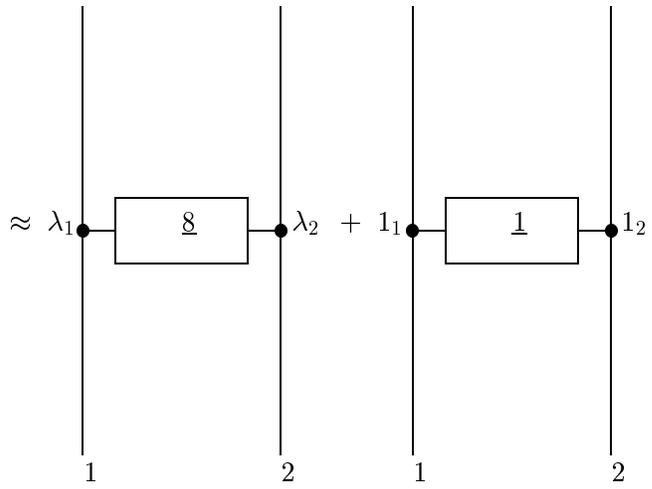
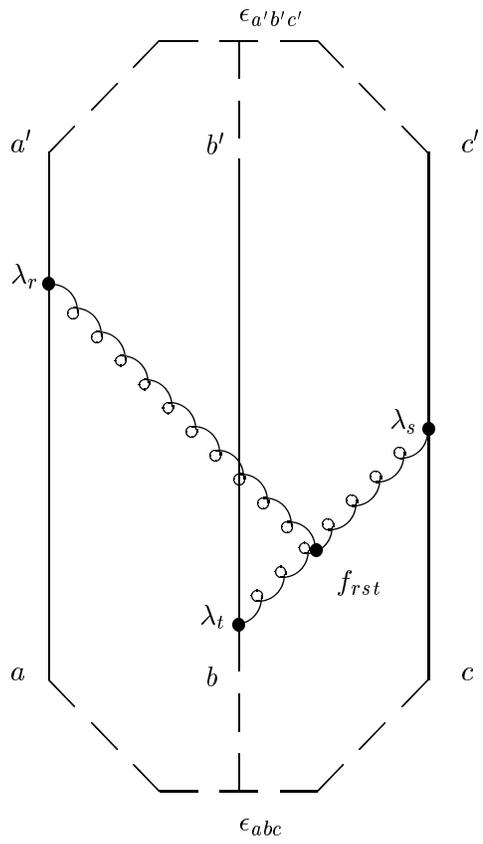
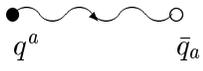
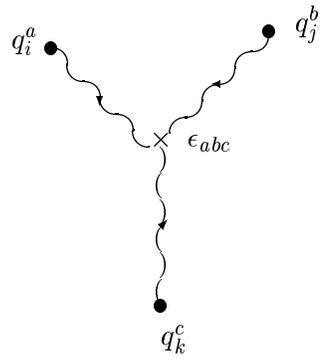
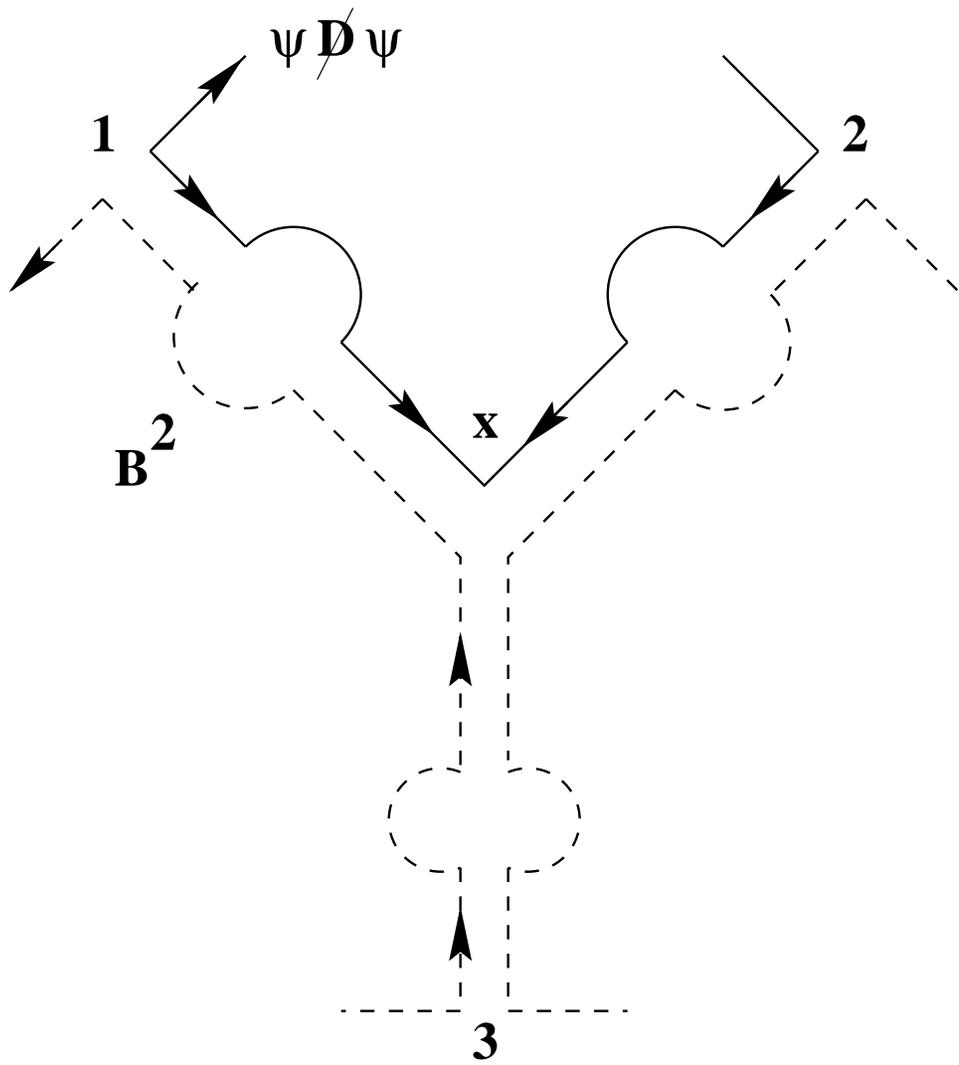
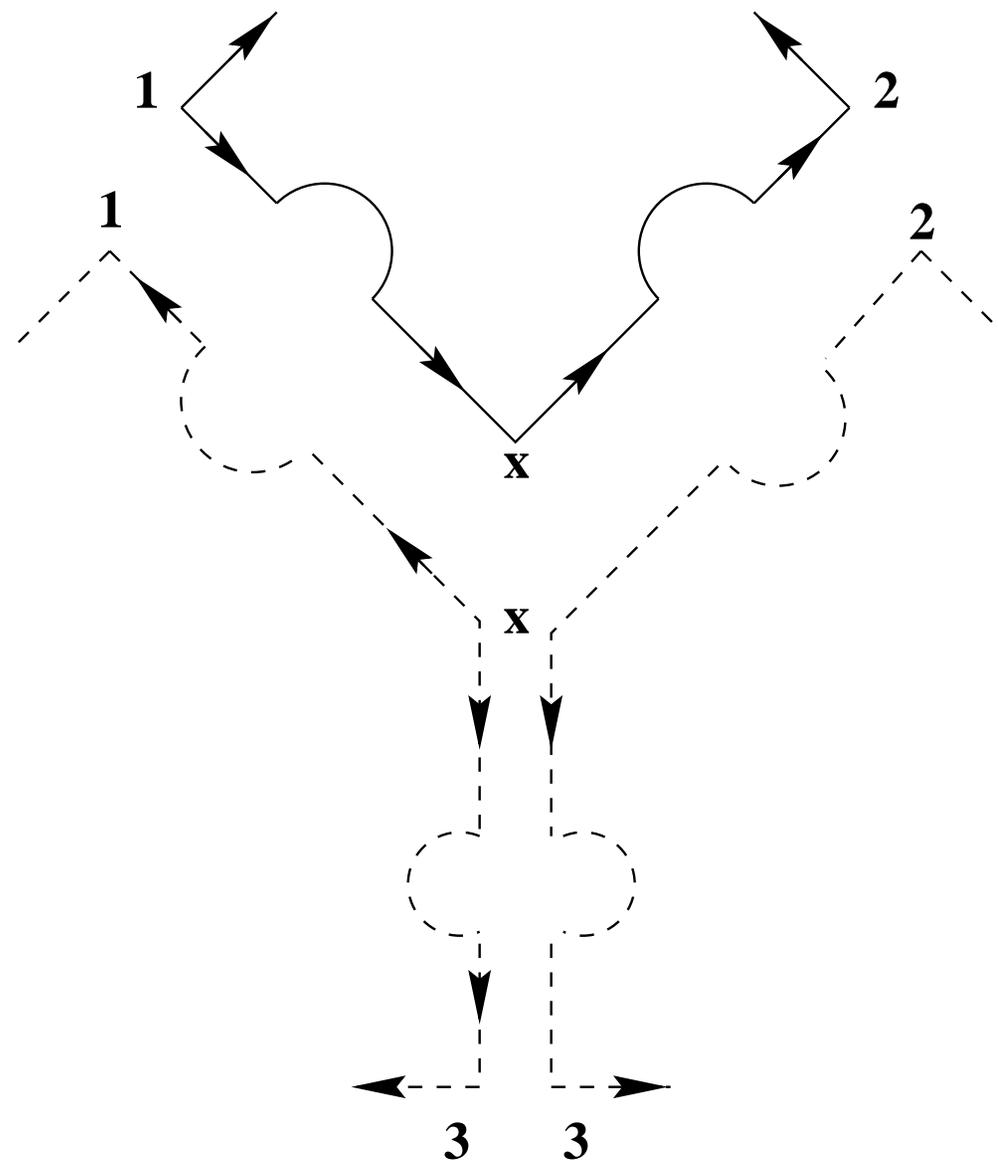
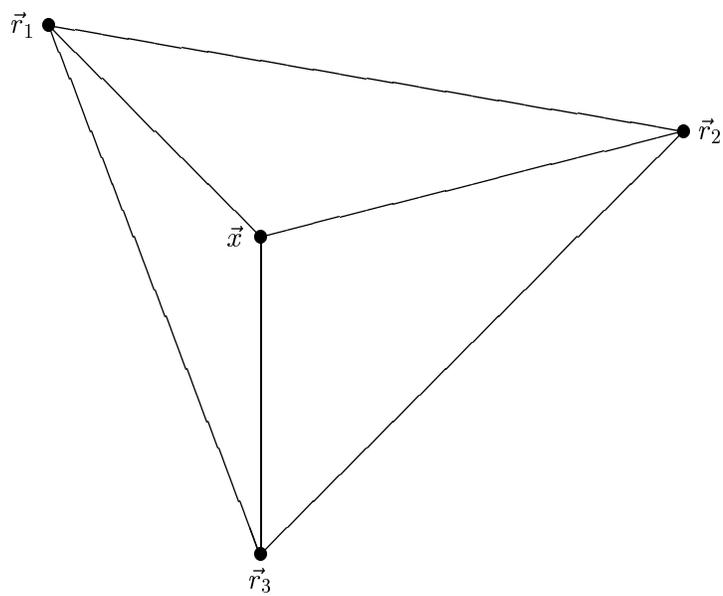
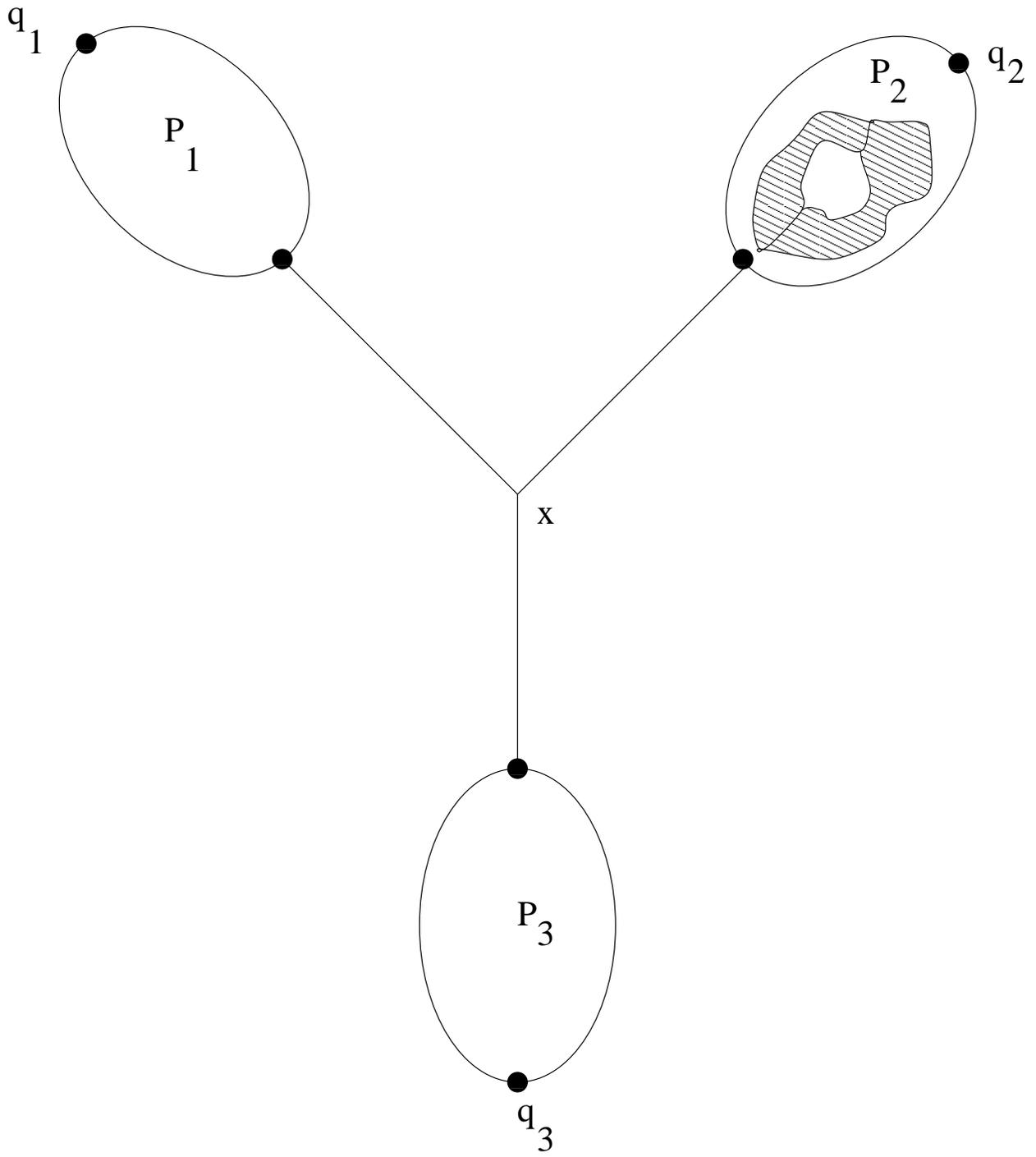
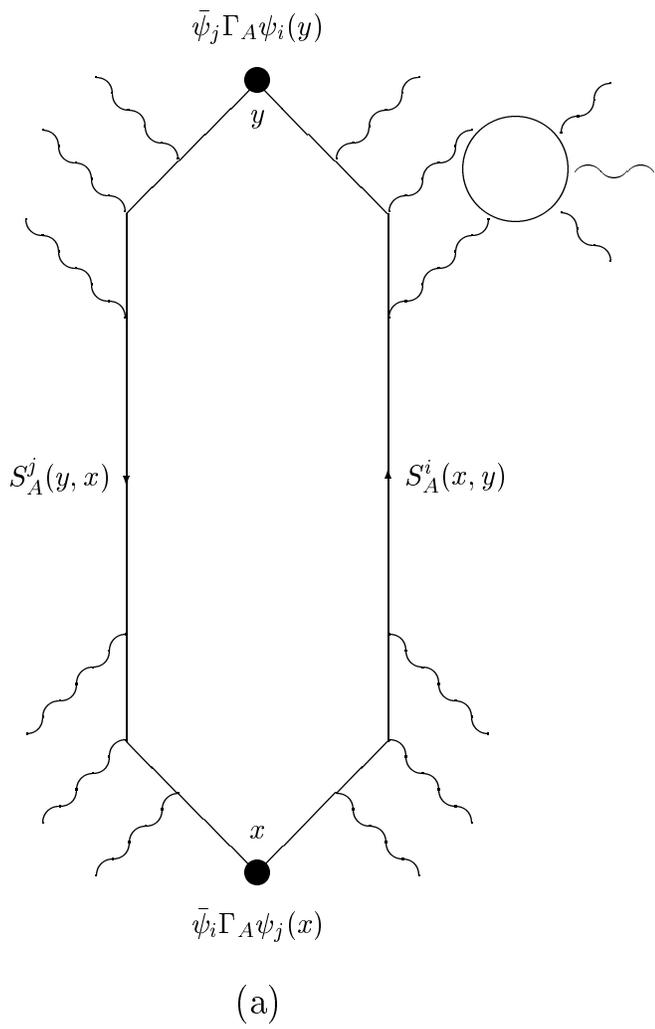
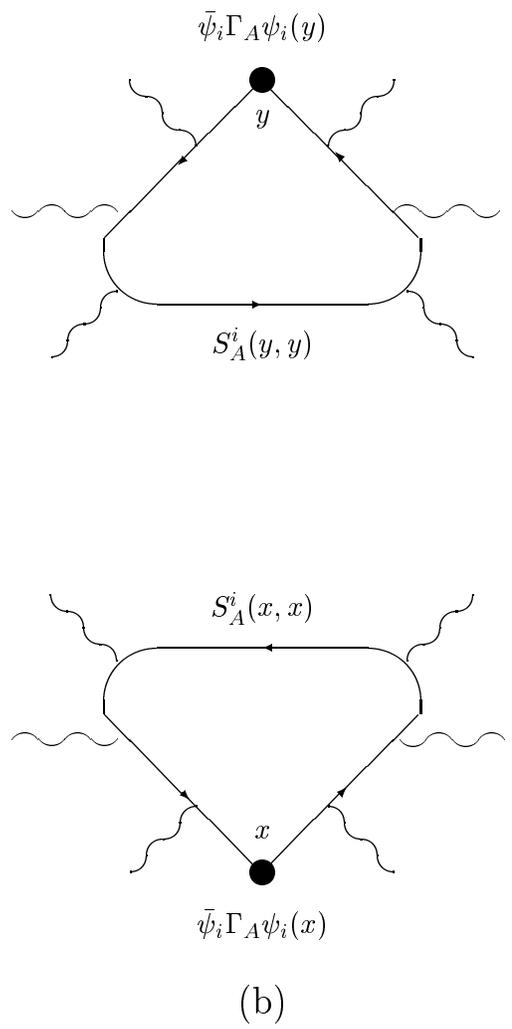
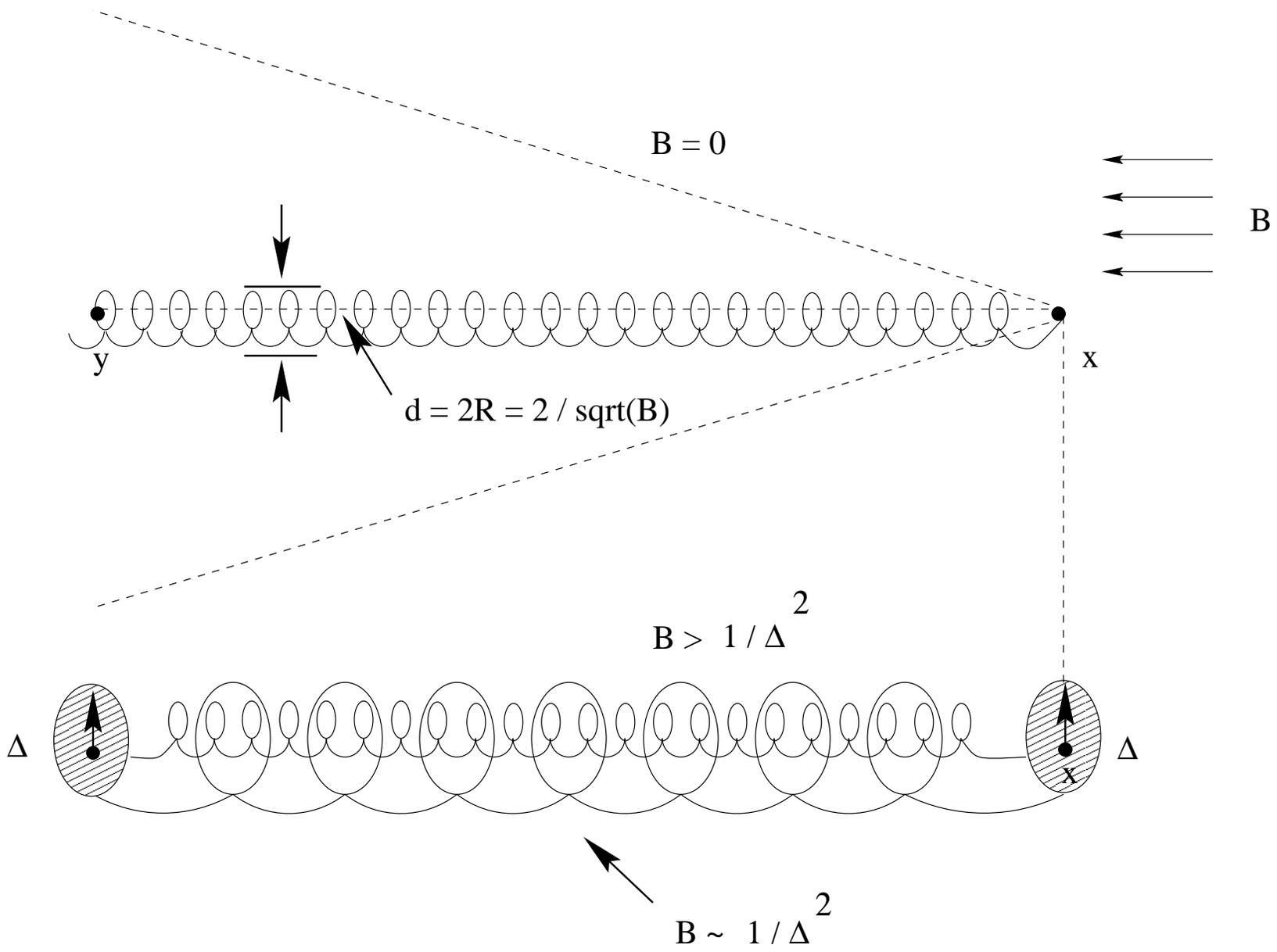
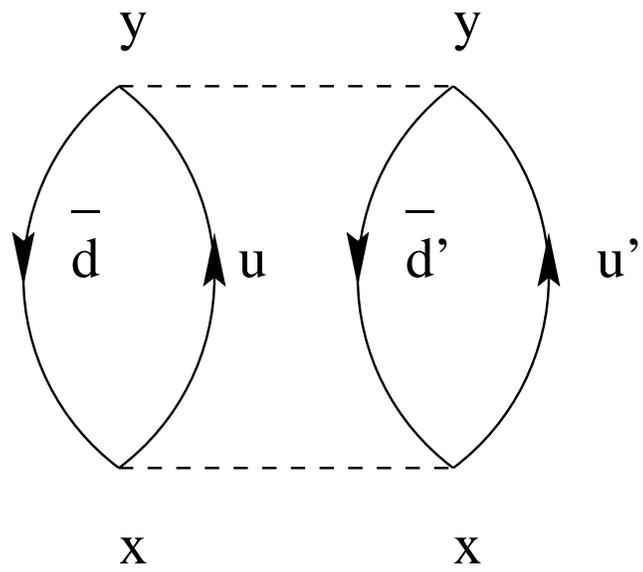
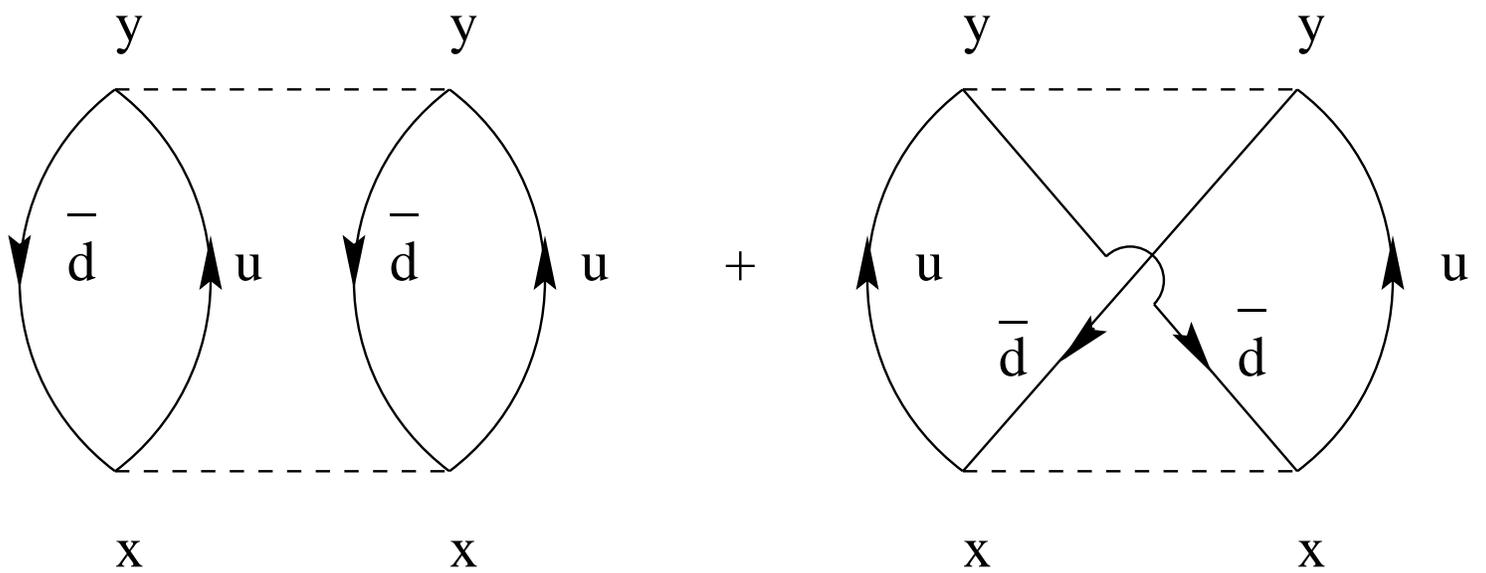
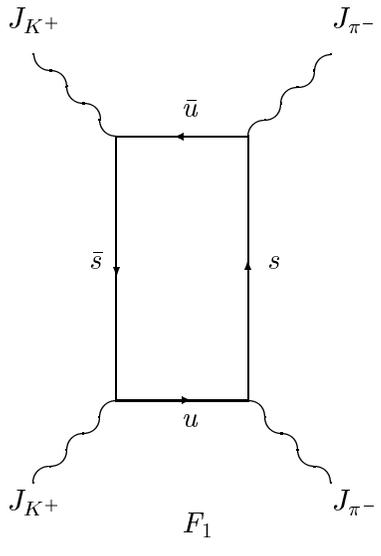
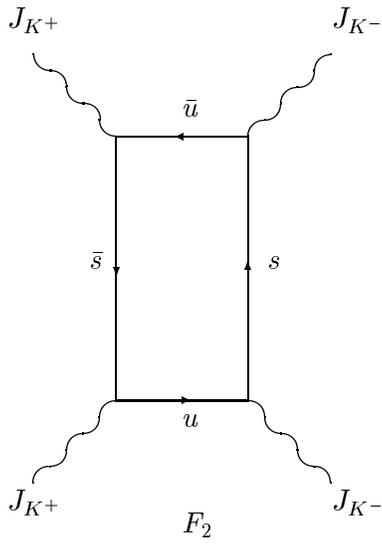
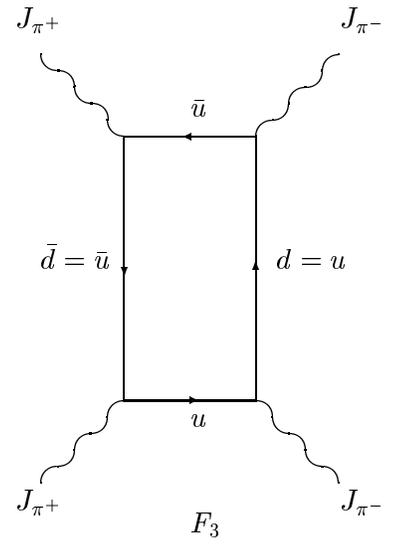
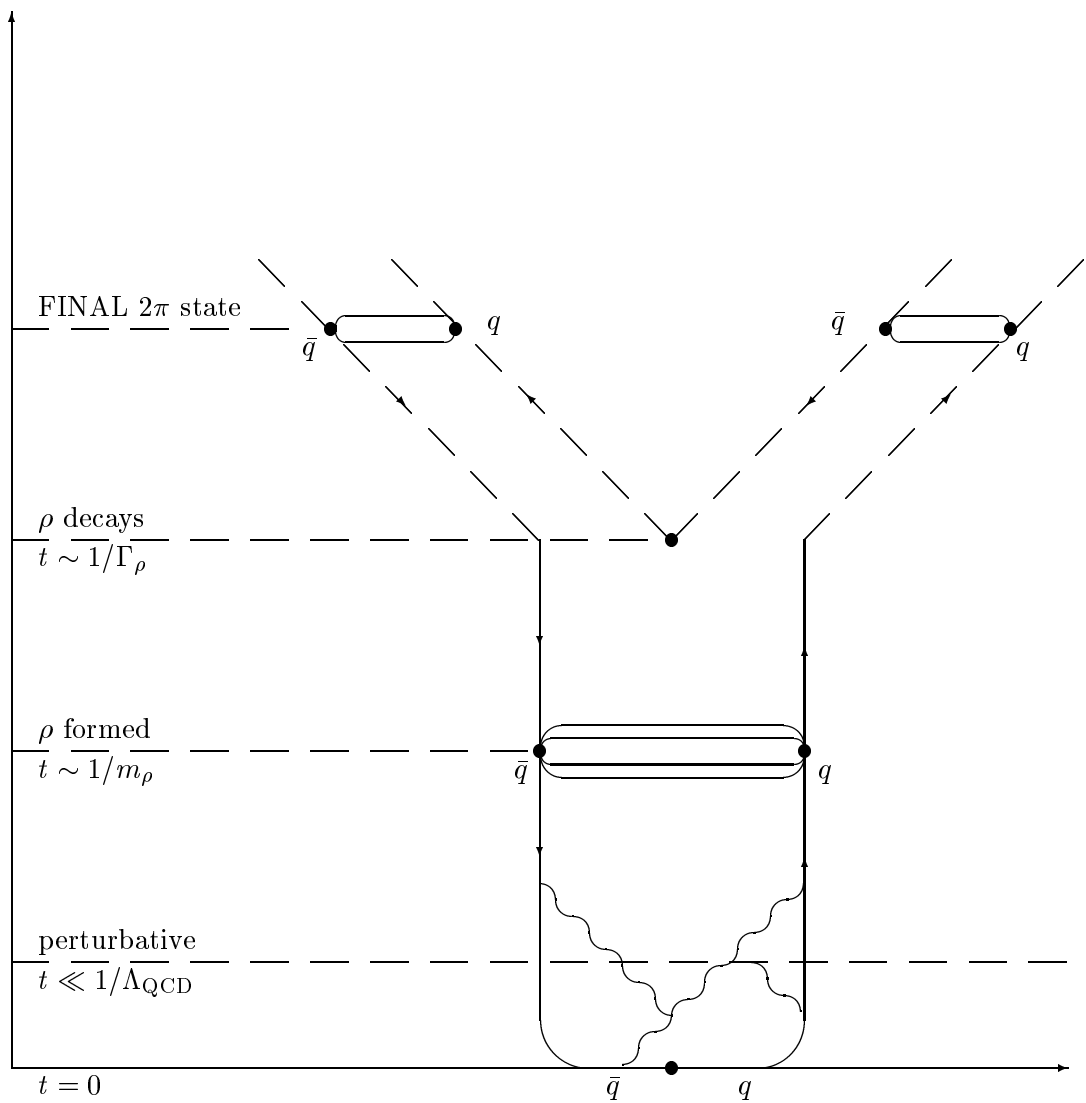
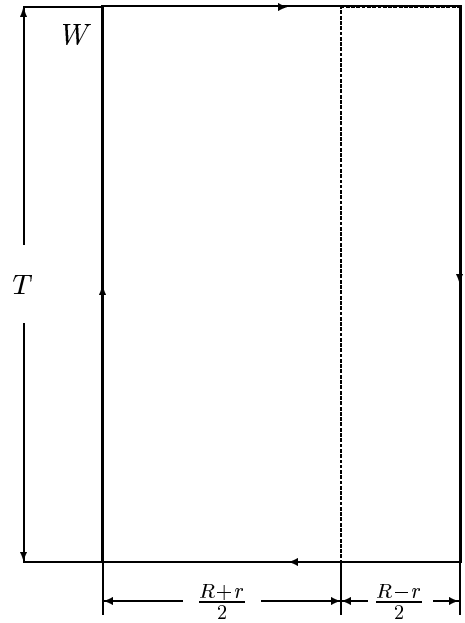
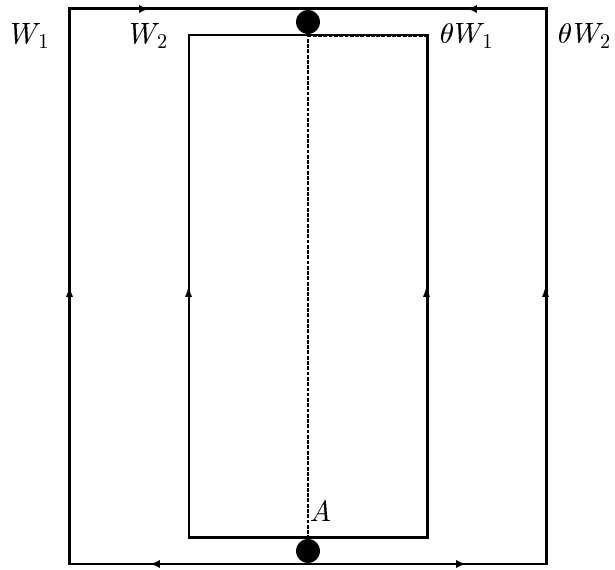
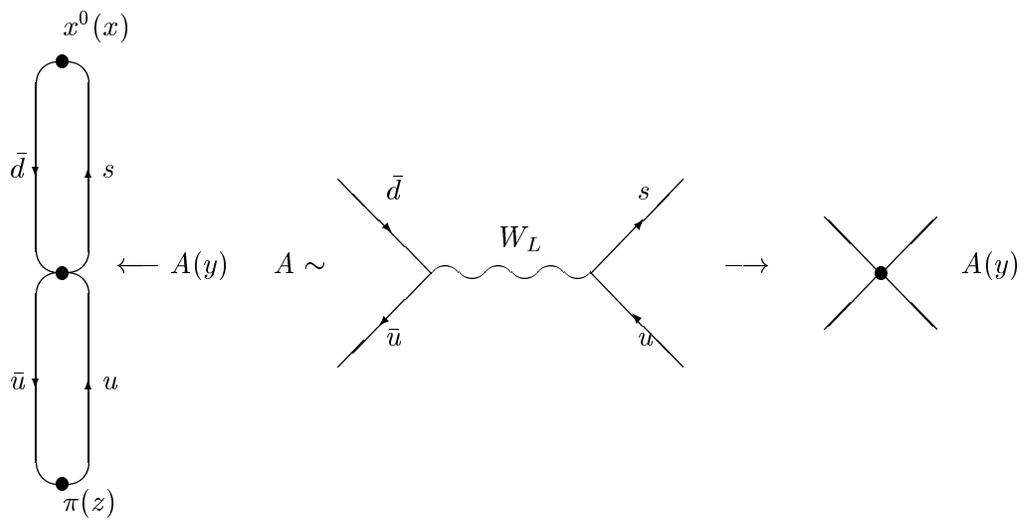
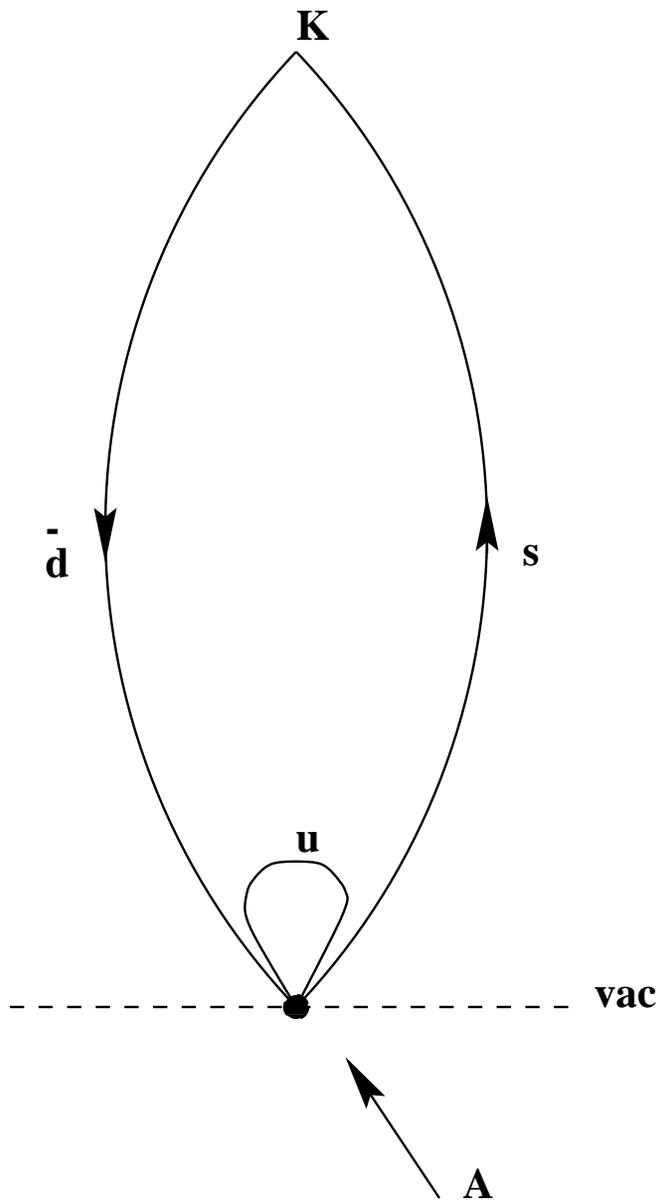
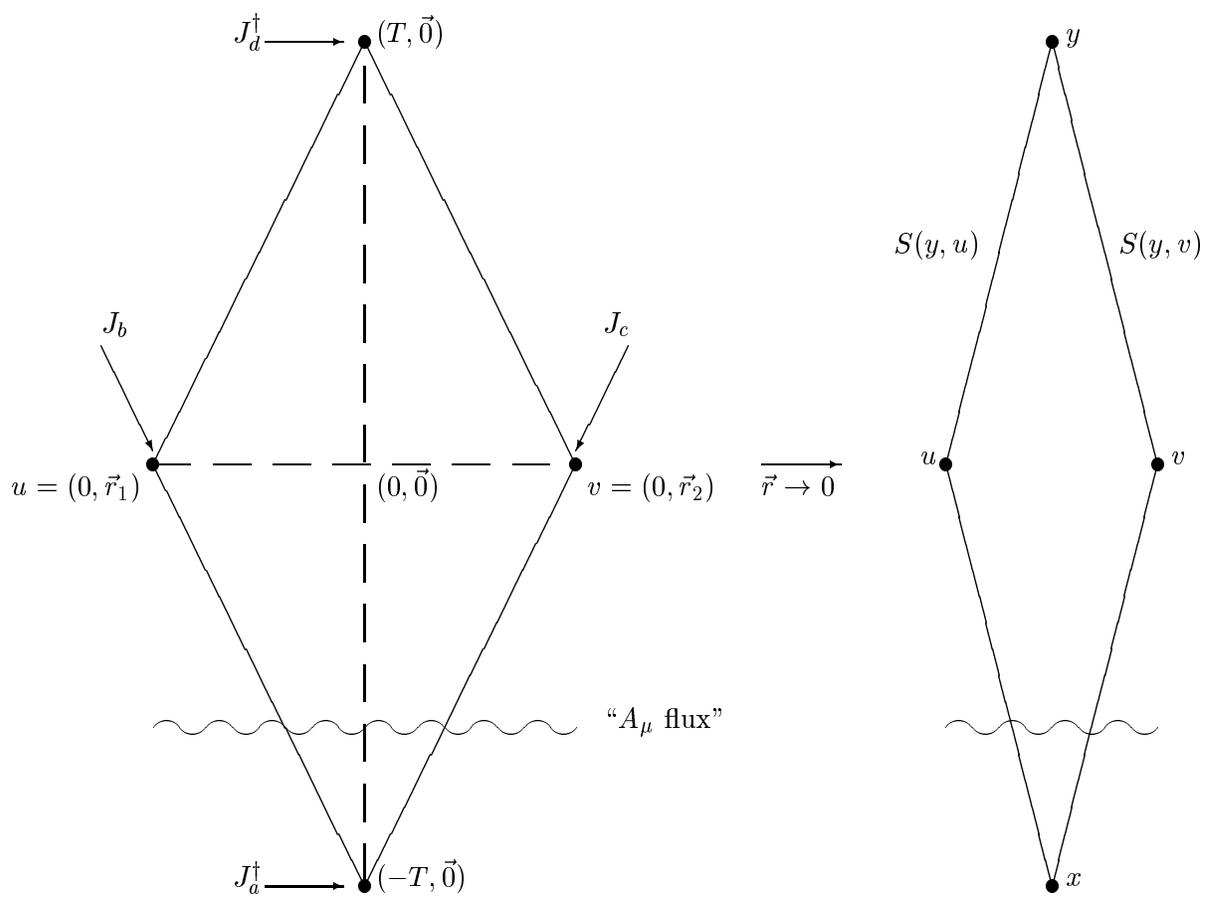
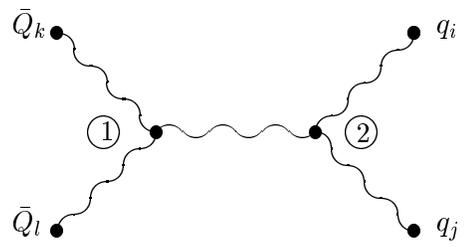
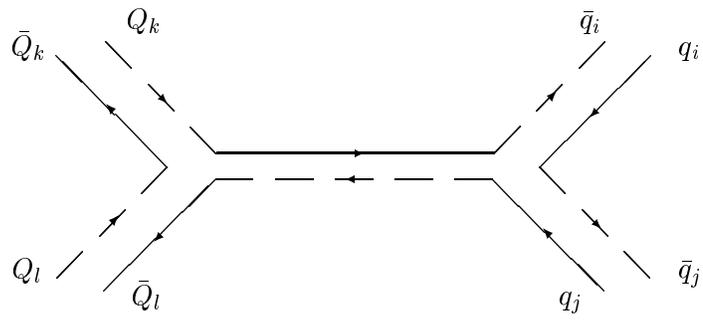
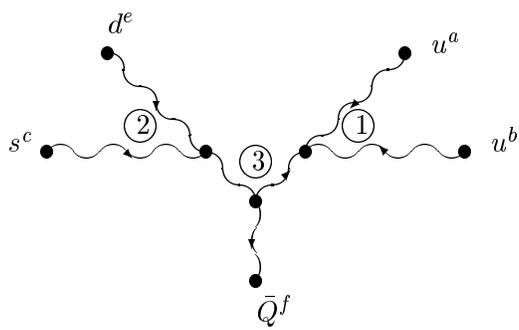
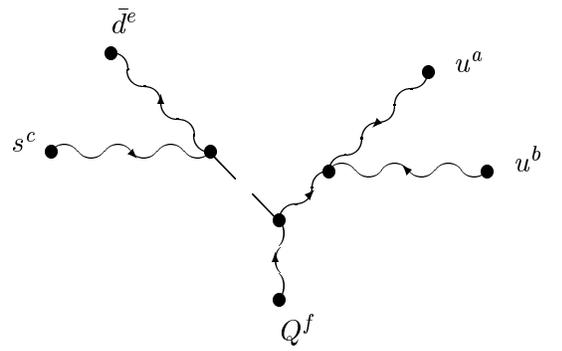
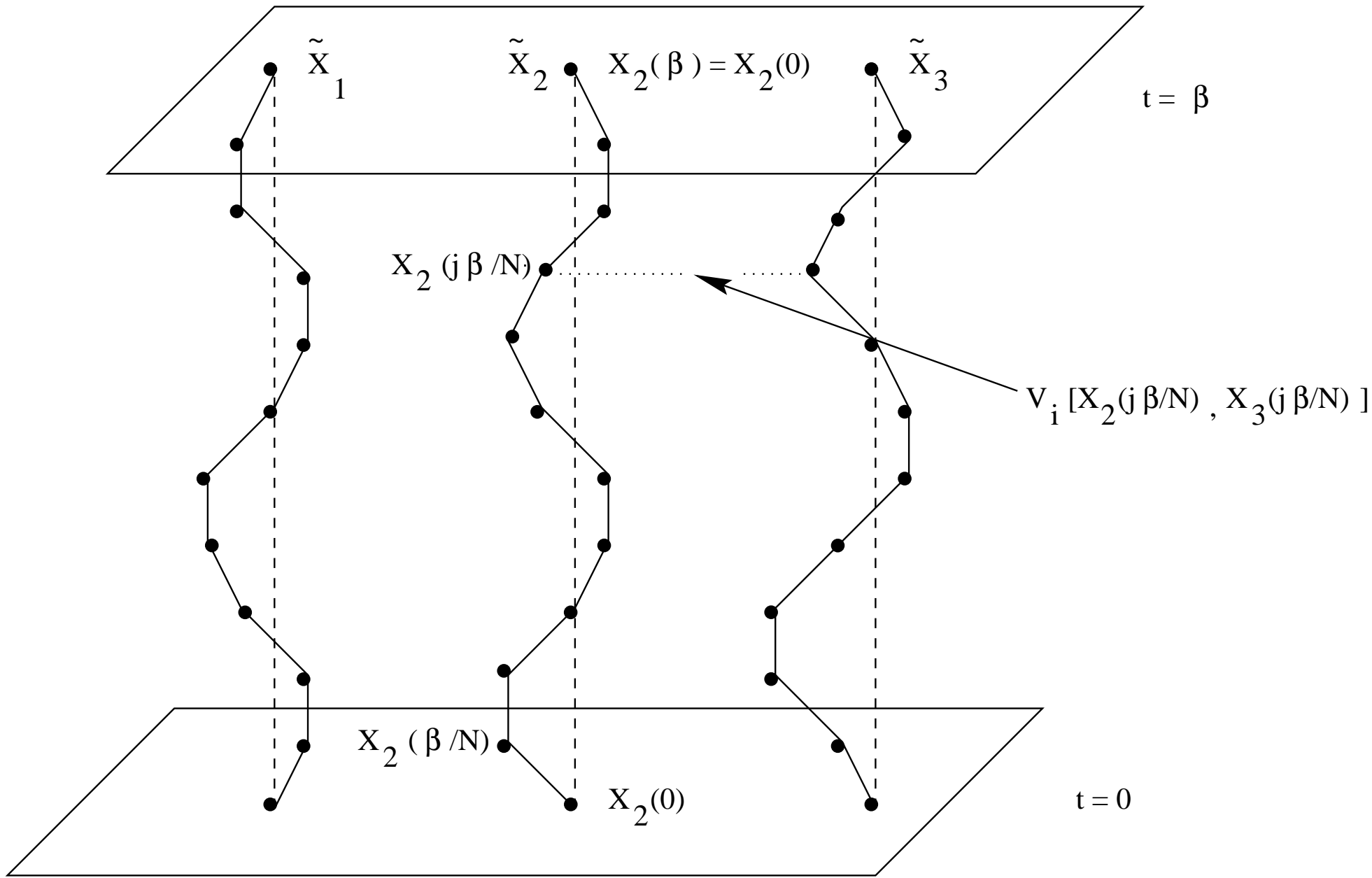
\end{document}